\documentclass[iop,apj]{emulateapj}
\usepackage[
        pdfencoding=auto,%
        pdftitle={Modules for Experiments in Stellar Astrophysics},%
        pdfauthor={Bill Paxton},%
        pdfstartview=FitV,%
        colorlinks=true,%
        linkcolor=blue,%
        citecolor=black, %
        urlcolor=blue]{hyperref}
\usepackage{txfonts}
\usepackage{bm}
\usepackage[usenames,dvipsnames]{color}

\providecommand\icarus{{\rmfamily Icarus}}
\newcommand\actaa{{\rmfamily Acta Astron.}}
\newcommand{\code}[1]{\texttt{#1}}

\newcommand{\mesa}{\code{MESA}}
\newcommand{\MESA}{\mesa}
\newcommand{\STERN}{STERN}
\newcommand{\ADIPLS}{\code{ADIPLS}}
\newcommand{\DSEP}{DSEP}

\newcommand{\const}{\code{const}}
\newcommand{\chem}{\code{chem}}
\newcommand{\diffusion}{\code{diffusion}}

\newcommand{\mesastar}{\mesa~\code{star}}
\newcommand{\MESAstar}{\mesastar}

\newcommand{\kap}{\code{kap}}
\newcommand{\eos}{\code{eos}}

\newcommand{\atm}{\code{atm}}

\newcommand{\mlt}{\code{mlt}}

\newcommand{\net}{\code{net}}
\newcommand{\neu}{\code{neu}}

\newcommand{\reaclib}{\code{reaclib}}

\newcommand{\sdk}{\code{SDK}}
\newcommand{\SDK}{\sdk}

\newcommand{\dif}{\ensuremath{\mathrm{d}}}
\newcommand{\D}{{\mathrm d}}
\newcommand{\DD}{{\,\D\!\!\;}}

\newcommand{\ppll}[2]{\left(\frac{\partial\ln #1}{\partial\ln
      #2}\right)_{\rho,T,\{X_{j\neq i}\}}}
\newcommand{\ddl}[2]{\frac{{\rm d}\ln #1}{{\rm d}\ln #2}}

\newcommand{\dxdycz}[3]{{\Brak{\frac{\partial{#1}}{\partial{#2}}}_{{#3}}}}

\newcommand{\av}[1]{{\langle{#1}\rangle}}

\newcommand{\abs}[1]{{\left|{#1}\right|}}

\newcommand{\nuclei}[2]{\ensuremath{\mathrm{^{#1}#2}}}

\newcommand{\proton}{\ensuremath{p}}
\newcommand{\pt}{\proton}
\newcommand{\photon}{\ensuremath{\gamma}}

\newcommand{\helium}[1][4]{\nuclei{#1}{He}}

\newcommand{\carbon}[1][12]{\nuclei{#1}{C}}
\newcommand{\nitrogen}[1][14]{\nuclei{#1}{N}}
\newcommand{\oxygen}[1][16]{\nuclei{#1}{O}}

\newcommand{\titanium}[1][48]{\nuclei{#1}{Ti}}

\newcommand{\chromium}[1][52]{\nuclei{#1}{Cr}}

\newcommand{\copernicum}[1][285]{\nuclei{#1}{Cn}}

\newcommand{\unitspace}{\ensuremath{\,}}
\newcommand{\usp}{\unitspace}
\newcommand{\numberspace}{\ensuremath{\;}}
\newcommand{\nsp}{\numberspace}
\newcommand{\unitstyle}[1]{\ensuremath{\mathrm{#1}}}
\newcommand{\power}[2]{\ensuremath{{#1}^{#2}}}
\newcommand{\ee}[1]{\ensuremath{\times 10^{#1}}}

\newcommand{\centi}{\unitstyle{c}}

\newcommand{\Mega}{\unitstyle{M}}
\newcommand{\Giga}{\unitstyle{G}}

\newcommand{\meter}{\unitstyle{m}}

\newcommand{\second}{\unitstyle{s}}

\newcommand{\Kelvin}{\unitstyle{K}}
\newcommand{\K}{\Kelvin}  

\newcommand{\cm}{\centi\meter}
\newcommand{\gram}{\unitstyle{g}}

\newcommand{\grampercc}{\gram\usp\power{\cm}{-3}} 
\newcommand{\squarecmpergram}{\power{\cm}{2}\usp\power{\gram}{-1}} 

\newcommand{\erg}{\unitstyle{ergs}} 
\newcommand{\ergs}{\erg}

\newcommand{\ergssecond}{\erg\unitspace\second}

\newcommand{\Msun}{\ensuremath{\unitstyle{M}_\odot}}
\newcommand{\Lsun}{\ensuremath{\unitstyle{L}_{\odot}}}
\newcommand{\Rsun}{\ensuremath{\unitstyle{R}_{\odot}}}

\newcommand{\Myr}{\Mega\yr}
\newcommand{\Gyr}{\Giga\yr}

\newcommand{\Msunyr}{\Msun\,\power{\yr}{-1}}
\newcommand{\MJ}{\ensuremath{\mathrm{M_J}}}
\newcommand{\RJ}{\ensuremath{\mathrm{R_J}}}
\newcommand{\Mearth}{\ensuremath{\mathrm{M_{\oplus}}}}
\newcommand{\AU}{\unitstyle{AU}}

\newcommand{\yr}{\unitstyle{yr}}        
\newcommand{\kms}{\ensuremath{\mathrm{km}\,\second^{-1}}}

\newcommand{\Tabff}[1]{{\ref{tab:#1}}}
\newcommand{\Tab}[1]{{Table~\Tabff{#1}}}

\newcommand{\isofont}[1]{{\mathrm{#1}}}
\newcommand{\isomass}[1]{{\ensuremath{\isofont{^{#1}}}}}
\newcommand{\isocharge}[1]{{\ensuremath{\isofont{_{#1}}}}}
\newcommand{\isotope}[3]{{\ensuremath{\isocharge{#1}\isomass{#2}\isofont{#3}}}}

\newcommand{\I}[2]{{\isotope{}{#1}{#2}}}

\newcommand{\Ep}[1]{{\ensuremath{10^{#1}}}}
\newcommand{\E}[1]{{\ensuremath{\powersep\Ep{#1}}}}

\newcommand{\powersep}{\times}
\newcommand{\Brak}[1]{{\left({#1}\right)}}
\newcommand{\SBrak}[1]{{\left[{#1}\right]}}

\newcommand{\epsnuc}{\ensuremath{\epsilon_{\mathrm{nuc}}}}	
\newcommand{\epsgrav}{\ensuremath{\epsilon_{\mathrm{grav}}}} 
\newcommand{\epsnu}{\ensuremath{\epsilon_{\mathrm{\nu}}}} 
\newcommand{\Teff}{\ensuremath{T_{\!\mathrm{eff}}}}	
\newcommand{\teff}{\Teff}
\newcommand{\Ledd}{\ensuremath{L_{\mathrm{Edd}}}} 
\newcommand{\logg}{\ensuremath{\log g}}	
\newcommand{\Tc}{\ensuremath{T_{\mathrm{\!c}}}} 
\newcommand{\Pc}{\ensuremath{P_{\mathrm{\!c}}}} 
\newcommand{\rhoc}{\ensuremath{\rho_{\mathrm{c}}}} 
\newcommand{\CP}{\ensuremath{C_{\!P}}} 
\newcommand{\Mdot}{\ensuremath{\dot{M}}} 
\newcommand{\Mc}{\ensuremath{M_{\rm c}}} 
\newcommand{\Mm}{\ensuremath{M_{\rm m}}} 
\newcommand{\Rc}{\ensuremath{R_{\rm c}}} 
\newcommand{\Lc}{\ensuremath{L_{\rm c}}} 
\newcommand{\Lacc}{\ensuremath{L_{\rm acc}}} 

\newcommand{\kappath}{\ensuremath{\kappa_{\mathrm{th}}}} 
\newcommand{\kappav}{\ensuremath{\kappa_{\mathrm{v}}}} 

\newcommand{\nB}{\ensuremath{n_{\mathrm{B}}}}	
\newcommand{\alphaMLT}{\ensuremath{\alpha_{\mathrm{MLT}}}}	
\newcommand{\chirho}{\ensuremath{\chi_{\rho}}}	
\newcommand{\chiT}{\ensuremath{\chi_{\raisebox{-2pt}{$\scriptstyle T$}}}}	
\newcommand{\Gammaone}{\ensuremath{\Gamma_{\!1}}} 
\newcommand{\Dov}{\ensuremath{D_{\mathrm{ov}}}}	
\newcommand{\nablaad}{\ensuremath{\nabla_{\!\mathrm{ad}}}}	
\newcommand{\nablarad}{\ensuremath{\nabla_{\!\mathrm{rad}}}}	
\newcommand{\nablaT}{\ensuremath{\nabla_{\!T}}}	
\newcommand{\nablaL}{\ensuremath{\nabla_{\mathrm{\!L}}}}	
\newcommand{\scaleheight}{\ensuremath{\lambda_P}}	
\newcommand{\Pgas}{\ensuremath{P_{\!\!\mathrm{gas}}}}	
\newcommand{\timestep}{\ensuremath{\delta t}} 
\newcommand{\Prad}{\ensuremath{P_{\!\!\mathrm{rad}}}}	
\newcommand{\Lrad}{\ensuremath{L_{\mathrm{rad}}}}	
\newcommand{\tderiv}[3]{\ensuremath{\left(\frac{\partial #1}{\partial #2}\right)_{#3}}} 
\newcommand{\Lrho}{\ensuremath{L_{\mathrm{inv}}}}	
\newcommand{\Lonset}{\ensuremath{L_{\mathrm{onset}}}}	
\newcommand{\Fconv}{\ensuremath{F_{\!\mathrm{conv}}}}		
\newcommand{\Frad}{\ensuremath{F_{\!\mathrm{rad}}}}	
\newcommand{\supernab}{\ensuremath{\delta_\nabla}}  
\newcommand{\superthresh}{\ensuremath{\delta_{\nabla,\mathrm{thresh}}}}  
\newcommand{\fsuper}{\ensuremath{f_\nabla}} 
\newcommand{\asuper}{\ensuremath{\alpha_\nabla}}  
\newcommand{\asupert}{\ensuremath{\widetilde{\asuper}}} 
\newcommand{\lambdamax}{\ensuremath{\lambda_{\max}}} 
\newcommand{\betamin}{\ensuremath{\beta_{\min}}} 

\newcommand{\alphasc}{\ensuremath{\alpha_{\mathrm{sc}}}} 
\newcommand{\alphath}{\ensuremath{\alpha_{\mathrm{th}}}} 
\newcommand{\Dth}{\ensuremath{D_{\mathrm{th}}}} 

\newcommand{\EFc}{\ensuremath{E_{\mathrm{F,c}}}}  
\newcommand{\kB}{\ensuremath{k_\mathrm{B}}} 
\newcommand{\NA}{\ensuremath{N_\mathrm{\!A}}} 
\newcommand{\mb}{\ensuremath{m_\mathrm{u}}} 
\newcommand{\sigmaSB}{\ensuremath{\sigma_\mathrm{\!SB}}} 

\newcommand{\veq}{\ensuremath{\varv_{\mathrm{eq}}}} 
\newcommand{\veqi}{\ensuremath{\varv_{\mathrm{eq,ini}}}}
\newcommand{\Om}{\ensuremath{\Omega}}  
\newcommand{\Omc}{\ensuremath{\Om_{\mathrm{crit}}}} 
\newcommand{\om}{\ensuremath{\omega}}  
\newcommand{\tkh}{\ensuremath{\tau_{\mathrm{KH}}}} 
\newcommand{\LP}{{L_{\mathrm{P}}}} 
\newcommand{\VP}{{V_{\mathrm{P}}}}
\newcommand{\SP}{{S_{\!\mathrm{P}}}} 
\newcommand{\rP}{{r_{\mathrm{P}}}}
\newcommand{\mP}{{m_{\mathrm{P}}}} 
\newcommand{\fP}{{f_{\mathrm{P}}}}
\newcommand{\fT}{{f_{\mathrm{T}}}}

\newcommand{\numax}{\ensuremath{\nu_{\mathrm{max}}}} 
\newcommand{\dnu}{\ensuremath{\Delta\nu}}  
\newcommand{\fov}{\ensuremath{f_{\mathrm{ov}}}} 
\newcommand{\cs}{\ensuremath{c_{\rm s}}} 
\newcommand{\Slamb}{\ensuremath{S_{\!\ell}}} 

\newcommand{\mesaone}{Paper~I} 
\newcommand{\bvfreq}{Brunt-V\"ais\"al\"a frequency}
\newcommand{\bvv}{Brunt-V\"ais\"al\"a}

\newlength{\apjcolwidth}
\newlength{\figwidth}
\newlength{\doublewide}
\newlength{\twoupwidth}
\newlength{\twoupsep}
\begin{document}
\title{Modules for Experiments in Stellar Astrophysics (MESA): 
Planets, Oscillations, Rotation, and Massive Stars }
\author{%
Bill Paxton\altaffilmark{1},
Matteo Cantiello\altaffilmark{1},
Phil Arras\altaffilmark{2},
Lars Bildsten\altaffilmark{1,3},
Edward F. Brown\altaffilmark{4},
Aaron Dotter\altaffilmark{5},
Christopher Mankovich\altaffilmark{3},
M. H. Montgomery\altaffilmark{6},
Dennis Stello\altaffilmark{7},
F. X. Timmes\altaffilmark{8},
and
Richard Townsend\altaffilmark{9}
}
\altaffiltext{1}{Kavli Institute for Theoretical Physics, University of California, Santa Barbara, CA 93106, USA}
\altaffiltext{2}{Department of Astronomy, University of Virginia, P.O. Box 400325, Charlottesville, VA 22904-4325, USA}
\altaffiltext{3}{Department of Physics, University of California, Santa Barbara, CA 93106, USA}
\altaffiltext{4}{Department of Physics and Astronomy, National Superconducting Cyclotron Laboratory, and Joint Institute for Nuclear Astrophysics, Michigan State University, East Lansing, MI 48864, USA}
\altaffiltext{5}{ Research School of Astronomy and Astrophysics, The Australian National University, Weston, ACT 2611, Australia}
\altaffiltext{6}{Department of Astronomy and McDonald Observatory, University of Texas, Austin, TX 78712, USA}
\altaffiltext{7}{Sydney Institute for Astronomy (SIfA), School of Physics, University of Sydney, NSW 2006, Australia}
\altaffiltext{8}{School of Earth and Space Exploration, Arizona State University, Tempe, AZ 85287, USA}
\altaffiltext{9}{Department of Astronomy, University of Wisconsin-Madison, Madison, WI 53706, USA}
\email{matteo@kitp.ucsb.edu}

\shortauthors{Paxton et al.}
\shorttitle{Modules for Experiments in Stellar Astrophysics (MESA)}

\journalinfo{The Astrophysical Journal Supplement Series}
\slugcomment{to appear in The Astrophysical Journal Supplement Series}

\begin{abstract}
We substantially update the capabilities of the open source software
package Modules for Experiments in Stellar Astrophysics (\MESA), and
its one-dimensional stellar evolution module, \MESAstar. Improvements 
in \MESAstar's ability to model the evolution of giant planets now
extends its applicability 
down to masses as low as one-tenth that of Jupiter. The dramatic
improvement in asteroseismology enabled by the space-based \emph{Kepler} and
\emph{CoRoT} missions motivates our full coupling of the ADIPLS adiabatic
pulsation code with \MESAstar. This also motivates a numerical
recasting of the Ledoux criterion that is more easily implemented when many nuclei
are present at non-negligible abundances.  This impacts the way in
which \MESAstar \ calculates semi-convective and thermohaline mixing. 
We exhibit the evolution of $3\textrm{--}8\nsp\Msun$ stars through the end of core
He burning, the onset of He thermal pulses, and arrival on the white
dwarf cooling sequence.
We implement diffusion of angular momentum and chemical abundances that enable calculations of rotating-star models, which we
compare thoroughly with earlier work. 
We introduce a new treatment of radiation-dominated envelopes that
allows the uninterrupted evolution of  massive
stars to core collapse. This
enables the generation of new sets of supernovae, long gamma-ray burst, and pair-instability progenitor models. 
We substantially modify the way in which \MESAstar\ 
solves the fully coupled stellar structure and composition equations, and we show how this has improved the scaling of
\MESA's calculational speed on multi-core processors. Updates to
the modules for equation of state, opacity, nuclear reaction rates,
and atmospheric boundary conditions are also provided. 
We describe the MESA Software Development Kit (\code{SDK}) that
packages all the required components needed to form a unified,
maintained, and well-validated build environment for \MESA. We also
highlight a few tools developed by the community for rapid
visualization of \MESAstar\ results. 
\end{abstract}


\maketitle
\tableofcontents

\section{Introduction}\label{s.introduction}

As the most commonly observed objects, stars remain at the forefront
of astrophysical research. Advances in optical detector technology,
computer processing power, and data storage capability have enabled
new sky surveys \citep[e.g., the Sloan Digital Sky
  Survey;][]{York2000The-Sloan-Digit}; triggered many new optical
transient surveys, such as the Palomar Transient Factory
\citep{Law2009The-Palomar-Tra} and Pan-STARRS1
\citep{Kaiser2010The-Pan-STARRS-}; and allowed for space missions
\citep[e.g., \emph{Kepler};][]{Koch2010Kepler-Mission-} that
continuously monitor more than 100,000 stars. The stellar discoveries
from these surveys include revelations about rare stars, unusual
explosive outcomes, and remarkably complex binaries. The immediate
future holds tremendous promise, as both the space-based survey
\emph{Gaia}
\citep{de-Bruijne2012Science-perform,Liu2012The-expected-pe} and the
ground based Large Synoptic Survey Telescope
\citep[LSST;][]{Ivezic2008LSST:-from-Scie} come to fruition.

These developments have created a new demand for a
reliable and publicly available research and education tool in
computational stellar astrophysics. We introduced the open source community
tool \mesa\ \citep[][hereafter \mesaone]{Paxton2010Modules-for-Exp}
to meet these new demands. This first ``instrument'' paper described
the design, implementation, and realm of validity of \mesa \ modules for numerics,
microphysics, and macrophysics, and introduced the stellar evolution
module, \MESAstar. We presented a multitude of tests and code
comparisons that served as our initial verification and demonstrated
\MESAstar's initial capabilities.  Since \mesaone, \mesa\ has
attracted over 500 registered users, witnessed over 5,000 downloads
from \url{http://mesa.sourceforge.net/}, started an annual Summer School
program, and provided a portal (\url{http://mesastar.org}) for the community to openly share
knowledge (e.g., the specific settings for a published \MESAstar\
run), codes, and publications.

This paper describes the major new \mesa\ capabilities for modeling
giant planets, asteroseismology, and the treatment of rotation and evolution of massive stars.  We also describe
numerous advances since \mesaone.  These include the  incorporation of composition
gradients in the determination of convective mixing and additional
verification for evolution of intermediate mass stars and the white dwarfs they create.

Our improvements to \MESAstar\ for gas giant planets
were motivated by the dramatic growth in this field. Over 800
exoplanets have been confirmed, and their study has prompted
enormous progress in our understanding of the formation and migration
of giant planets, and of the importance of factors such as stellar
mass \citep{laughlin_2004_aa, alibert_2011_aa,boss_2011_aa},
composition \citep{fischer_2005_aa,young_2012_aa}, and binarity
\citep{patience_2002_aa,mugrauer_2009_aa,roell_2012_aa}. Puzzles
remain, though, both in our solar system and in the studies of the
plethora of these newly discovered exoplanets, including the
characteristics of the planet-hosting stars and the interiors,
atmospheres, surface gravities, temperatures, and compositions of the
planets \citep[e.g.,][]{udry_2007_aa,seager_2010_aa}. 
Many of these variations can now be numerically explored, as
can the incorporation of an inert core in an otherwise regular gas
giant and the impact of irradiation.

The ability to infer stellar properties (e.g., mass, radius, internal
state, and rotation) from measurements of the radial and non-radial
oscillation modes has been dramatically improved by two space-based
optical telescopes (Convection Rotation and Planetary Transits, \emph{CoRoT};
\citealt{Baglin2009CoRoT:-Descript} and \emph{Kepler}; \citealt{Borucki2009KEPLER:-Search-}). The high
cadences and precision (often better than ten parts per million) reveal
and accurately measure multitudes of oscillation frequencies for over 10,000 stars, substantially raising the need for accurate and
efficient computations of stellar mode frequencies and the resulting 
eigenfunctions. The intrinsic flexibility of \MESAstar\ allows for
the exploration of model-space required to precisely infer stellar
properties from the observed frequencies. 

An important new addition to \mesa\ is the incorporation of stellar
rotation and  magnetic fields in radiative regions.  As stars are not
solid bodies, they undergo radial differential rotation
\citep{thompson_2003_aa,balbus_2012_aa} and also rotate at different
angular velocities at different latitudes
\citep{ruediger_1998_aa,bonanno_2007_aa,kuker_2011_aa}.  These
rotational shears have a significant impact on the evolution of the
stellar magnetic field. Despite the resulting 3D nature of magnetism
and rotation, the stellar evolution community has come a long way in
understanding stars with 1D simulations
\citep{meynet_1997_aa,langer_1999_aa,maeder_2000_aa,heger_2000_aa,
  hirschi_2004_aa, cantiello_2010_aa}, thus motivating our need to
fully incorporate rotation within \mesa. The new flexibility in
angular momentum transport mechanisms allows for numerical exploration
of alternate rotational outcomes should the observations (e.g.,
asteroseismology) require it.

The paper is outlined as follows. Section \ref{s.planets} describes the new
capability of \mesa\ to evolve models of giant planets, while 
\S\ref{s.astroseismology} discusses the new asteroseismology capabilities. 
The \mesa\ implementation of composition gradients in stellar
interiors and their impact on convective mixing is described in 
\S\ref{s.mixing}. The status of the evolution of intermediate mass  stars and the \MESAstar\ construction and
evolution of white dwarfs is described in \S\ref{s.agb-wd}. 
The new capabilities for evolving rotating stars is described in 
\S\ref{s.rotation}. The onset of near Eddington luminosities and
radiation pressure dominance in the envelopes of evolving massive
stars  has been a challenge for many stellar evolution codes ever
since the realization of the iron opacity bump at $\log T\approx 5.3$ 
\citep{Iglesias1992Spin-orbit-inte}. We discuss in \S\ref{s.massive} the resulting
improvements for evolving massive stars. 
This allows for the  uninterrupted evolution of rotating massive stars to the onset of core collapse. 
We conclude
in \S\ref{s.conclusions} by highlighting where additional improvements
to \mesa\ are likely to occur in the near future. 
Appendix \ref{s.input-physics} describes the many improvements to the
physics modules since \mesaone; Appendix \ref{s.nuts-and-bolts}
presents ``nuts and bolts'' information on the primary components of
evolution calculations; and Appendix \ref{s.mesasdk} presents the
\mesa\ Software Development Kit (\SDK). 
All of our symbols are defined in Table~\ref{t.list-of-symbols}.  We denote components of \mesa, such as modules and routines, in Courier font, e.g., \code{evolve\_star}. 

{\LongTables
\begin{deluxetable}{cll}
	\tablecolumns{3}
	\tablecaption{Variable Index.\label{t.list-of-symbols}}
	\tablewidth{\apjcolwidth}
	\tablehead{\colhead{Name} & \colhead{Description} & \colhead{First Appears}}
	\startdata
	$A$				& atomic mass number		&  \ref{s.chem}  \\
	$\Delta_i$		& mass excess of the $i$th isotope			& \ref{s.chem} \\
	$\eta$ & wind mass loss coefficient & \ref{s.completing-evolution}\\
	$F_\star$ & day-side flux incident on an irradiated planet & \ref{sec:irradiation} \\
	$\Gamma$			& Coulomb coupling parameter			&  \ref{s.completing-evolution}  \\
	$i$   & specific moment of inertia & \ref{s.nuts-rotation}\\
	$\kappa$	& opacity	& \ref{s.construction-models} \\
	$L$ & stellar luminosity    & \ref{s.astero} \\
	$m$				& Lagrangian mass coordinate		& \ref{s.construction-models} \\
	$M$			& stellar mass		&  \ref{s.construction-models} \\
	$N$ & \bvfreq & \ref{s.ledoux}\\
	$n_i$				& number density of the $i$th isotope		& \ref{s.chem} \\
	$\nu$   & turbulent viscosity & \ref{s.nuts-rotation}\\
	$r$       & radial coordinate                                & \ref{sec:irradiation} \\
	$R$     & total stellar radius    & \ref{s.construction-models} \\
	$\rho$			& baryon mass density				& \ref{s.chem} \\
	$S$ & specific entropy             & \ref{s.construction-models} \\
	$\Sigma$ & mass column & \ref{sec:irradiation} \\
	$\Sigma_\star$ & depth for heating from irradiation & \ref{sec:irradiation} \\
	$\tau$			& optical depth			&  \ref{s.completing-evolution} \\
	$w_{c}$ & magnitude of changes during a timestep & \ref{s.timestep-controls} \\
	$w_{t}$ & target value for $w_{c}$ & \ref{s.timestep-controls} \\
	$W$				& atomic weight			& \ref{s.chem}   \\
	$X$				  & H mass fraction				&  \ref{s.astero}  \\
	$X_i$				& baryon mass fraction of the $i$th isotope					& \ref{s.ledoux} \\
	$Y$         & He mass fraction      & \ref{s.planets} \\
	$Y_{e}$			& electrons per baryon ($\bar{Z}$/$\bar{A}$)			&   \ref{s.chem} \\
	$Y_i$				& abundance of the $i$th isotope							& \ref{s.chem} \\
	$Z$         & metallicity           & \ref{s.planets} \\
	$Z$				& atomic number					&  \ref{s.chem}  \\
	\alphaMLT &  mixing length parameter & \ref{s.planets-no-core} \\
	\alphasc &  semiconvection efficiency parameter & \ref{s.semiconvection} \\
	\alphath &  thermohaline efficiency parameter & \ref{s.thermohaline} \\
	\asuper &  smoothing parameter for MLT++ & \ref{s.superadiabatic} \\
	\asupert &  MLT++ parameter used in construction of \asuper & \ref{s.superadiabatic} \\
	\betamin &  $ \min(P/\Pgas)$ & \ref{s.superadiabatic} \\
	\chirho &  $(\partial\ln P/\partial\ln\rho)_T$ & \ref{s.ledoux} \\
	\chiT &  $(\partial\ln P/\partial\ln T)_{\rho}$ & \ref{s.ledoux} \\
	\CP &  specific heat at constant pressure & \ref{s.semiconvection} \\
	\cs &  adiabatic sound speed & \ref{s.astroseismology} \\
	\dnu &  large frequency separation of pulsation modes & \ref{s.astero} \\
	\Dov &  overshoot diffusion coefficient & \ref{s.ledoux} \\
	\Dth &  thermohaline diffusion coefficient & \ref{s.thermohaline} \\
	\EFc &  Fermi energy at center & \ref{s.planets-no-core} \\
	\epsgrav &  gravitational heating rate & \ref{s.completing-evolution} \\
	\epsnuc &  nuclear heating rate & \ref{s.reactions} \\
	\epsnu &  neutrino loss rate & \ref{s.rotation-modifications} \\
	\Fconv &  convective flux & \ref{s.massive-evol} \\
	\fov &  convective overshoot parameter & \ref{s.astero} \\
	\Frad &  radiative flux & \ref{s.massive-evol} \\
	\fsuper &  reduction factor for $\supernab$ & \ref{s.superadiabatic} \\
	\Gammaone &  $ (\partial\ln P/\partial \ln\rho)_S$ & \ref{s.astroseismology} \\
	\kappath &  opacity for thermal radiation orig.\ in planet & \ref{sec:irradiation} \\
	\kappav &  opacity for irradiation from star & \ref{sec:irradiation} \\
	\kB &  Boltzmann constant & \ref{s.planets-no-core} \\
	\Lacc &  accretion luminosity & \ref{s.compressional} \\
	\lambdamax &  $ \max(\Lrad/\Ledd)$ & \ref{s.superadiabatic} \\
	\Lc &  core luminosity & \ref{s.planets-cores} \\
	\Ledd &  Eddington Luminosity & \ref{s.rotation-mass-loss} \\
	\logg &  log surface gravity & \ref{s.atmospheres} \\
	\Lonset &  luminosity at which the onset of convection occurs & \ref{s.massive-evol} \\
	\Lrad &  radiative luminosity & \ref{s.massive-evol} \\
	\Lrho &  luminosity at which a density inversion occurs & \ref{s.massive-evol} \\
	\mb &  atomic mass unit & \ref{s.chem} \\
	\Mc &  core mass & \ref{s.planets-cores} \\
	\Mdot &  mass-loss rate & \ref{s.compressional} \\
	\Mm &  modeled mass & \ref{s.mesh-controls} \\
	\NA &  Avogadro number & \ref{s.planets-no-core} \\
	\nablaad &  adiabatic temperature gradient & \ref{s.ledoux} \\
	\nablaL &  Ledoux criterion & \ref{s.semiconvection} \\
	\nablarad &  radiative temperature gradient & \ref{s.ledoux} \\
	\nablaT &  actual temperature gradient & \ref{s.ledoux} \\
	\nB &  baryon density & \ref{s.chem} \\
	\numax &  frequency of maximum power & \ref{s.astero} \\
	\Om &  surface angular velocity & \ref{s.rotation-mass-loss} \\
	\om &   angular velocity & \ref{s.rotation-modifications} \\
	\Omc &  surface critical angular velocity & \ref{s.rotation-mass-loss} \\
	\Pc &  central pressure & \ref{s.construction-models} \\
	\Pgas &  gas pressure & \ref{s.compressional} \\
	\Prad &  radiation pressure & \ref{s.compressional} \\
	\Rc &  core radius & \ref{s.planets-cores} \\
	\rhoc &  central density & \ref{s.construction-models} \\
	\scaleheight &  pressure scale height & \ref{s.ledoux} \\
	\sigmaSB &  Stefan-Boltzmann constant & \ref{s.construction-models} \\
	\Slamb &  Lamb frequency & \ref{s.astroseismology} \\
	\supernab &  superadiabaticity, $\nablaT-\nablaad$ & \ref{s.superadiabatic} \\
	\superthresh &  controls when MLT++ is applied & \ref{s.superadiabatic} \\
	\Tc &  central temperature & \ref{s.construction-models} \\
	\Teff &  effective temperature & \ref{s.construction-models} \\
	\timestep &  numerical timestep & \ref{s.compressional} \\
	\tkh &  thermal (Kelvin-Helmholtz) timescale & \ref{s.rotation-mass-loss} \\
	\veq &  equatorial velocity & \ref{s.rotation} \\
	\enddata
\end{deluxetable}
}

\section{Giant Planets and Low-Mass Stars}\label{s.planets} 

Evolutionary models of giant planets and low-mass stars differ from their higher-mass stellar counterparts
in both the microphysics needed to describe the interior and the role of stellar irradiation 
in the outer boundary condition. For masses $M \la 84\,\MJ$,
hydrogen burning is insufficient to prevent cooling and contraction. Deuterium 
burning can briefly slow the cooling for $M \ga 13\,\MJ$, where 
$\MJ=9.54 \times 10^{-4}\,\Msun$ is Jupiter's mass, but has a negligible influence 
on the cooling for smaller masses. Hence nuclear burning can be ignored in the planetary
mass regime.

For hydrogen-helium rich objects with $M \gg \MJ$, an ideal gas equation
of state (EOS), with arbitrary degeneracy, is a good approximation while for  $M \la \MJ$ particle interactions play an important role. Specifically, 
pressure ionization of hydrogen at $\rho \simeq 1\,\grampercc$ and $T \simeq 10^4\,\Kelvin$
causes a sudden change from a H$_2$-dominated phase to an ionized phase. 
\MESA\ employs the \citet{1995ApJS...99..713S} equation of state (SCVH EOS), smoothly interpolated from the low to high pressure phase,
for this complicated region of parameter space where thermal, Fermi, and electrostatic
energies may all be comparable. 
The  SCVH EOS includes pressure ionization of hydrogen, but not helium. The temperature range covered by the tables is $2.10 < \log T({\rm K}) < 7.06$, and the pressure ranges from $\log P\,({\rm dyne\ cm^{-2}}) = 4$ to a maximum value $19$ dependent on the temperature. Smooth interpolation to other EOS occurs near the SCVH boundaries (for more details see Paper I).
At the low temperatures in planetary atmospheres,
abundant species such as CNO atoms will be in molecular form, and may 
condense into clouds.  \MESA\ does not follow the transition from atomic to molecular
form for these species in the EOS---they are currently included by increasing the helium abundance from
$Y$ to $Y+Z$ when calling the SCVH EOS. \MESA\ does, however, include the effect of molecules in the Rosseland opacities.
Currently, the \citet{Ferguson:2005} and \citet{Freedman:2008} tables, which include the opacity from molecules,
but ignore condensates, are available. 

Lastly, for planets in close-in orbits about their parent star,
the external irradiation flux may be orders of magnitude larger than the cooling flux from the planet's interior. 
This may dramatically increase the surface temperature and affect the outer boundary condition. \MESA\ now 
implements several options for this surface heating, including the flexibility to include user-supplied prescriptions. 

In the following subsections, we discuss a new \MESA\ module that creates initial models in the planetary mass
range $M \simeq 0.1\textrm{--}10\,\MJ$, and present a suite of evolutionary calculations. We discuss 
how surface irradiation may be included, as well as an inert core at
the center of the planet. 
We also show what \MESAstar\ yields for the 
mass-radius relation for sub-solar mass stars in \S  \ref{s.mass-radius}.

\subsection{ Construction of Starting Models }\label{s.construction-models}

For stellar mass objects, the \code{pre\_ms\_model} routine constructs
pre-main-sequence (PMS) models
assuming $L(r) \propto m$, where $L(r)$ is the luminosity at radius $r$, by iterating on the starting conditions at the center
to find a model with a given $M$ and central temperature $\Tc$.   This PMS routine works well for $M \ga 0.03\,\Msun$, but
lower masses may not converge when the guess for central
density $\rhoc$ and luminosity are not close enough to the (unknown) true values.
As a result, it is difficult and time consuming to create models with $M<0.03 \nsp\Msun$
using the same routine for giant planets as for stars.

A new routine called \code{create\_initial\_model}  builds a model of given $M$ and radius $R$ using an adiabatic temperature
profile. Given the central pressure \Pc\ and specific entropy $S$, the equation of hydrostatic
balance is integrated outward, and the temperature at each step determined from the equation
of state using $T=T(P,S)$. The values of \Pc\ and $S$ are iterated to attain the desired
$M$ and $R$. The luminosity profile is then derived treating $S$ as constant in space for the fully convective
planet (e.g., \citealt{1998ApJ...497..253U}), so
\begin{eqnarray}
\int_0^m \dif m'\ T(m') \frac{\dif S}{\dif t} & \simeq & \frac{\dif S}{\dif t} \int_0^m \dif m' T(m') = - L(m).
\end{eqnarray}
The luminosity at the surface, $L(M)$, is estimated using the radius $R$ and temperature $\Teff$ at the $\tau=\kappa P/g=2/3$ point as $L(M)=4\pi R^2 \sigmaSB \Teff^4$. Given $L(M)$, the luminosity at interior points is found by
\begin{eqnarray}
L(m) & = & L(M) \left( \frac{ \int_0^m \dif m'\ T(m') }{ \int_0^M \dif m'\ T(m')  } \right).
\end{eqnarray}
This procedure works well for $M$ down to $\sim 0.1\,\MJ$ and over a range of initial radii, allowing the user to choose either a  $\sim 1\, \RJ$ radius appropriate for a cold planet, to radii $\sim 2\textrm{--}3\,\RJ$ appropriate for young or inflated planets (e.g., \citealt{2007ApJ...655..541M}).  Here $\RJ=7.192
\times 10^9\,\cm$ is the equatorial radius of Jupiter.

\subsection{ Evolutionary Calculations }\label{s.planets-no-core}
\begin{figure}[htbp]
\centering
\includegraphics[width=\figwidth]{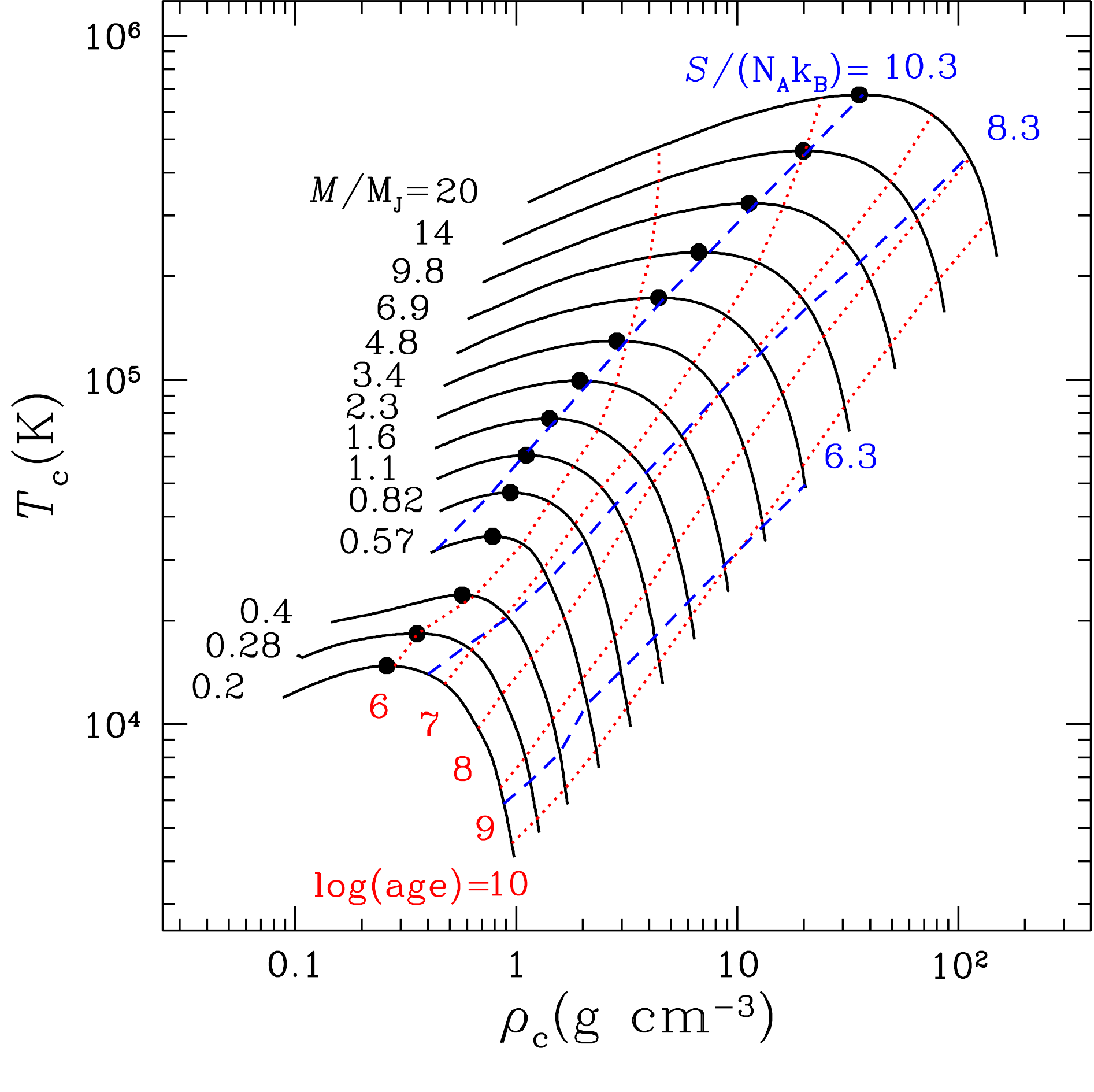}
\caption{
The solid black lines show \Tc\ versus \rhoc\ during the evolution. Each line is labeled on the left
by the mass in units of \MJ. The dotted red lines show constant values of 
$\log(\rm age[yr])$, labeled at the base of each line. The blue dashed lines
show fixed values of 
$S/(\NA\kB) $, labeled at the top of
each line. The large black dots show the position of maximum \Tc\ along the evolutionary track.
}
\label{fig:Tc_vs_rhoc_isoage_isoent}
\end{figure}

\begin{figure}[htbp]
\centering
\includegraphics[width=\figwidth]{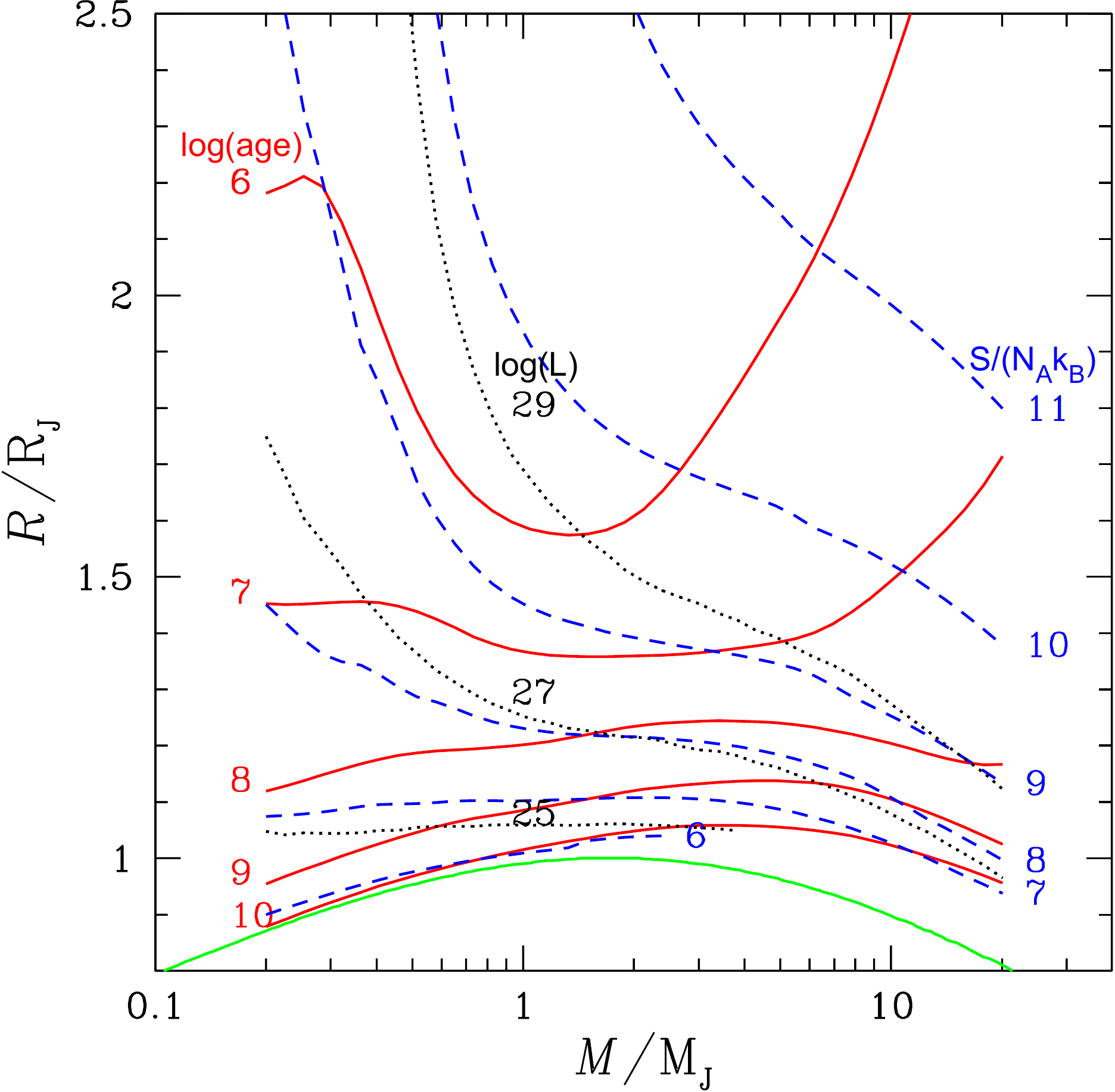}
\caption{
Radius versus mass iso-contours from a suite of evolutionary calculations. The solid red lines show 
$R/\RJ$ versus $M/\MJ$ at fixed values of $\log(\rm age[yr])$, labeled on the left of each curve.
The dashed blue curves are for fixed entropy, with each curve labeled by $S/(\NA\kB) $ on the right.
The dotted black curves are for fixed luminosity, with each curve labeled by $\log(L[\rm erg\ s^{-1}])$
above $M=1\nsp\MJ$. The green curve at the bottom is the $T=0$ $M\textrm{-}R$ relation from 
\citet{1969ApJ...158..809Z} for a solar mixture of H and He.
}
\label{fig:R_vs_M_iso_tsl}
\end{figure}

Figures \ref{fig:Tc_vs_rhoc_isoage_isoent} and \ref{fig:R_vs_M_iso_tsl} show evolutionary
calculations for models with  masses $M=0.2\textrm{--}20\,\MJ$.
All models were evolved for $20$ Gyr.
The initial models from \code{create\_initial\_model} had a large radius $R=5\,\RJ$.
The other parameters used are $Y=0.27$, $Z=0.02$ and $\alphaMLT=2$. 
The opacity and EOS tables used are \code{eos\_file\_prefix} = \code{mesa}, \code{kappa\_file\_prefix} = \code{gs98} and 
\code{kappa\_lowT\_prefix} = \code{lowT\_Freedman11}. The atmosphere model is \code{which\_atm\_option} = \code{simple\_photosphere}.

Figure \ref{fig:Tc_vs_rhoc_isoage_isoent} is a low mass extension of Figure 16 from 
\mesaone, showing evolution in the \rhoc-\Tc\ plane.
Each track (solid black curve) is labeled on the left by the planet's mass, and evolution goes from left to right. Initially the planet is non-degenerate and
contraction increases both $\rhoc \propto R^{-3}$ and 
$\Tc \propto R^{-1} \propto \rhoc^{1/3}$. A maximum
\Tc\ is reached when $\kB \Tc \sim \EFc$, where \EFc\ is the electron Fermi energy at the center, beyond which \rhoc\ approaches a constant as \Tc\ 
decreases further.
Ignoring Coulomb interactions in the EOS, $S$ is a function of the electron degeneracy parameter $\mu_{\rm e}/\kB T$,
where $\mu_{\rm e}$ is the electron chemical potential
and all models
should have maximum $\kB \Tc \sim\EFc$ at the same $S$. The line labeled $S/(\NA\kB) =10.3$ 
indeed coincides with maximum \Tc\ down to $M \simeq 1\,\MJ$, but at smaller masses
where non-ideal effects are more important, maximum \Tc\ occurs when $S/(\NA\kB)  < 10.3$. Also shown in Figure \ref{fig:Tc_vs_rhoc_isoage_isoent} are lines of constant 
age, shown as dotted red lines, and labeled on the bottom of the plot.
 
The same evolutionary calculations are used in Figure \ref{fig:R_vs_M_iso_tsl} to show radius versus mass at fixed values of age, entropy or luminosity. At late times, or low entropy and luminosity, the radius
approaches the zero-temperature value (green curve; \citealt{1969ApJ...158..809Z}) for which thermal support is insignificant.
The maximum radius occurs where gravitational and Coulomb energies, per ion, are comparable. The solid red lines, labeled by age on the left, show that contraction down to $R \simeq 1.5\,\RJ$ is rapid, taking less than 10 Myr for $M \la 10\,\MJ$. This initial rapid cooling phase occurs because the initial luminosity is orders of magnitude higher than the luminosity around one Gyr. This can been seen in the black dotted contours of constant $\log(L)$, where $L$ is larger by a factor of 100 for $R=1.3\,\RJ$ and $10^4$ for $R=1.7\,\RJ$, as compared to $R=1.1\,\RJ$.
The blue dashed lines show contours of constant entropy, labeled on the right by $S/(\NA\kB)$.

\subsection{ Implementation of Inert Cores }\label{s.planets-cores}

In the core accretion model of planet formation (e.g., \citealt{1996Icar..124...62P,
2005Icar..179..415H}), a rock/ice
core is first assembled. Once this core grows to $\sim\!10\,\Mearth$, where $\Mearth$ denotes an Earth mass, it can initiate rapid accretion of nebular gas,
which could then dominate the mass of the planet. For studies of
planetary radii, a central core composed of high mean molecular weight 
material can 
decrease the radius of the planet by a significant amount ($\simeq 0.1\textrm{--}0.2\,\RJ$). 
The \mesastar\ inert core feature allows one
to add a core of specified mass \Mc\ and radius \Rc, or
more conveniently, density \rhoc. A luminosity 
\Lc\ may also be specified, although the high mean molecular
weight of the core, as compared to the overlying H/He envelope,
implies that even large cores will tend to have small heat content
\citep{2006ApJ...642..495F}. This inert core is not presently
evolved in any way, and changes in \Pc\
during evolution are neglected as $R$ changes.
While cores of mass $\la 10\textrm{--}20\,\Mearth$ are
commonly used for modeling solar system giants
(e.g. \citealt{2005AREPS..33..493G}), the large masses and small
radii of some exoplanets may imply far larger core masses (e.g. HD
149026; \citealt{2005ApJ...633..465S}). In addition, Neptune-like
planets with smaller ratios of envelope to core masses may be modeled with
\MESA\ \citep{2013arXiv1303.3899O}.

\subsection{ Irradiation }\label{sec:irradiation}

\begin{figure}[htbp]
\centering
\includegraphics[width=\figwidth]{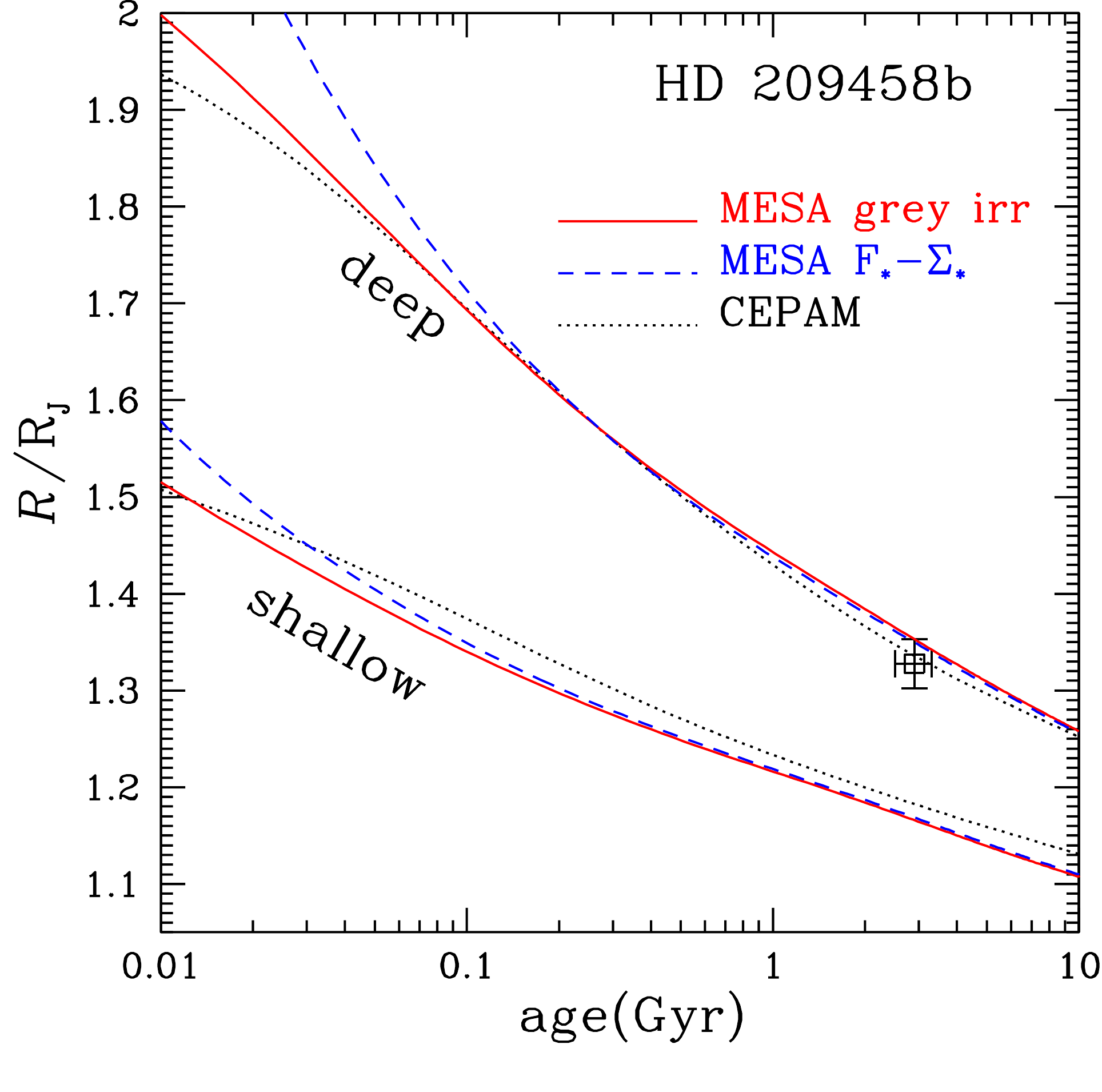}
\caption{
Radius versus age for the planet HD 209458b. The solid red lines are for \MESA, using the
\code{grey\_irradiated} atmosphere model. The dotted black lines show the CEPAM code results.
The dashed blue lines show the \MESA\ calculation using the 
$F_\star$-$\Sigma_\star$ surface heat source.  The data point with error bars is the observed value of the 
radius for HD 209458b quoted in \citet{2010A&A...520A..27G}. 
The two sets of curves are deep heating (upper three curves) and shallow heating (lower three curves).}
\label{fig:Rph_vs_age_compare_to_cepam_guillot2010_fig9}
\end{figure}

Surface heating by stellar irradiation changes the boundary condition for
the planet's cooling and contraction.  This modifies the planetary radius versus age for 
exoplanets at orbital separation $\la 0.1\,\AU$.  \MESA\ provides several ways to
implement surface heating with varying degrees of 
fidelity to the true solution. These presently include:
\begin{list}{}
\item a) An energy generation rate $\epsilon = F_\star/4\Sigma_\star$ applied
in the outer mass column $\Sigma \leq \Sigma_\star$. Here $F_\star$ is the day-side
flux from the star, and $\Sigma ( r ) = \int_r^R \dif r' \rho(r')$ is the mass column. 
In steady-state, this generates
an outward flux $F_\star/4$, which is meant to simulate the angle-averaged flux
over the planetary surface. This model implicitly assumes that day-night heat transport is efficient, and at the depths of interest the temperature is uniform over the surface. 
The parameters $F_\star$ and $\Sigma_\star$ are specified 
through the user-specified variables \code{irradiation\_flux} and \code{column\_depth\_for\_irradiation},
making
this the simplest method to use. This heating mechanism represents
absorption of stellar optical radiation well below the
photosphere of the planet's thermal radiation and gives rise
to greenhouse heating of the atmosphere where $\epsilon \neq 0$.

\item b) \MESA's \code{grey\_irradiated} atmosphere model (see also \S \ref{s.atmospheres}) 
implements the angle-averaged temperature profile of
\citet{2010A&A...520A..27G}. This approximate solution to the transfer equation
assumes two frequency bands: optical radiation from the star (with user-specified 
opacity $\kappav$) and thermal radiation originating in the planet 
(with user-specified opacity $\kappath$). 
The temperature profile is derived
using the Eddington approximation, assuming an external flux 
from the star as well as a  flux from the planetary interior. 
While the \citet{2010A&A...520A..27G} model implemented in \MESA\ uses a single temperature as a function of depth, it is derived allowing for local temperature variations over the surface which are then averaged over angle. This temperature profile is shown to be valid in the presence of horizontal heat transport by fluid motions. 
This is the only \MESA\ atmosphere model that uses pressure instead of optical depth to determine the surface boundary condition. As this 
pressure may be relatively deep in the atmosphere, a correction
to the radius may be required to give either the vertical thermal
photosphere, or the optical photosphere in transit along a chord.
Lastly, the \code{relax\_irradiation} routine improves initial convergence by providing a starting
model closer to the irradiated one.

 \item c) Finally, \MESA\ allows user-specified heating
functions (e.g., $F_\star$-$\Sigma_\star$ surface heating) or atmosphere models (e.g.,
\code{grey\_irradiated}). User-supplied routines may be easily implemented by using the \code{other\_energy}
module.
\end{list}

Figure~\ref{fig:Rph_vs_age_compare_to_cepam_guillot2010_fig9} shows radius versus age for the planet HD 209458b
\citep{2010A&A...520A..27G}. The two groupings of lines are for different
heating depths, and within each grouping of lines, there are three calculations: \MESA\ using \code{grey\_irradiated}
surface boundary condition (solid red line), \MESA\ using the $F_\star$-$\Sigma_\star$ surface heating profile
(dashed blue line), and CEPAM \citep{1995A&AS..109..109G} 
using the same grey irradiated boundary condition (dotted black line; kindly provided by Tristan
Guillot).
The lower curves
, corresponding to shallow heating,
use fiducial values
$(\kappath,\kappav)=(10^{-2},6\times 10^{-3})\ 
{\rm cm^2\ g^{-1}}$ and give a model radius significantly smaller than the observed radius.
The upper curves
, corresponding to deep heating,
use $(\kappath,\kappav)=(10^{-2},6\times 10^{-4})\,\squarecmpergram$, yielding significantly hotter temperatures deep in the surface radiative
zone, which slow the cooling enough to agree with the observed radius. 
The choice $\Sigma_\star = 2/\kappav$ gives agreement between the grey irradiated
and $F_\star$-$\Sigma_\star$ methods, where the factor of 2 accounts for the fact that the 
grey irradiated boundary condition has
some heating below $\Sigma = 1/\kappav$. The radii 
are at the $\tau_{\rm th}=2/3$ photosphere for a vertical path into the atmosphere. 

The agreement between all three methods is excellent, at the 1\textrm{--}2\%
level after 100\,\Myr. 
The remaining discrepancy between the \MESA\ and CEPAM grey irradiated results are likely
due to different opacity tables, with the \MESA\ result using an update of \citet{Freedman:2008}  (Freedman 2011, priv.\ comm.) while the CEPAM run uses
the \citet{2001ApJ...556..357A} COND table. The differences at ages $\la 100\,\Myr$ are due
to different starting conditions. The CEPAM calculation
started with initial radius $2\,\RJ$, whereas the MESA calculations started with $5\,\RJ$. 
The \MESA\ grey irradiated and $F_\star$-$\Sigma_\star$ calculations  differ
at $\lesssim 100\,\Myr$, likely because the former has a fixed thermal opacity while the latter allows the  opacity to change.

\subsection{Low-Mass Main Sequence Stars}\label{s.mass-radius}

Most of $\MESAstar$'s capability to evolve low-mass ($M<2\,\Msun$) stars was demonstrated in Section 7.1 of \mesaone. \MESA\ 
has seen use in the asteroseismology of helium core flashing stars
\citep{2012ApJ...744L...6B}
and the discovery of a new instability from the onset of \helium[3] burning
\citep{2012ApJ...751...98V}. We expect the
future use of \MESAstar\  for asteroseismic investigations of these stars to be substantial (see \S \ref{s.astroseismology}).

\begin{figure}[htbp]
\centering
\includegraphics[width=\figwidth]{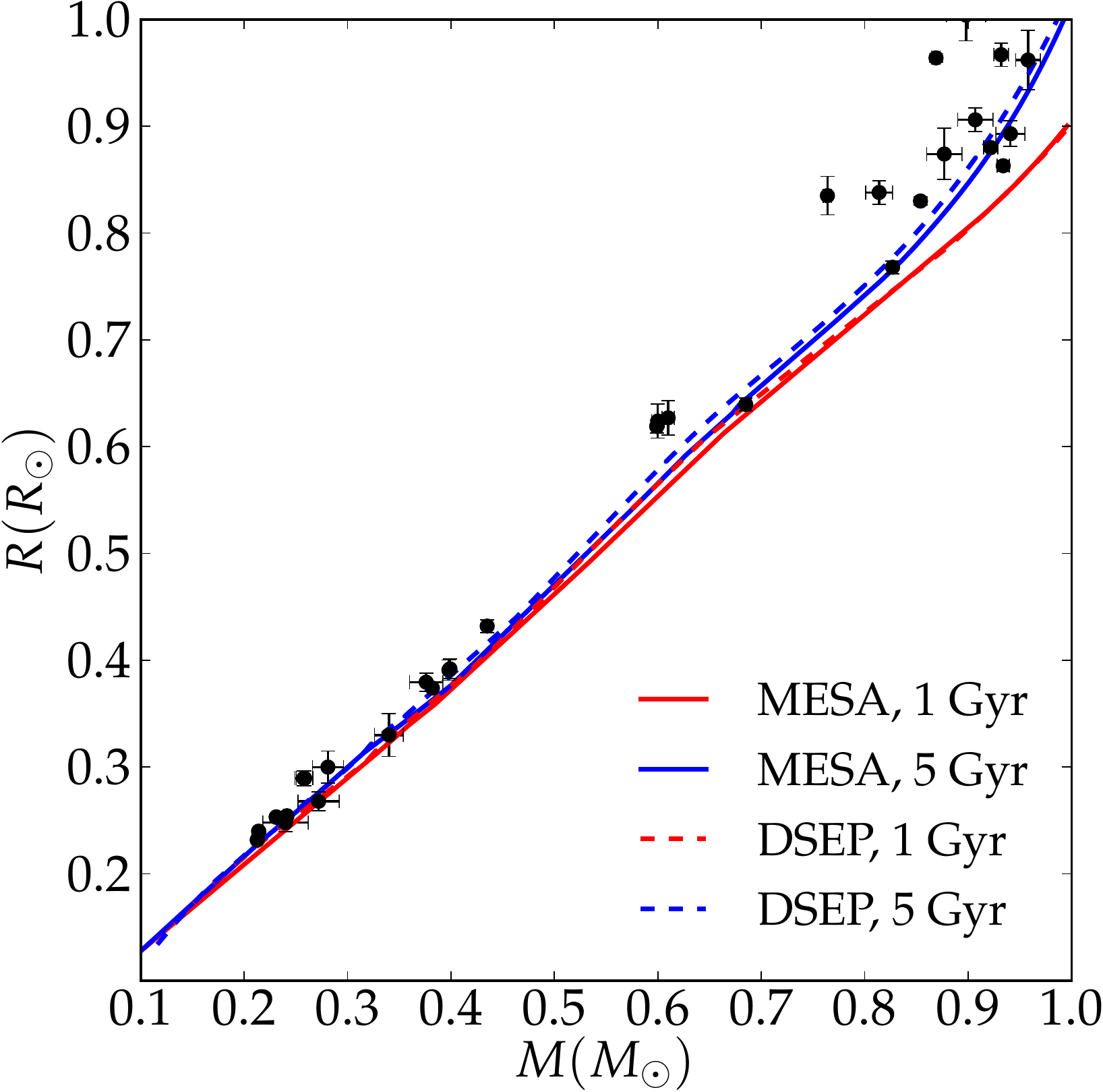}
\caption{ 
Stellar isochrones at solar
  composition spanning 0.1 to $1\nsp \Msun$ from \MESAstar\ (solid lines)
  and \citet[][dashed lines]{2008ApJS..178...89D} in the mass-radius
  plane. The data points plotted are the same as shown by
  \citet{2012arXiv1211.1068B}. }
\label{fig:stellar_M_R}
\end{figure}

The derivation of accurate planetary radii based on transits requires
accurate radii of the host stars; this motivates \MESAstar\ investigations
of low-mass stars \citep{2011ApJ...739L..49L}.  Figure \ref{fig:stellar_M_R} shows 1 and 5
Gyr isochrones at solar composition ($Y=0.27, Z=0.019$) from
\MESAstar\ (solid lines) and \citet[][dashed
 lines]{2008ApJS..178...89D} in the mass-radius diagram.  Data points
shown in Figure \ref{fig:stellar_M_R} are taken from
\citet{2010A&ARv..18...67T}, \citet{2011Sci...331..562C},
\citet{2011ApJ...742..123I}, and \citet{2012arXiv1211.1068B}.  This
figure is a reproduction of the upper panel of Figure 11 from
\citet{2012arXiv1211.1068B}.  Figure \ref{fig:stellar_M_R} indicates
that \MESAstar\ is capable of producing mass-radius relations for main
sequence stars that are consistent with other widely-used models as
well as observational data. The \MESAstar\ models were computed using,
as much as possible, the same physical assumptions as the models used by
\citet{2008ApJS..178...89D}. The main difference is the equation of
state, for which \citet{2008ApJS..178...89D} used
FreeEOS\footnote{http://freeeos.sourceforge.net} and \MESAstar\ uses a
combination of the OPAL \citep{OPAL2002} and SCVH EOS for thermodynamic
parameters relevant to this diagram.

\section{Asteroseismology} \label{s.astroseismology}

With its highly configurable output options, and its ability to
calculate asteroseismic variables,
\mesastar\ can readily produce models suitable for use with a
range of oscillation codes.  In addition to its own text output files, \MESA\ can produce outputs in formats 
widely
used by stellar oscillation codes, such as \code{fgong} and \code{osc} \citep{Monteiro09}. 

\begin{figure*}[htbp]
\centering{\includegraphics[width=\twoupwidth]{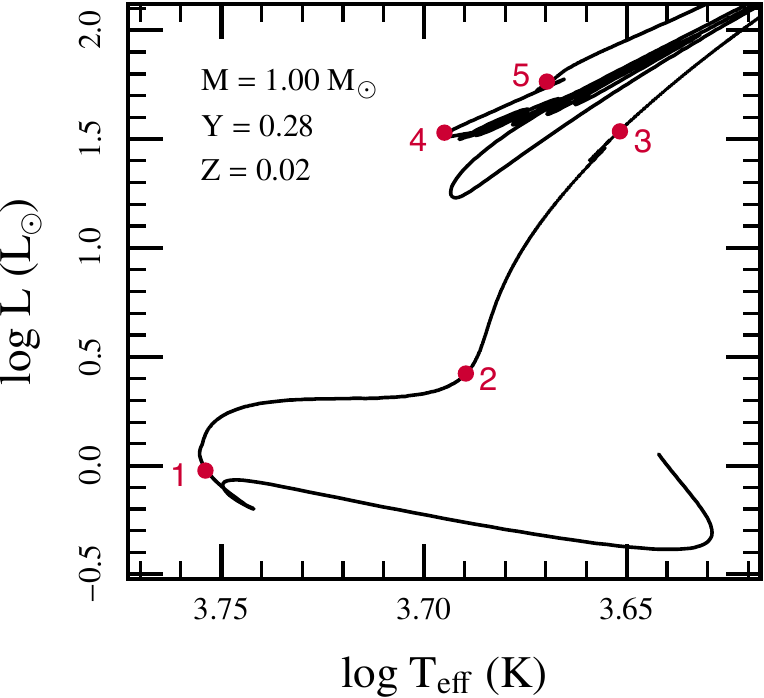}\hspace{\twoupsep}
\includegraphics[width=\twoupwidth]{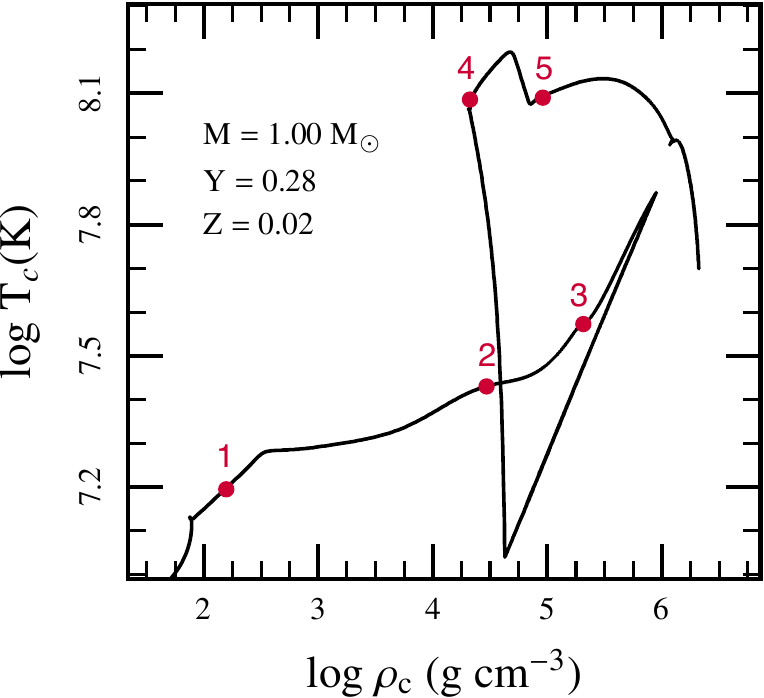}}
\caption{\label{astero:1M_pre_ms_to_wd} 
Hertzsprung-Russell diagram and $\Tc$-$\rhoc$ evolution of a $1\,\Msun$
model evolving from pre-main sequence to the white dwarf cooling
sequence. The number labels denote selected models, for which we show internal profiles in  Fig.~\ref{astero:profiles}.}
\end{figure*}

\begin{figure*}[htbp]
\centering{
\includegraphics[height=0.8\textheight]{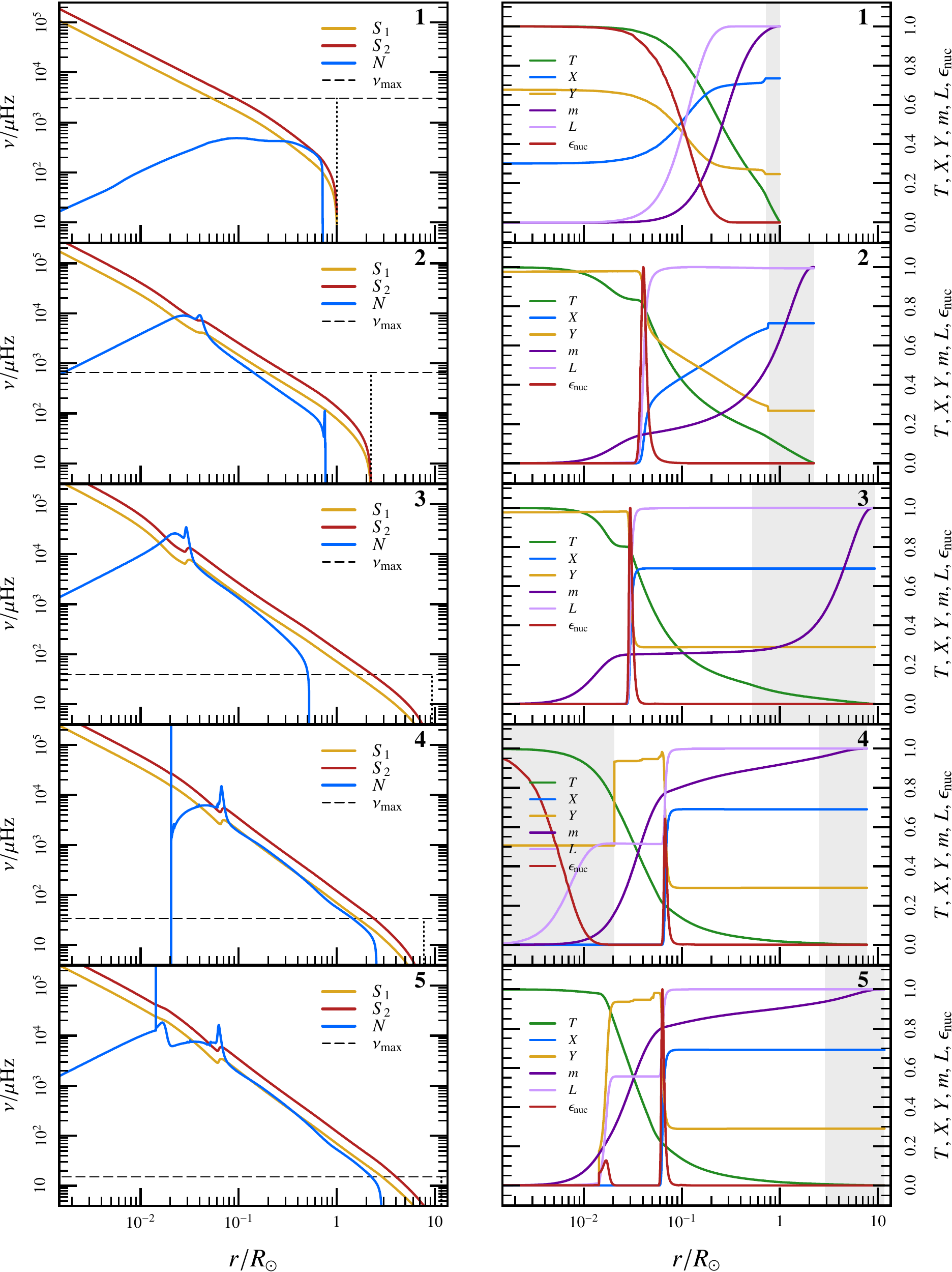}
}
\caption{\label{astero:profiles} 
Internal structure of the five points (indicated by the numbers in
each panel) marked in Figure~\ref{astero:1M_pre_ms_to_wd}. The left
panel for each point shows $N$ and $\Slamb$ for harmonic degrees
$\ell=1$ and 2.  The dashed line indicates the frequency of maximum
power $\nu_{\mathrm{max}}$ of the stochastically excited solar-like
modes. The vertical dotted lines mark the radius of the model.  Right
panels show temperature, hydrogen and helium mass fractions, mass,
luminosity, and the nuclear energy generation rate. Grey areas mark
convective regions according to the Schwarzschild criterion.
}   
\end{figure*}

In Figure~\ref{astero:1M_pre_ms_to_wd}
we show the evolution of a $1\usp\Msun$ model in the
Hertzsprung-Russell Diagram (HRD) and in \Tc-\rhoc\ space. These were evolved following the test 
case found in \code{1M\_pre\_ms\_to\_wd}, which was modified to include diffusion. This runs without user
intervention from pre-main sequence to white dwarf.  To demonstrate the changing stellar structure as
the model evolves from the main sequence to post helium-core burning on the
Asymptotic Giant Branch (AGB), we show in Figure~\ref{astero:profiles}
some of the fundamental quantities extracted from the corresponding
\code{profile.data} files for the models marked in
Figure~\ref{astero:1M_pre_ms_to_wd}. These include the Lamb and \bvv\ frequencies defined respectively by
\begin{eqnarray}
\label{lambfreq}
\Slamb^{2} &=& \frac{\ell\left(\ell+1\right)\cs^{2}}{r^{2}}, \\
N^{2} &=& \frac{g}{r} \left[ 
    \frac{1}{\Gammaone}\frac{{\dif} \ln P}{{\dif} \ln r}
    -\frac{{\dif} \ln \rho}{{\dif} \ln r}\right],
\label{N1}
\end{eqnarray}
where \cs\ is the adiabatic sound speed and $\ell$ is the spherical harmonic degree.

%
%

\subsection{The Solar Sound Speed Profile}\label{s.solar}

\begin{figure}[htbp]
\centering{\includegraphics[width=\figwidth]{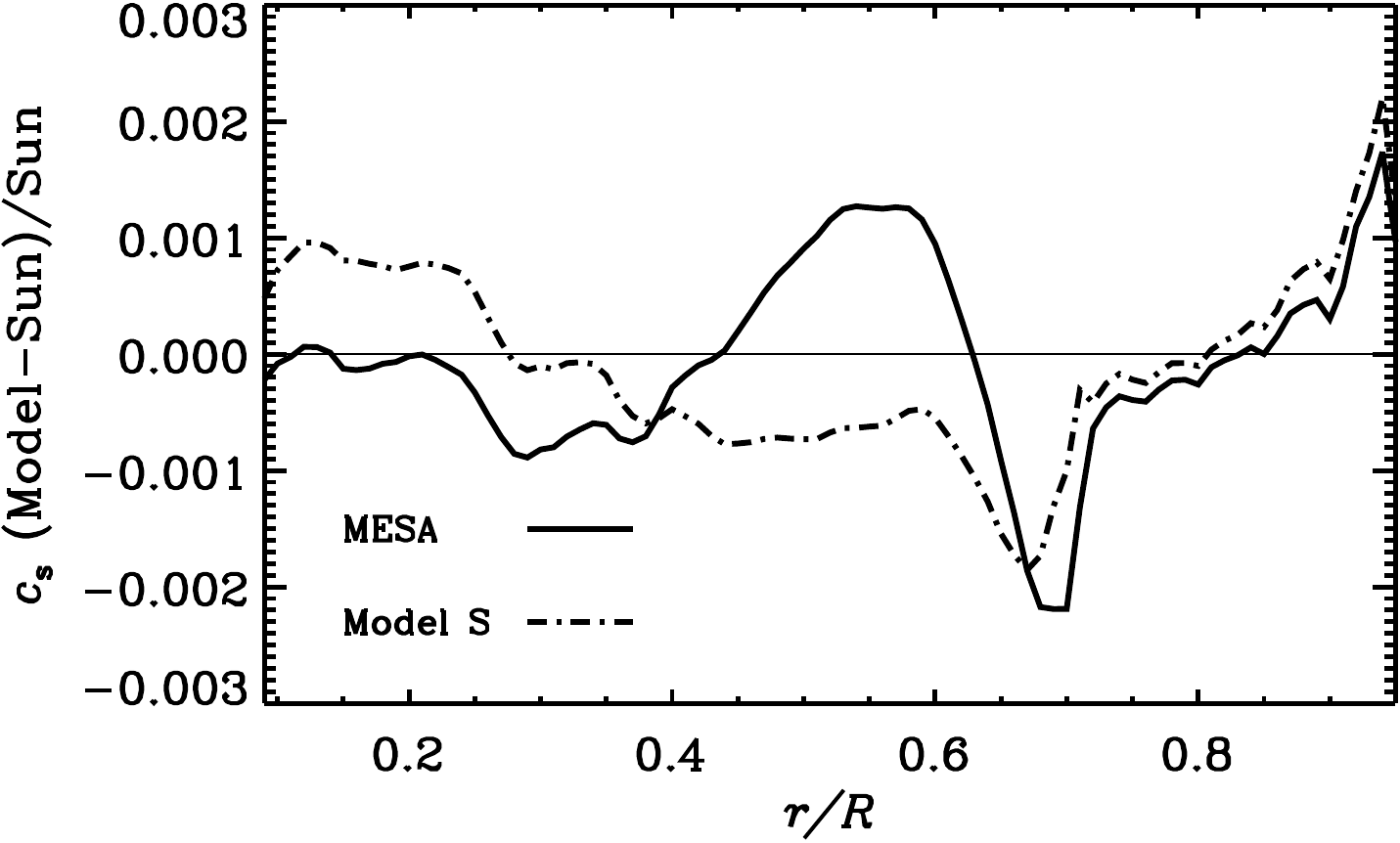}}
\caption{\label{astero:solar} 
Comparison of the
difference between the helioseismically-inferred sound speed profile
\citep{Bahcall98} of a \MESAstar\ model and Model
S \citep{ModelS}.}  
\end{figure}

The seismic properties of the Sun provide a test of stellar
evolution models, and an opportunity to calibrate $\alphaMLT$ for any particular set of 
input physics and other assumptions. The \MESAstar\ test case \code{solar\_calibration}
produces
a calibrated Standard Solar Model.
Figure \ref{astero:solar} shows the difference between the
helioseismically-inferred solar sound speed profile and this model. We also show ``Model S'' from \citet{ModelS}. Both models
employ comparable input physics and assume solar abundances from  \citet{GN93} and \citet{GS98}.
One clear improvement since \mesaone\ is a smoother 
sound speed profile at small $r/R$, which is primarily due to improvements in the \diffusion\ module. 
This is particularly important for 
asteroseismology, where sharp features in the sound speed profile can
influence the stellar oscillation frequencies. 
The results are based on the solar calibration test case compiled with the 
GNU Fortran compiler version 4.7.2 on Mac OS X 10.7.5; Appendix \ref{s.compiler_OS}
 provides information about 
how the solar calibration results may depend on different operating systems and 
compilers.

\subsection{New Asteroseismic Capabilities in MESA}\label{s.astero}

 The ``astero'' extension to \mesastar\ 
implements an integrated approach that passes results automatically
between \mesastar\ and the new \mesa\ module based on the adiabatic code ADIPLS \citep[][June 2011 release]{ADIPLS}.  The \MESA\ module \ADIPLS\ also supports independent use for post-processing, including the calculation of pulsation frequencies.

This astero extension enables calculation of selected pulsation frequencies by \MESAstar\ during the evolution of the model. This allows fitting to the observations
that can include spectroscopic constraints (e.g., [Fe/H] and \teff),
asteroseismic constraints, such as the large frequency separation
(\dnu) and the frequency of maximum power (\numax), and even
individual frequencies.
A variety of approaches for finding a best-fitting model are available,
including grid searches and automatic $\chi^2$ minimization by the
Hooke-Jeeves algorithm \citep{HookeJeeves61}
or by the
``Bound Optimization BY Quadratic Approximation'' \citep[BOBYQA;][]{powell_2009_aa}
technique.
These searches are user controlled through a number of parameter
bounds and step sizes.  Users also have full control over the relative
weight assigned to the seismic and spectroscopic parts of the $\chi^{2}$
statistic. 

For the automated $\chi^2$ minimization, astero will
evolve a pre-main sequence model from a user defined starting point,
and find the best match along that single evolutionary track. The code
then recalculates the track, again initiated at the pre-main
sequence, with different initial parameters such as mass, composition,
mixing length parameter and overshoot, and repeats 
until the lowest $\chi^2$ has been found.

Calculating specific mode frequencies is computationally intensive. Hence, a number of options exist to
improve the efficiency of the minimization when individual frequencies
are included.  Bounds can be established on stellar parameters (e.g.,
\teff, central H mass fraction, \dnu), so that \ADIPLS\ is invoked
only when the model falls within these bounds.  This enables certain
evolutionary stages to be skipped when other observational diagnostics
rule them out---if a star is known to be a red giant, for instance,
there is no sense in invoking \ADIPLS\ when models are on the main
sequence.  The large frequency separation, \dnu, of the model is
calculated as the inverse of the sound travel time through the star,
\dnu\ $=[2 \int \dif r/\cs]^{-1}$
\citep{Tassoul80,Gough86}.  There is also the option 
to derive \dnu\ using simple solar scaling: \dnu\ $\propto (M/R^3)^{0.5}$
\citep{KjeldsenBedding95}. To obtain
\numax, \MESA\ scales the solar value with the acoustic cut-off
frequency: \numax\ $\propto g/\sqrt{T_{\mathrm{eff}}}$ \citep{Brown91,KjeldsenBedding95}.

Moreover, hierarchical approaches to the frequency fitting can be
selected, saving large amounts of computational time.  In one case the
radial modes are first calculated, and only when they match reasonably
well are the non-radial mode frequencies derived and included in the
$\chi^{2}$. This is particularly beneficial for red giants where the
calculation of the non-radial frequencies is extremely time
consuming.  
Another example is when the time steps in the stellar evolution calculations are 
too large to find an accurate minimum of $\chi^{2}$.
Hence, as a further option to increase
efficiency while attaining accuracy, the time steps can be set to
automatically reduce when the model comes close to the ``target box'' of
the observational constraints. As for other modules used in \mesastar,
astero offers a range of graphical outputs including an \'echelle diagram
where the fitting process can be followed in real time. 

There is also an option for including corrections to the model
frequencies on-the-fly to compensate for the inadequate modelling of the near
surface layers of the star.  The effect, known as the ``surface
term,'' is seen as a frequency dependent offset between the modelled
and observed acoustic frequencies of the Sun
\citep[e.g.][]{Christensen-DalsgaardThomson97}.  The offset increases
towards higher frequencies and is well described by a power law
\citep{Kjeldsen08}.  
\mesastar\ follows the approach described by \citet{Kjeldsen08} for
correcting the surface term.

To illustrate the performance of
astero, we show here a fit to the star HD49385. The input frequencies
and the spectroscopic constraints are from \citet{Deheuvels10}.  We first
ran a wide-range grid search over $M$, $\alphaMLT$, [Fe/H], and $Y$, including only [Fe/H], \teff, and \dnu\ as observational constraints.  The results of this initial search
guided our starting parameters and ranges for the next automatic
$\chi^2$ minimization.  We first compare our grid results
with those of the RADIUS grid search routine \citep{Stello09}, which is
based on a grid of ASTEC models \citep{ASTEC} and find agreement within
uncertainties. 

We then include the individual oscillation frequencies and
use the Hooke-Jeeves algorithm for the $\chi^2$ minimization.
Model frequencies were corrected for the surface term, and the part of the
$\chi^2$ coming from the frequencies was
given 2/3 of the weight in the final $\chi^2$, similar to that
used by \citet{Metcalfe12}.   
To ensure we adequately sample the parameter space, we initiate the search at 
several initial values within a broad range. By starting the
search from multiple initial values, we aim to reduce the chance of
ending up in a local minimum, which could potentially provide unphysical
results, such as the spuriously low helium abundances reported by
\citet{Mathur12}.  Current developments in astero further seeks to
overcome such problems and improve the robustness of the results 
by including frequency ratios \citep{RoxburghVorontsov03,SilvaAguirre13} in the $\chi^2$
minimization. 

Each ``Hooke'' search generates several stellar evolution tracks, each with a best $\chi^2$ value. We
then combine the data from about 1400 tracks  to estimate the
1-$\sigma$ uncertainties in the varied parameters following the approach by
\citet{Deheuvels10}.  
The lowest (reduced) $\chi^2$ value we obtained was 2.4 with a few tens of
models in the 2.4--4.0 range, which all fit the
frequencies similarly well.   
Among these models there are two families of results, one of
which has slightly lower [Fe/H] and $Y$, and a slightly increased value for the spectroscopic part of the
$\chi^2$.  

\begin{figure}[htbp]
\centering{\includegraphics[scale=0.8]{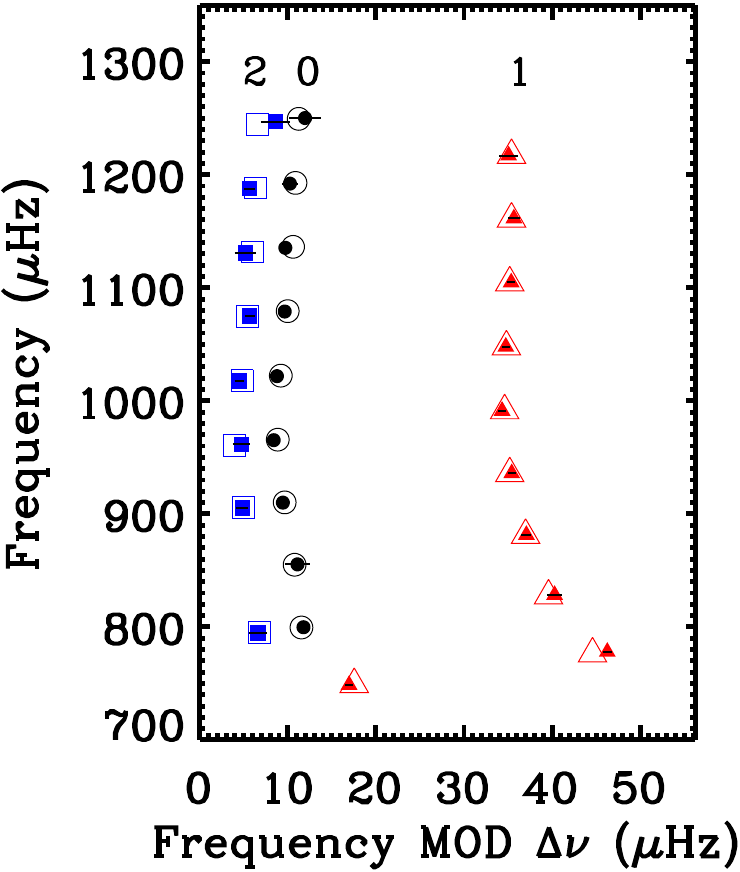}}
\caption{\label{astero:echelle} 
\'Echelle diagram of the oscillation frequencies of the
  subgiant HD49385. Observed frequencies are shown with filled symbols as
  blue squares ($\ell=2$), black circles ($\ell=0$), and red triangles ($\ell=1$),
  and the matched model frequencies are shown with open symbols.
  Black horizontal lines indicates 1-$\sigma$ error bars.}
\end{figure}

\begin{figure}[htbp]
\centering{\includegraphics[width=0.8\hsize]{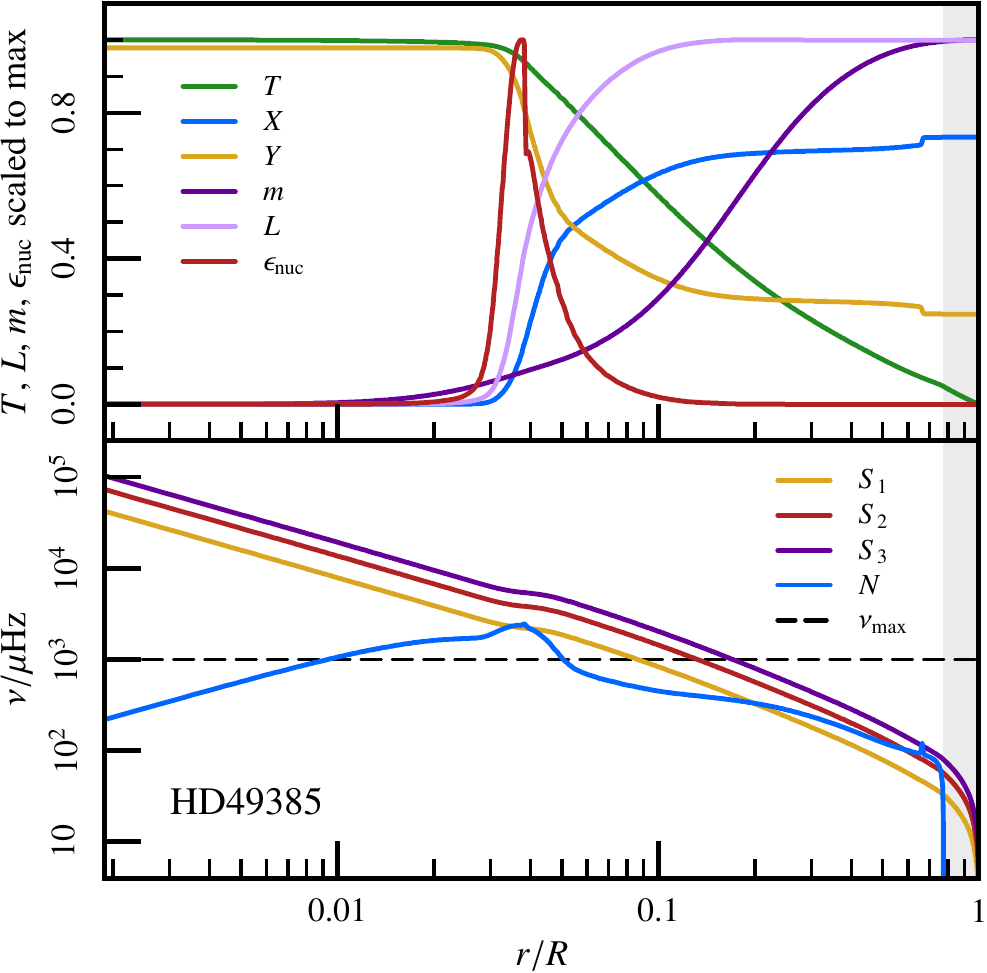}}
\caption{\label{astero:hd49385profile} 
Same format as in
Figure~\ref{astero:profiles}, but for the best-fitting model of HD49385
(see also Table~\ref{t.astero:hd49385}).
}   
\end{figure}

The comparison of the observed and modeled frequencies for the realization with
the lowest $\chi^2$ is shown in the
\'echelle diagram format in Figure~\ref{astero:echelle}.  A plot of the
internal structure including the \bvv\ and Lamb frequencies is shown in
Figure~\ref{astero:hd49385profile}, and the parameters of
the model are listed in Table~\ref{t.astero:hd49385}.  We set $\fov=0.015$
and use the GN98 solar abundances.
Our results can be best compared to those
listed as ``low $\alpha_{\mathrm{ov}}$'' and ``GN93''
in Table 4 of \citet{DeheuvelsMichel11} and agree within the uncertainties.

\begin{deluxetable}{rr}
\tablecolumns{2}
\tablewidth{0pc}
\tablecaption{Properties of best fitting model to HD49385\label{t.astero:hd49385}}
\tablehead{\colhead{Quantity}&\colhead{Value}}
\startdata
$M/\Msun$ & $1.30\pm0.04$\\
$R/\Rsun$ & $1.972\pm0.016$\\
$L/\Lsun$ & $4.9\pm0.4$\\
$\log g$  & $3.962\pm0.003$ \\
$\Teff$/K & $6115\pm125$\\
Age/Gyr   & $4.1\pm0.4$\\
$\alpha_{\mathrm{MLT}}$ & $1.9\pm0.1$\\
$[\mathrm{Fe/H}]_i$     &  $0.15\pm0.04$\\ 
$[\mathrm{Fe/H}]_s$\tablenotemark{a}     &  $0.063$\\
$Y_\mathrm{initial}$    & $0.29\pm0.02$  \\ 
$Z_\mathrm{initial}$    & $0.0222$ \\
$\chi^2$  & $2.40$ \\
\enddata
\tablenotetext{a}{
$[\mathrm{Fe/H}]_s$ is the
log of the ratio of the surface ($Z$/$X$) relative to the solar value of
0.02293.} 
\end{deluxetable}

\subsection{The Effect of Composition Gradients on the Brunt-V\"ais\"al\"a Frequency} \label{s.ledoux}

Including the effect of composition gradients in the
  calculation of the \bvv\ frequency is important for two reasons.
  First, it is necessary for implementing the Ledoux criterion for
  convection, which is used to determine the chemical mixing and
  convective heat transport in a region (see
  \S~\ref{s.semiconvection}).  Second, a smooth and accurate
  method for calculating $N^{2}$ is crucial for studies of g-mode
  pulsation in stars.  In a highly degenerate environment, the
pressure is nearly independent of temperature, and $P \propto
\rho^{\Gammaone}$, so from eq.~(\ref{N1}) we see that $N^2$ depends on
the difference of two large and nearly equal quantities. This can lead
to a loss of precision and a noisy $N^2$. To eliminate this problem,
$N^{2}$ is re-written into a form that depends on the difference of
the adiabatic and true temperature gradients and on the composition
gradient:
\begin{equation}
  N^2 = \frac{g^2\rho}{P} \frac{\chiT}{\chirho} \left(\nablaad-\nablaT + B\right).
  \label{bled}
\end{equation}
The term $B$ explicitly takes into account the effect of composition
gradients and is commonly called the Ledoux term
\citep[e.g.,][]{Unno89,Brassard91}.  For the general case of an
$N$-component plasma with mass fractions $\{X_i\}$, the standard
formula for $B$ is \citep[e.g.,][]{Unno89}
\begin{equation}
  B = -\frac{1}{\chiT} \sum_{i=1}^{N-1} \ppll{P}{X_i}\ddl{X_i}{P}.
  \label{bsum}
\end{equation}
Since $\sum_{i=1}^N X_i = 1$, one of the
mass fractions can be eliminated, so that the sum in
eq.~(\ref{bsum}) runs from 1 to $N-1$. We note that the partial
derivatives in eq.~(\ref{bsum}) hold all the \{$X_j$\} constant except
for $X_i$ and $X_N$, where $X_N$ is varied so as to maintain
$\sum_{i=1}^N X_i = 1$.  

\begin{figure}[htbp]
\centering{
  \includegraphics[width=\figwidth]{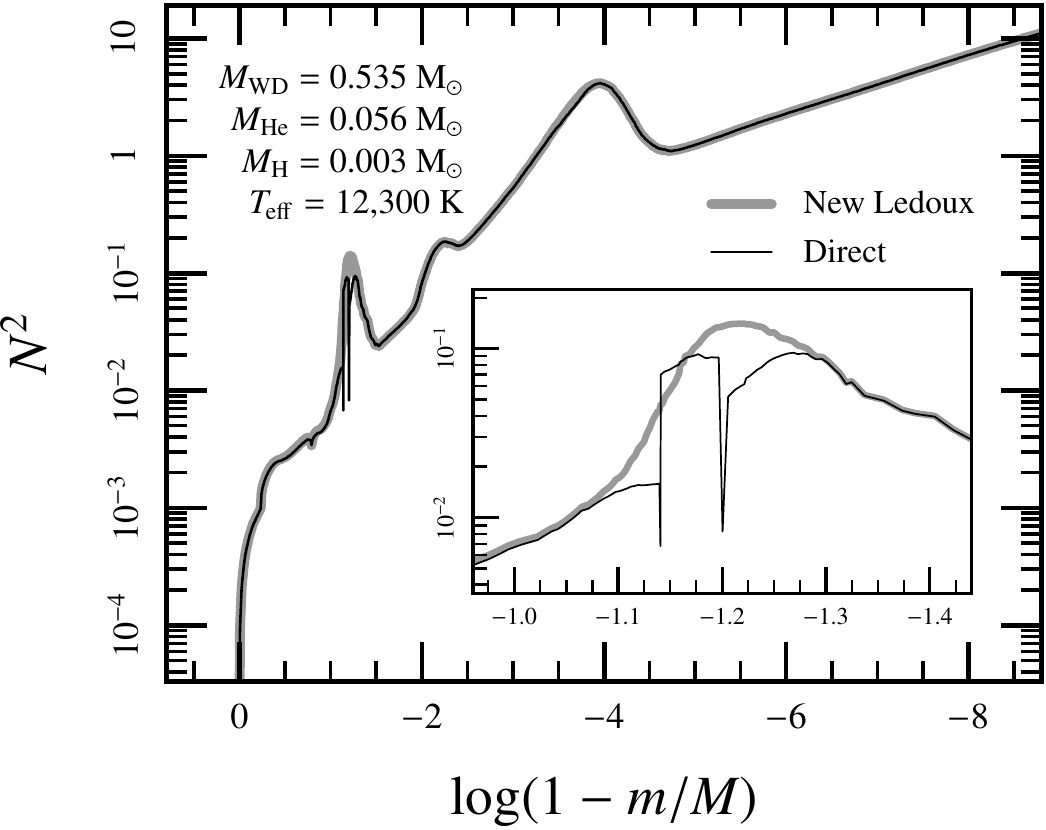}
}
\caption{\label{BVdiff} 
  A comparison of the new Ledoux prescription for $N^2$
  (eqs.~\ref{bled} and \ref{new}) versus the direct numerical
  calculation (eq.~\ref{N1}). This calculation is for a $0.535
  \,\Msun$ white dwarf model at $\Teff = 12,300\,\Kelvin$.  }
\end{figure}

Although eq.~(\ref{bsum}) is correct as written, we have developed a
new, formally-equivalent prescription that is both
numerically robust and simpler to implement. We define a new Ledoux
term by taking a directional derivative along the radial composition gradient in
the stellar model,
\begin{equation}
 B \equiv -\frac{1}{\chiT} \lim_{\delta \ln P \to 0} \frac{
   \ln P(\rho,T,\vec{X}+ (\dif \vec{X}/\dif \ln P)\,\delta \ln P) -
   \ln P(\rho,T,\vec{X}) }{\delta \ln P}.
 \label{bdef}
\end{equation}
The implementation of the above derivative typically involves the use
of quantities on neighboring mesh points. Using the subscript $k$ to
denote the value of a given quantity on the $k$th mesh point, we
therefore have
\begin{equation}
 B = -\frac{1}{\chiT} \frac{
   \ln P(\rho_k,T_k,\vec{X}_{k+1}) -
   \ln P(\rho_k,T_k,\vec{X}_{k})}{\ln P_{k+1} - \ln P_k}.
 \label{new}
\end{equation}
This is the form of the Ledoux term that is implemented in \mesa\ and
we term it the ``New Ledoux'' formulation. Since \mesa\ ensures that
$\sum_{i=1}^N X_i = 1$ at each mesh point, this condition does not
have to be separately enforced. This formulation requires just one
numerical difference along $\vec{X}$ that is consistent with the
stellar model and equation of state.
Because \mesa's EOS does not directly supply the partial derivatives
required for the formulation in eq.~\ref{bsum}, an implementation of
that method would suffer in both accuracy and efficiency from having
to do a large number of numerical differences. \citet{Brassard91}
dealt with a similar problem by using a restricted form of
eq.~\ref{bsum} that included only the helium composition
gradient. They showed that for cases where their restricted form
applied, it gave significantly better numerical results than an
implementation of eq.~\ref{N1} based on finite differences.
Figure~\ref{BVdiff} shows that our New Ledoux prescription (grey heavy
curve) retains their good results compared to eq.~\ref{N1} (thin black curve)
while extending the applicability to cases that cannot be dealt with
using only helium gradients.

\section{Mixing Mechanisms Involving Composition
  Gradients}\label{s.mixing}

We described the implementation of mixing-length theory (MLT) in
\mesaone, including the allowance for overshoot beyond the boundaries
of the convective zones as determined by the standard Schwarzschild
condition, $\nablarad>\nablaad$. Overshooting is
implemented via an exponential decay of the
convective diffusion coefficient beyond the boundary of full
convection, following \citet{2000A&A...360..952H}:
\begin{equation}\label{eq:overshoot}
\Dov = D_{\rm conv, 0}\exp\left(-\frac{2\Delta r}{\fov\,\scaleheight}\right),
\end{equation}  where $D_{\rm conv,0}$ is the diffusion coefficient at the convective border, 
$\Delta r$ is the distance from the start of overshoot, and $\scaleheight$ is the local pressure scale
height. The user-adjusted dimensionless parameter $\fov$ then determines the
extent of the overshooting region.  \MESA\ also allows for the
adoption of a step-function overshooting model, where the mixing
region extends a distance $\fov\scaleheight$ beyond the convective
boundary with a constant specified diffusion coefficient.

In \mesaone\, we did not implement the influence of composition gradients
on mixing and the resulting diffusion coefficients when
instabilities are operative. The description of how \MESAstar\ calculates
the Ledoux criterion is in \S \ref{s.ledoux}. In this section, we
describe the implementation of mixing due to composition
gradients in stellar interiors. 

We refer to \S\ref{s.nuts-free-parameters} 
for a discussion of the free parameters involved in the implementation of these mixing mechanisms.

\subsection{Semiconvection}\label{s.semiconvection}

Semiconvection refers to mixing in regions unstable to
Schwarzschild but stable to Ledoux, that is
\begin{equation}\label{eq:ledoux_stable}
	\nablaad<\nablaT<\nablaL,
\end{equation}
where $\nablaL$ is the sum of the adiabatic gradient and the Brunt
composition gradient term (see eqs.~[\ref{bled}] and
  [\ref{new}]),
\begin{equation}
	\nablaL\equiv\nablaad+B.
\end{equation}
Once $\nablaL$ is calculated, regions satisfying equation
(\ref{eq:ledoux_stable}) undergo mixing via a time-dependent diffusive
process with a diffusion coefficient calculated by the \mlt\, module
following \citet{1983A&A...126..207L},
\begin{equation}\label{eq:semi_diffusion_coeff}
	D_{sc}=\alphasc\left(\frac{K}{6\CP\rho}\right)\frac{\nablaT-\nablaad}{\nablaL-\nablaT},
\end{equation}
where $K=4acT^3/(3\kappa\rho)$ is the radiative conductivity, \CP\ is
the specific heat at constant pressure, and \alphasc\ a dimensionless efficiency parameter.
See \S\ref{s.nuts-free-parameters} for a discussion of the range of values for $\alphasc$.

\begin{figure}[htbp]
\centering
\includegraphics[width=\figwidth]{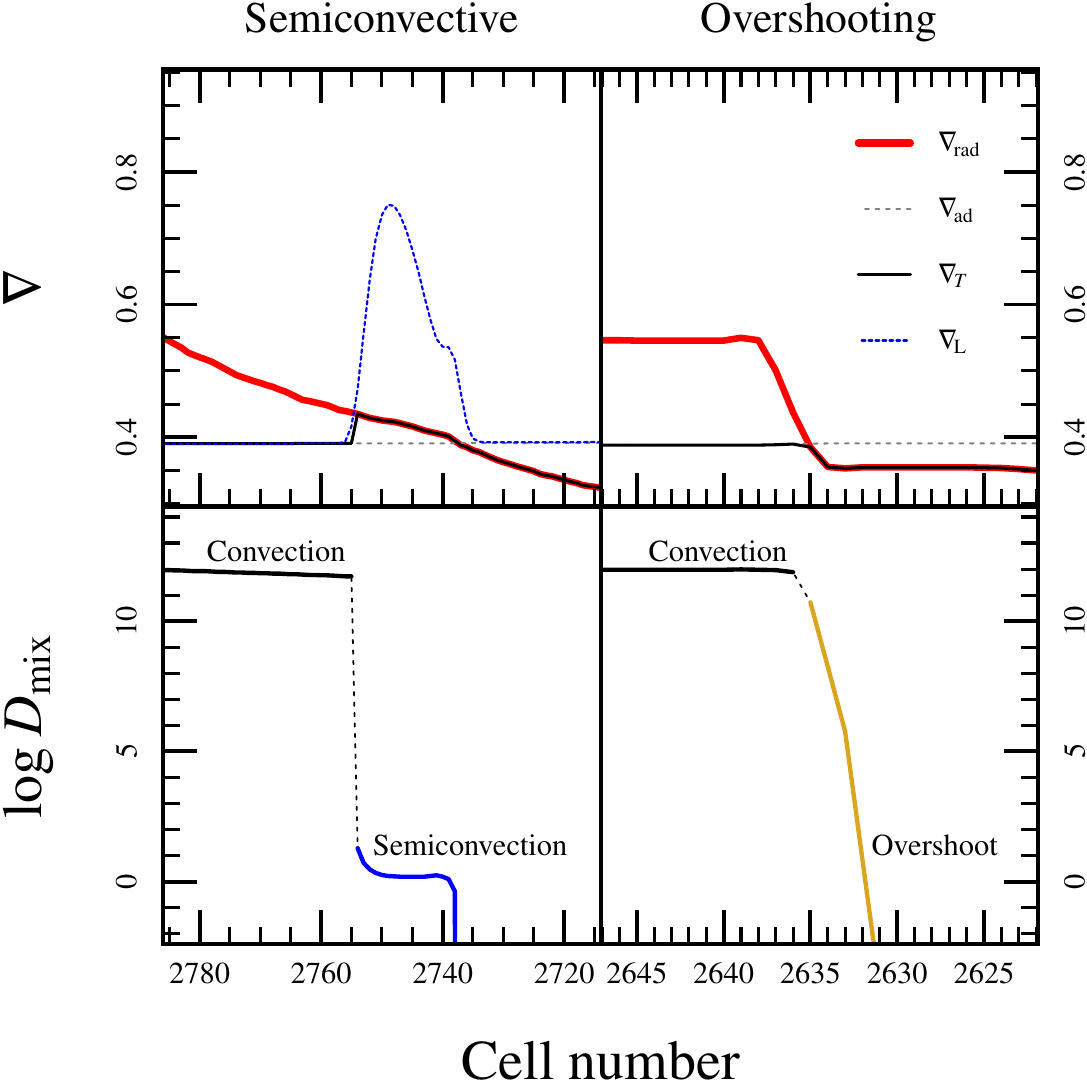}
\caption{\label{3M_grad_dmix}
 Sample profiles of semiconvective (left) and exponentially overshooting (right)
 $3\Msun$ models undergoing core helium burning.  Top panels show
 the radiative, adiabatic, temperature, and Ledoux gradients that
 determine mixing boundaries and diffusion coefficients.
 Bottom panels show the resulting diffusion
 coefficients for energy and chemical transport. In either case, a
 thin dotted line spanning a single intermediate cell joins the
 convective and semiconvective/overshoot curves.  This is intended
 merely as a guide for the eye, as diffusion coefficients are  defined
 only at the two boundaries of a cell.  In particular, diffusion for
 this intermediate cell is governed by convection at its interior
 boundary and semiconvection/overshoot at the exterior.  The
 semiconvective model shown here was run with $\alphasc=0.01$; the
 exponentially overshooting model with $\fov=10^{-5}$.  The profiles are taken at
 the points marked in Figure \ref{3M_he_burn_mixing}.}
\end{figure}

We stress that semiconvection and overshooting have distinct
implementations in \MESA.  Both are time-dependent diffusive
processes. As an example, in Figure \ref{3M_grad_dmix} we display
profiles of thermodynamic gradients and their resulting diffusion
coefficients during core helium burning in a semiconvective model with
$\alphasc=0.01$ and in an exponentially overshooting model with
$\fov=10^{-5}$.

\subsection{Thermohaline Mixing}\label{s.thermohaline}

\begin{figure*}[htbp]
\centering{
\includegraphics[angle=0,width=\twoupwidth]{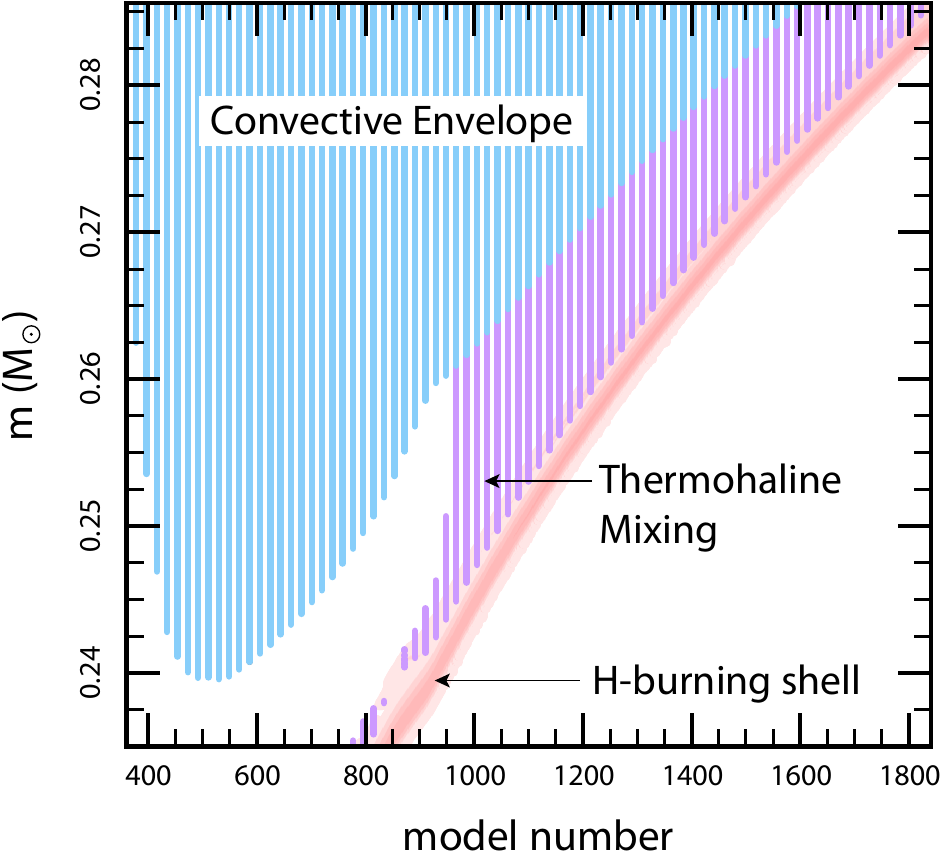}\hspace{\twoupsep}
\includegraphics[angle=0,width=\twoupwidth]{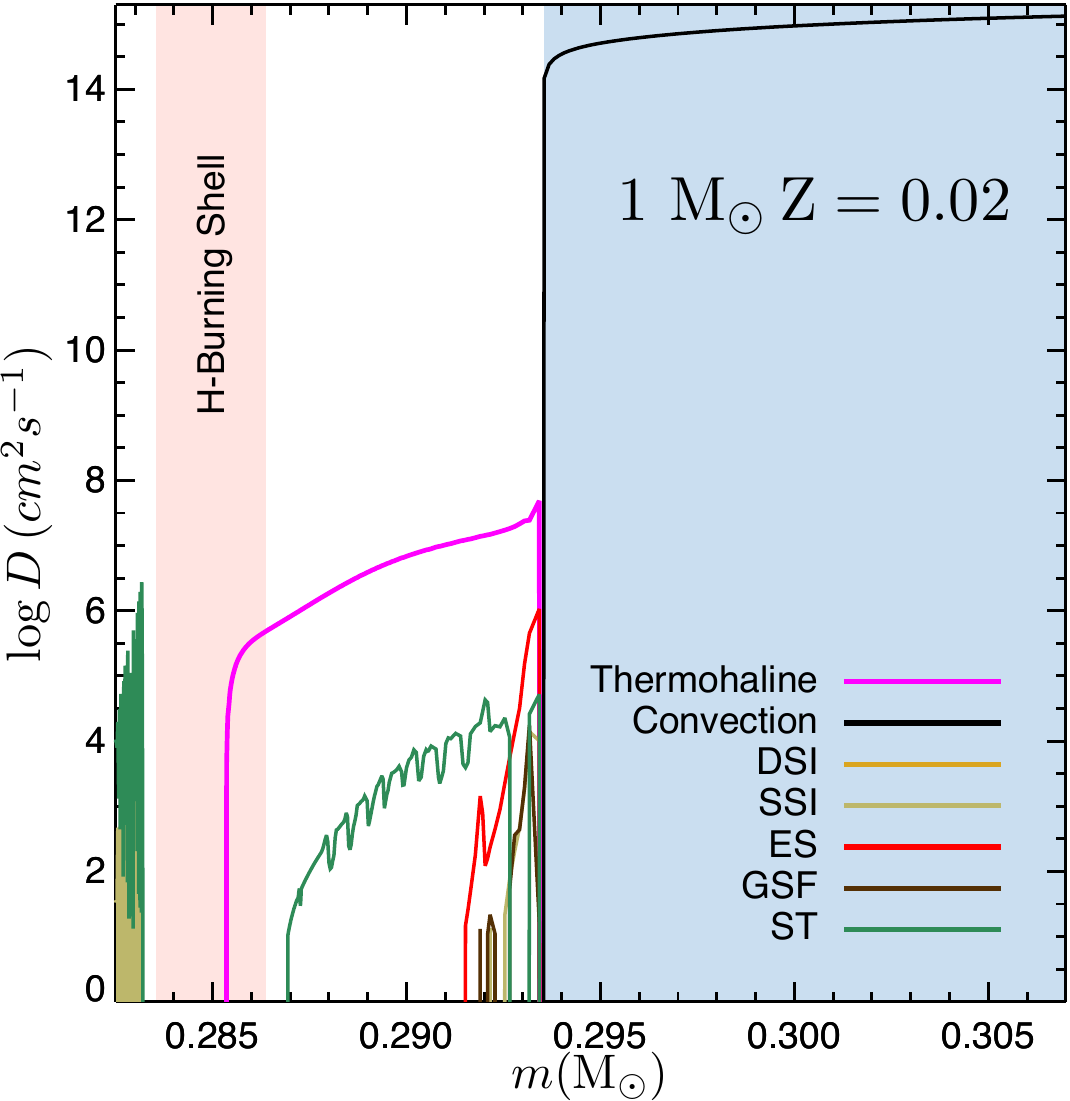}
}
\caption{Thermohaline mixing during the RGB phase of a $Z=0.02$, 1\,\Msun\ model, initially rotating with an equatorial velocity of $10\,\kms$ and adopting  
$\alphath = 2$.
In the left panel a Kippenhahn diagram shows, in mass coordinate and as function of model number, the locations 
of the retreating convective envelope (blue), of the H-burning shell (red) and of the thermohaline mixing region (magenta).
The right panel shows diffusion coefficient profiles extracted at model number 1849, which is the last model shown in the Kippenhahn plot.  
The H-burning shell and the convective envelope are shaded in red and blue, respectively. 
Thermohaline mixing (magenta line) transports chemicals between the burning shell
and the convective envelope. Also shown are the diffusion coefficients
resulting from Eddington-Sweet circulation (ES), 
magnetic torques by dynamo generated fields (ST), Dynamical Shear (DSI), Secular Shear (SSI) and Goldreich-Schubert-Fricke (GSF) instability (see \S \ref{s.rotation} 
for details). 
 \label{f.thermohaline} }
\end{figure*}

Thermohaline mixing arises in the presence of an inversion of the mean molecular
weight in regions that are formally stable against convection
according to the Ledoux criterion,
\begin{equation}
\nablaT - \nablaad \le B \le 0,
\end{equation}
In \MESA\ thermohaline mixing is treated in a diffusion
approximation, with a diffusion coefficient motivated by the 
linear stability analysis of \citet{Ulrich:1972} and
\citet{Kippenhahn:1980}
\begin{equation}
  \Dth = \alphath\; \frac{3K}{2\,\rho\, \CP}\,
  \frac{B}{(\nablaT-\nablaad)}. \label{coefficient}
\end{equation}
The quantity \alphath\ is a dimensionless efficiency parameter. 
In the linear analysis it depends on the aspect ratio of the blobs/fingers arising from the instability.  In the case of salt fingers
such a value is calibrated using laboratory experiments in water
\citep[e.g.][]{Krishnamurti:2003}, where the fingers have an
aspect ratio of $\approx 5$.  In the stellar case the value of this
parameter is vexatious
\citep[e.g.][]{Charbonnel:2007,Denissenkov:2008,cantiello_2010_aa,Wachlin:2011},
with recent 2D and 3D hydrodynamical calculations pointing toward a
much reduced value of \alphath\ relative to the salt
fingers case \citep{denissenkov_2010_aa,Traxler:2011,Brown:2013}.  Figure
\ref{f.thermohaline} shows a calculation including the effects of
thermohaline mixing  during the RGB phase of a $1\Msun$ star after
the luminosity bump \citep[e.g.][]{Charbonnel:2007,cantiello_2010_aa}. For this calculation
a value $\alphath = 2$ has been adopted, 
but see \S\ref{s.nuts-free-parameters} for a discussion of the range of options.

\subsection{Impact of Mixing on Convective Core Hydrogen and Helium Burning}\label{s.mixing-core-burning}

The duration of the hydrogen and helium
core burning depends on the extent of the convective
core, so we focus here on exhibiting the \MESA\ capabilities during
these phases. As we noted above, there are many physical effects that
change the size of the convective core, such as semiconvection,
overshooting, and rotation-induced mixing.  For example, the
Schwarzschild criterion implies larger cores than the
Ledoux criterion, but when using Ledoux alone, the region above the
convective boundary is overstable and so semiconvection occurs (see \S 
\ref{s.semiconvection}).

\begin{figure}[htbp]
\centering
\includegraphics[width=\figwidth]{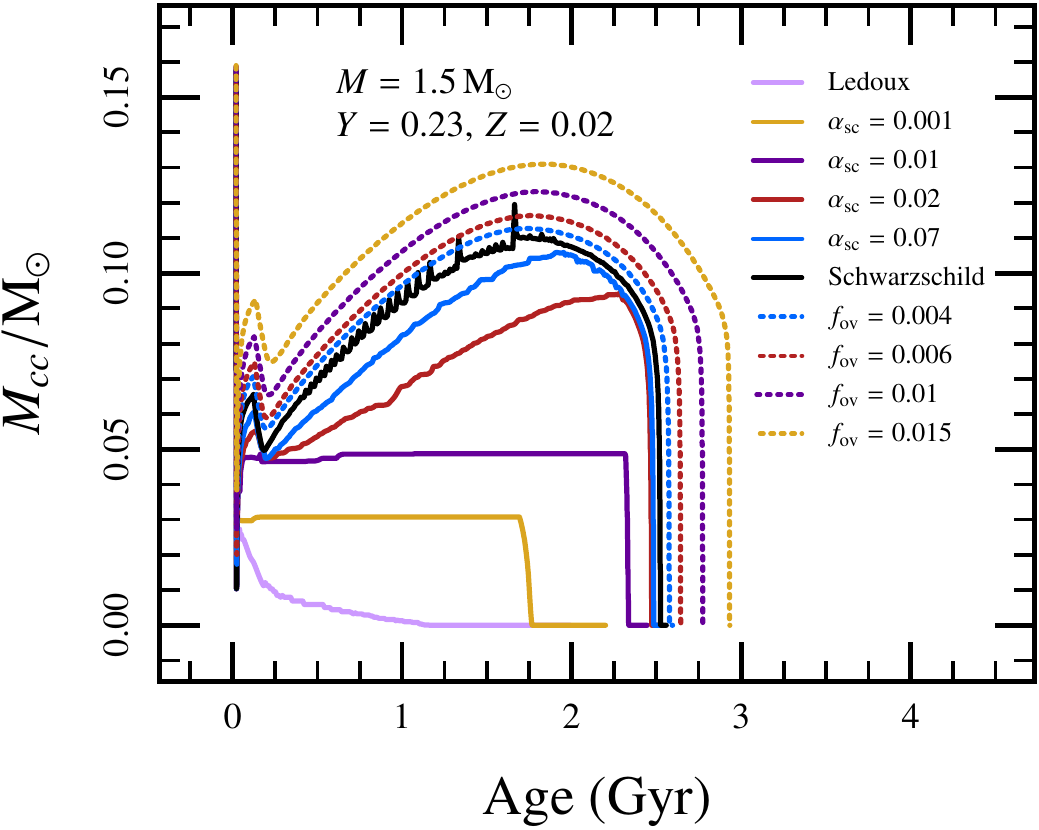}
\caption{\label{1p5M_main_sequence_Mcc}
History of convective core extent during the
main sequence for a non-rotating $1.5\,\Msun$ star with various mixing options.  The plot shows the boundary of convection not including the extent of semiconvection or overshooting.}
\end{figure}

\begin{figure}[htbp]
\centering
\includegraphics[width=\figwidth]{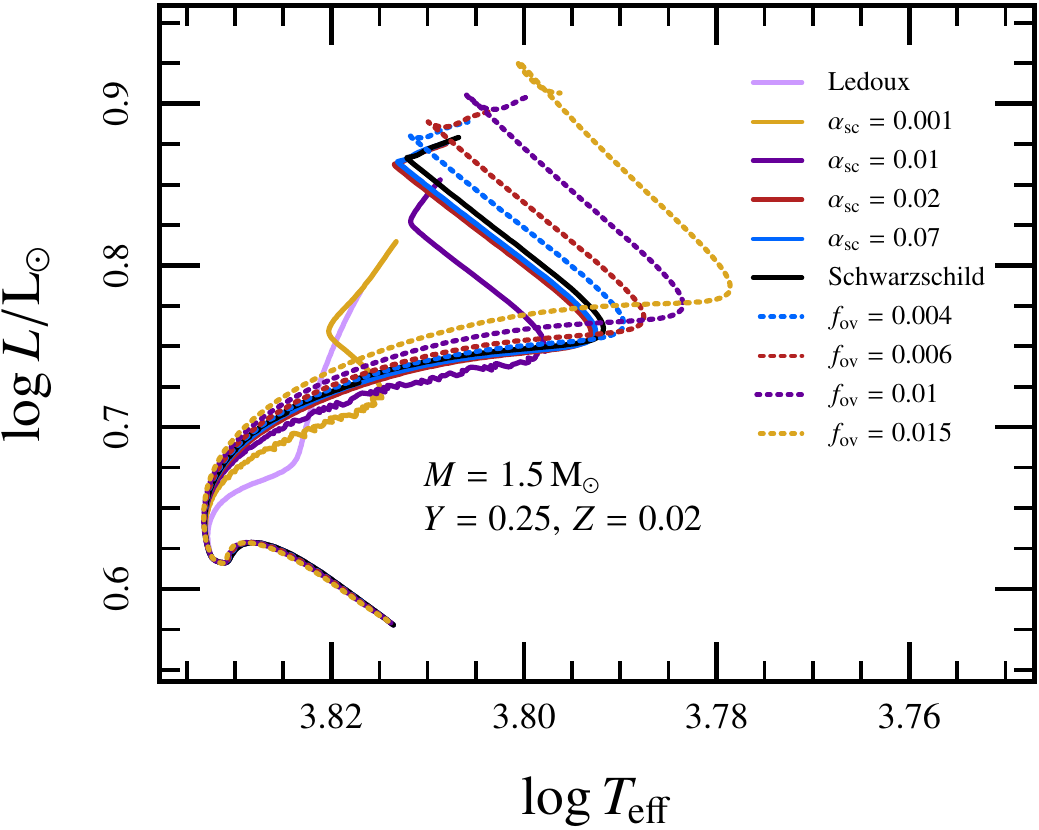}
\caption{\label{1p5M_main_sequence_hrd}
The HR diagram for the non-rotating $1.5\,\Msun$ star with various mixing options.  Tracks are displayed from ZAMS until depletion of core hydrogen to $X=10^{-5}$.
}
\end{figure}

We evolved a non-rotating $1.5\,\Msun$ star with $(Y,Z)=(0.23,0.02)$ through
central hydrogen burning using Ledoux, Ledoux
plus semiconvection, Schwarzschild, and Schwarzschild plus overshoot. As is evident in Figure
\ref{1p5M_main_sequence_Mcc}, this set of physical processes leads to a
large range of convective core masses and thereby main
sequence lifetimes. For the parameters explored we found that
overshooting increases the lifetime by a factor
$\lesssim$1.2 for Schwarzschild and $\lesssim$2.5 for Ledoux. Figure \ref{1p5M_main_sequence_hrd} shows an HR diagram for each
of the $1.5\,\Msun$ models undergoing core hydrogen burning, showing
the impact of convective core extent on main-sequence turnoff
morphology.

We also evolved a non-rotating $3\,\Msun$ star with $(Y,Z)=(0.25,0.02)$ through
central helium burning.   Overshooting extends the burning
lifetime by a factor $\la$1.6 for Schwarzschild  and $\la$2.8 for Ledoux.  Although this lengthening of the
core burning phase is always true of convective overshoot, we find
that the extension of the overshoot and convective regions is sensitive to the temporal resolution adopted. With sufficiently large values of
\fov\ the upper boundary develops oscillatory behavior which can also affect the lifetime.
This behavior also occurs with the
step-function implementation of overshoot.  This instability is not seen in overshoot during hydrogen burning and has yet to be studied in detail.

\begin{figure}[htbp]
\centering
\includegraphics[width=\figwidth]{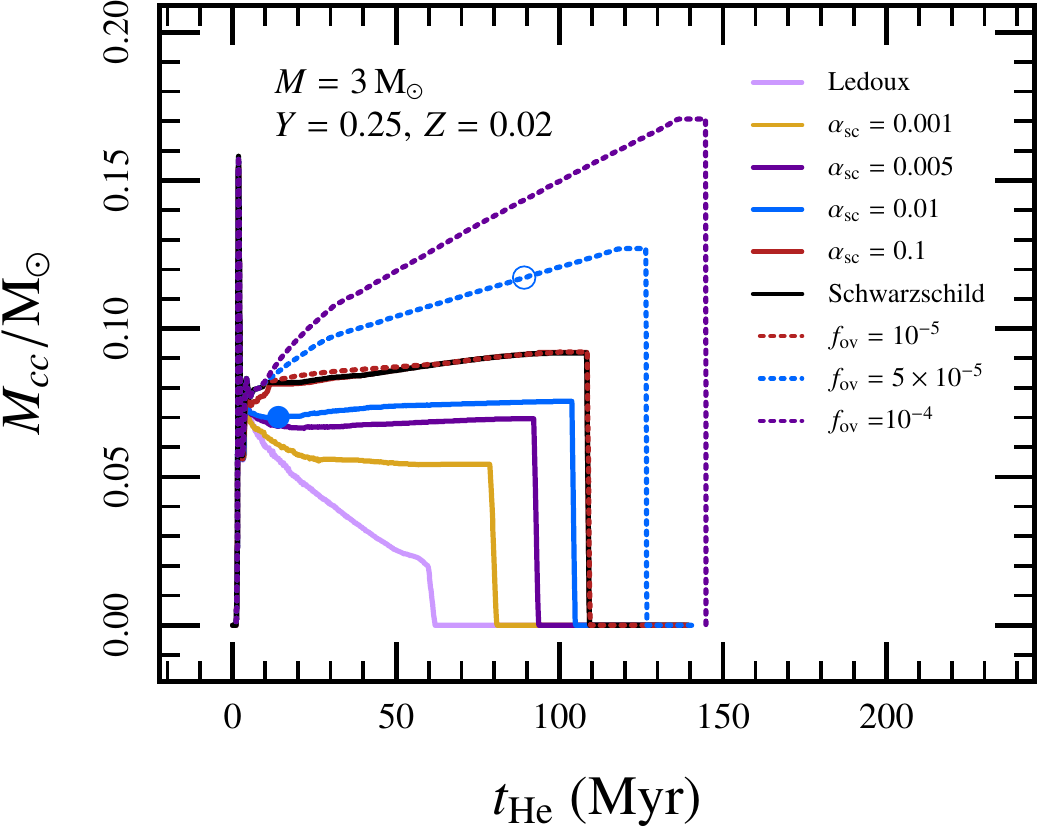}
\caption{\label{3M_he_burn_mixing}
History of convective core extent during the
core helium burning phase for a non-rotating $3\,\Msun$ star with various mixing options,   as in Figure~\ref{1p5M_main_sequence_Mcc}. Time is measured relative to the onset of the convective core burning.
Efficient semiconvection ($\alphasc=0.01$) and inefficient overshooting ($\fov=10^{-5}$) coincide with the pure Schwarzschild model.  The filled (open) circle indicates the time for which we display a profile detailing semiconvection (overshooting) in Figure \ref{3M_grad_dmix}.
}
\end{figure}

\section{Evolution beyond the Main Sequence and White Dwarfs}\label{s.agb-wd}

Extending the verification of \mesaone, we now compare to other available codes for intermediate-mass stars, $3\textrm{--}8\nsp\Msun$.  We describe the techniques used by \MESAstar\ to evolve stars through the AGB phase to the white dwarf cooling sequence.
We also demonstrate
how \MESAstar\ incorporates compressional heating from accretion.

\subsection{Code Comparisons during Helium Core Burning}\label{s.core-he}

We start by comparing the results of \MESAstar\ to those from the
Dartmouth Stellar Evolution Program (\DSEP; \citealt{2008ApJS..178...89D}) for stars with
$M = 3\textrm{--}8\,\Msun$. In both cases, the models were
evolved from the pre-main sequence to the depletion of helium in their
cores. For completeness, the \MESAstar\ models were further evolved 
 to the occurrence of the first helium thermal pulse.

All models have an initial composition $Y=0.272$, $Z=0.02$, and no
mass loss or rotation was included. The boundaries of mixing zones are
determined by the Schwarzschild criterion with
$\alphaMLT=2$. In order to compare the codes, we do not
allow overshooting or semiconvection. We adopt the \citet{Kunz:2002}
rate for \carbon[12]$(\alpha,\photon)$\oxygen[16]
 and the \citet{Imbriani:2004}
rate for \nitrogen[14]$(\proton,\photon)$\oxygen[15]; for all other rates we use the
NACRE compilation \citep{Angulo:1999}. We use the OPAL Type 2
opacity tables (\citealt{1993ApJ...412..752I}) to account for the
carbon- and oxygen-enhanced opacities during helium burning.

\begin{figure}[htbp]
\centering
\includegraphics[width=\figwidth]{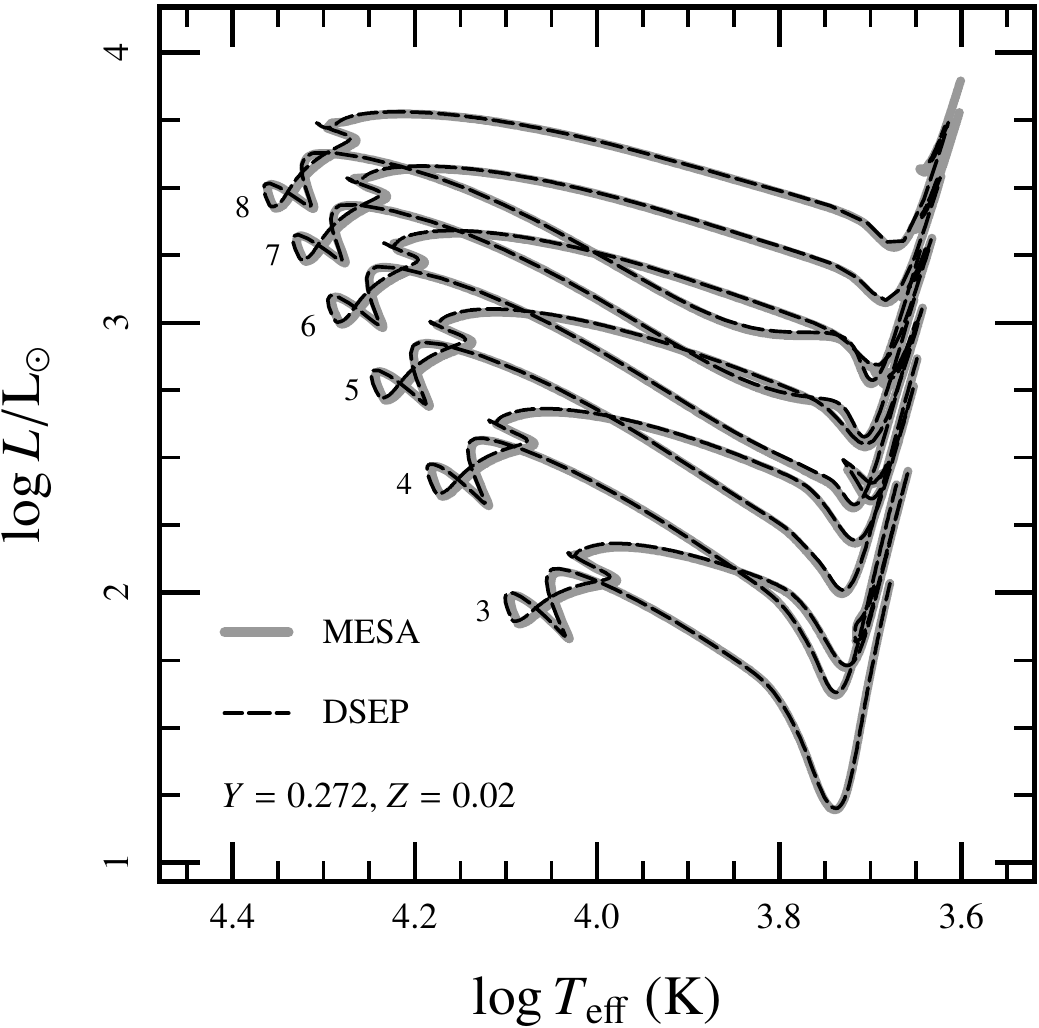}
\caption{\label{fig:mesa_vs_dsep_HRD}
Hertzsprung-Russell
 diagram for evolution of $3\textrm{--}8\,\Msun$ stars from the pre-main sequence
 through core helium depletion. Models are from \MESA\  (thick grey lines)
and \DSEP\ (dashed black lines).  Each curve is labeled with its corresponding initial mass in solar units.}
\end{figure}

\begin{figure}[htbp]
\centering
\includegraphics[width=\figwidth]{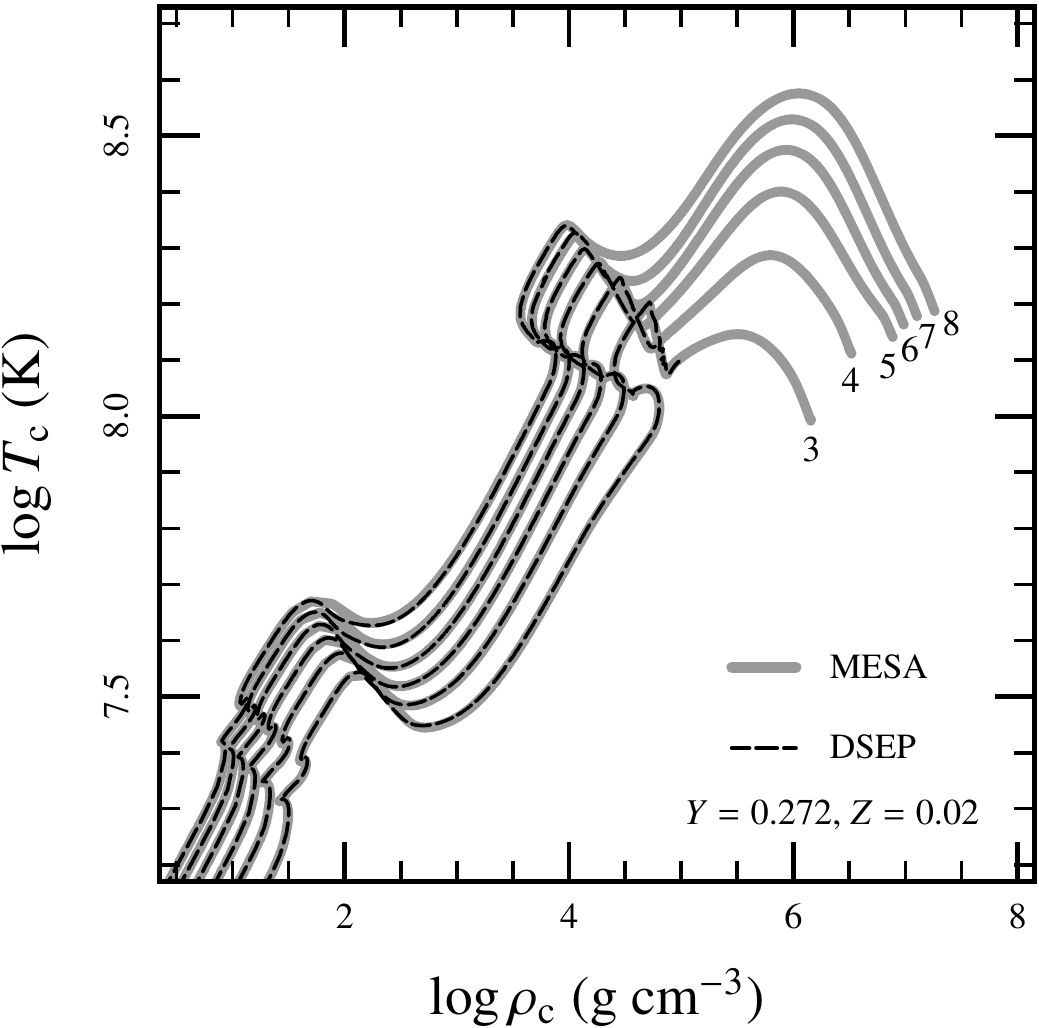}
\caption{\label{fig:mesa_vs_dsep_tc-rhoc}
Same as
  Fig. \ref{fig:mesa_vs_dsep_HRD}, but in the \Tc-\rhoc\ plane.  The
  \MESA\  models (thick grey lines) are evolved until the occurrence
  of the first thermal pulse.}
\end{figure}

\begin{figure}[htbp]
\centering
\includegraphics[width=\figwidth]{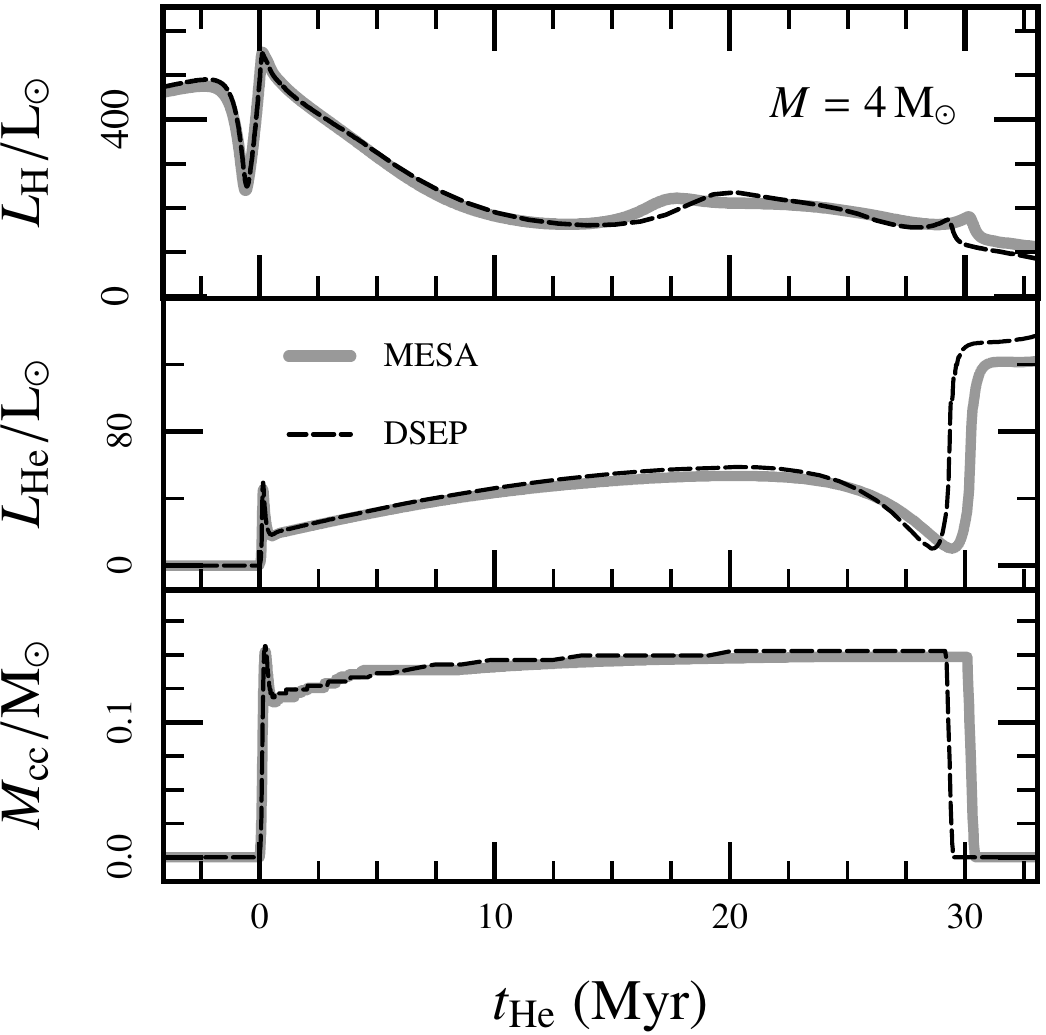}
\caption{\label{fig:mesa_vs_dsep_4M}
History of hydrogen burning
luminosity (top), helium-burning luminosity (center), and convective core extent (bottom) during the
core helium burning phase for the $4\,\Msun$ models. Time is measured
relative to the onset of the convective core.}
\end{figure}

\begin{figure}[htbp]
\centering
\includegraphics[width=\figwidth]{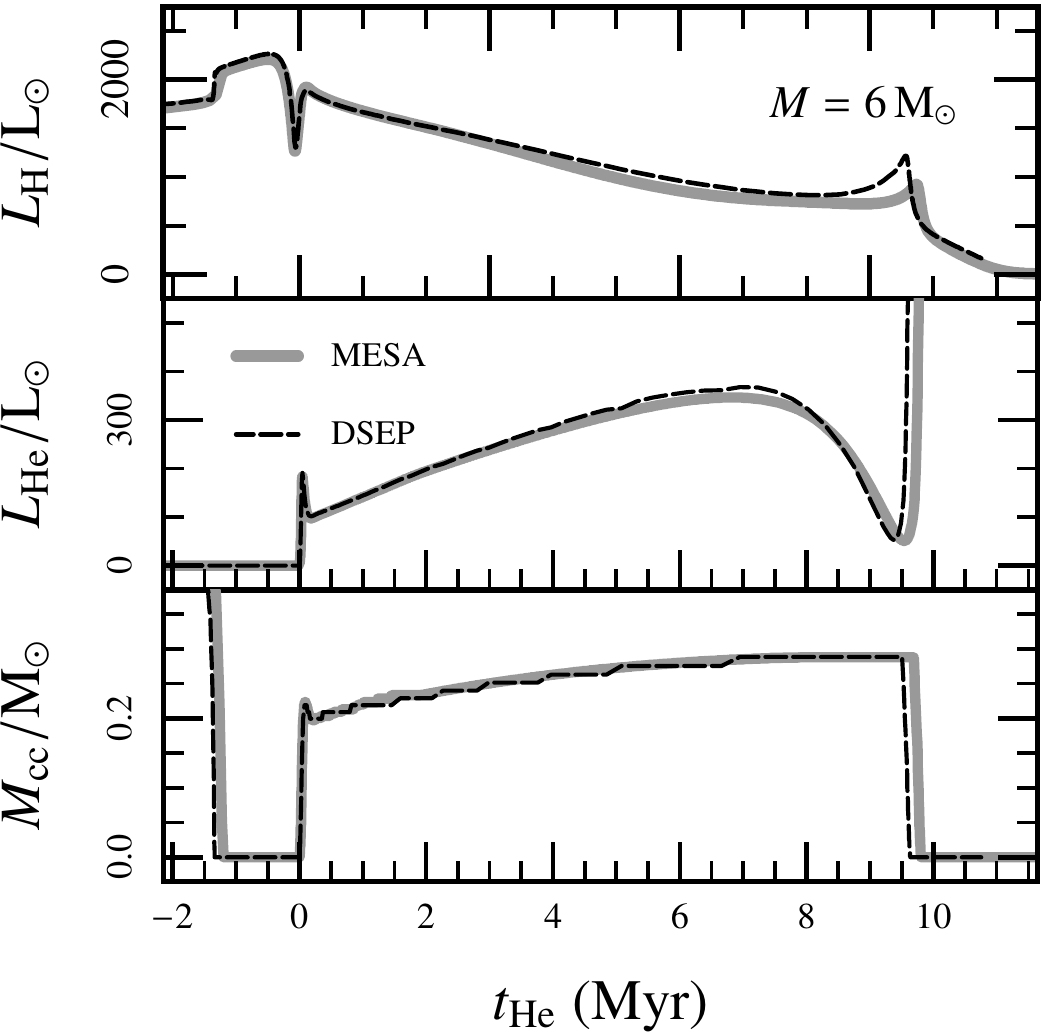}
\caption{\label{fig:mesa_vs_dsep_6M}
History of hydrogen burning
  luminosity (top), helium-burning luminosity (center), and
  convective core extent (bottom) during the
core helium burning phase for the $6\,\Msun$ models.  Time is measured
relative to the onset of the
convective core.}
\end{figure}

The resulting  tracks in the HR diagram of Figure \ref{fig:mesa_vs_dsep_HRD} and
the evolution in the \Tc-\rhoc\ plane
of Figure \ref{fig:mesa_vs_dsep_tc-rhoc} show excellent
agreement between the codes.
Figures \ref{fig:mesa_vs_dsep_4M} and \ref{fig:mesa_vs_dsep_6M} show
the hydrogen-burning luminosity, the helium-burning luminosity, and
the extent of the convective core during convective helium core
burning for a $4\,\Msun$ model (Fig.~\ref{fig:mesa_vs_dsep_4M}) and
a $6\,\Msun$ model (Fig.~\ref{fig:mesa_vs_dsep_6M}).  Table~\ref{mesa_dsep_table} gives a
summary of the core hydrogen burning lifetime, the core helium burning lifetime, the final extent of the convective core during central helium burning, and the final carbon mass fraction $X_{\mathrm{C}}$
in the core for each model. For the \MESA \ models, we also show the
maximum extent of the convective core during central hydrogen burning, the mass of the helium core before helium ignition, and the mass of the C/O core at the time of the first helium thermal pulse.

\begin{deluxetable*}{rrrrrrrrrrrrr}
\tablewidth{0pc}
\tablecolumns{13}
\tablecaption{\label{mesa_dsep_table}Properties of the $3$--$8\,\Msun$ evolution (masses in solar units).  Selected quantities are also shown from \DSEP\ for comparison.}
\tablehead{%
  & \multicolumn{7}{c}{\MESA} && \multicolumn{4}{c}{\DSEP}\\ \cline{2-8}\cline{10-13}
  \colhead{$M/\Msun$} &
  \colhead{$\Delta t_{\rm H}$\tablenotemark{(1)}} &
  \colhead{$M_{\mathrm{cc}}^{\rm max}$\tablenotemark{(2)}} & 
  \colhead{$M_{\rm core}^{\rm He}$\tablenotemark{(3)}} &
  \colhead{$\Delta t_{\rm He}$\tablenotemark{(4)}} & 
  \colhead{$M_{\rm cc}^{f}$\tablenotemark{(5)}} &
  \colhead{$X_{\rm C}$\tablenotemark{(6)}} & 
  \colhead{$M_{\rm core}^{\rm CO}$\tablenotemark{(7)}} &
  & 
  \colhead{$\Delta t_{\rm H}$\tablenotemark{(1)}} &
  \colhead{$\Delta t_{\rm He}$\tablenotemark{(4)}} & 
  \colhead{$M_{\mathrm{cc}}^f$\tablenotemark{(5)}} & 
  \colhead{$X_{\rm C}$\tablenotemark{(6)}}
 }
\startdata
3.0	&	320.6	&	0.69	&	0.36	&	83.59	&	0.097	&	0.426	&	0.466	&&	312.0	&	80.81	&	0.098	&	0.456\\
4.0	&	152.7	&	1.01	&	0.47	&	29.78	&	0.149	&	0.490	&	0.667	&&	147.3	&	28.91	&	0.153	&	0.516\\
5.0	&	85.61	&	1.34	&	0.59	&	15.52	&	0.214	&	0.511	&	0.827	&&	84.75	&	15.19	&	0.210	&	0.507\\
6.0	&	55.98	&	1.68	&	0.72	&	9.62	&	0.288	&	0.514	&	0.870	&&	55.41	&	9.61	&	0.289	&	0.505\\
7.0	&	39.91	&	2.03	&	0.86	&	6.51	&	0.375	&	0.511	&	0.915	&&	39.69	&	6.79	&	0.401	&	0.454\\
8.0	&	30.42	&	2.40	&	1.02	&	4.67	&	0.480	&	0.504	&	0.966	&&	30.26	&	4.71	&	0.482	&	0.515\\
\enddata
\tablenotetext{(1)}{Central H burning lifetime (Myr)}
\tablenotetext{(2)}{Maximum extent of the convective core during core H burning }
\tablenotetext{(3)}{Mass of the He core before central He ignition }
\tablenotetext{(4)}{Central He burning lifetime (Myr)}
\tablenotetext{(5)}{Stable final extent of the Schwarzschild convective core during core He burning }
\tablenotetext{(6)}{Central mass fraction of \I{12}{C} at the end of core He burning }
\tablenotetext{(7)}{Mass of the C/O core at the time of the first thermal pulse }
\end{deluxetable*}

We close with an additional comparison of the helium core burning
phase of a $M=3\,\Msun$, $Z=0.02$ model computed by \MESA\ to that of
\citet{2003ApJ...583..878S}. Both models were evolved using the
\citet{Kunz:2002} rate for \carbon[12]$(\alpha,\photon)$\oxygen[16].  The
results for \MESAstar\ are a helium core burning lifetime of 83.6~Myr and
final C/O mass fractions of $X_{\mathrm{C}} = 0.43$, $X_{\mathrm{O}} =
0.55$; \citet{2003ApJ...583..878S} find a lifetime of 88~Myr and
$X_{\mathrm{C}} = 0.42$, $X_{\mathrm{O}} = 0.56$.

\subsection{Making and Cooling White Dwarfs}\label{s.completing-evolution}

In the previous section, we discussed the evolution of
$3\textrm{--}8\,\Msun$ stars up to the occurrence of the first He
thermal pulse. In \mesaone \ we showed detailed comparisons of the
evolution of a $2\,\Msun$ star to the EVOL code \citep{herwig_2004_aa}, exhibiting the
ability of \MESAstar\ to calculate multiple helium shell pulses. We now illustrate the final evolution of
intermediate-mass stars, and how to construct white dwarfs (WDs) by using winds.

We evolve $3\,\Msun$, $5\,\Msun$, and $7\,\Msun$ stars from the ZAMS
using the test suite case
\code{make\_co\_wd}.  This makes use of RGB
mass loss following \citet{1975psae.book..229R} with an efficiency parameter $\eta=0.5$ and
AGB mass loss following \citet{1995A&A...297..727B} using $\eta=0.1$
until the occurrence of the first helium shell flash.  At that time, an
increased Bloecker $\eta=5$ is adopted to allow only a small number of
thermal pulses before the wind mass loss eliminates the envelope. Such
intervention allows \MESAstar \ to make a high-mass WD. To avoid shortening of timesteps due to
radiation-dominated envelopes, these cases also use the MLT++ capability described in \S \ref{s.superadiabatic}.

\begin{figure}[htbp]
\centering
\includegraphics[width=\figwidth]{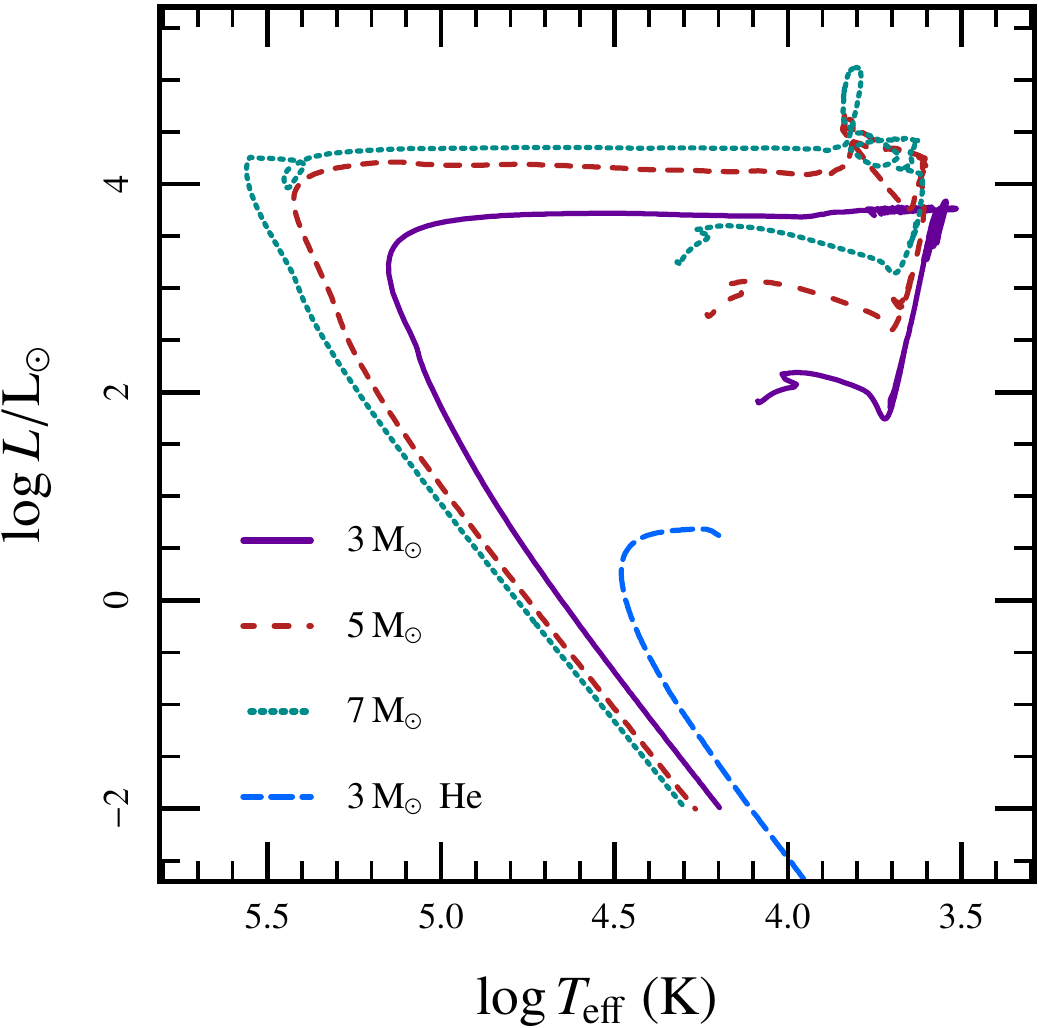}
\caption{\label{fig:357_zams_to_wd_HR}
Evolution
  of $3$, $5$ and $7\,\Msun$ models from zero-age main sequence to
  cooling white dwarfs.  A Bloecker mass loss strips
  the stars of their envelopes on the thermally pulsing AGB to make the three C/O white dwarfs. The single $0.32\,\Msun$ He white dwarf was made with mass loss after the hydrogen main sequence for the $3\,\Msun$ model was completed.
}
\end{figure}

Figure \ref{fig:357_zams_to_wd_HR} shows the resulting tracks on the
HR diagram.  The $3\,\Msun$ star underwent eight thermal pulses after
the enhancement of Bloecker winds, while the $5\,\Msun$ and
$7\,\Msun$ stars lost their envelopes so quickly that thermal pulses
were immediately halted.  The $5\,\Msun$ star ended up as an
$M=0.844\,\Msun$ C/O WD with a helium shell of thickness $M_{\rm
  He}=0.009\,\Msun$ and a hydrogen envelope of $M_{\rm
  H}=2.3\times10^{-5}\,\Msun$. Note that the C/O WD mass is only
slightly larger than the C/O mass at the first thermal pulse
($0.827\,\Msun$) reported in Table \ref{mesa_dsep_table}.

\begin{figure}[htbp]
\centering
\includegraphics[width=\figwidth]{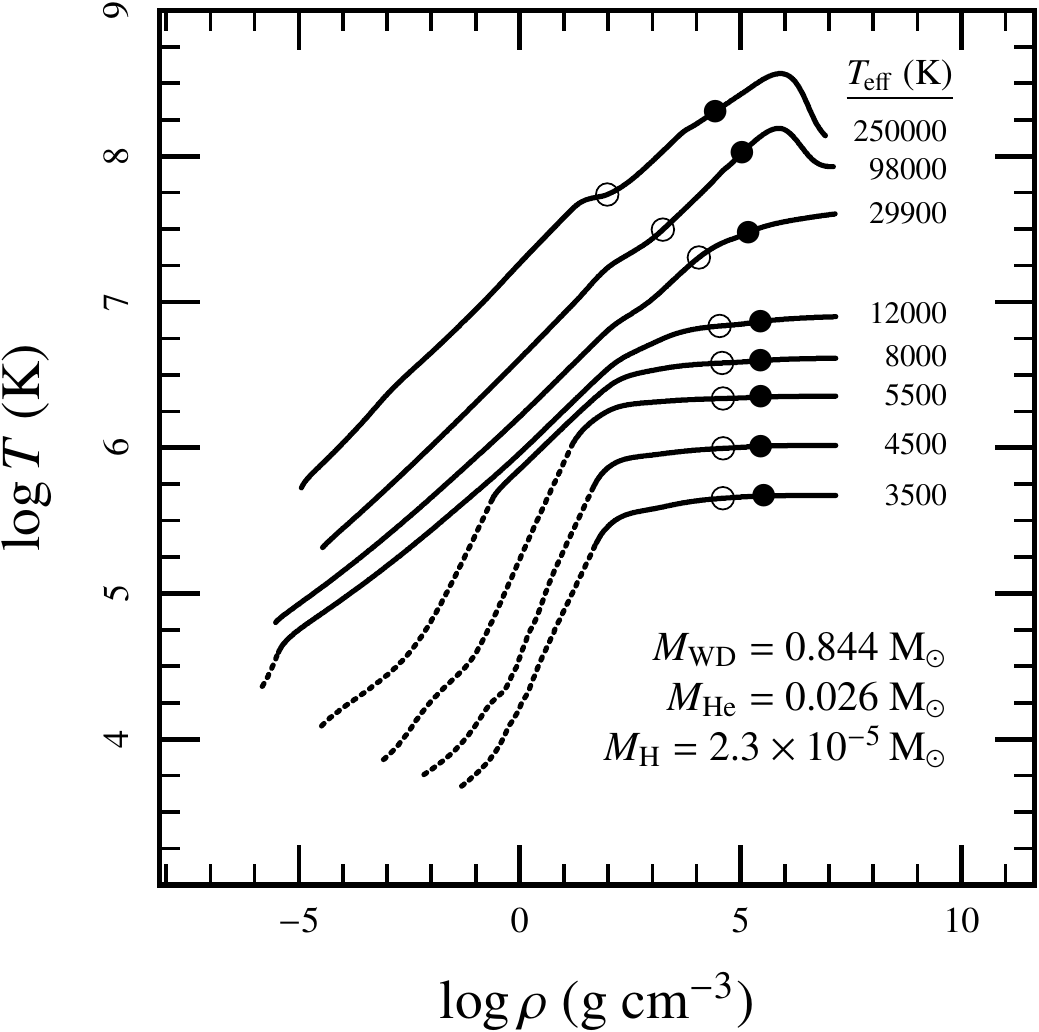}
\caption{\label{fig:co_wd_from_5M_TRho}
Profiles
  in $\log T$-$\log\rho$ space of the cooling $0.844\,\Msun$
  C/O white dwarf evolved from a $5\,\Msun$ progenitor.
  Each model is labeled to the right by \Teff.  The outermost point of the model is at $\tau=25$.
  Dotted curves denote convective regions.  Going toward the interior, open
  circles designate the transition into the helium-rich shell, and
  filled circles designate the transition into the C/O
  core.}
\end{figure}

\begin{figure}[htbp]
\centering
\includegraphics[width=\figwidth]{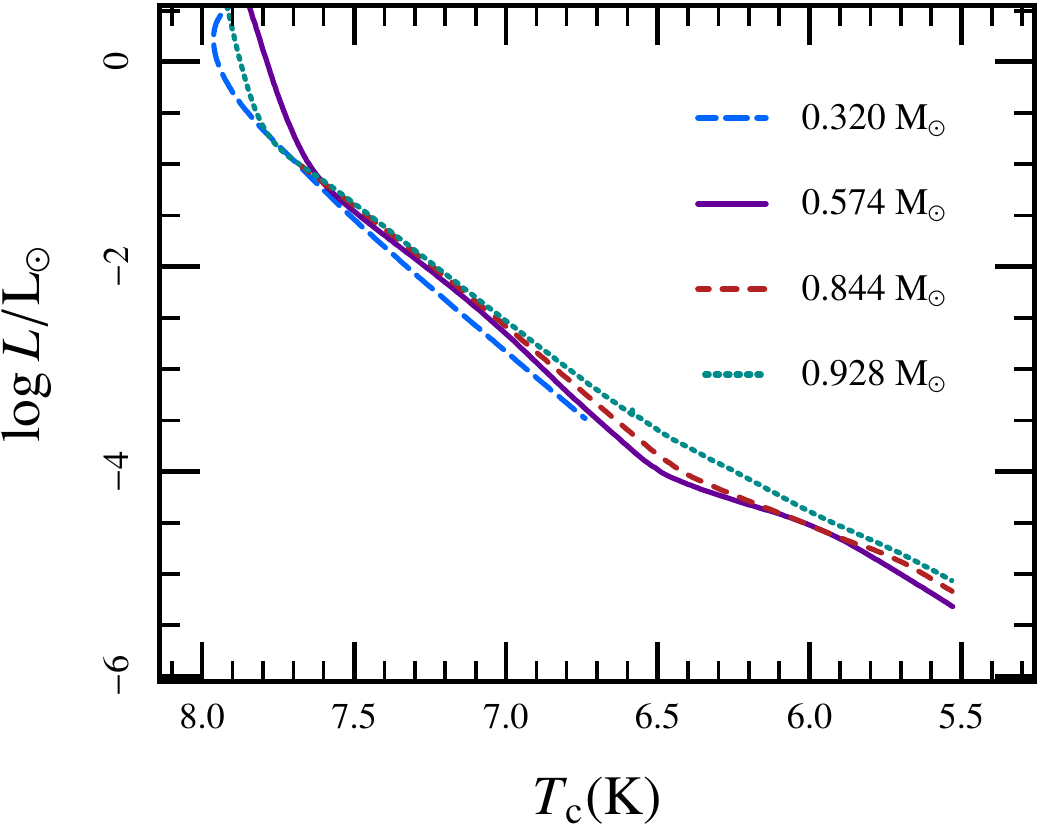}
\caption{\label{fig:357_cool_L_Tc}
Surface
  luminosity as a function of central temperature for the cooling $0.32$,
  $0.574$, $0.844$, and $0.928\,\Msun$ WDs evolved from $3$, $5$,
  and $7\,\Msun$ progenitors. }
\end{figure}

After removal of the envelope, the evolution of the white dwarf is continued through its cooling phase past solidification.  We include  gravitational settling and chemical diffusion of the outermost
layers. Figure
\ref{fig:co_wd_from_5M_TRho} shows $T$--$\rho$ profiles taken at
various effective temperatures during the cooling of the
$M=0.844\,\Msun$ C/O WD made from the $5\,\Msun$ star. The growing
depth of the convection zone is shown by the dashed line, and the
open circles designate the H/He transition, while the filled circles
denote the He/CO transition. Figure~\ref{fig:357_cool_L_Tc} illustrates
the resulting $L$-\Tc\ relation as these models cool.

\begin{figure}[htbp]
\centering
\includegraphics[width=\figwidth]{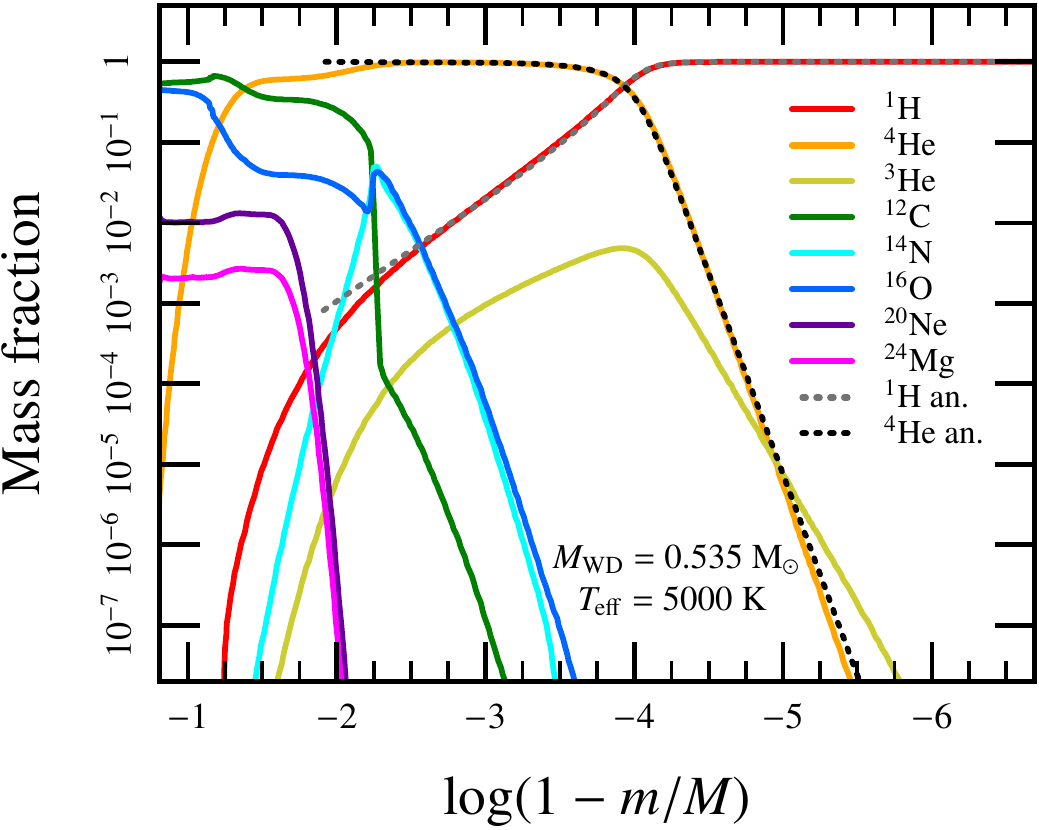}
\caption{\label{WDdiffuse} 
  A comparison of time-dependent diffusion calculations for a
  $M=0.535\,\Msun$ WD with $M_H=5.9\times
    10^{-5}\,\Msun$ and $M_{\rm He}=1.0\times
  10^{-2}\,\Msun$ with \MESAstar\ (solid lines) to those assuming
  diffusive equilibrium and an ideal gas equation of state (dashed
  lines).}
\end{figure}

The test suite case \code{wd\_diffusion} uses the implementation of
diffusion described in \mesaone\ to evolve a WD of mass
$0.535\,\Msun$ until the  $M_{\rm
  H}=5.9\times10^{-5}\,\Msun$ hydrogen layer and the $M_{\rm
  He}=1.0\times10^{-2}\,\Msun$ helium layer approach diffusive
equilibrium.  At this point, the WD has an effective temperature of
$\Teff \approx 5,000\,\Kelvin$.  We show the resulting abundance
profiles in Figure~\ref{WDdiffuse}, and, for comparison, the abundance
profiles derived from the analytic form for diffusive
equilibrium \citep[eq.~(22) of][]{Althaus03}. This formula is
  obtained by integrating equation (A.5) of \citet{Arcoragi80} and
  assuming an ideal gas equation of state and complete ionization of
  both species.

\begin{figure}[htbp]
\centering{\includegraphics[width=\figwidth]{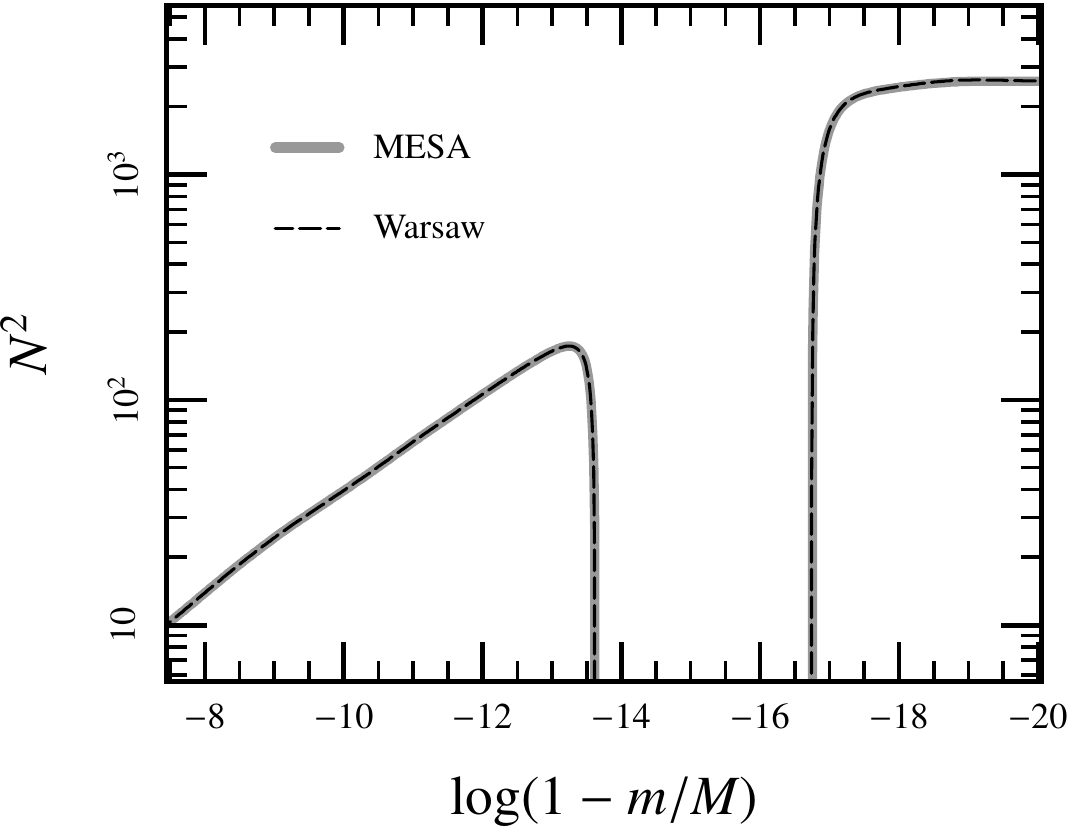}}
\caption{\label{WDconv} 
A comparison of the
  \bvfreq\ calculated with \mesa\ (solid grey line) to that using the Warsaw
  envelope code \code{pig35.f} (dashed line) for the same WD in Figure \ref{WDdiffuse},
 but at a cooler $\Teff=11,354\,\Kelvin$. }
\end{figure}

The specific treatment of convection can also impact WD evolution.  In
\mesaone, \mesa\ used the \citet{Cox68} prescription for convection
as its default convective MLT, with the
optional extension of \citet{Henyey65}.
Since \mesaone, we have added support for the formulations of
\citet{Bohm-Vitense58}, \citet{bohm71}, and \citet{Mihalas78}. In
particular, the B\"ohm \& Cassinelli prescription, often referred to
as ``ML2,'' is frequently employed in WD studies \citep[e.g.,][]{Bergeron:1995}.  In
Figure~\ref{WDconv} we show a comparison of the \bvfreq\ calculated
with \mesa\ to that using the Warsaw envelope code
\citep{Paczynski69,Paczynski70,Pamyatnykh99},
assuming the ML2 prescription. This
is the same WD as in Figure \ref{WDdiffuse}, but now at a lower
$\Teff=11,354\,\Kelvin$. To more accurately integrate these opaque but thin
layers, we reduce $\tau$ at the  boundary of the model by a factor of 1000 from its photospheric value of $2/3$. This calculation is a sensitive test of the envelope
integrations because $N^{2}$ is a derivative of the envelope
structure. The two codes give indistinguishable results for this case
and all other cases that we have calculated.

\begin{figure}[htbp]
\centering{
  \includegraphics[width=\figwidth]{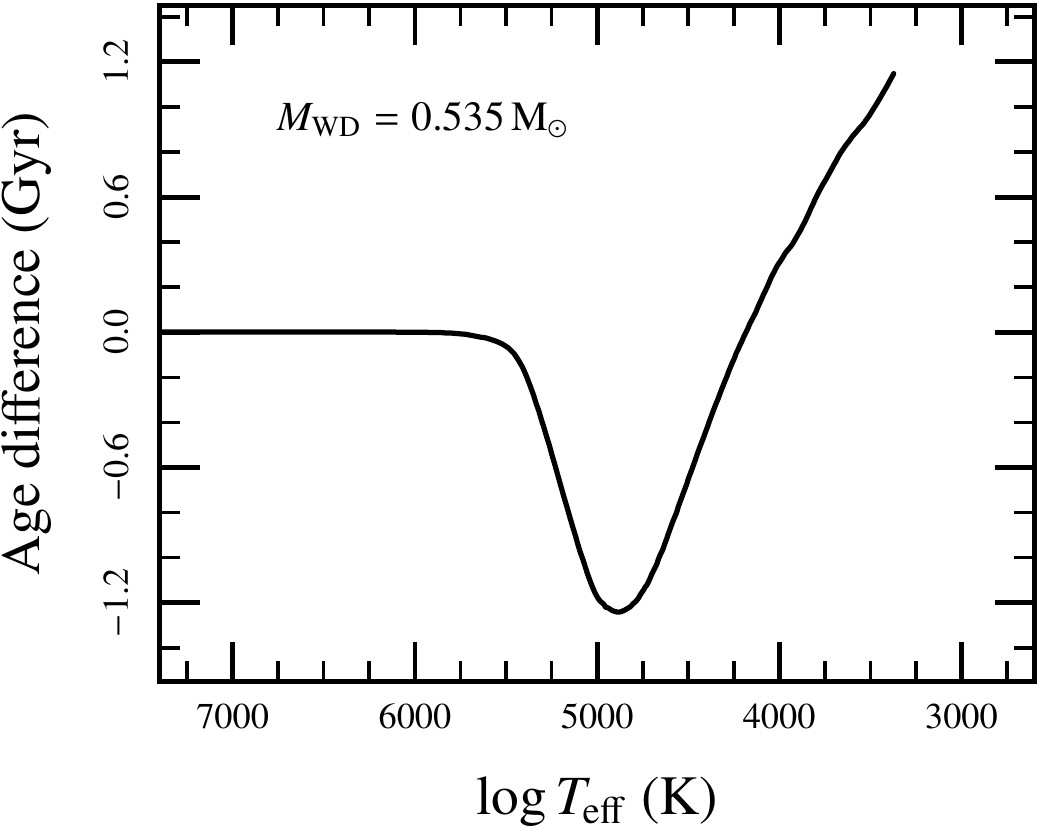}
}
\caption{\label{age_diff} 
The age difference
  (non-grey minus grey) in Gyr as a function of \Teff.  
}
\end{figure}

\mesa\ now includes atmospheric tables based on the non-grey model
atmospheres for hydrogen-atmosphere WDs \citep{Rohrmann01,Rohrmann12},
spanning the following range of parameters: $2,000\nsp\K \leq
\Teff \leq 40,000\nsp\K$ and $5.5 \leq \logg \leq 9.5$.  Such an
approach is necessary at $\Teff\la 6000\,\Kelvin$, where WDs develop
deeper convection zones.  When the convection zone comes in contact
with the degenerate, nearly isothermal core, energy is able to flow
out of the core much more efficiently. The use of non-grey atmosphere
models results in shallower convection zones, so this convective
coupling of the core and envelope is delayed. For reliable cooling
ages, we therefore recommend using non-grey atmospheres when $\Teff
\la 6000\,\Kelvin$.  Figure~\ref{age_diff} demonstrates the impact of
non-grey atmospheres with the $0.535\,\Msun$ WD, which was cooled with
and without the non-grey atmosphere.

\mesa\ currently treats crystallization by employing the
\citet{Potekhin10} EOS (PC EOS).  The PC EOS is callable
  for arbitrary mixtures of chemical species and for densities with
  $\log \rho \ge 2.8$; it is applicable in the domains of
  non-degenerate and degenerate, non-relativistic and relativistic
  electrons, weakly and strongly coupled Coulomb liquids, and
  classical and quantum Coulomb crystals.  The phase transition is
first-order, so the PC EOS exhibits a latent heat between the solid and
liquid phases, i.e., the entropy and internal energy both experience
finite jumps. This energy is included in \MESAstar\ models of cooling
white dwarfs through the gravitational source term in the energy
equation,
\begin{equation}   \label{epsent}
  \epsgrav \equiv -T\frac{\dif S}{\dif t}.
\end{equation}
This form for \epsgrav\ replaces  the default one
(see eq.~(\ref{eq:eps_grav_lnd}) below) in cells where $\Gamma\geq160$ ($\Gamma$ is the Coulomb coupling parameter), and is smoothly interpolated with the default form in
cells where $130\leq\Gamma<160$.  The PC EOS uses the criterion $\Gamma = 175$ to determine
crystallization, but it is straightforward to include explicit
crystallization curves for C/O and other mixtures
\citep[e.g.,][]{Schneider12,Medin10}. For example, using the
parameters of the model in Figure~\ref{WDdiffuse}, the age difference
at late times ($\Teff < 3,500\,\Kelvin$) between a model with and
without the latent heat of crystallization is $\approx 0.8\,\Gyr$; a
slightly larger value would be obtained using the phase diagram of
\citet{Schneider12}. \mesa\ does not currently treat
phase separation of different chemical species upon crystallization.

Low mass WDs ($M \la 0.4\, \Msun$) with helium cores and hydrogen
envelopes may be produced in binary systems when the envelope is
stripped by the companion as the primary evolves up the giant branch
\citep[][and references therein]{iben_1991_aa}.
He-core WDs of mass $M
\simeq 0.4\textrm{--}0.5 \,\Msun$ may also be produced through strong RGB winds
(D'Cruz et al.\ 1996), although we do not discuss this possibility
further here.

Here we discuss the prescription for stripping the envelope used in the test case
\code{make\_he\_wd}.  The first step is to evolve a star,
$M= 3.0 \, \Msun$ in this example, from the PMS until a He core of the correct size has
been made. The remnant total mass is determined by the
mass interior to where the H abundance has dropped below a
preset value, for example, $X_{\rm H}=0.1$, moving in from the
surface.  Next, the routine \code{relax\_mass} is used to remove mass
from the model until it has the desired remnant mass.  After the
initial remnant has been constructed, diffusion can then be turned on
to allow an outer H layer to form.  After this
stage, normal evolution of the WD occurs, as shown in Figures \ref{fig:357_zams_to_wd_HR} and \ref{fig:357_cool_L_Tc}.

\subsection{Compressional Heating and Accretion} \label{s.compressional}

Accretion onto stars occurs in many contexts and requires special treatment for the outermost layers added in each timestep. In particular, a special evaluation of the $\epsgrav=-T\,\D S/\D t$ term is required for fluid parcels that were not present in the previous timestep.  Prior to addressing that subtlety, we restate (as discussed in \S 6.2 of \mesaone) that \MESAstar \ calculates \epsgrav\ of
eq.~(\ref{epsent}) in terms of the local thermodynamic variables ($T$
and $\rho$) used by \MESA,
\begin{equation}\label{eq:eps_grav_lnd}
\epsgrav = -\CP T\left[\left(1-\nablaad\,\chiT\right)\frac{\dif \ln
    T}{\dif t}-\nablaad\,\chirho\frac{\dif \ln\rho}{\dif t}\right].
\end{equation}
\MESAstar\ takes the quantities in this equation as provided by \code{eos}, and computes the Lagrangian time derivatives to find
\epsgrav.  \MESAstar\ can alternatively work under the assumption that
$P=\Pgas+\Prad$,
in which case \MESAstar\ treats \Pgas\ rather than $\rho$ as its basic variable (see \S
\ref{s.solve-burn-and-mix} for a discussion). In that case,
\begin{equation}\label{eq:eps_grav_prho}
\epsgrav=- \CP T\left[ 
  \left(1-4\nablaad \frac{\Prad}{P}\right) \frac{\dif \ln T}{\dif t} -
  \nablaad \frac{\Pgas}{P}\frac{\dif \ln\Pgas}{\dif t} \right].
\end{equation}
Either formulation can be used deep within the star, as long as the
location is safely removed from any phase transition. \mesaone\, described the validation of these formulations.

We now turn to the complication which arises when \epsgrav\ needs to be evaluated in material that was not present in the previous timestep.  Defining the envelope mass coordinate $\Delta M\equiv M-m$, we need to resolve the entropy for $\Delta M< \delta
M=\Mdot\,\timestep$, as the explicit Lagrangian time derivatives of
eqs.~(\ref{eq:eps_grav_lnd}) and (\ref{eq:eps_grav_prho}) cannot be
numerically evaluated. Since there can be important physics that needs to be resolved
for these mass shells for $\Delta M\ll \delta M$, an approximation must be
derived that allows for accurate modeling of the star's outermost
layers without having to result to a dramatic shortening of \timestep.

The luminosity
$\Lacc=GM\Mdot/R$ from the accretion shock (or boundary layer)
goes outwards and does not determine the entropy of the material as it
becomes part of the hydrostatically adjusting star.  Rather, the
entropy of the material at $\Delta M\ll \delta M$ is
determined by the the transport of $L$ (\citealt{1977PASJ...29..765N}; \citealt{1982ApJ...253..798N};
\citealt{2004ApJ...600..390T}). Consider such an outermost layer,
where there are two relevant timescales, the thermal time, $t_{\rm
  th}=C_PT\Delta M/L$, and the local accretion time, $t_{\rm
  acc}=\Delta M/\Mdot$. In nearly all relevant cases, the ratio
$t_{\rm th}/t_{\rm acc}=\CP T\Mdot/L\ll 1$; this implies that the fluid
element adjusts its temperature to that needed to transport the
stellar luminosity from deep within. This simplifies $\epsgrav$ in that part of the star (following
\citealt{2004ApJ...600..390T}) to
\begin{equation}\label{eq:eps_grav_thin_radiative}
\epsgrav = \frac{C_PTGm\Mdot}{4\pi r^4P}(\nablaad-\nablaT),
\end{equation}
enabling accurate modeling within \mesastar \ of nearly all fluid
elements that become part of the star during each timestep, many of
which have envelope mass coordinates $\Delta M\ll \Mdot \timestep$. 

We give an explicit example of this thin-shell radiative calculation
of \epsgrav\ in a C/O white dwarf accreting hydrogen-rich material and
undergoing classical nova (CN) cycles.  We present two models
accreting at rates of $\Mdot=10^{-11}\,\Msunyr$ and
$10^{-10}\,\Msunyr$.  Both cases were evolved from a $0.6\,\Msun$
starting model with $\Tc=10^7\,\Kelvin$ which had
undergone a few flashes while accreting at
$\Mdot=10^{-11}\,\Msunyr$. The accreted material has solar-like
composition $X=0.70$, $Y=0.26$, and $Z=0.04$
where the metal mass fractions are taken from
\citet{2003ApJ...591.1220L}.

\begin{figure}[htbp]
\centering
\includegraphics[width=\figwidth]{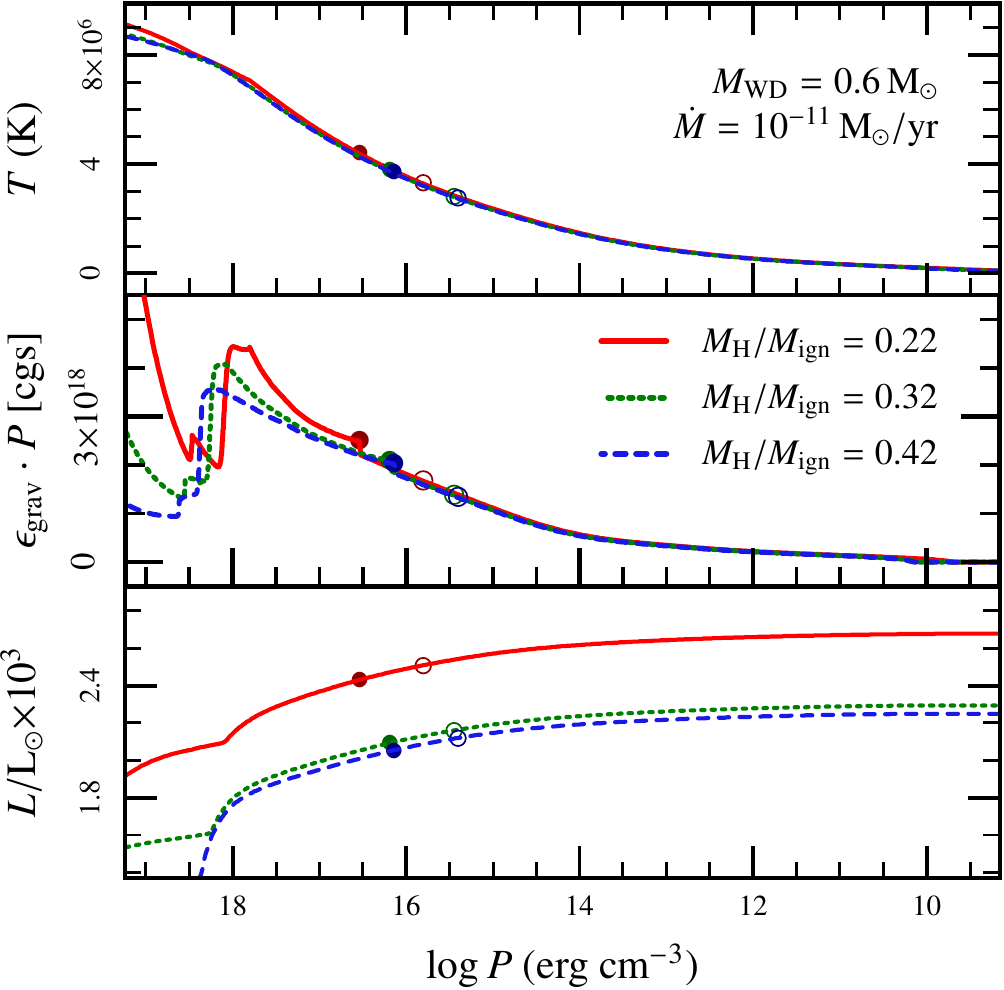}
\caption{\label{co_wd_accrete_1d-11} 
Envelope
  profiles as a function of pressure of the accreting white dwarf for
  three instants during the mass accumulation phase;
  $\Mdot=10^{-11}\,\Msunyr$ model.  The top panel shows temperature, the central panel
  shows the gravitational energy release rate, and the bottom shows the luminosity.
  Material to the right of the open circle is newly
  accreted.  The code treats material to the right of the filled circle using the thin-shell
  radiative calculation of \epsgrav.}
\end{figure}

\begin{figure}[htbp]
\centering
\includegraphics[width=\figwidth]{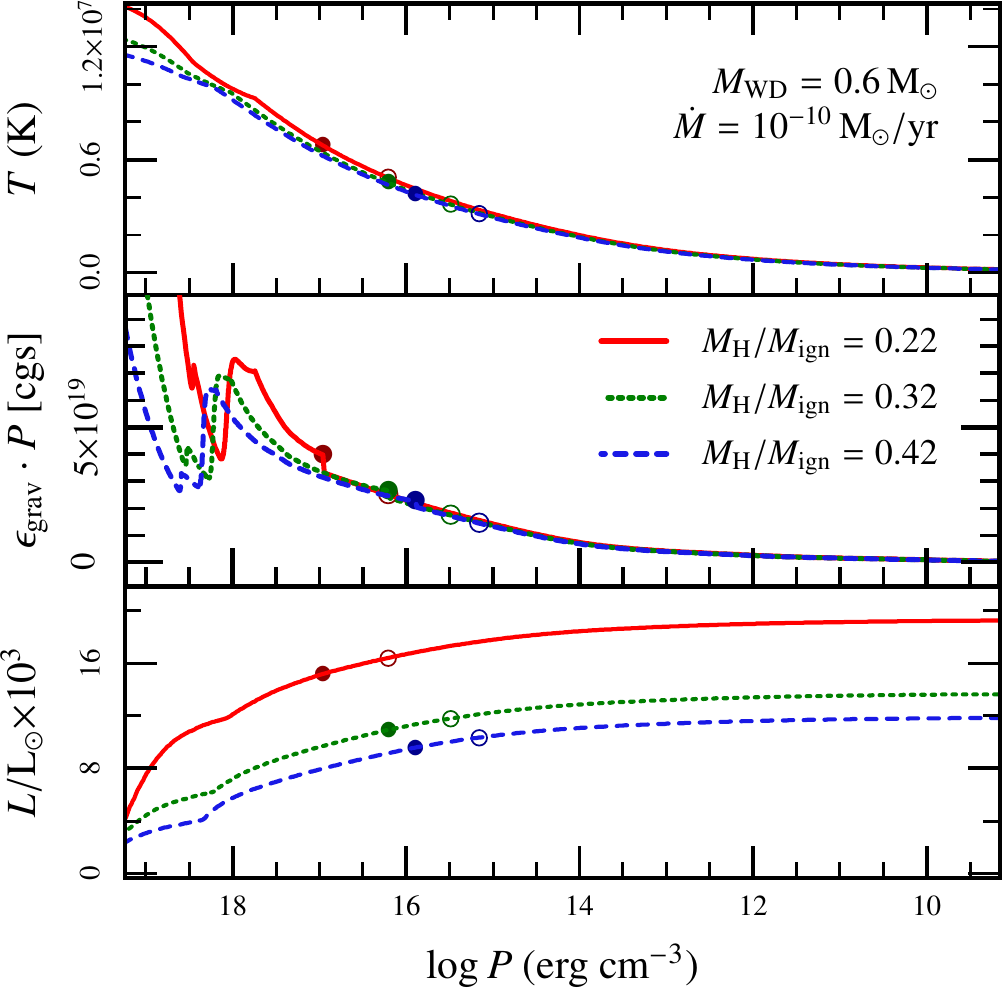}
\caption{\label{co_wd_accrete_1d-10} 
Same as
  Fig. \ref{co_wd_accrete_1d-11}, but for model accreting
  $10^{-10}\,\Msunyr$.}
\end{figure}

Profiles of the envelope during the mass accumulation phase between CN
outbursts for the two accretion rates are displayed in Figures
\ref{co_wd_accrete_1d-11} and \ref{co_wd_accrete_1d-10}. Each line
represents a different time in the accumulation cycle up to the
unstable ignition, when the hydrogen mass reaches $M_{\rm H}=M_{\rm
  ign}$. All material at pressures smaller than that shown by the open
circle is new to the model in that timestep (e.g., it has $\Delta M<
\delta M$) and employs the modified \epsgrav\ of
eq.~(\ref{eq:eps_grav_thin_radiative}). This highlights the
significance of this approximation as it allows \mesastar \ to
calculate material properties at $\Delta M\sim 10^{-8}\,\delta M$. The solid
points show where \epsgrav\ switches to the explicit form employing
the Lagrangian time derivatives, such as eq.~(\ref{eq:eps_grav_lnd}).

The middle panel shows $\epsgrav P\propto\epsgrav \Delta M$, which reflects the contribution
of \epsgrav\ to the outward luminosity. The discontinuity of \epsgrav\ at the solid
point reflects the error associated with the abrupt transition in the
calculational approach. The substantially larger luminosity of the
early ($M_{\rm H}/M_{\rm ign}=0.22$) stages is due to the ongoing transfer of heat
from the previous outburst. The near-discontinuous drop in
\epsgrav\ occurs at the base of the hydrogen-rich envelope, and
reflects the jump in composition from the accreted material to the
nearly pure \helium\ layer. The expected amplitude of the jump in
\epsgrav\ depends on both the composition jump and the local degree of
electron degeneracy (see Appendix B of \citealt{2004ApJ...600..390T} for a discussion).

\section{Rotation}\label{s.rotation}
A star's rotational energy is usually a small fraction of the
gravitational energy: for the Sun it is $\sim 10^{-5}$ and for a  $25\,\Msun$ star rotating with a typical equatorial velocity $\veq=200\,\kms$ on the
main sequence it is $\sim 0.04$.
Therefore the effects on the stellar hydrostatic
equilibrium are marginal, with the exception of stars close to critical rotation (see \S \ref{s.rotation-mass-loss}).
Even in the case of a small perturbation to hydrostatic
equilibrium, rotation induces a modification to the star's thermal
equilibrium \citep{vonzeipel:1924}. Together with the
emergence of rotationally-induced dynamical and secular instabilities,
this can significantly affect the evolution of stars
\citep{maeder_2000_aa}. Due to the destabilizing effect of increasing
radiation pressure,  rotation is particularly important in massive stars \citep[see,
  e.g.,][]{Heger:2000,Meynet:2000}.  Moreover, the final fate
of a massive star depends chiefly on the relative importance of
rotation during its evolution
\citep[e.g.,][]{Heger:2000,hirschi_2004_aa,Heger:2005,Yoon:2006,Woosley:2006,Ekstrom:2012,Georgy:2012,Langer:2012}.

Here we describe the implementation of rotation in \MESAstar. 
We briefly discuss the modification to the stellar structure equations and the inclusion of rotationally- and  
magnetically-induced mixing. Magnetic fields generated by differential rotation in radiative regions have been implemented 
following the work of \citet{Spruit:2002} and in the same fashion as in \citet{Petrovic:2005} and \citet{Heger:2005}.
Rotationally enhanced mass loss is also discussed.

We compare rotating massive-star models calculated with \MESAstar\ to previous calculations performed with KEPLER \citep{Heger:2005}.
We also directly compare runs from \MESAstar\ and \STERN\ \citep{Petrovic:2005,Yoon:2005,Brott:2011}.
The purpose of these tests is to verify  our implementation of rotation, which is derived from \STERN.
We do not compare to codes that have a different implementation of rotation \citep[e.g.,][]{hirschi_2004_aa,Ekstrom:2012,Georgy:2012,Potter:2012a,Potter:2012}.  Although beyond the scope of this paper,  such comparisons are critical when coupled to observations of the effects of rotation in stars \citep[e.g.,][]{Hunter:2007,Evans:2011}  including asteroseismology \citep{Beck:2012,Mosser:2012}. 

\subsection{Implementation of Shellular Rotation}\label{s.rotation-modifications}
Stellar structure deviates from spherical symmetry in the presence of rotation.
While the structure is inherently three-dimensional, it suffices
to solve the stellar structure equations in one dimension if the angular velocity, $\omega$, is constant over isobars 
\citep[the so-called shellular approximation; see, e.g.,][]{meynet_1997_aa}. 
This is expected in the presence of strong anisotropic turbulence acting along isobars. 
In radiative regions such turbulence is a consequence of differential rotation \citep{Zahn:1992} 
and efficiently erases gradients along isobars and enforces shellular rotation \citep{meynet_1997_aa}.
Turbulence in the vertical direction (i.e., perpendicular to the isobars)  is much weaker due to the stabilizing effect of stratification.
In \MESAstar\ we adopt the shellular approximation \citep{meynet_1997_aa} and calculate the modification to the stellar 
equations due to centrifugal acceleration \citep{Kipp:1970,Endal:1976}.
\subsubsection{Stellar Structure}
An isobar with volume $\VP$ and surface area $\SP$ deviates from spherical symmetry in the presence of rotation.
However one can retain a 1D approximation by re-defining the radius
coordinate as the radius of a sphere containing the same volume
$\VP=4\pi \rP^3/3$, allowing an equation of continuity in the usual
form
\begin{equation}
  \dxdycz{\mP}{\rP}{t}=4\pi\rP^2\rho
  \;,
\end{equation}
with $\rho$ being the density and $\mP$ the mass enclosed by $\SP$.
The energy equation also retains its usual, non-rotating form
\begin{equation}
  \dxdycz{\LP}{\mP}{t}=\epsnuc-\epsnu+\epsgrav
  \;,
\end{equation}
where $\LP$ is the rate of energy flow through the equipotential surface $\SP$.
Then the next step is to define mean values for the quantities varying on isobars,
\begin{equation}
\av{\;\cdot\;} \equiv \frac{1}{\SP} \oint_{\SP}\cdot\;\;\D\sigma\;,
\end{equation}
where $\DD\sigma$ is an isobaric surface area element. 
The equation of momentum balance can be written as
\begin{equation}
  \dxdycz{P}{\mP}{t}=
     -\frac{G \mP}{4\pi \rP^4} \fP
     -\frac{1}{4\pi \rP^2}\dxdycz{^2 \rP}{t^2}{\mP}
  \;,
\end{equation}
where $P$ is the pressure, $G$ is the gravitational constant and $t$ the
time. The last term in the equation is the inertia term. 
Rotation enters the momentum equation through the quantity $\fP$
\begin{equation}
  \fP \equiv \frac{4\pi \rP^4}{G\mP\SP} 
               \av{g^{-1}}^{-1}
  \;,
\end{equation}
where $g\equiv\abs{\vec{g}}$, with $g$ the effective gravitational acceleration ($\vec{g}$ is normal to $\SP$). 
Then the radiative temperature gradient becomes
\begin{equation}
  \dxdycz{\ln T}{\ln P}{t} = 
	\frac{3\kappa}{16\pi a c G}
	\frac{P}{T^4}
	\frac{\LP}{\mP}
 	\frac{\fT}{\fP}
	\SBrak{1+\frac{\rP^2}{G \mP \fP}
              \dxdycz{^2 \rP}{t^2}{\mP}}^{\!-1}
\;,
\end{equation}
with  $a$ the radiation
constant, $\kappa$ is the opacity, $T$ the temperature and $\LP$, the energy flux through $\SP$.  
The last factor on the right hand side accounts for inertia, and
\begin{equation}
\fT \equiv \Brak{\frac{4\pi \rP^2}{\SP}}^{\!2} 
             \Brak{\av{g}\av{g^{-1}}}^{-1}
  \;.
\end{equation}
In rotating models the  values of  $\fT$ and $\fP$  differ from 1 mostly in the outer stellar layers.
Limits to the minimum values of $\fT$ and $\fP$ are set in the code (default values are 0.95 and 0.75, respectively). 
This prevents numerical instabilities in models approaching critical rotation ($\Om/\Omc=1$, see~\ref{s.rotation-mass-loss}). 
In such cases the outer layers greatly deviates from spherical symmetry and the results from 1D calculations should be
considered particularly uncertain.


\subsubsection{Mixing and angular momentum transport}
Transport of angular momentum and chemicals
due to rotationally-induced instabilities is implemented in a diffusion approximation 
\citep[e.g.,][]{Endal:1978,Pinsonneault:1989,Heger:2000}. 
This choice has also been adopted by other stellar evolution codes  (e.g., KEPLER, \citealp{Heger:2000}; STERN, \citealp{Yoon:2005}). We stress that this is not the only possibility,
and other groups have implemented a diffusion-advection approach (e.g., GENEVA, \citealp{Eggenberger:2008};  RoSE, \citealp{Potter:2012a}).  The RoSE code can switch between the two different implementations.  
The two approaches are equivalent for the transport of chemicals. Potentially large differences can arise, however, for the transport of angular momentum.
A detailed description of the advection-diffusion equation for angular momentum  is given in \citet{Zahn:1992} and \citet{Maeder:1998}.

In \MESAstar\ the turbulent viscosity $\nu$ is determined as the sum of the diffusion coefficients for convection, semiconvection, and rotationally-induced instabilities.
In convective regions, the very large diffusion coefficient implies that the  rotation law is not far from solid body. 
This is a very common assumption in stellar evolution codes \citep[e.g.][]{Pinsonneault:1989,Heger:2000,Eggenberger:2008};  
note however that helioseismology has clearly shown this is not the case for the solar convection zone \citep[e.g.][]{Brown:1989,Thompson:1996,Schou:1998}.
\MESAstar\ calculates diffusion coefficients for five different rotationally-induced mixing processes: 
dynamical shear instability, Solberg-H{\o}iland instability, secular shear instability, Eddington-Sweet circulation, and the Goldreich-Schubert-Fricke instability.
See \citet{Heger:2000} for a detailed description of the physics of the different instabilities and the calculation of the respective diffusion coefficients. These enter the angular momentum and abundance diffusion equations 
that are solved at each timestep (see \S\ref{s.nuts-rotation}).

\subsection{Internal Magnetic Fields}\label{s.magnetic-fields}
It has been suggested that differential rotation in the radiative layers of a star can amplify a seed magnetic field.
Such a dynamo process has been proposed by \citet[][Spruit-Tayler dynamo]{Spruit:2002}; a theoretical debate on this 
is still ongoing \citep{Braithwaite:2006,Zahn:2007,Denissenkov:2007}.
From the observational point of view, pure hydrodynamic models fail to predict the solar core rotation 
\citep[e.g., ][]{Pinsonneault:1989}, with the exception of models that include transport of angular momentum by gravity waves 
\citep{Charbonnel:2005}. Models that include the Spruit-Tayler dynamo can reproduce the flat rotation profile of the Sun. 
Note however that these have difficulty explaining the core-envelope 
decoupling observed in low-mass, young cluster stars \citep{Denissenkov:2010}.
On the other hand, observations of  the final spins of both WDs and neutron stars \citep{Heger:2005,suijs:08} suggest that angular momentum transport with an efficiency similar
to the torques provided by the Spruit-Tayler dynamo operates. 
Models that only include angular momentum transport through rotational instabilities do not produce the core-envelope 
ratio of angular velocity observed through the splitting of mixed modes in red giant stars \citep{Eggenberger:2012}.

\MESAstar\  accounts for transport by magnetic fields of angular momentum and chemicals due to the Spruit-Tayler dynamo.
We refer to \citet{Spruit:2002} for a description of the physics of the dynamo loop and to \citet{Maeder:2003}, \citet{Maeder:2004} and \citet{Heger:2005}  for a discussion of its inclusion in stellar evolution codes. 
We implement the Spruit-Tayler dynamo in \MESAstar\ following KEPLER \citep{Heger:2005} and STERN \citep{Petrovic:2005}.
  
\subsection{Surface Magnetic Fields}\label{s.magnetic-fields-surface}
Rotating stars that have a significant outer convective zone
can produce surface magnetic fields through a dynamo \citep[See e.g.,][for a review on astrophysical dynamos]{Brandenburg:2005}. 
This is the case for low-mass main sequence stars below about 1.5~\Msun, and
observationally the break in the rotation properties around this mass 
is attributed to the presence of magnetized stellar winds \citep[e.g.,][]{Schatzman:1962,Kawaler:1988}.
Note that dynamo action in a subsurface convective layer is in principle possible also in early-type stars \citep{Cantiello:2009,Cantiello:2011}.
Surface magnetic fields can also be of fossil origin, as is usually discussed in the context of Ap stars \citep{Braithwaite:2004}.
Whatever the origin of surface magnetic fields, these are expected to couple to the wind mass-loss and, if strong enough, produce  magnetic braking \citep[e.g.,][]{Weber:1967,uddoula:2002,Meynet:2011}. 
Such magnetic braking has been directly observed in the case of the main sequence massive star $\sigma$-Ori E \citep{Townsend:2010}.
Here we do not include the physics of magnetic braking, as we only consider the evolution of stars without surface magnetic fields.

\subsection{Rotationally-Enhanced Mass Loss}\label{s.rotation-mass-loss}
We include the rotational modification to the wind mass loss rate \citep{Friend:1986,Langer:1998,Heger:1998,Maeder:2000}.
Similar to other codes \citep[e.g.,][]{Heger:2000,Brott:2011,Potter:2012}, in \MESAstar\ 
the stellar mass loss is enhanced as the rotation rate increases according to the prescription
\begin{equation}\label{mdotomega}
  \Mdot\left(\Om\right) = \Mdot(0)\,\left(\frac{1}{1-\Om/\Omc}\right)^{\xi},
\end{equation} 
where $\Om$ is the value of the surface angular velocity and $\Omc$ is the critical angular velocity at the surface.
 This last quantity is defined as $\Omc^2=(1-L/\Ledd)\, GM/R^3$, where $\Ledd =  4 \pi c G  M /\kappa$ is calculated as a mass-weighted average in a user-specified optical depth range (default value $\tau \in [1-100]$).
In \MESAstar\ the default value for the exponent $\xi$ is 0.43 \citep{Langer:1998}.  
Other implementations of rotationally enhanced mass loss can be found in \citet{Maeder:2000} and \citet{Georgy:2011}.

For stars approaching $\Om/\Omc=1$, the mass loss calculated using 
equation (\ref{mdotomega}) diverges. Notice that luminous stars can approach this limit without having to rotate very rapidly as $\Omc \rightarrow 0$ when $L/\Ledd \rightarrow 1$. 
Following \citet{Yoon:2010a} we limit the mass loss timescale
to the thermal timescale of the star $\tkh$
\begin{equation}\label{mdotomegatkh}
  \Mdot= \textrm{min}\,\left[ \Mdot(\Om) \,, f\,\frac{M}{\tkh} \right]
\end{equation}
where $f$ is an efficiency factor of order unity (default value is $f = 0.3$).

\subsection{Initial models}\label{s.initial-models}
In all the rotating models presented in this paper, rotation is initialized by imposing
a solid body rotation law on the zero-age main sequence (ZAMS, $L = L_{\rm{nuc}}$).
In these massive stars this is motivated by the presence of rotationally-induced angular momentum transport 
during the pre-main sequence evolution. This alone is able to enforce a state of close-to-rigid rotation by the time the star reaches the 
ZAMS \citep{Heger:2000}. Overall initial solid body rotation is a common choice in stellar evolution codes, 
but other rotational laws are certainly possible.

\subsection{Test Cases: 15~\Msun\ and 25~\Msun}\label{s.test-cases}

As a first test we  initialize a 15\Msun\ model with $Z=0.02$ and 
initial equatorial rotational velocity $\veq=200\,\kms$ and run two calculations:
\begin{itemize}
\item 15MAG includes the effects of rotation and Spruit-Tayler magnetic fields on both the transport of chemicals and angular momentum.  
\item 15ROT includes only the effect of rotation on both the transport of chemicals and angular momentum;
\end{itemize}
The initial conditions have been calibrated  to match as closely as possible the  KEPLER $15\nsp\Msun$ models \citep{Heger:2005}.
Moreover, we  directly compare the \MESAstar\ models with calculations from STERN \citep[see e.g.,][]{Yoon:2005,Yoon:2006}.
In particular we adopt a value of $f_c = 1/30$ for the ratio of the turbulent viscosity to the diffusion coefficient  and a value $f_{\mu} = 0.1$ for the sensitivity to 
$\mu$-gradients  \citep[see][for a discussion of these calibration parameters]{Heger:2000}. The Ledoux criterion is used for the treatment of convective boundaries, together with semiconvection ($\alphasc=1$). 
We use $\alphaMLT=1.6$,  mass loss as in \citet{Yoon:2006} with rotational enhancement as described in \S~\ref{s.rotation-mass-loss}.

\begin{figure*}[htbp]
\centering{
\includegraphics[angle=0,width=\twoupwidth]{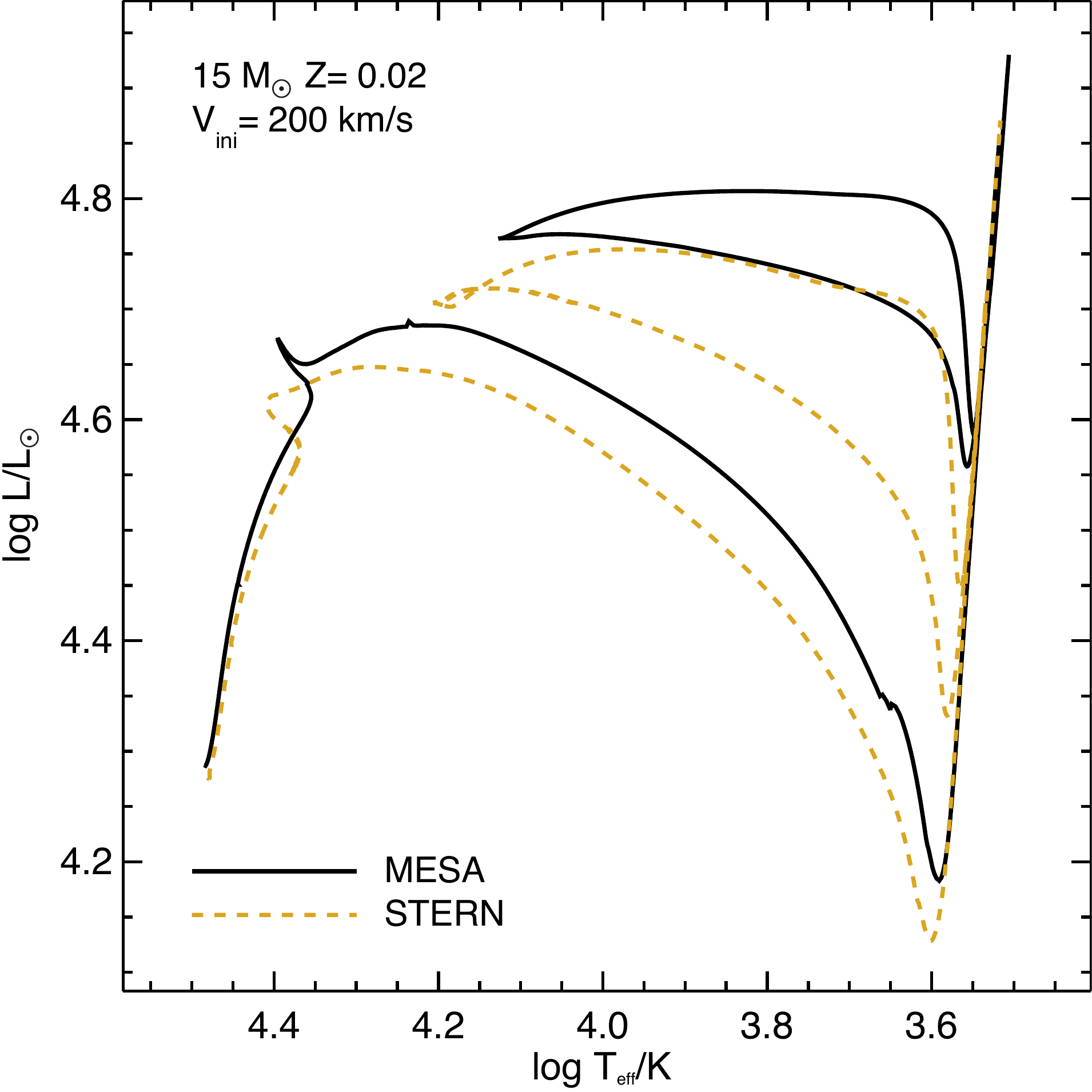}\hspace{\twoupsep}
\includegraphics[angle=0,width=\twoupwidth]{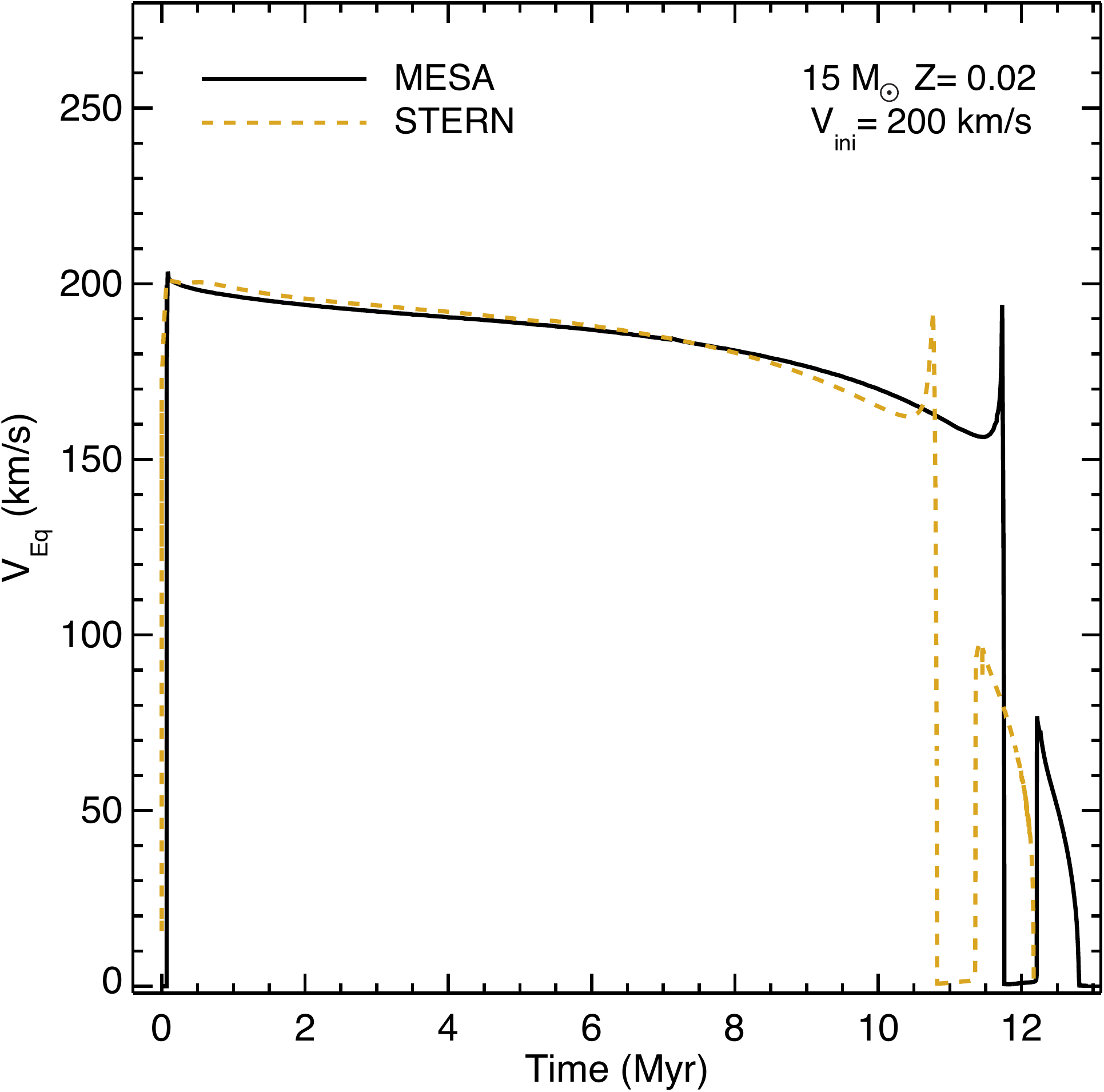} 
}
\caption{ Comparison of evolutionary tracks (left) and equatorial rotational velocities (right) for a $15\,\Msun$ model with $Z=0.02$ rotating initially 
with  $\veq = 200\,\kms$ (15MAG). The solid black line shows \MESAstar\ results, and the dashed gold line shows the STERN calculations.
\label{fig:hrdfig15}}
\end{figure*}

\begin{figure*}[htbp]
\centering{
\includegraphics[angle=0,width=\twoupwidth]{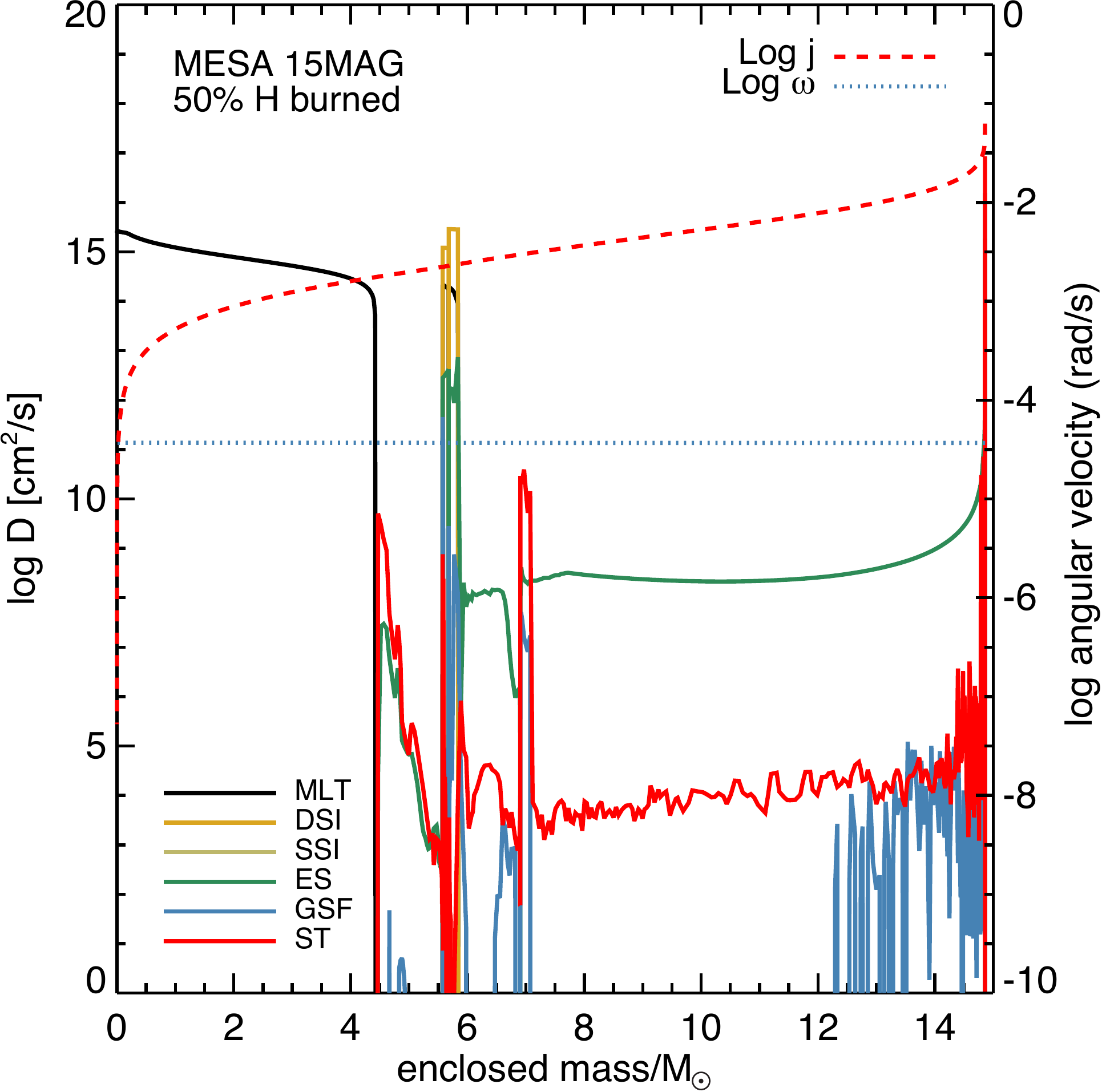}\hspace{\twoupsep}
\includegraphics[angle=0,width=\twoupwidth]{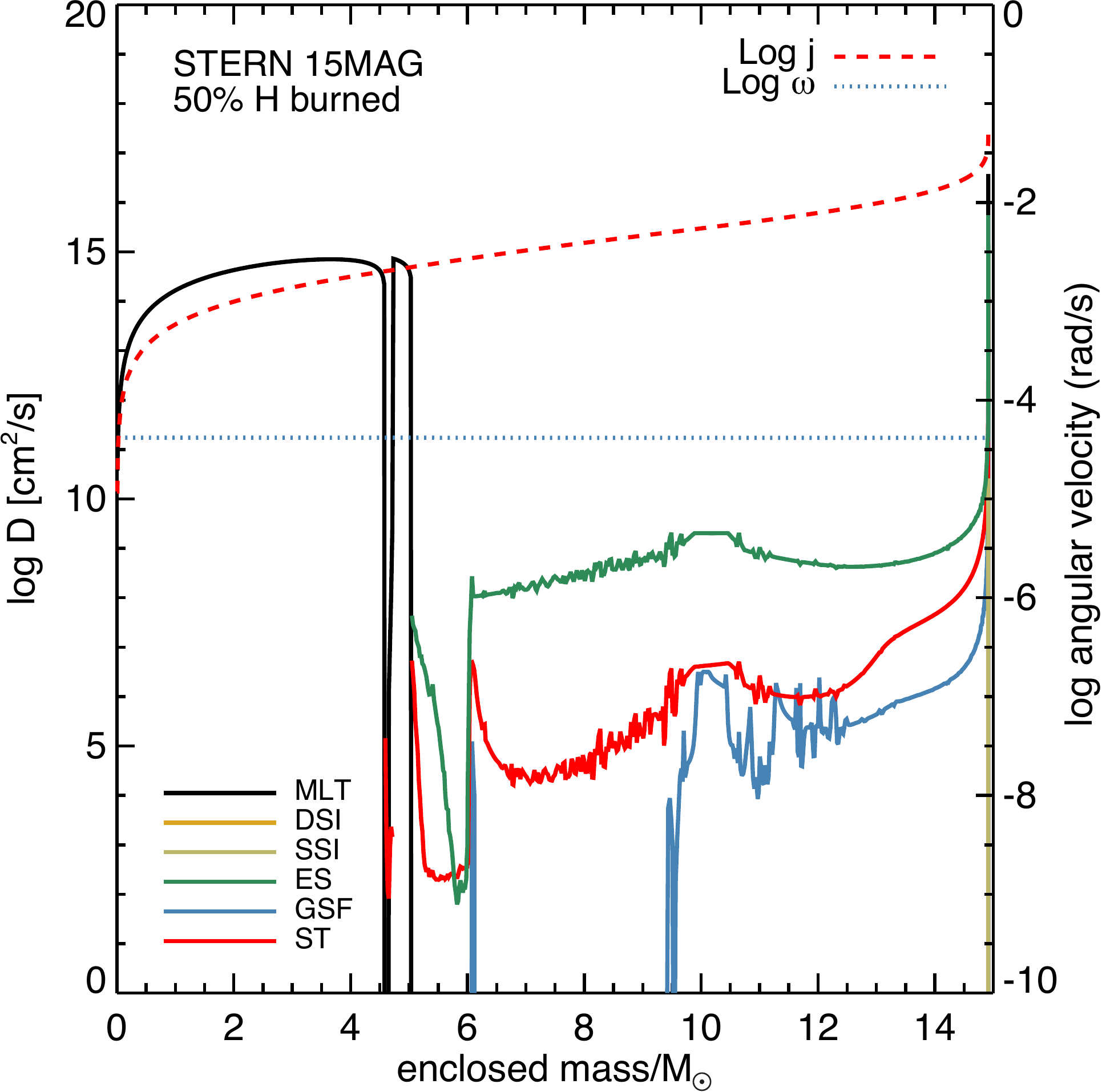}
}
\caption{ Same as Fig.~\protect\ref{fig:hrdfig15}.
As function of mass coordinate we plot the values of the diffusion coefficient for convection (MLT), Eddington-Sweet circulation (ES), 
magnetic torques by dynamo generated fields (ST), Dynamical Shear (DSI), Secular Shear (SSI) and Goldreich-Schubert-Fricke (GSF) instability. Following STERN, we turn off the Solberg-H{\o}iland instability (SH)  for this comparison. This does not affect the results, as the diffusion coefficient 
for SH is usually smaller than the ones for ES and ST. 
The values of the specific angular momentum $j$ and the angular velocity $\omega$ are also plotted.
Left panel shows the results using \MESAstar, while the right panel shows analogous STERN calculations.
\label{fig:msfig15}}
\end{figure*}

In Fig.~\ref{fig:hrdfig15} we show the evolutionary track and  the evolution of surface equatorial rotational velocity for the 15MAG  model.
Results of a similar calculation using STERN are shown as a dashed curve. The two results are in  excellent agreement. Small differences in 
luminosity and lifetimes are not unexpected, as we have only matched the physics of rotation between the two calculations and not other ingredients.
Values for the diffusion coefficients for rotationally induced mixing and magnetic torques during the main sequence of 15MAG are shown in Fig.~\ref{fig:msfig15}.
The comparison reveals a very good agreement. Both stars are kept in solid-body rotation during the main sequence
by the efficient transport of angular momentum provided by the Eddington-Sweet circulation and Spruit-Tayler magnetic fields. 

\begin{figure*}[htbp]
\centering
\begin{tabular}{cc}
\includegraphics[angle=0,width=\twoupwidth]{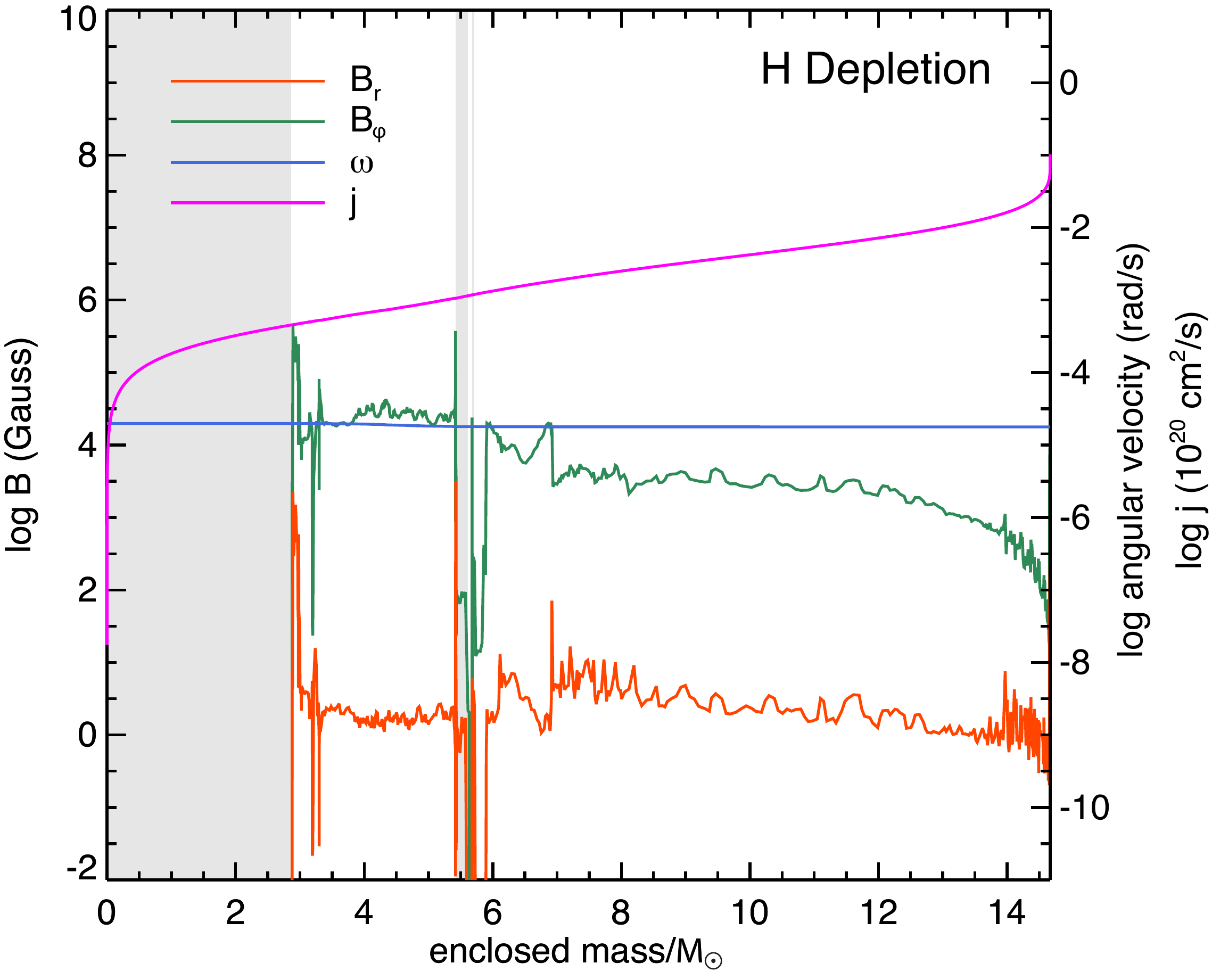} &
\includegraphics[angle=0,width=\twoupwidth]{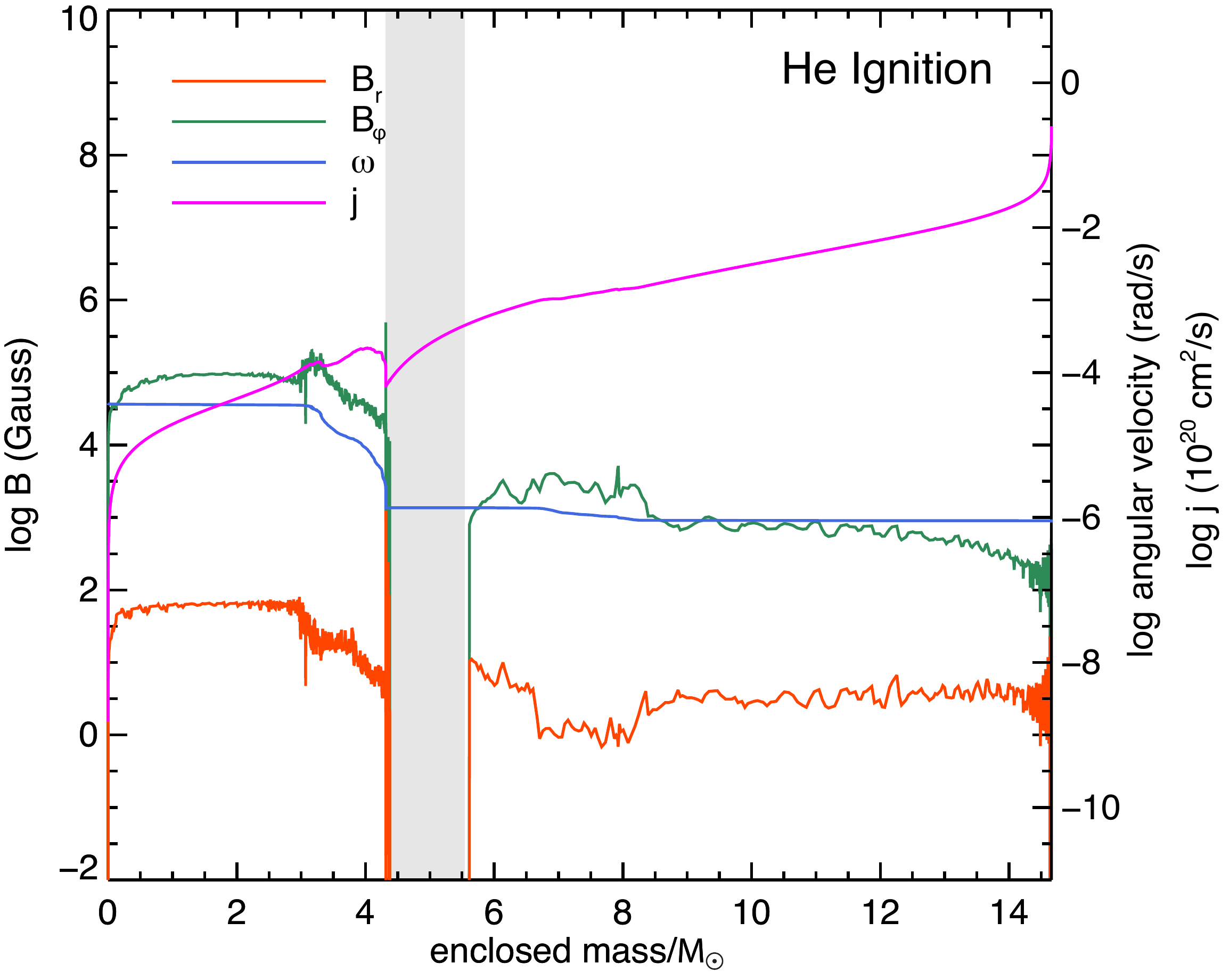} \\
\includegraphics[angle=0,width=\twoupwidth]{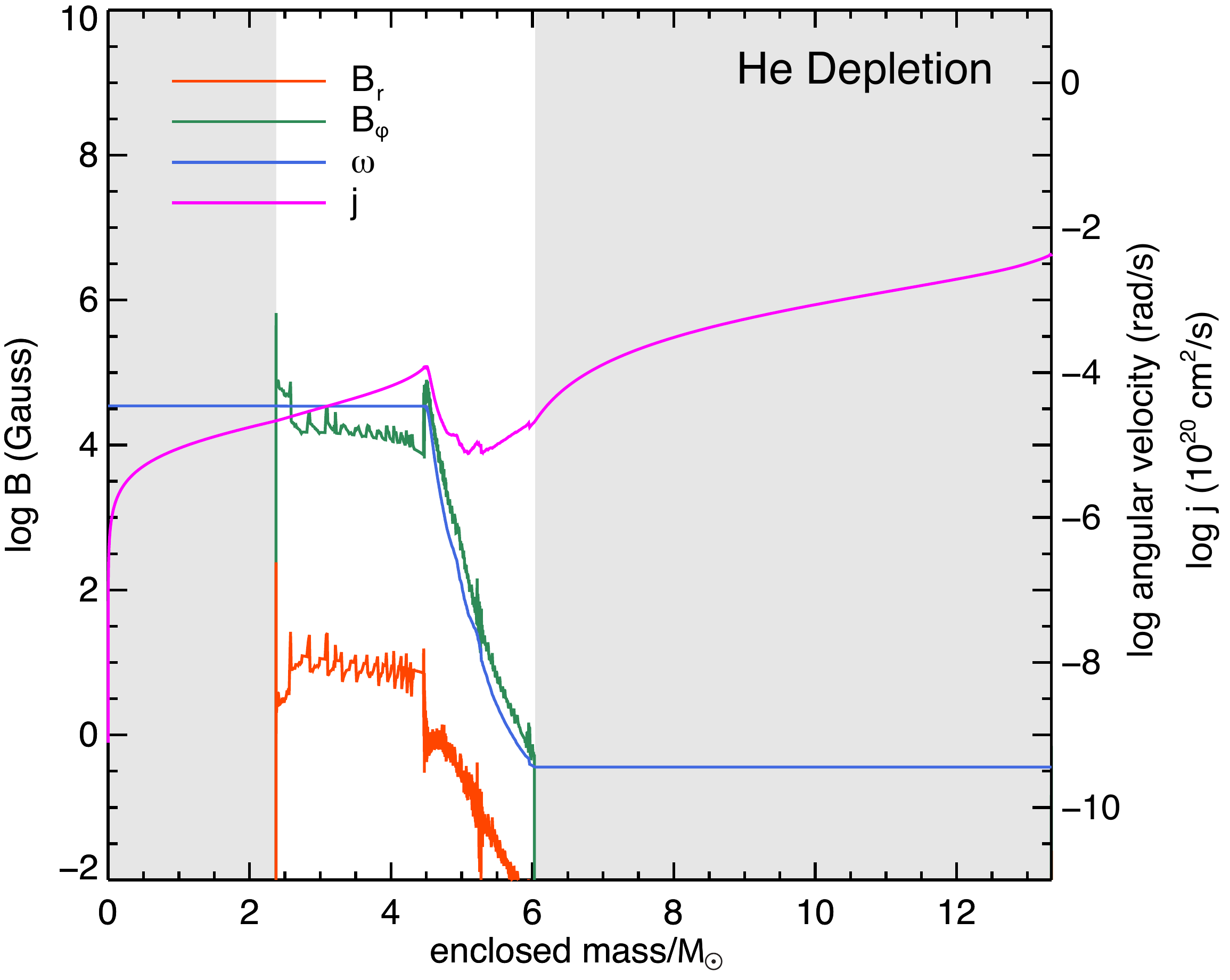} &
\includegraphics[angle=0,width=\twoupwidth]{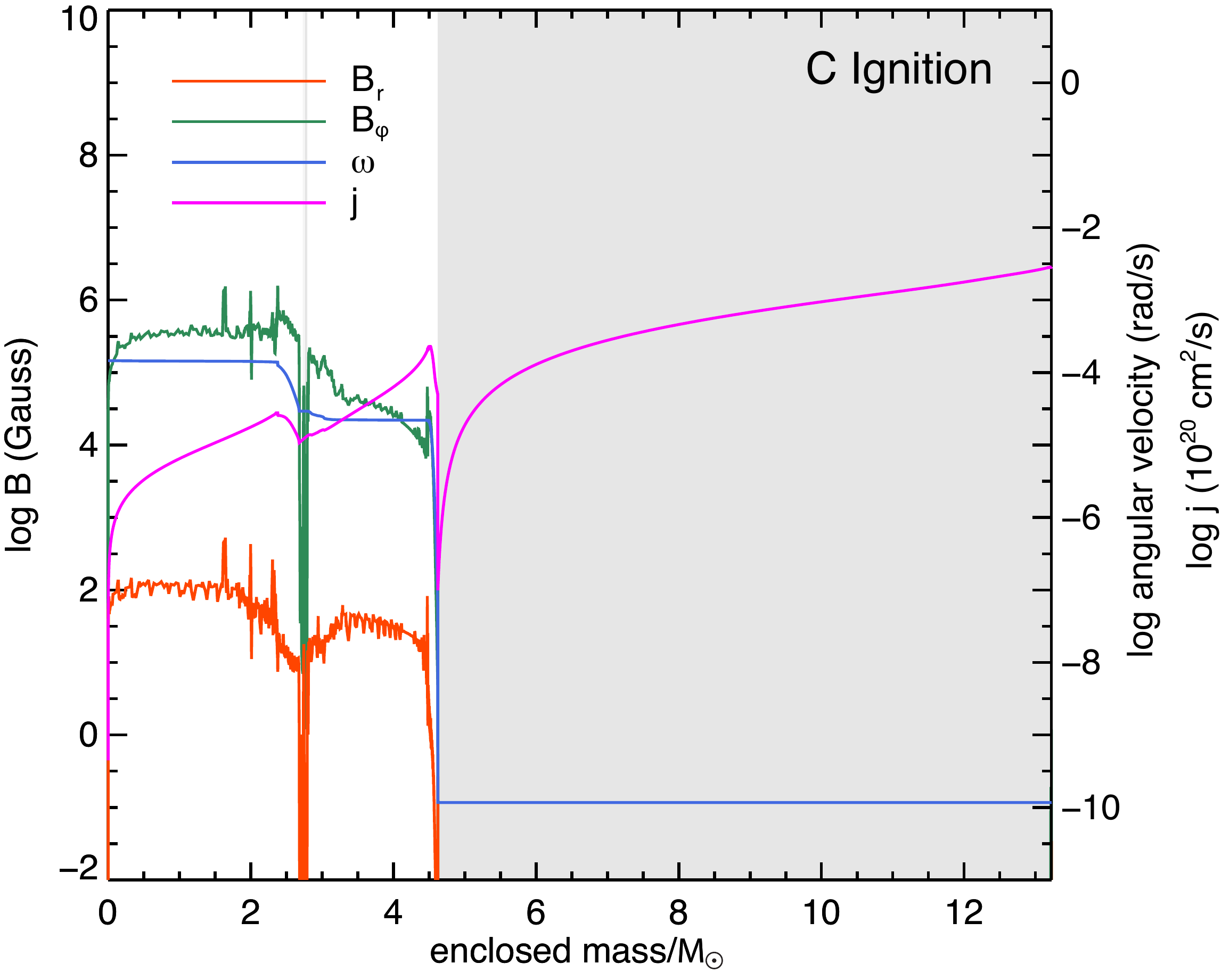}
\end{tabular}
\caption{ Magnetic field structure and angular momentum distribution for model 15MAG at different evolutionary stages (see Table~\ref{tab:15evo}). The curves show profiles for specific angular momentum ($j$), angular velocity ($\omega$), azimuthal and radial components of magnetic field ($B_{\phi}$ and $B_{r}$).
 The shaded regions represent convective parts of the star. 
Compare with Fig.\ 1 in
\citet{Heger:2005}. \label{fig:bfig15}}
\end{figure*}

\begin{figure*}[htbp]
\centering{
\includegraphics[angle=0,width=\twoupwidth]{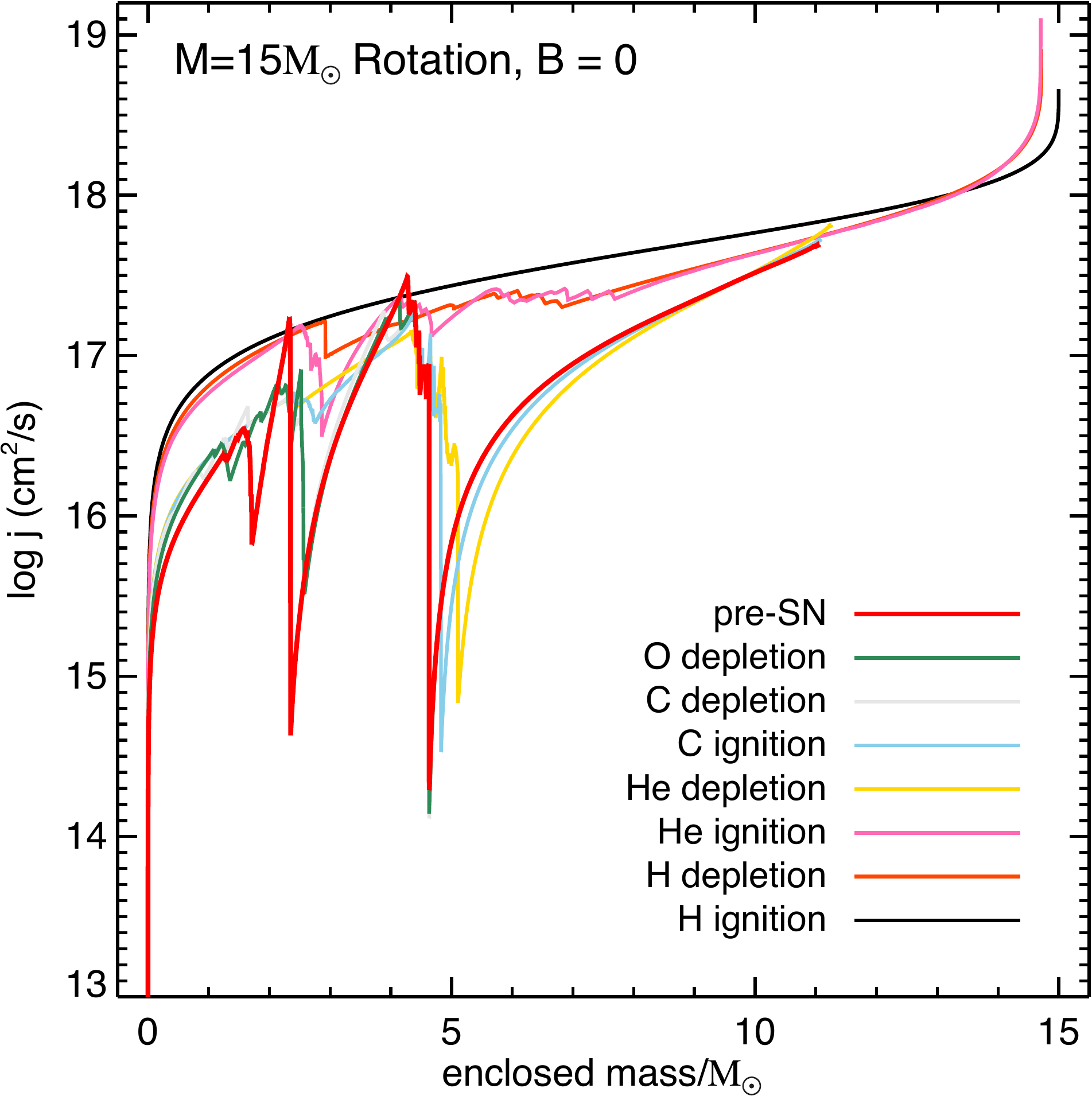}\hspace{\twoupsep}
\includegraphics[angle=0,width=\twoupwidth]{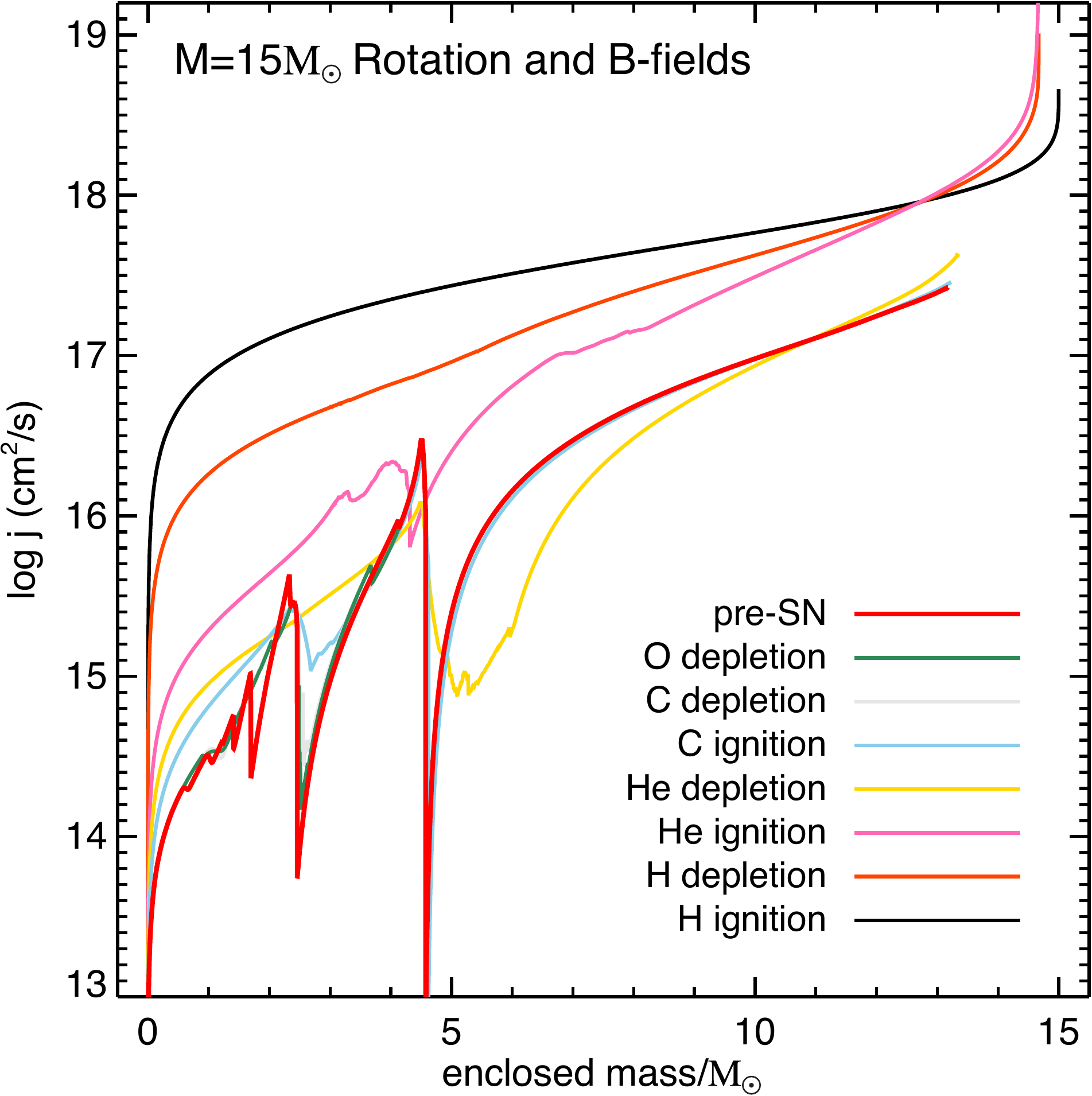}
}
\caption{Specific angular momentum distribution  at different evolutionary stage for 15MAG and 15ROT.  See \Tab{15evo} for
the definitions of these times.  Compare with Fig.~2 in
\citet{Heger:2005}. \label{fig:jfig15}}
\end{figure*}

The amplitude and location of the azimuthal ($B_{\phi}$) and radial ($B_{r}$) components of the magnetic fields during different phases of the evolution of 15MAG are shown in Fig.~\ref{fig:bfig15}.
As expected, these fields are generated only in radiative regions of the star and  $B_{\phi} > B_{r}$ \citep{Spruit:2002}. As the star evolves away from the main 
sequence its structure departs from solid-body rotation with the core rotating faster than the envelope. During this stage the role of magnetic fields is very important in transporting
 angular momentum from the core to the envelope. The effect can be seen in Fig.~\ref{fig:jfig15}, which shows the evolution of the internal specific angular momentum in
  models 15ROT and 15MAG. The presence of magnetic torques  results in a dramatic spin-down of the core of 15MAG with respect to 15ROT (see also Table~\ref{tab:15evo}).
These results are in very good agreement with the ones obtained by STERN and KEPLER.

\begin{table*}
\caption{\label{tab:15evo}Evolution of Angular Momentum at Fiducial Mass Coordinates for a $Z=0.02$, $15\,\Msun$ star initially rotating with $\veq$ = 200\,\kms\ with (15MAG) and without (15ROT) the inclusion of magnetic fields. 
}
\smallskip\centering{
\begin{tabular}{llcccccc}
\hline\hline\noalign{\smallskip}
Evolution Stage & & \multicolumn{3}{c}{\hrulefill 15MAG\hrulefill} 
& \multicolumn{3}{c}{\hrulefill 15ROT\hrulefill} 
\\  
& & J(1.5) & J(2.5) & J(3.5) 
& J(1.5) & J(2.5) & J(3.5)   \\
\hline\noalign{\smallskip}


ZAMS  &MESA      & 1.82\E{50} & 4.38\E{50} & 7.90\E{50} & 1.82\E{50} & 4.38\E{50} & 7.90\E{50} \\
      &KEPLER$^a$     & 1.75\E{50} & 4.20\E{50} & 7.62\E{50} & 2.30\E{50} & 5.53\E{50} & 1.00\E{51} \\
      &STERN$^b$     & 1.76\E{50} & 4.27\E{50} & 7.74\E{50} & 1.76\E{50} & 4.28\E{50} & 7.76\E{50} \\ 
H-burn$^c$ &MESA  & 1.25\E{50} & 3.03\E{50} & 5.51\E{50} & 1.64\E{50} & 3.99\E{50} & 7.26\E{50} \\
            &KEPLER     & 1.31\E{50} & 3.19\E{50} & 5.83\E{50} & 1.51\E{50} & 3.68\E{50} & 6.72\E{50} \\
            &STERN     & 1.21\E{50} & 2.96\E{50} & 5.40\E{50} & 1.62\E{50} & 3.97\E{50} & 7.25\E{50} \\
H-dep$^d$ &MESA  & 4.32\E{49} & 1.08\E{50} & 2.03\E{50} & 1.54\E{50} & 3.86\E{50} & 6.44\E{50} \\          
          &KEPLER     & 5.02\E{49} & 1.26\E{50} & 2.37\E{50} & 1.36\E{50} & 3.41\E{50} & 6.37\E{50} \\
          &STERN  & 4.81\E{49} & 1.21\E{50} & 2.29\E{50} & 1.48\E{50} & 3.74\E{50} & 6.99\E{50} \\
He-ign$^e$ &MESA  & 4.56\E{48} & 1.36\E{49} & 3.46\E{49} & 1.37\E{50} & 3.63\E{50} & 5.35\E{50} \\          
           &KEPLER  & 4.25\E{48} & 1.21\E{49} & 2.57\E{49} & 1.16\E{50} & 2.98\E{50} & 4.87\E{50} \\
&STERN  & 4.10\E{48} & 1.16\E{49} & 3.25\E{49} & 1.33\E{50} & 3.47\E{50} & 6.36\E{50} \\
He-burn$^f$ &MESA & 2.71\E{48} & 7.23\E{48} & 1.52\E{49} & 7.48\E{49} & 1.98\E{50} & 3.93\E{50} \\          
           &KEPLER & 2.85\E{48} & 7.84\E{48} & 1.83\E{49} & 7.06\E{49} & 1.85\E{50} & 3.86\E{50} \\
&STERN & 3.30\E{48} & 8.57\E{48} & 1.87\E{49} & 8.46\E{49} & 2.16\E{50} & 4.39\E{50} \\
He-dep$^g$ &MESA & 2.10\E{48} & 5.65\E{48} & 1.22\E{49} & 5.40\E{49} & 1.44\E{50} & 2.81\E{50} \\   
           &KEPLER & 2.23\E{48} & 5.95\E{48} & 1.21\E{49} & 4.72\E{49} & 1.26\E{50} & 2.52\E{50} \\
&STERN & 2.70\E{48} & 7.17\E{48} & 1.51\E{49} & 6.80\E{49} & 1.75\E{50} & 3.41\E{50} \\
C-ign$^h$  &MESA & 1.54\E{48} & 5.21\E{48} & 8.89\E{48} & 5.40\E{49} & 1.44\E{50} & 2.58\E{50} \\         
           &KEPLER & 1.88\E{48} & 5.52\E{48} & 1.12\E{49} & 4.69\E{49} & 1.26\E{50} & 2.46\E{50} \\
&STERN & 1.56\E{48} & 5.58\E{48} & 1.04\E{49} & 5.85\E{49} & 1.59\E{50} & 2.79\E{50} \\
C-dep$^i$ &MESA  & 7.54\E{47} & 3.84\E{48} & 6.71\E{48} & 5.11\E{49} & 1.39\E{50} & 2.09\E{50} \\        
          &KEPLER  & 8.00\E{47} & 3.26\E{48} & 9.08\E{48} & 4.06\E{49} & 1.25\E{50} & 2.24\E{50} \\
 &STERN  & 9.04\E{47} & 4.48\E{48} & 9.33\E{48} & 5.04\E{49} & 1.56\E{50} & 2.61\E{50} \\
 
O-dep$^j$ &MESA  & 7.52\E{47} & 3.71\E{48} & 6.41\E{48} & 4.61\E{49} & 1.37\E{50} & 1.97\E{50} \\
          &KEPLER  & 7.85\E{47} & 3.19\E{48} & 8.43\E{48} & 3.94\E{49} & 1.20\E{50} & 1.99\E{50} \\

Si-dep$^k$ &MESA  & 7.28\E{47} & 3.64\E{48} & 5.90\E{48} & 4.03\E{49} & 1.22\E{50} & 1.76\E{50} \\
          &KEPLER  & 7.76\E{47} & 3.05\E{48} & 7.23\E{48} & 3.75\E{49} & 1.16\E{50} & 1.95\E{50} \\
\hline
\end{tabular}}
\smallskip\\
NOTE: 
$^a$ Results from Table~1 of \citet{Heger:2005};
$^b$ See e.g., \citet{Petrovic:2005,Yoon:2005,Yoon:2006};
$^c$40\,\% central hydrogen mass fraction; 
$^d$1\,\% hydrogen left in the core; 
$^e$1\,\% helium burnt;  
$^f$50\% central helium mass fraction; 
$^g$1\,\% helium left in the core; 
$^h$central temperature of 5\E8\,\K; 
$^i$central temperature of 1.2\E9\,\K; 
$^j$central oxygen mass fraction drops below 5\,\%;
$^k$central Si mass fraction drops below \Ep{-4}; 
\end{table*}

\begin{figure*}[htbp]
\centering{
\includegraphics[angle=0,width=\twoupwidth]{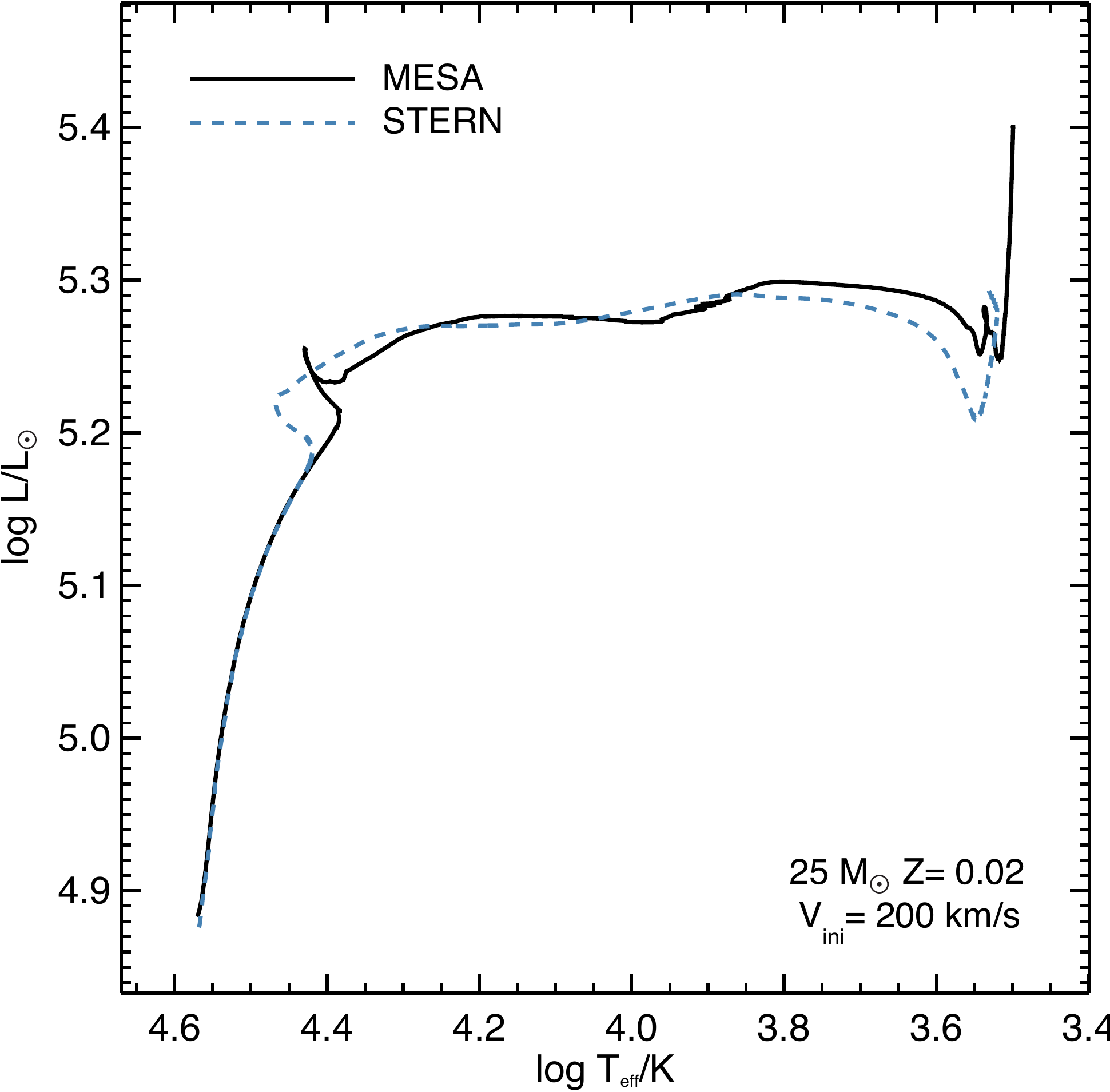}\hspace{\twoupsep}
\includegraphics[angle=0,width=\twoupwidth]{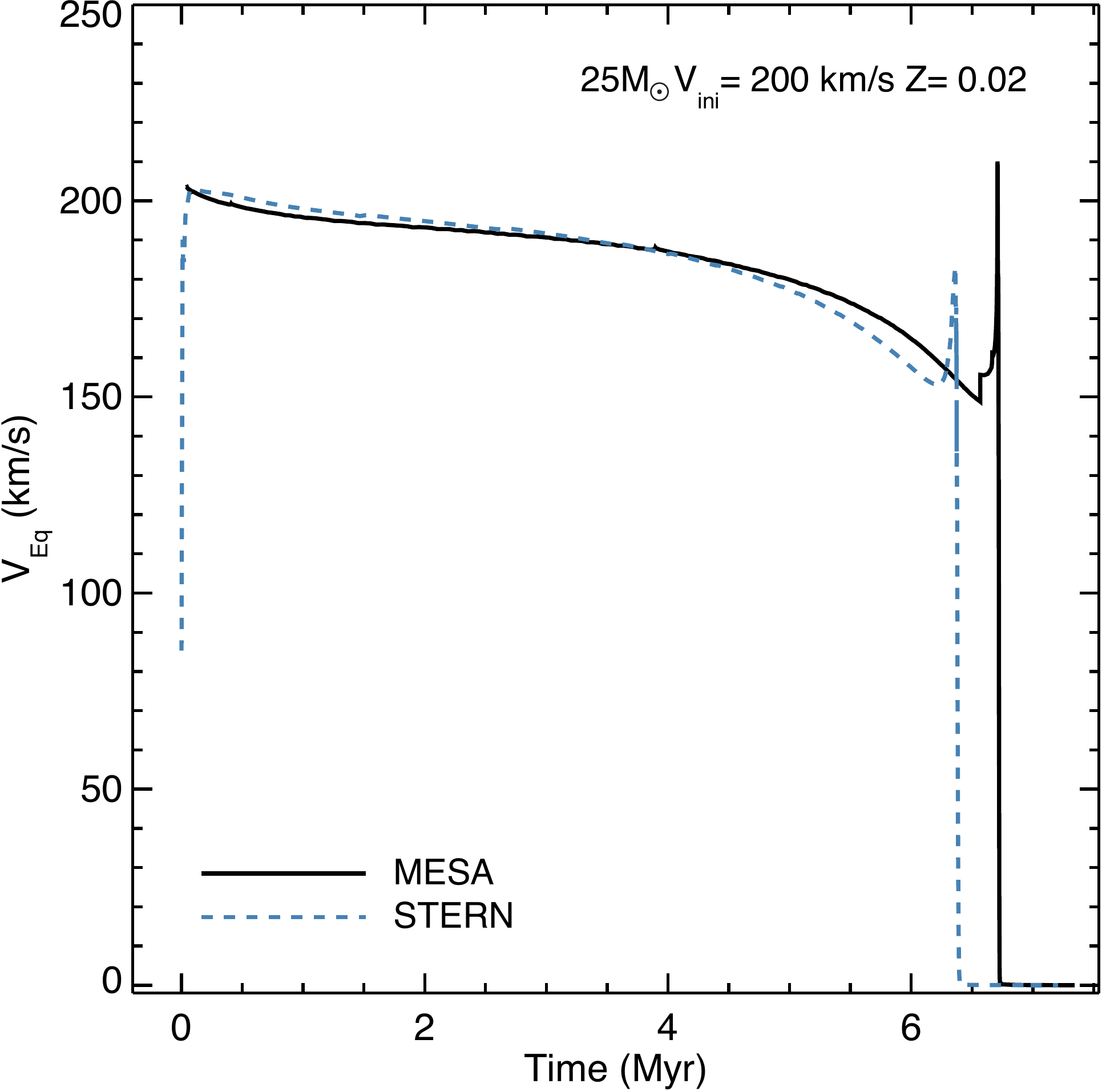}
}
\caption{ Same as Fig.~\ref{fig:hrdfig15}, except for the 25MAG model.
\label{fig:hrdfig25}}
\end{figure*}

\begin{figure*}[htbp]
\centering{
\includegraphics[angle=0,width=\twoupwidth]{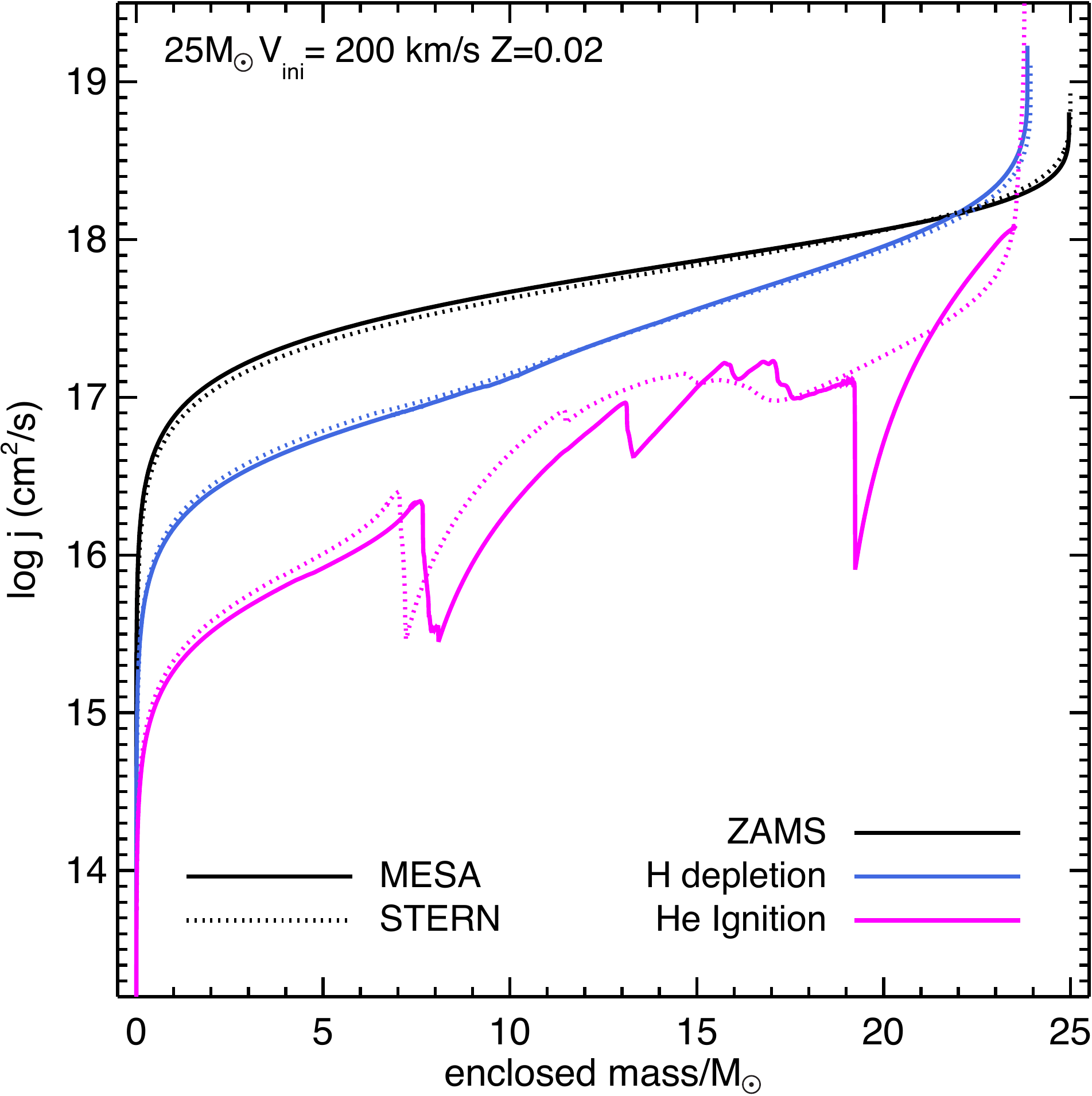}\hspace{\twoupsep}
\includegraphics[angle=0,width=\twoupwidth]{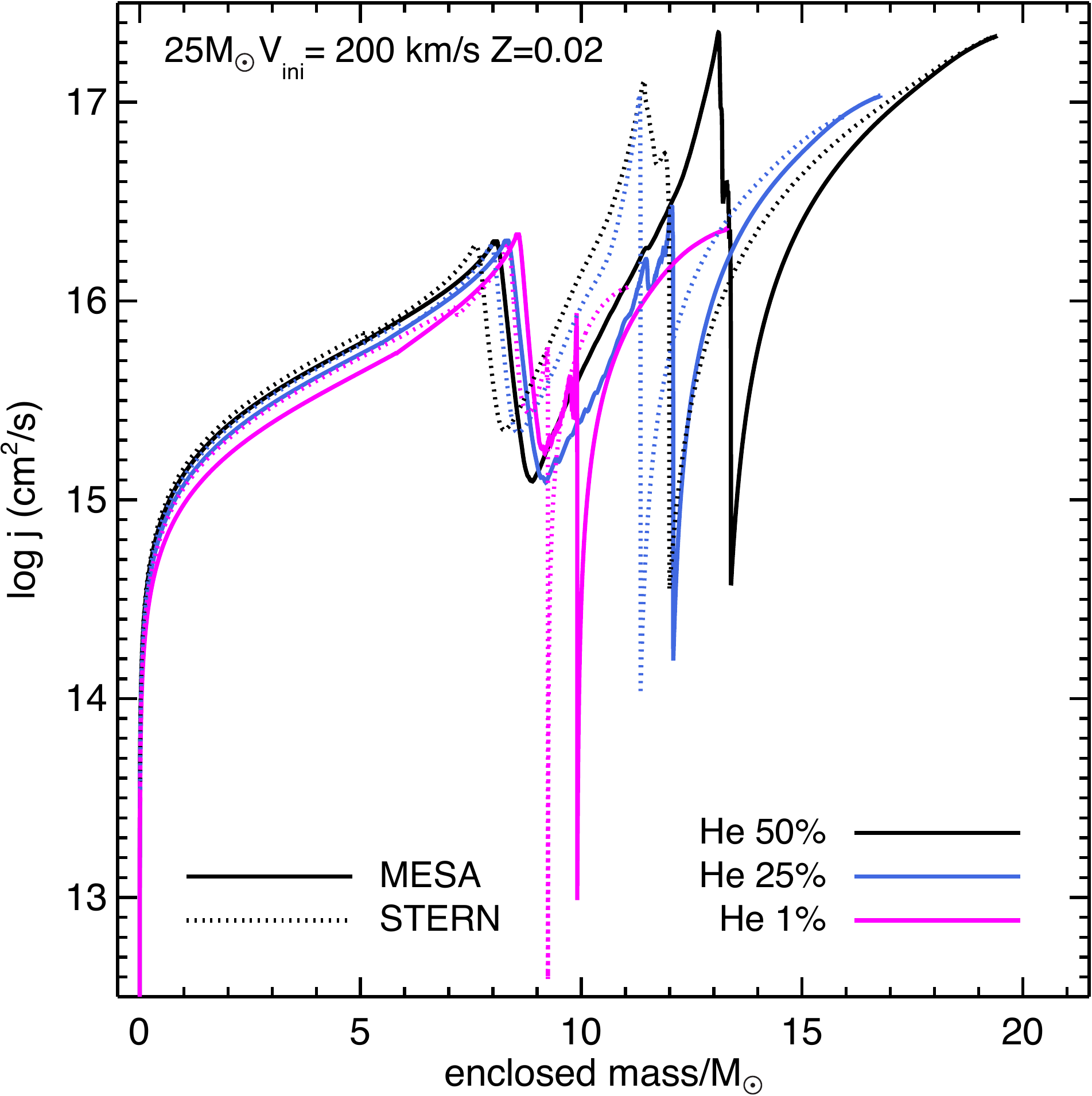} 
}
\caption{ Evolution of internal specific angular momentum for the 25MAG model. Solid lines show \MESAstar\ result, while dashed lines refer to STERN. Left panel shows the evolution from zero age main sequence to He ignition.
Right panel shows the evolution during core He-burning (from 50\% of He in the core to He depletion). Notice the different axis range in the two plots.
\label{fig:j25comparison}}
\end{figure*}

\begin{figure}[htbp]
\center
\includegraphics[width=\figwidth]{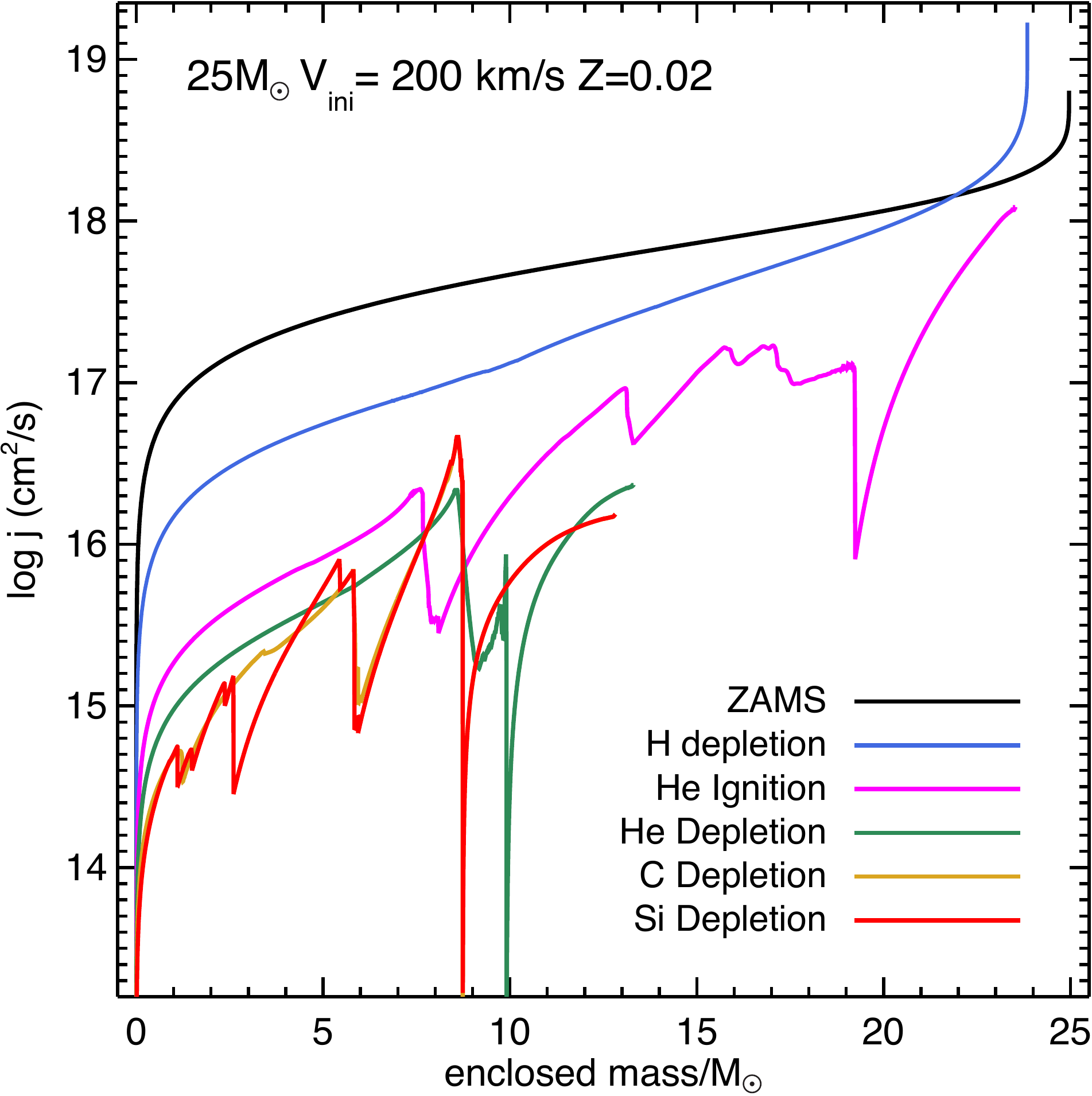} 
\caption{ Evolution to Si-depletion of the internal specific angular momentum for the 25MAG model.
\label{fig:j25all}}
\end{figure}

As a second test, we now evolve a $25\nsp\Msun$ model (25MAG) with the same physics as in 15MAG. 
Figure~\ref{fig:hrdfig25} directly compare results with calculations performed with STERN.
In Fig.~\ref{fig:j25comparison} we show a detailed comparison of the evolution of the internal specific angular momentum profile.
We find a very good quantitative agreement between \MESAstar\ and STERN down to He depletion in the core.
The timescale for nuclear burning decreases substantially after He-burning and becomes shorter than the angular momentum transport timescale after C depletion.
Thus only minor changes in the final angular momentum content of the stellar core are expected after this stage. 
Figure \ref{fig:j25all} shows the full  evolution of the specific angular momentum profile of the \MESAstar\ calculation from ZAMS to Si exhaustion.

\subsection{Rapidly Rotating Massive Stars}\label{s.rapid}

\MESAstar\ can calculate the evolution to core collapse of rapidly rotating massive stars.
Rotational instabilities can be efficient enough to erase the compositional gradients built by nuclear burning. 
In such cases the model never develops a compositional stratification and remains almost completely mixed throughout its evolution \citep{Maeder:1987a,Yoon:2005,Woosley:2006}.
This process leads to a bifurcation in the HR-diagram, with stars above a certain mass and rotation rate becoming more luminous and hotter.
The threshold required for this bifurcation depends mostly on the initial mass of the star \citep{Yoon:2006}.  Metallicity also plays an important role, as angular momentum is lost through line-driven stellar winds, with mass-loss rates depending on the metallicity at the stellar surface \citep{Vink:2001}.
For the calculations in this section, we adopt the same mass-loss prescription of \citet{Yoon:2006}.

\begin{figure*}[htbp]
\centering{
\includegraphics[angle=0,width=\twoupwidth]{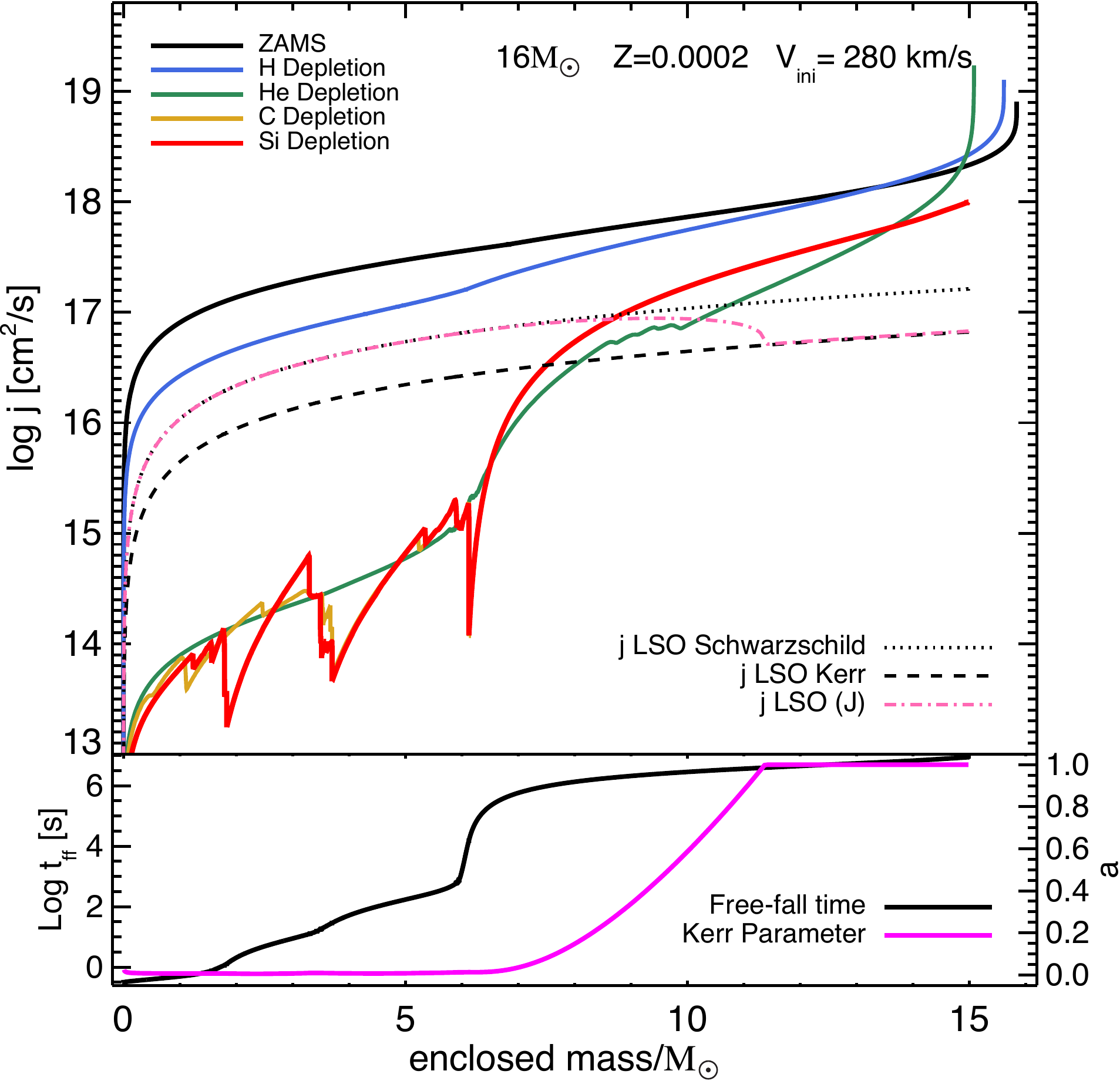}\hspace{\twoupsep}
\includegraphics[angle=0,width=\twoupwidth]{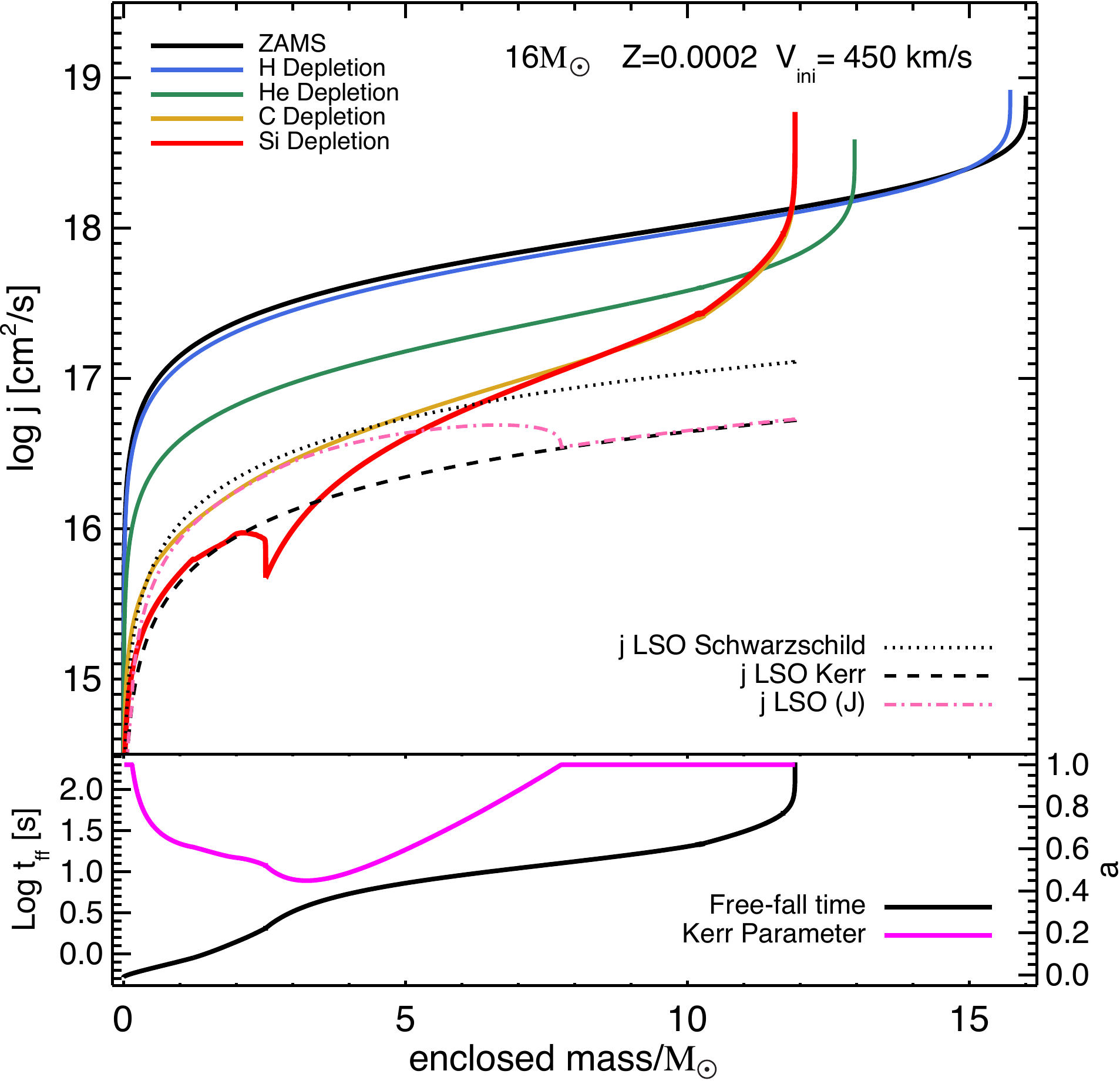} 
}
\caption{ Specific angular momentum distribution for the two
$16\,\Msun$ models.
In the top panels, the solid curves show the distribution of specific angular
momentum at different evolutionary stages.  The other curves in the top panel show the specific angular momentum of the last stable orbit around a Schwarzschild black hole, a maximally rotating
Kerr black hole ($a=1$), and a black hole with a Kerr parameter corresponding to the angular momentum content of the stellar progenitor at that mass coordinate.
The bottom panels show the free-fall time at the relative mass coordinate at the end of Si-burning. Notice the different ranges of the y-axis.
These models can be compared to the calculations of \citet{Woosley:2006}, in particular their models 16SG and 16TI respectively. \label{fig:jfig16ff}}
\end{figure*}

Figure~\ref{fig:jfig16ff} shows the evolution of two $16\,\Msun$ models at metallicity $Z=0.0002$ with rotation initialized at the ZAMS.
One model is rotating very rapidly, with $\veq=450\,\kms$ (corresponding to $\Om/\Omc=0.55$ and  $J=3.23\times10^{52}\,\ergssecond$), while the other rotates at $\veq=280\,\kms$ (corresponding to $\Om/\Omc=0.39$ and $J=2.52\times10^{52}\,\ergssecond$).
The model with $\Om/\Omc=0.55$ avoids the core-envelope structure and  becomes a compact Wolf-Rayet star. 
The absence of a RSG phase eliminates the large magnetic torques from an extended envelope.
The evolution of the internal profile of specific angular momentum in the two models clarify this point: the model with $\Om/\Omc=0.39$ becomes a RSG, and the core spins down rapidly. When it reaches core-collapse its structure is extended, as implied by the large free-fall timescale shown in the left panel of Fig.~\ref{fig:jfig16ff}.
As a consequence, there is not enough angular momentum in its core to build an accretion disk around a newly formed compact object. This model is expected to produce a Type~IIP supernova. 
On the contrary, the model with $\Om/\Omc=0.55$ is compact  (the free-fall timescale is on the order of seconds, right panel of Fig.~\ref{fig:jfig16ff}) and has enough angular momentum to produce an accretion disk around the central compact object.
Therefore this model is a candidate progenitor for a long gamma-ray burst \citep{Woosley:1993}. This last calculation can be directly compared to the KEPLER model 16TI in \citet{Woosley:2006}.  

\begin{figure}[htbp]
\newcommand{\panelwidth}{\figwidth}
\centering
\includegraphics[angle=0,width=\panelwidth]{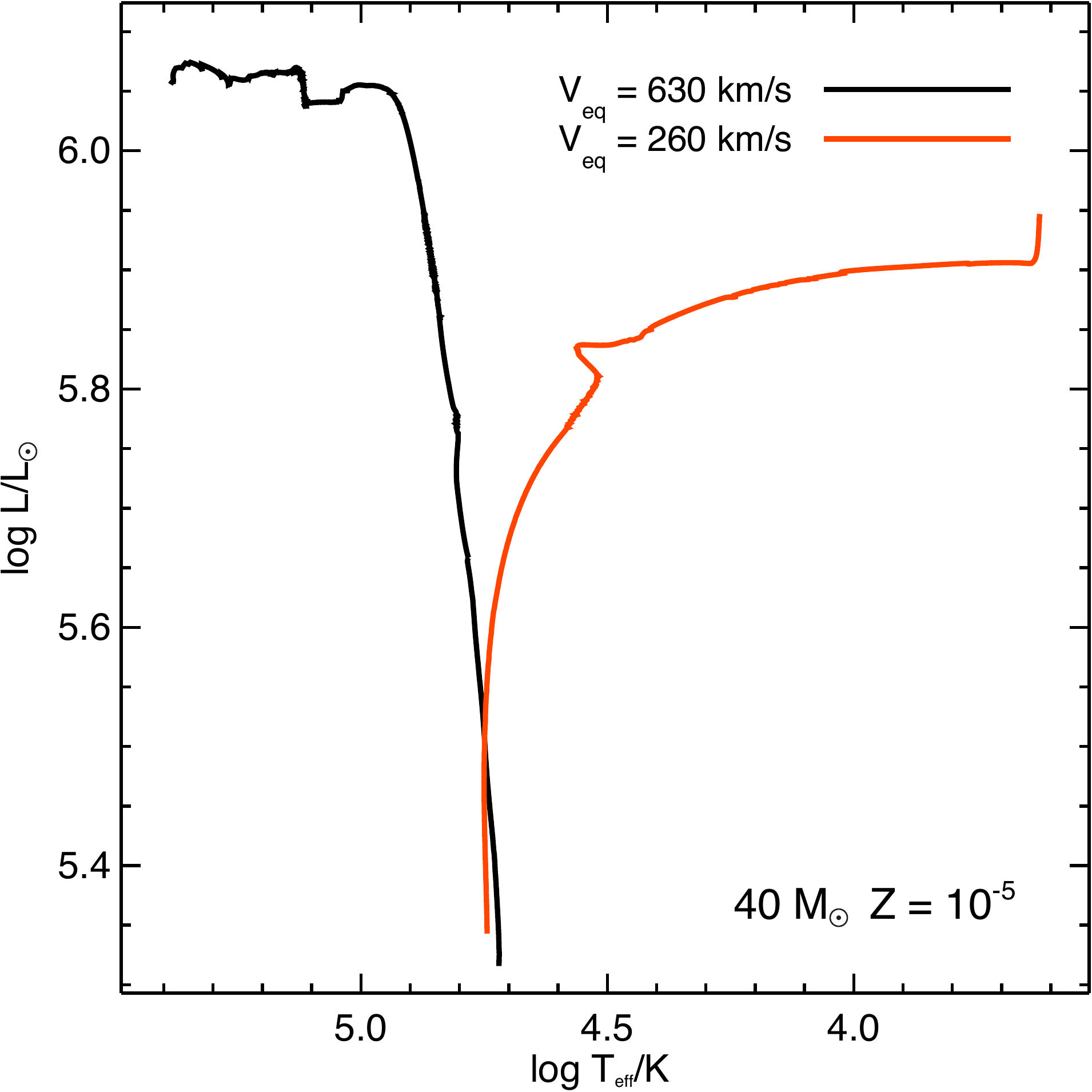}
\caption{Evolution in the HR diagram for two rotating $40\,\Msun$ models at $Z=10^{-5}$. The slower rotating model evolves 
toward the red part of the HRD;  the other model evolves 
toward the blue part of the HRD. The internal evolution of the angular momentum is shown in Fig.~\ref{fig:jfig40}.
This can be compared to Fig.~2 of \citet{Yoon:2005}.
\label{fig:hrd40}}
\end{figure}

\begin{figure*}[htbp]
\centering{
\includegraphics[angle=0,width=\twoupwidth]{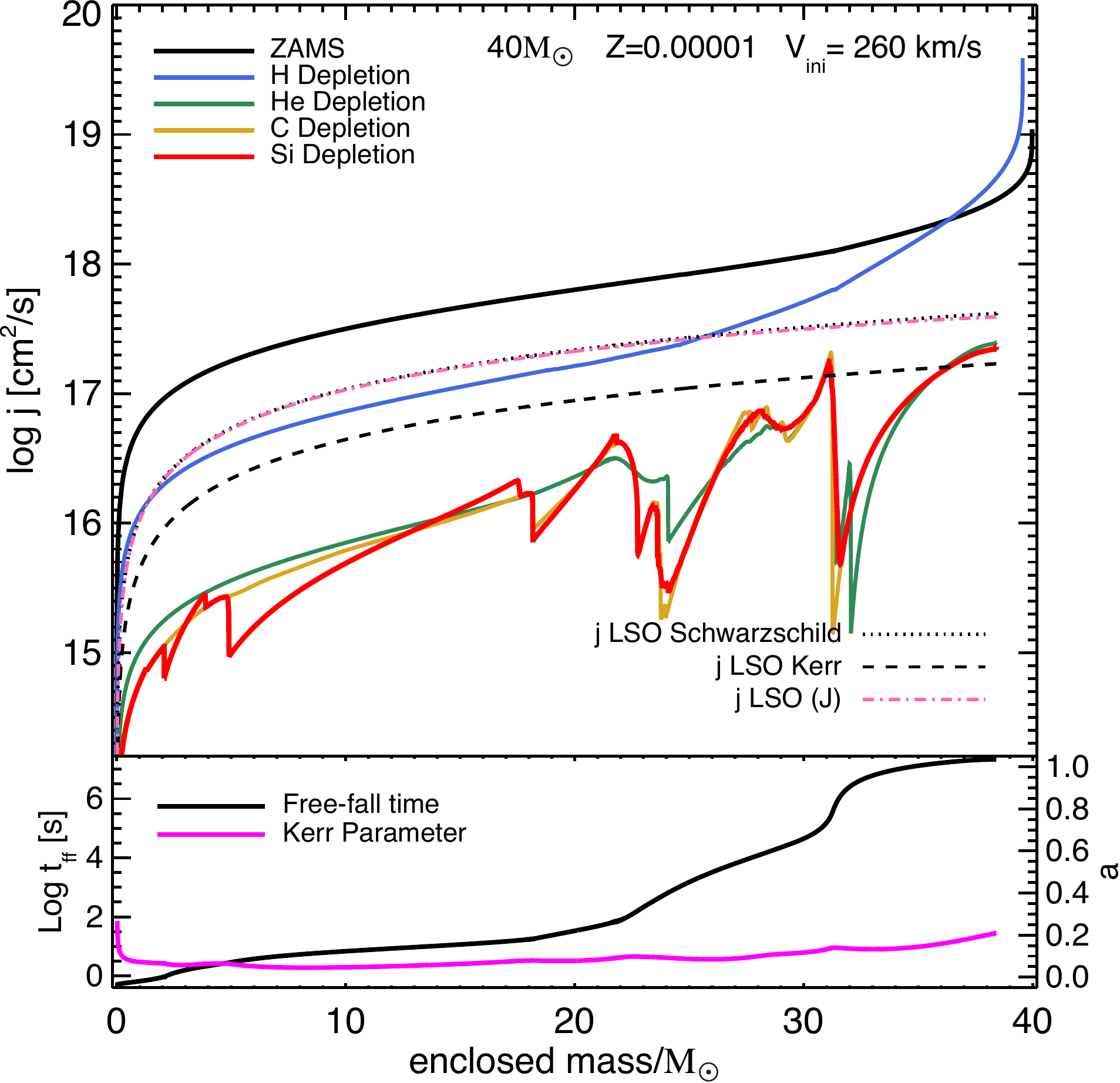}\hspace{\twoupsep}
\includegraphics[angle=0,width=\twoupwidth]{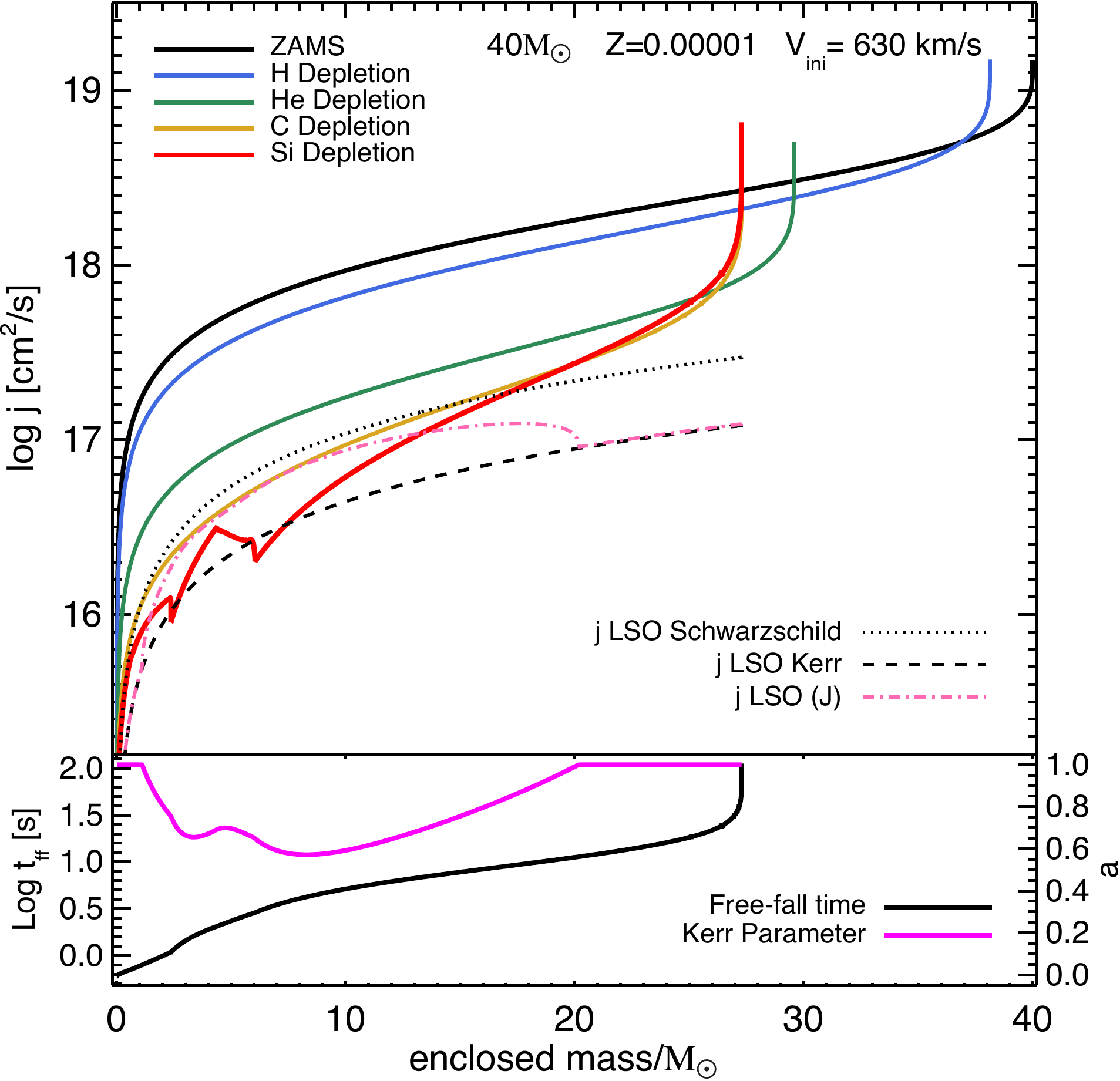} 
}
\caption{ Specific angular momentum distribution for two
$40\,\Msun$ models at $Z=10^{-5}$ with $\veq= 260\,\kms$ (top panel) and $630\,\kms$ (bottom panel).
Lines are showing the distribution of specific angular momentum at different stages of the evolution, together with the specific angular momentum of the last stable orbit around a Schwarzschild black hole, a maximally rotating
Kerr black hole ($a=1$) and a black hole with a Kerr parameter corresponding to the angular momentum content of the stellar progenitor at that mass coordinate. Note the different ranges of the y-axis.
The evolutionary tracks for these models are shown in Fig.~\protect\ref{fig:hrd40}. These calculations should be compared to Fig.~5 of \citet{Yoon:2005}. \label{fig:jfig40}}
\end{figure*}

We further test \MESA\ capabilities by evolving  two rotating $40\,\Msun$ models at $Z=10^{-5}$.
One model is initialized at the ZAMS with $\veq= 260\,\kms$, while the other has  $\veq=630\,\kms$.
The results of these calculations can be compared with the models shown in \citet{Yoon:2005}.
Figure~\ref{fig:hrd40} shows that for the more rapidly rotating model, rotational mixing (mainly due to the Eddington-Sweet circulation) is large enough that the star evolves blueward in the HR-diagram.
This evolution results in a compact configuration and enough angular momentum to fulfill the requirements of the collapsar scenario for long gamma-ray bursts, as shown in Fig.~\ref{fig:jfig40} (right panel). 
On the other hand, the slower rotating model becomes a RSG and loses most of its core angular momentum, as shown in Fig.~\ref{fig:jfig40} (left panel).

\section{Massive Stellar Evolution}\label{s.massive}
Modeling massive stars is numerically difficult. One problem is they develop loosely bound, radiation pressure dominated envelopes that can cause density and gas pressure inversions. 
Indeed, very massive stars are observed to suffer sporadic ``eruptions'' of extreme mass loss (i.e., the Luminous Blue Variables), and the tendency to form inversions has been speculatively mentioned as playing a role in such episodes \citep[see][and references therein]{Humphreys1994The-luminous-bl}. 
 This environment poses a physical and numerical challenge that all stellar evolution codes must address to evolve massive stars past the main sequence. In this section we discuss \MESAstar's capability to evolve rotating massive stars from their zero age main sequence to core-collapse.  

\subsection{Evolution of Massive Stars with \MESA}\label{s.massive-evol}

Previous computations with \MESAstar\ found these envelopes to be numerically (and probably physically) unstable.
This is a known issue in the literature \citep[e.g.,][]{Maeder:1987}, which  reveals the limitations of the 1D treatment of late phases of evolution of massive stars.
 The evolution of stars with radiation-dominated envelopes can require prohibitively short timesteps in \MESAstar\ if the standard mixing length theory is adopted. 
This problem usually appears during the evolution of high mass and/or high metallicity stars after hydrogen-core burning and prevents evolution to core collapse.
We discuss in \ref{s.superadiabatic} our treatment of superadiabatic convection in these envelopes, which allows uninterrupted evolution,  from ZAMS to core collapse.  

\begin{figure*}[htbp]
\centering \includegraphics[width=\doublewide]{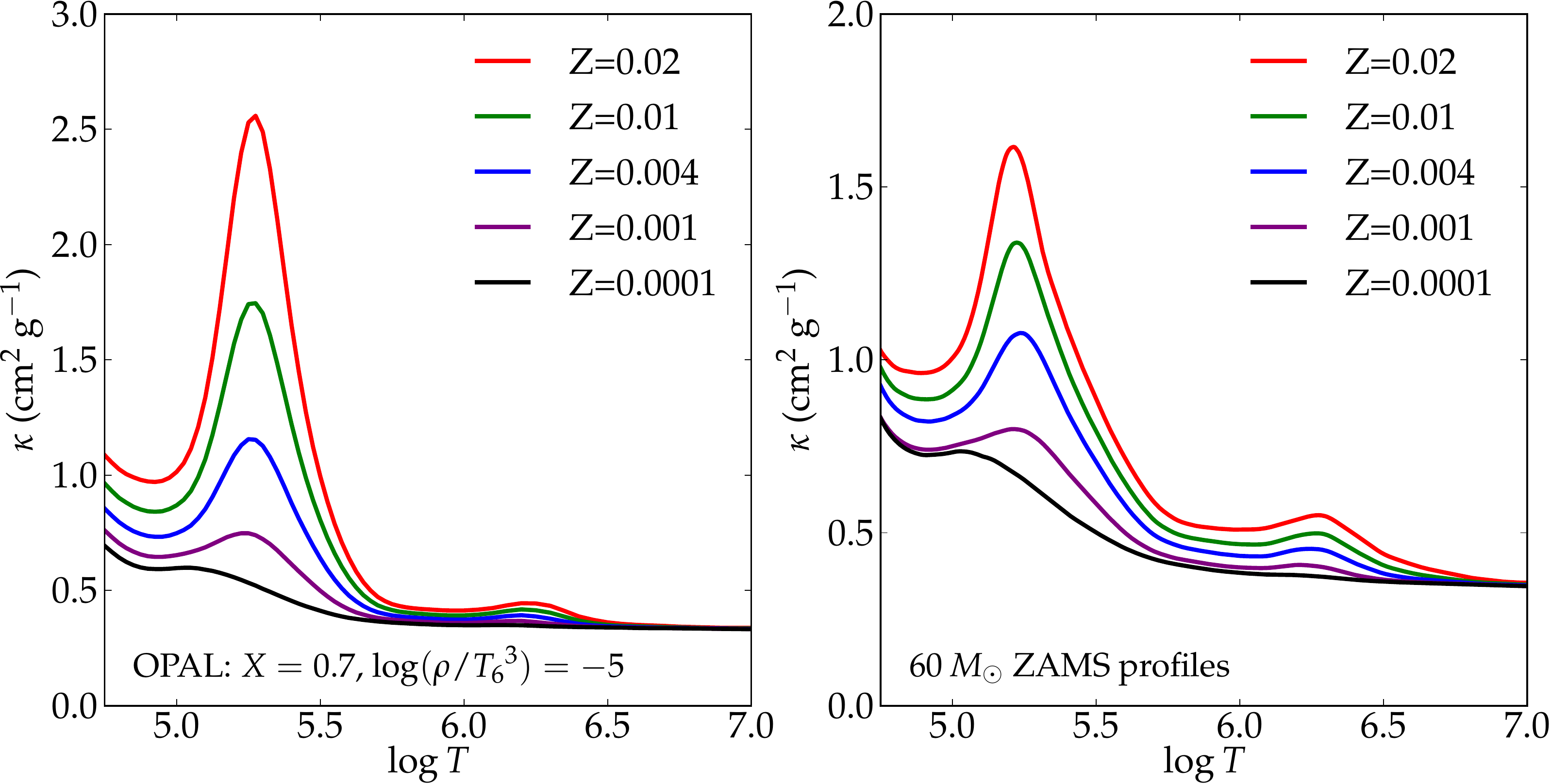}
\caption{\emph{Left:} A plot of the OPAL opacity data for five $Z$ values at $X=0.7$ and $\log (\rho/{T_6}^3) = -5$.
These curves show the increase in the iron opacity bumps at $\log T \approx 5.3$ and 6.3 as $Z$ 
increases from $10^{-4}$ to 0.02. \emph{Right:} The opacity-temperature profiles of $60\,\Msun$ ZAMS models for the same $Z$ values.}
\label{opacity}
\end{figure*}

Since it is relevant to later discussions we start with a plot of the OPAL opacity data \citep{OPAL1996}
and $60\,\Msun$ ZAMS models in Figure \ref{opacity}. The plot is inspired by
Figure 1 of \citet{Cantiello:2009}. The left-hand panel of Figure \ref{opacity} shows the OPAL data 
for five different $Z$ values at constant $X=0.7$ and $\log (\rho/{T_6}^3)=-5$, where  $T_6$ is the temperature in units of $10^{6}\,\Kelvin$. The right-hand panel shows the
opacity  profiles of five $60\,\Msun$ ZAMS models for the same five $Z$ values.
 The model profiles exhibit the same general behavior in the 
opacity-temperature profile as the raw opacity data. Of particular importance are the iron opacity bumps that 
occur at  $\log T \approx 5.3$ and 6.3. These bumps cause both the local radiation pressure to dominate and the luminosity to approach the Eddington luminosity \Ledd. 

Where both the pressure is dominated by radiation and $\Lrad$ approaches \Ledd, specific conditions can be reached that cause convection and inversions in density and gas pressure. To define the conditions under which these occur, we follow the discussion of \citet{Joss1973On-the-critical}, going from high to low $\Lrad$. We assume that $\dif T/\dif r < 0$, $\dif P/\dif r < 0$, and that the inertial terms in the momentum equation are small. First, we establish a condition for the occurrence of an inversion in the gas pressure \Pgas.
Recasting the equation for the temperature gradient gives
\begin{equation}
\Lrad = -\frac{4\pi r^{2} c}{\rho\kappa}\frac{\dif\Prad}{\dif r},
\end{equation}
and using the equation of hydrostatic equilibrium, one obtains
\begin{equation}\label{e.dPraddP}
\frac{\dif\Prad}{\dif P} = \frac{\Lrad}{\Ledd}.
\end{equation}
Writing $\dif \Pgas/\dif r$ = $\dif (P-\Prad)/\dif r$ and using equation~(\ref{e.dPraddP}) and the fact that both $\Prad$ and $P$ monotonically decrease with $r$, one obtains
\begin{equation}
\frac{\dif\Pgas}{\dif r}
	= \left(\frac{\dif\Prad}{\dif r}\right)\left[\frac{\Ledd}{\Lrad}-1\right].
\label{e.dPgasdr}
\end{equation}
Since $\dif\Prad/\dif r < 0$, equation~(\ref{e.dPgasdr}) implies that for $\Lrad > \Ledd$, the gas pressure gradient will increase outward, $\dif\Pgas/\dif r > 0$, as shown by \citet{Joss1973On-the-critical}.

The next step is to establish the condition for a density inversion to occur. Writing the gas equation of state as $\Pgas = \Pgas(\rho, \Prad)$ gives
\begin{equation}\label{e.dPgasdr2}
\frac{\dif\Pgas}{\dif r} = \tderiv{\Pgas}{\rho}{\Prad}\frac{\dif\rho}{\dif r} + \tderiv{\Pgas}{\Prad}{\rho}\frac{\dif\Prad}{\dif r}.
\end{equation}
Solving eq.~(\ref{e.dPgasdr2}) for $\dif\rho/\dif r$ and using eq.~(\ref{e.dPgasdr}) eliminates $\dif\Pgas/\dif r$. Gas equations of state have $(\partial \Pgas/\partial\rho)_{T} > 0$, so that for $\dif \rho/\dif r > 0$ (a density inversion), one must have
\[ \left(\frac{\dif\Prad}{\dif r}\right) \left[ \frac{\Ledd}{\Lrad} - 1 - \tderiv{\Pgas}{\Prad}{\rho}\right] > 0. \]
Recognizing that $\dif\Prad/\dif r < 0$, we find that a density inversion occurs when 
\begin{equation}\label{e.luminosity-density-inversion}
\frac{\Lrad}{\Ledd} > \frac{\Lrho}{\Ledd} \equiv \left[ 1 + \tderiv{\Pgas}{\Prad}{\rho} \right]^{-1}.
\end{equation}
This equation is identical to eq.~(8) of \citet{Joss1973On-the-critical}. Since under conditions of interest $(\partial\Pgas/\partial\Prad)_{\rho} >0$, we have $\Lrho < \Ledd$. For $\Lrho < \Lrad < \Ledd$, a density inversion will occur even though $\dif\Pgas/\dif r < 0$. 

Next, we shall consider the luminosity \Lonset\ at which convection occurs.  In a convective region, the entropy is either constant or declining with radius.  Hence, convection will occur once
\begin{equation}
\frac{\dif \ln\Prad}{\dif\ln P} > \tderiv{\ln\Prad}{\ln P}{s};
\end{equation}
using equation~(\ref{e.dPraddP}) and solving for the luminosity, we find that  convection starts once
\begin{equation}\label{e.convection-onset}
\frac{\Lrad}{\Ledd} > \frac{\Lonset}{\Ledd} \equiv \left(1 - \frac{\Pgas}{P}\right) \tderiv{\ln\Prad}{\ln P}{s}.
\end{equation}
Equation~(\ref{e.convection-onset}) corresponds to eq.~(9) of \citet{Joss1973On-the-critical}.  As argued in that paper, entropy decreases as density increases; therefore a density inversion implies a superadiabatic gradient, and as a result, $\Lonset < \Lrho$.  This can be shown explicitly for a chemically homogenous mixture of an ideal gas and radiation.  For such a mixture, equation~(\ref{e.luminosity-density-inversion}) becomes
\begin{equation}\label{e.Lrho-ideal}
 \frac{\Lrad}{\Ledd} > \frac{\Lrho}{\Ledd} = \left[ \frac{1-\Pgas/P}{ 1 - 3 \Pgas/4P} \right],
\end{equation}
and equation (\ref{e.convection-onset}) becomes
\begin{equation}\label{e.Lonset-ideal}
\frac{\Lrad}{\Ledd} > \frac{\Lonset}{\Ledd} = \frac{8 (1-\Pgas/P) (4-3\Pgas/P)}{32-24 \Pgas/P + 3(\Pgas/P)^{2} },
\end{equation}
allowing one to show that $\Lonset < \Lrho$.  At high luminosities where the gas becomes radiation-dominated, however, the difference between \Lonset\ and \Lrho\ becomes small.  Expanding equations~(\ref{e.Lrho-ideal}) and (\ref{e.Lonset-ideal}) for $\Pgas/P \ll 1$ gives $\Lrho-\Lonset \approx (3/4) \times (\Pgas/P)\times\Ledd$.
For such high-luminosity, radiation-dominated stars, a small inefficiency in convection is sufficient to drive a density inversion.

We now demonstrate that such inefficient convection can arise in the convective, radiation-dominated, envelopes of massive stars. In order of magnitude the convective and radiative fluxes are, respectively,
$\Fconv \sim \rho c_{s}^{3}\left(\nablaT - \nablaad\right)^{3/2}$ and
$\Frad \sim c\Prad/\tau$.  To carry the flux, we need $\Fconv\sim\Frad$; equating and substituting $\rho c_{s}^{2} \sim P\sim\Pgas$, we arrive at an expression that sets the level of superadiabaticity,
\begin{equation}\label{e.gradient-diff}
(\nablaT - \nablaad)^{3/2} \sim \frac{c}{\cs} \frac{\Prad}{\Pgas}\tau^{-1}.
\end{equation}
Under typical conditions in massive star envelopes, $c/\cs \sim 10^{4}$ at the iron opacity bump, but at this location, $\tau$ is not large enough to prevent the superadiabaticity from  triggering a density inversion.

\begin{figure}[htbp]
\centering{\includegraphics[width=\figwidth]{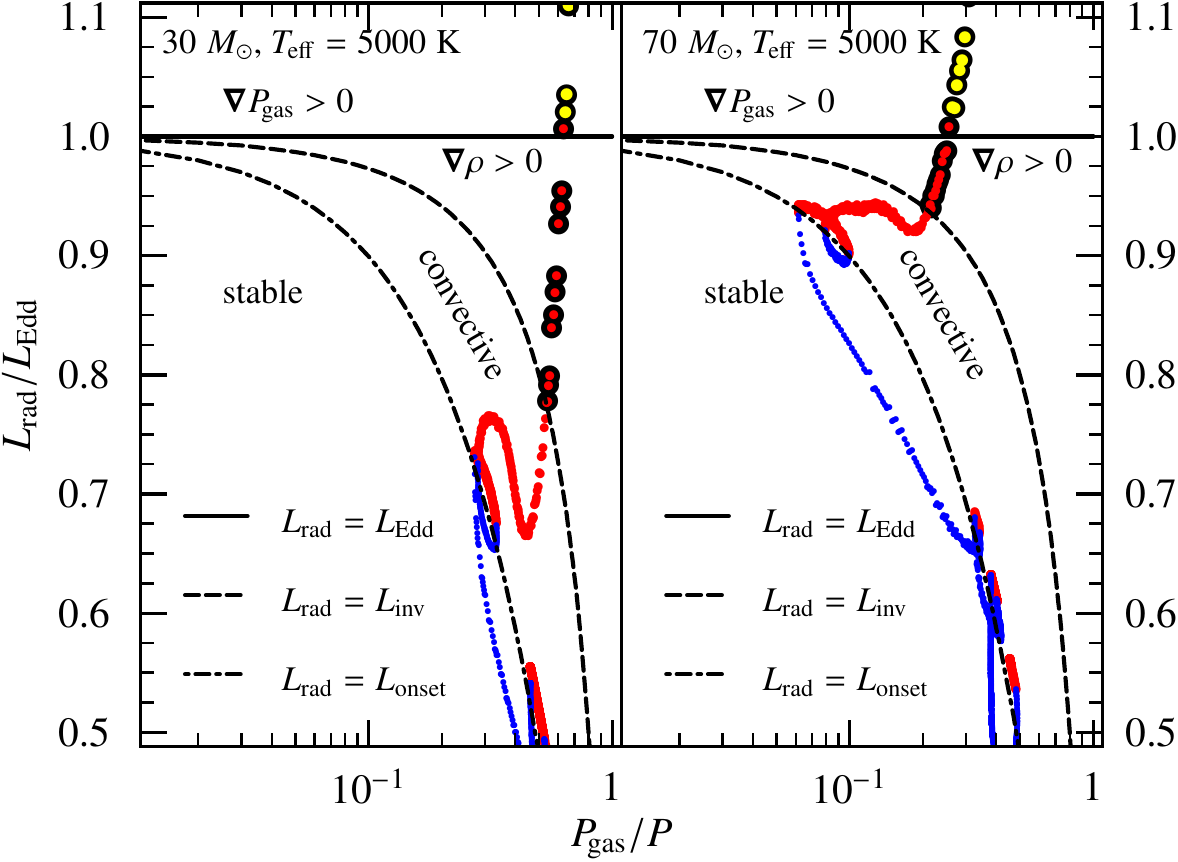}}
\caption{\label{f.Lcrit} 
The critical luminosities $\Lrad = \Lonset$ (eq.~[\ref{e.Lonset-ideal}], dot-dashed line),  $\Lrad = \Lrho$ (eq.~[\ref{e.Lrho-ideal}], dashed line), and $\Lrad = \Ledd$ (solid line) as a function of $\Pgas/P$ for an ideal gas-radiation mixture. Compare this with Fig.~1 of \citet{Joss1973On-the-critical}.  For $\Lrad < \Lonset$, the gas is convectively stable; for $\Lonset < \Lrad < \Lrho$, the gas is convective; for $\Lrho < \Lrad < \Ledd$, the density is inverted, $\dif\rho/\dif r > 0$; and for $\Ledd < \Lrad$, the gas pressure is inverted, $\dif\Pgas/\dif r > 0$.  Overlaid on the plots are the profiles from a $30\,\Msun$ (left panel) and a $70\,\Msun$ (right panel) model with $Z = 0.02$: blue dots indicate zones that are radiative; red dots indicate $\nablarad > \nablaad$; dots with a black border have a density inversion; and the yellow dots with black borders indicate a gas pressure inversion. As the profile moves out from the stellar center it traces out the points on the plot from bottom to top. Only a part of the model profiles are visible in the plot. The calculations correspond to the first crossing of the Hertzsprung gap when $\teff= 5000\,\Kelvin$.}
\end{figure}

The lines in Fig.~\ref{f.Lcrit} show these luminosity conditions as a function of $\Pgas/P$, and reveal that as the stellar conditions become radiation dominated, there is only a small gap  between a convective model that is adiabatically stratified  and a model with a density inversion. This corresponds to the region between the curves $\Lrad=\Lonset$ (dot-dashed line) and $\Lrad = \Lrho$ (dashed line).
The gas pressure does not invert until $L > \Ledd$, which in Fig.~\ref{f.Lcrit} is the region above the solid horizontal line.  We show profiles from a $30\,\Msun$ (left panel) and a $70\,\Msun$ model (right panel).  These are from the first crossing of the Hertzsprung gap when $\Teff = 5000\,\Kelvin$.  Each dot corresponds to a zone in the calculation; as the profile moves outward from center to surface the traces go from bottom to top in the plot.  The blue dots indicate zones where the star is radiative; red indicates convection; a black border denotes a density inversion, $\dif\rho/\dif r > 0$; and yellow indicates a gas pressure inversion, $\dif\Pgas/\dif r > 0$. There is excellent agreement between the detailed \mesa\ evolutionary calculations and the analytical conditions (eq.~[\ref{e.Lrho-ideal}] and [\ref{e.Lonset-ideal}]).  The $70\,\Msun$ profile goes  into the low $\Pgas/P$, high $\Lrad/\Ledd$ regime. 

\begin{figure}[htbp]
\centering{\includegraphics[width=\apjcolwidth]{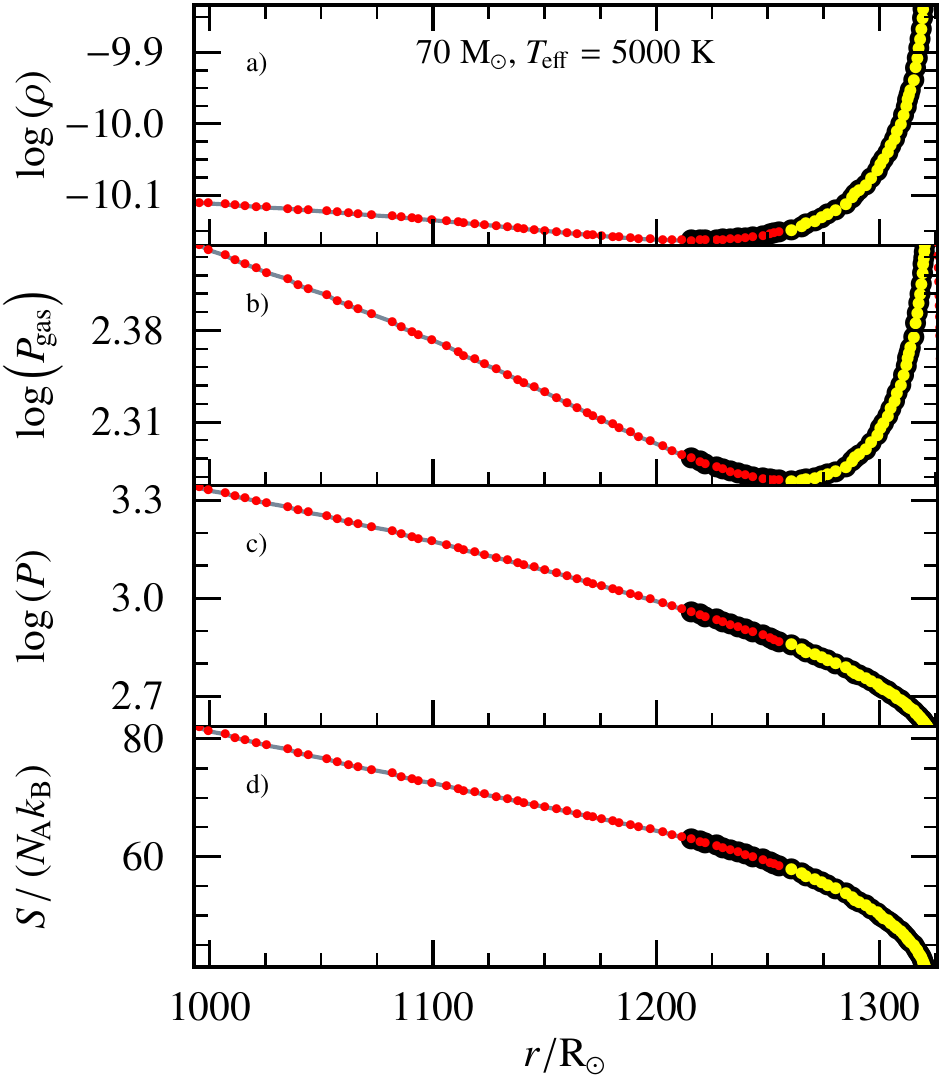}}
\caption{\label{f.rho-Pgas-S}%
The panels display, from top to bottom, the density, gas pressure, total pressure, and entropy as functions of radius for the $70\,\Msun$ model shown in Fig.~\protect\ref{f.Lcrit}. The range of radii is restricted to the region where density and gas pressure inversions develop.  Each zone is marked by a dot; a small red dot indicates convection with no predicted gas or gas pressure inversion ($\Lrad < \Lrho$); a large red dot with black border indicates a predicted density inversion but no gas pressure inversion ($\Lrho<\Lrad<\Ledd$); and a yellow dot with black border indicates a convective region with a predicted gas pressure inversion ($\Lrad > \Ledd$). The total pressure (panel c) is well-behaved at all radii. Note also the decrease in entropy (panel d): the region is superadiabatic.}
\end{figure}

Figure~\ref{f.rho-Pgas-S} displays the physical conditions in the $70\,\Msun$ model where the density and gas pressure inversions develop.  The panels display, from top to bottom, density, gas pressure, total pressure, and entropy, all as functions of radius.  The total radius is $R = 1330\,\Rsun$. Regions with $\nabla > \nablaad$ and $\Lrad < \Lrho < \Ledd$ are marked with a small red dot.  Regions where $\Lrho < \Lrad < \Ledd$ (cf.\ eq.~[\ref{e.Lrho-ideal}]) are marked with a large red dot with a black border. Regions where $\Lrad > \Ledd$ are marked with a large yellow dot with black border.  Although the pressure (panel c) is well-behaved in this superadiabatic (panel d) region, a density inversion does develop where $\Ledd > \Lrad > \Lrho$ (panel a) and a gas pressure inversion develops (panel b) where $\Lrad >\Ledd$, as predicted.  In this region the superadiabaticity $\nablaT-\nablaad > 10^{-2}$ and is greater than unity for $r/\Rsun \gtrsim 1300$. This is much larger than a typical value ($\sim 10^{-6}$) where convection is efficient and results in the entropy decreasing with $r$ as shown in panel c.

\subsection{Treatment of Superadiabatic Convection in Radiation-Dominated Regions}\label{s.superadiabatic}

In \MESAstar\ the superadiabatic gradient arising in radiation-dominated envelopes 
can force the adoption of prohibitively short timesteps. Energy is mostly transported by 
radiation, and the convective velocities resulting from MLT approach the sound speed.
The stability of such radiation-dominated envelopes has been discussed in the past, 
and is still a matter of debate \citep[see, e.g.,][]{Langer:1997,bisnovatyi-kogan-1999,Maeder:2009,Suarezmadrigal:2013}.
In this regime, the treatment of convective energy transport by MLT is admittedly out of its domain of applicability.
Hydrodynamical instabilities and the transport of energy from waves excited by near-sonic turbulent convection are 
important for energy transport, and three-dimensional hydrodynamical calculations are 
required to capture fully the complex physics occurring in this regime.

Here we develop a treatment of convection, known as MLT++, that reduces the superadiabaticity in some radiation-dominated convective regions. This treatment allows \MESAstar\ to calculate models of massive stars up to core collapse. 
For every model, \MESAstar\ computes the values of
\begin{equation}
\lambdamax\equiv \max\left(\frac{\Lrad}{\Ledd}\right) \quad \textrm{and}\quad  \betamin
\equiv\min\left(\frac{\Pgas}{P}\right).
\end{equation}
When $\betamin$ is small and $\lambdamax$ is large, and MLT yields a $\supernab > \superthresh$, we  artificially decrease the superadiabaticity, $\supernab \equiv \nablaT - \nablaad$, implied by MLT. 
The default of the user-specified parameter $\superthresh$ is sufficiently large, $\sim 10^{-3}$, so that convection is still inefficient.

\begin{figure}[htbp]
\centering{\includegraphics[width=\figwidth]{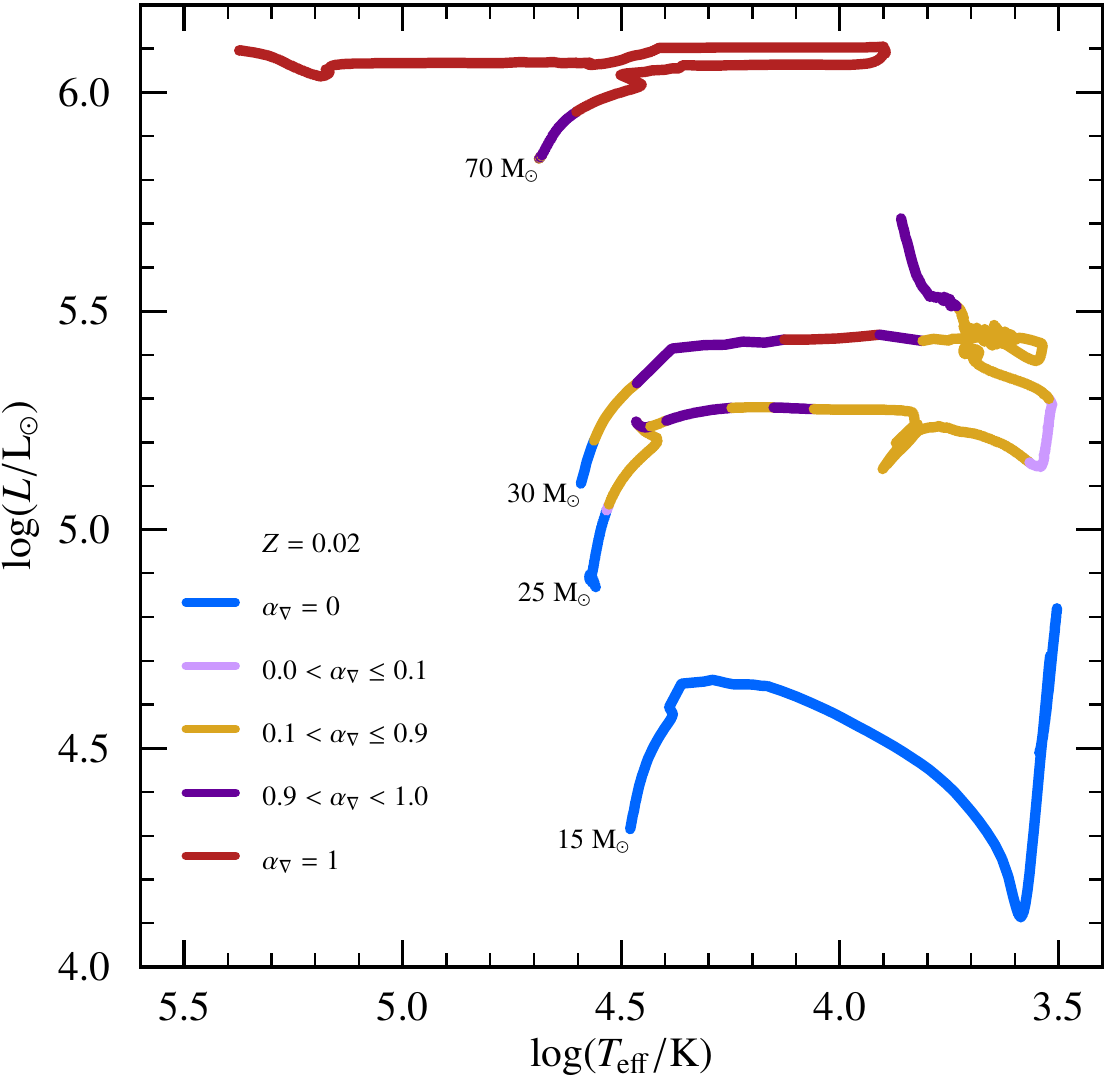}}
\caption{\label{f.mlt}HR diagram of 15, 25, 30, and $70\,\Msun$ models. The color indicates the value of $\asuper$ at that point in the star's evolution. For the $25\,\Msun$ and $30\,\Msun$ stars, there is a sharp spike in $\asuper$ as the star crosses the Hertzsprung gap followed by a sharp drop at the base of the red giant branch.  The $70\,\Msun$ model has $\asuper > 0.9$ for its entire evolution.}
\end{figure}

\MESAstar\ sets $\nablaT$ to reduce the $\supernab-\superthresh$ by a factor
$ \asuper\fsuper$, where $\fsuper$ is specified by the user, and  $\asuper$ is updated at each timestep to a linear combination of its previous value and a value $\asupert(\lambdamax,\betamin)$.
For large values of $\lambdamax$ and small values of $\betamin$, $\asupert\to1$; in typical usage, the transition happens where $\lambdamax\approx 0.5$ and $\betamin\approx 0.3$. For small values of $\lambdamax$ and large values of $\betamin$, $\asupert\to0$.  Thus \fsuper\ sets the maximum reduction of $\supernab-\superthresh$.  Figure~\ref{f.mlt} shows how \MESAstar\ turns on the reduction in \supernab\ as a star evolves.  Tracks in the HR diagram are shown for four stellar models: 15, 25, 30, and $70\,\Msun$.  The color of each line indicates the value of \asuper\ at each point.

Such a decrease of the temperature gradient reduces \Lrad\ and implies additional physical transport. Potential agents for the excess transport include waves
excited by turbulent convection \citep[see, e.g.,][]{Maeder:1987} and
radiative diffusion enhanced by porous clumping of the envelope
\citep[e.g.,][]{Owocki:2004}.  As these radiation-dominated envelopes
might be physically unstable, with a resulting strong enhancement of
mass loss, we caution that the results of any 1D stellar evolution
calculation for the late evolutionary phases of massive stars should
be considered highly uncertain.

\begin{figure*}[htbp]
\centering{
\includegraphics[angle=0,width=\twoupwidth]{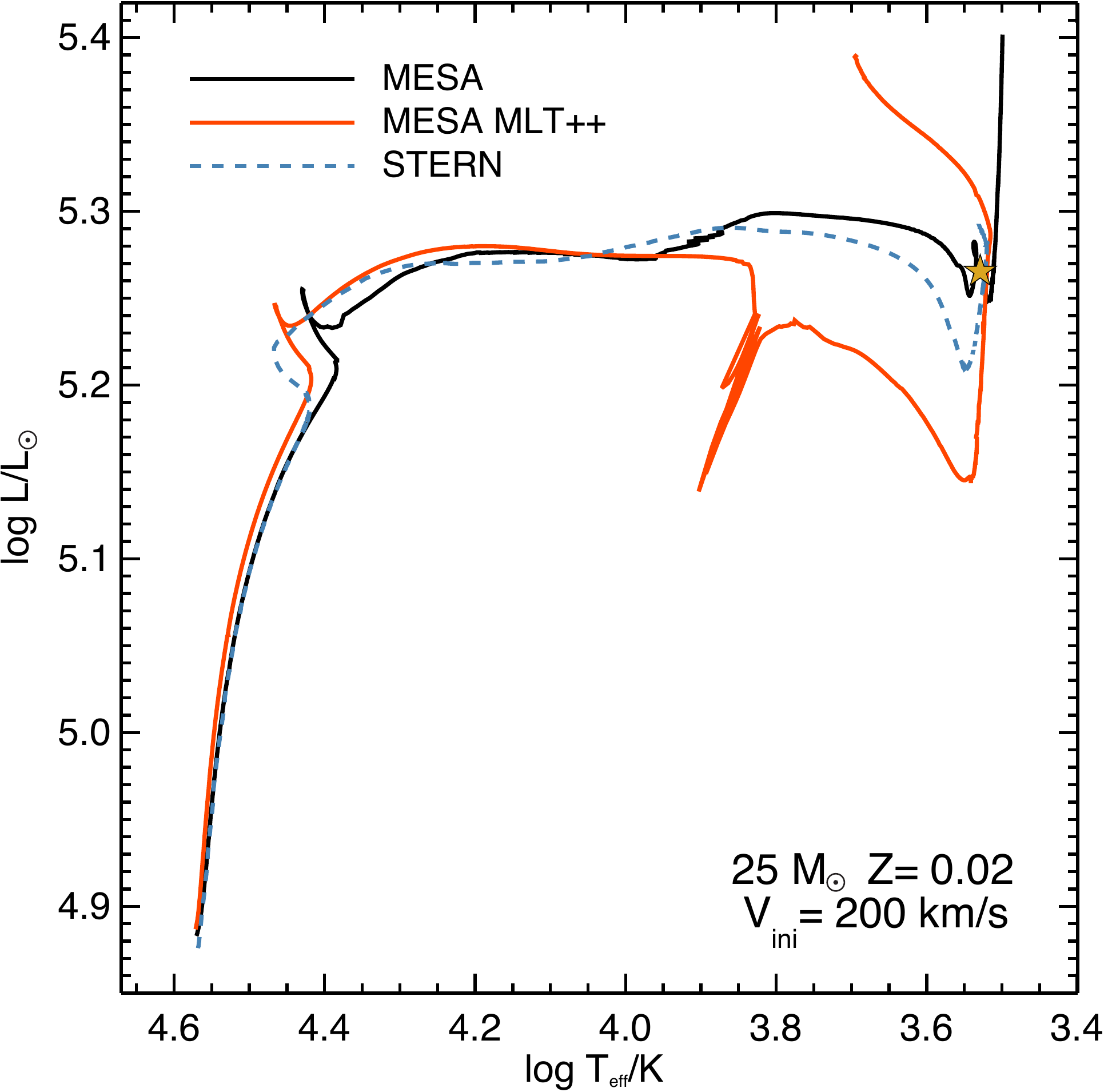}\hspace{\twoupsep}
\includegraphics[angle=0,width=\twoupwidth]{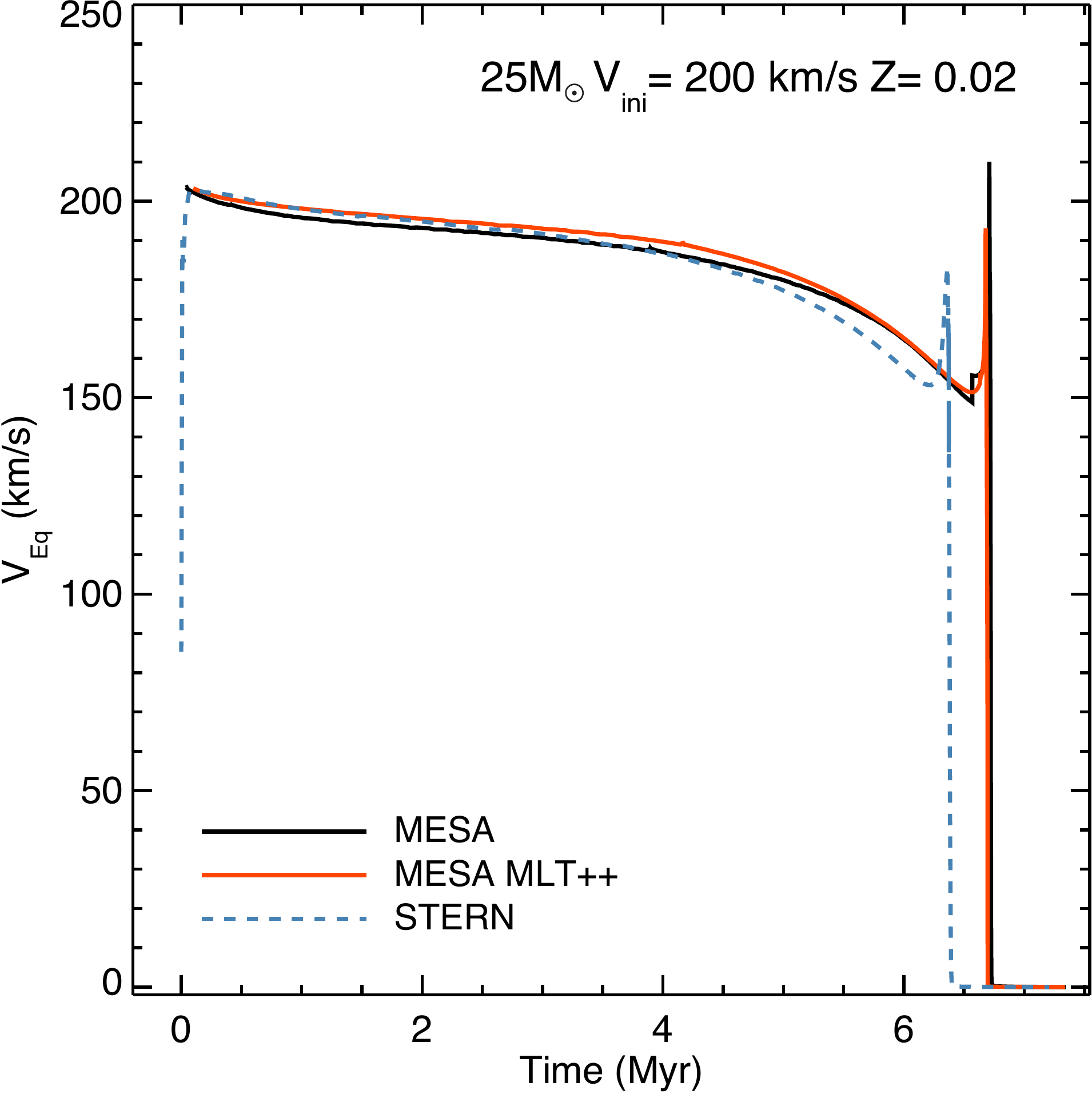} 
}
\caption{ Comparison of evolutionary tracks (left) and equatorial rotational velocity (right) for a $25\,\Msun$ model with $Z=0.02$ and $\veq= 200\,\kms$. The solid black lines show \MESAstar\ results with MLT (black) and MLT++ (orange), while the dashed blue line refers to STERN calculations.
The star symbol shows the location where we started the calculation for the RSG pulsations discussed in \S\ref{rsg-pulsation}.
\label{fig:hrdfig25plus}}
\end{figure*}

\begin{figure}[htbp]
\centering
\includegraphics[width=\figwidth]{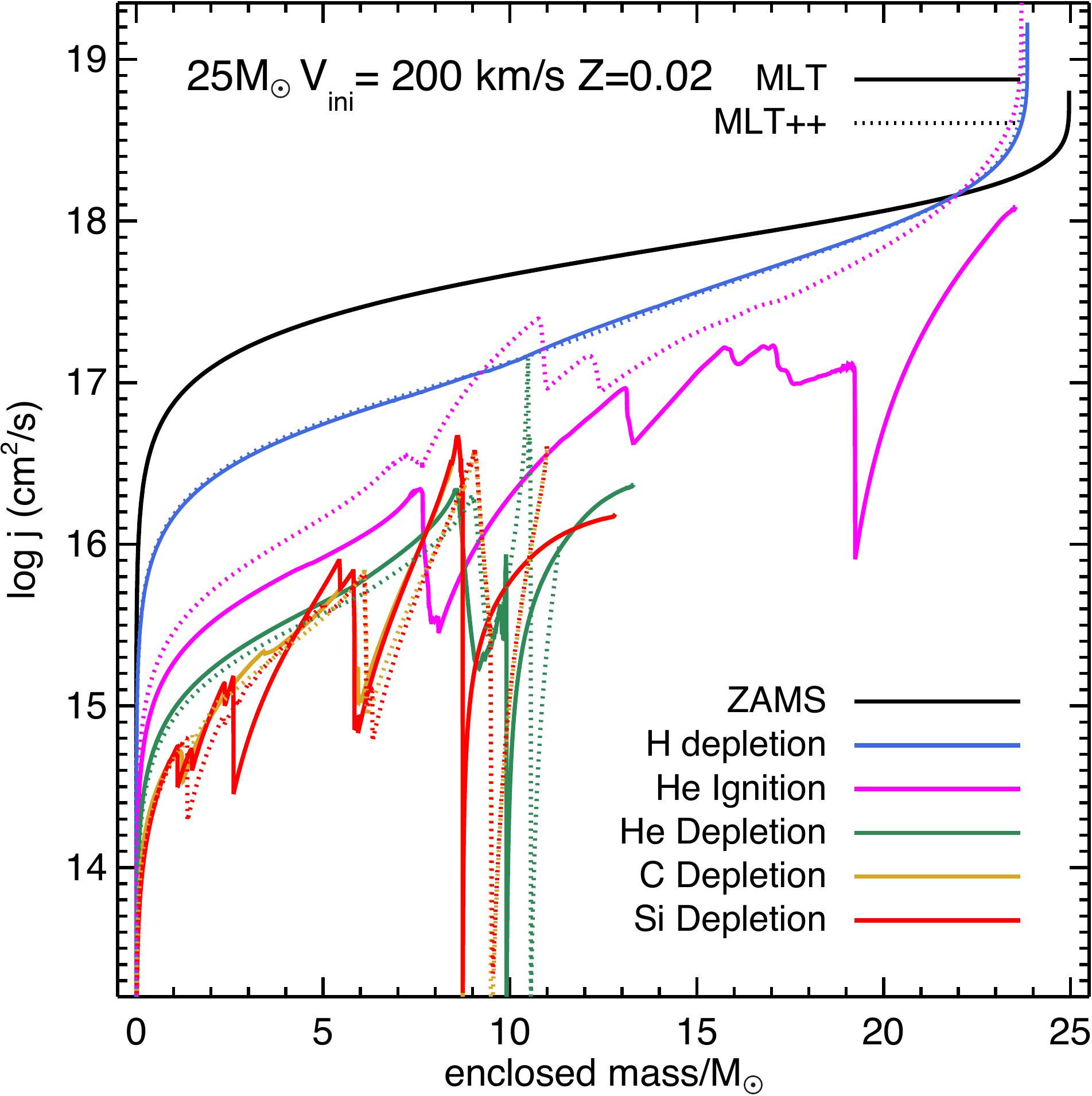} 
\caption{ Evolution of internal specific angular momentum for the two
$25\,\Msun$ models of Fig.~\protect\ref{fig:hrdfig25plus}.
The dashed lines show models calculated with MLT++. Due to different excursions in the HRD (see Fig.~\ref{fig:hrdfig15}) 
calculations with MLT and MLT++ end  with different final masses. There are no substantial changes, however, in the specific angular momentum content of the stellar cores.
\label{fig:j25allplus}}
\end{figure}

We now show a comparison of \MESAstar\ calculations of rotating massive stars
done with and without MLT++. 
We used the 25\,\Msun\ model described in \S\ref{s.test-cases}, which at $Z=0.02$ 
is around the upper mass limit that can converge using a reasonably short timestep without having to rely on the MLT++.
The most prominent difference between the calculations is the evolutionary track in the HR-diagram (Fig.~\ref{fig:hrdfig25plus}). 
This is not surprising, as MLT and MLT++ result in different efficiencies of  energy transport in radiation-dominated stellar envelopes.  The  sharp drop in $L$ for the MLT++ case is the result of a brief period of enhanced mass loss due to super-critical rotation.
The structure and the angular momentum content of the collapsing core are weakly dependent, however, on the choice of MLT vs.\ MLT++ (Fig.~\ref{fig:j25allplus}).

\subsection{Core-Collapse Progenitor Models}\label{cc-progenitors}
We evolve a grid of massive
stars initially rotating with $\Om/\Omc=0.2$.
The models have been initialized using solid body rotation.
Models with initial $M/\Msun=$ 30, 40, 50, 60, 70, 80, 90 and 100 have initial $Z= 0.02$, while
models with initial $M/\Msun=$ 120, 150, 250, 500 and 1000 have been initialized with $Z= 0.001$.
To calculate convective boundaries we adopt the Ledoux criterion including the impact of semiconvection (with $\alphasc=0.02$, see \S\ref{s.semiconvection}).
The transport of angular momentum and chemicals by rotational instabilities and magnetic torques is included and calibrated following
\citet{Heger:2000,Heger:2005} and \citet{Yoon:2005}. 
Wind mass-loss is been implemented following the recipe of \citet{Glebbeek:2009}.
For $\teff > 10^4\,\Kelvin$ and H-surface fraction $> 0.4$, the mass-loss prescription of \citet{Vink:2001} is used.
In the same temperature range, but when the H-surface fraction decreases below 0.4, \citet{Nugis:2000} determine the mass-loss rate.
At low temperatures ($\teff < 10^4\,\Kelvin$) the mass-loss rate of \citet{deJager:1988} is used. 

\begin{figure}[htbp]
\newcommand{\panelwidth}{\figwidth}
\centering
\includegraphics[angle=0,width=\panelwidth]{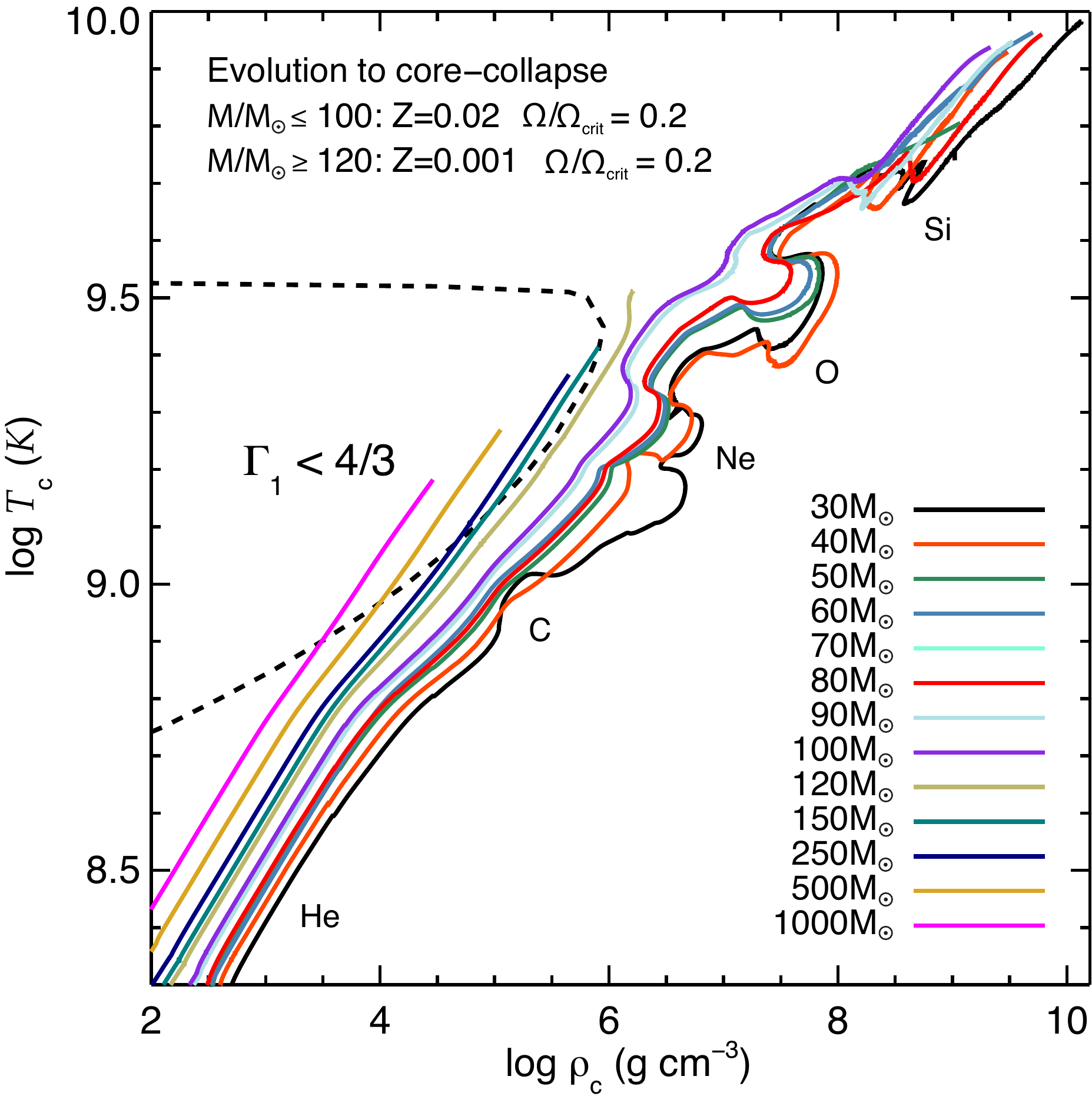}
\caption{ Evolution of \Tc\ and \rhoc\ in the massive rotating models.
The locations of core helium, carbon, neon, oxygen, and silicon burning are labeled. 
A dashed curve marks the electron-positron pair-instability region where
$\Gammaone < 4/3$.
All models are rotating initially at $20\%$ of critical rotation. 
The calculations include the effects of rotation and Spruit-Tayler magnetic fields as discussed in \S\ref{s.rotation}.
Models with initial mass $\le 100\,\Msun$ have initial metallicity $Z=0.02$, while models with mass $\ge 120\,\Msun$ have initial metallicity $Z=0.001$. 
The end of the line for each mass corresponds to the time of core-collapse, defined as when any part of the collapsing-core exceeds an in-fall velocity of 1000\,\kms.
The tracks for the $60\,\Msun$ and $70\,\Msun$ overlap in this plot. 
\label{fig:rhotmassive}}
\end{figure}

Figure~\ref{fig:rhotmassive} shows the central conditions of these massive rotating models.
For each model the calculation stops when any part of the collapsing core reaches an infall velocity of 1000\,\kms. 
Some of the initial and final properties are summarized in Table~\ref{tab:massive}.
These calculations are performed to reveal the new capabilities of \MESAstar. The values of the parameters for these calculations have not been calibrated against existing calculations or observations.

\begin{table*}
\caption{\label{tab:massive} Initial (ZAMS) and final (core-collapse) properties. }
\centering
\renewcommand{\arraystretch}{1.1}
\begin{tabular}{ccccc|ccc|ccccc}
\hline\hline
  $M_{\rm{ini}}$ &
   $Z_{\mathrm{ini}}$ &
  $\Om/\Omc$\tablenotemark{a} &
  $\veqi$\tablenotemark{b} &
  $J_{\mathrm{ini}}$\tablenotemark{c} & 
  $\Delta t$\tablenotemark{d}&
  $\Delta t_{\rm H}$\tablenotemark{e} &
  $\Delta t_{\rm He}$\tablenotemark{e} & 
  $M_{\rm f}$\tablenotemark{f}&
  $M_{\rm Fe}$\tablenotemark{g}&
  $J_{\mathrm{f}}$\tablenotemark{h} & 
  $J_{\mathrm{Fe}}$\tablenotemark{i}\\
  $[\Msun]$ &
   &
   &
  $[\kms]$ &
  $[\ergssecond]$ & 
  $[\Mega\yr]$ &
  $[\Mega\yr]$ &
  $[\Mega\yr]$ & 
  $[\Mega\yr]$ &
  $[\Mega\yr]$ &
  $[\ergssecond]$ & 
  $[\ergssecond]$\\
\hline
     30   &    0.020 &     0.20 &   129.69 & 3.28\E{52} &     6.30 &     5.87 &     0.36 &    17.77 &     1.41 & 2.87\E{50} & 1.03\E{48} \\
     40   &    0.020 &     0.20 &   122.86 & 4.87\E{52} &     5.06 &     4.71 &     0.31 &    19.37 &     1.81 & 3.77\E{50} & 1.61\E{48} \\
     50   &    0.020 &     0.20 &   112.02 & 6.30\E{52} &     4.41 &     4.08 &     0.29 &    25.04 &     1.38 & 5.39\E{50} & 1.09\E{48} \\
     60   &    0.020 &     0.20 &    98.37 & 7.34\E{52} &     4.04 &     3.66 &     0.35 &    22.88 &     1.76 & 7.81\E{50} & 2.76\E{48} \\
     70   &    0.020 &     0.20 &    78.76 & 7.53\E{52} &     3.90 &     3.57 &     0.29 &    26.19 &     1.75 & 5.30\E{50} & 1.54\E{48} \\
     80   &    0.020 &     0.20 &    50.10 & 5.88\E{52} &     3.70 &     3.38 &     0.29 &    29.20 &     1.78 & 6.16\E{50} & 1.44\E{48} \\
     90   &    0.020 &     0.20 &     2.27 & 3.57\E{52} &     3.10 &     2.80 &     0.27 &    44.90 &     1.71 & 4.39\E{50} & 5.23\E{47} \\
    100   &    0.020 &     0.20 &     2.34 & 3.91\E{51} &     2.98 &     2.69 &     0.26 &    49.02 &     1.92 & 5.50\E{50} & 6.58\E{47} \\
    120   &    0.001 &     0.20 &   145.41 & 2.93\E{53} &     3.26 &     2.99 &     0.23 &    79.38 &     --   & 4.79\E{51} & -- \\
    150   &    0.001 &     0.20 &   134.75 & 3.84\E{53} &     3.03 &     2.77 &     0.23 &    95.52 &     --   & 6.80\E{51} & -- \\
    250   &    0.001 &     0.20 &    69.30 & 4.39\E{53} &     2.56 &     2.32 &     0.21 &   167.49 &     --   & 9.13\E{51} & -- \\
    500   &    0.001 &     0.20 &     3.78 & 6.40\E{52} &     2.19 &     1.96 &     0.20 &   410.28 &     --   & 7.92\E{51} & -- \\
   1000   &    0.001 &     0.20 &     4.42 & 2.09\E{53} &     1.99 &     1.77 &     0.19 &   860.48 &     --   & 2.44\E{52} & -- \\
\hline
\end{tabular}
\tablenotetext{1}{ Initial rotation rate, see definition in \S.~\ref{s.rotation-mass-loss}.}
  \tablenotetext{2}{Initial equatorial rotational velocity.}
  \tablenotetext{3}{Total initial angular momentum.}
  \tablenotetext{4}{Stellar lifetime.}
  \tablenotetext{5}{Main sequence and core He-burning lifetimes. These are defined as the interval between onset of core burning and depletion of central hydrogen (or helium) to 1\% by mass.}
  \tablenotetext{6}{Final mass.}
  \tablenotetext{7}{Mass of the Iron core (if present).}
  \tablenotetext{8}{ Final total angular momentum.}
  \tablenotetext{9}{Final total angular momentum of the iron-core.}
\end{table*}

\subsection{Radial Instability of Red Supergiants}\label{rsg-pulsation}
Massive red supergiants (RSG) are unstable to radial pulsations
driven by the $\kappa$-mechanism in the hydrogen ionization zone.
Both linear and non-linear calculations show the occurrence of oscillations 
with the period and growth rate of the dominant fundamental mode increasing with 
$L/M$  \citep{Li:1994,Heger:1997,Yoon:2010}. The periods are of the order of years. As discussed by \citet{Yoon:2010} the occurrence of RSG pulsations can impact stellar mass-loss rates and modify the evolution 
of massive stars above a certain mass. 
We study the occurrence of RSG pulsations with \MESAstar\ and compare results with existing non-linear calculations.

\begin{figure}[htbp]
\newcommand{\panelwidth}{\figwidth}
\centering
\includegraphics[angle=0,width=\panelwidth]{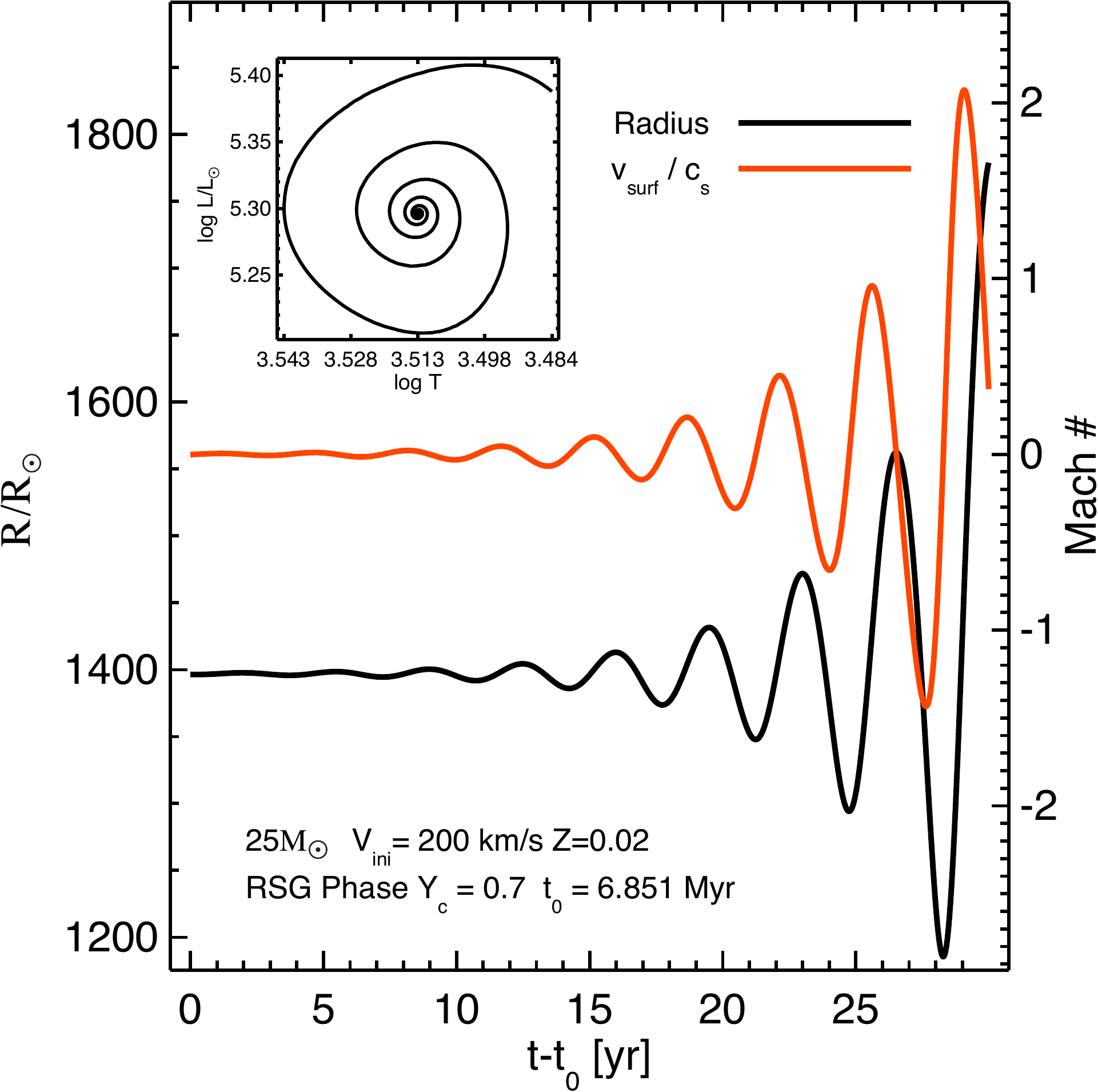}
\caption{Surface properties of a pulsating RSG. This is the same 25\Msun\ model discussed in Sec.~\ref{s.test-cases}, evolved from $t_0=6.851$~Myr (corresponding to $Y_c=0.7$, star symbol in Fig.~\ref{fig:hrdfig25}) with timesteps $\timestep \le 0.01\,\yr$.
The black line shows the evolution of the stellar radius, while the orange line shows the value of the surface radial velocity (in units of the local sound speed). The inset shows the corresponding evolution in
the HR-diagram. \label{fig:puls25}}
\end{figure}

In Fig.~\ref{fig:puls25} we show the capability of \MESAstar\ to exhibit radial oscillations in luminous RSGs.
We use the same $25\,\Msun$ rotating model discussed in \S~\ref{s.test-cases}, and we restart the calculation when the He mass fraction in the core is $Y_c = 0.7$.
For non-rotating RSG with $Z=0.02$, \citet{Yoon:2010} found pulsation periods in the range 1--$8\,\yr$.
To resolve the RSG pulsations we force the timestep to $< 0.01\,\yr$, much shorter than the usual timestep during He-burning ($\timestep \gtrsim 10^2\,\yr$,  see Appendix~\ref{s.timestep-controls}).  This explains why RSG pulsations are usually not found during the evolution of massive stars.
Before the code stops  due to the emergence of  supersonic radial velocities in the envelope, we find a pulsational period $\approx 4\,\yr$, in good agreement with the results of \citet{Yoon:2010}.

\section{Summary and Conclusions}\label{s.conclusions}

We have explained and, where possible, verified the improvements and
major new capabilities implemented in \MESA \ since the publication of
\mesaone.  These advancements include evolutionary modeling for
giant planets (\S\ref{s.planets}), tools for asteroseismology
(\S\ref{s.astroseismology}), implementation of composition gradients
in stellar interiors and their impact on convective mixing
(\S\ref{s.mixing}) , the evolution of intermediate mass stars and white
dwarfs (\S\ref{s.agb-wd}) the treatment of rotation during stellar
evolution (\S\ref{s.rotation}), addressing the onset of radiation
pressure dominance in the envelopes of evolving massive stars due to
the iron opacity bump, and evolving massive stars to the onset of core
collapse (\S\ref{s.massive}).  The enhancements include the
physics modules (Appendix \ref{s.input-physics}), the 
algorithms (Appendix \ref{s.nuts-and-bolts}), and the addition of a
\mesa\ Software Development Kit (Appendix \ref{s.mesasdk}). \MESAstar\
input files and related materials for all the figures are 
avaliable at \url{http://mesastar.org}.

These hitherto unpublished advancements have already enabled a number
of studies in planets
\citep[e.g.,][]{passy_2012_aa,huang_2012_aa,carlberg_2012_aa},
classical novae \citep{denissenkov_2013_aa}, 
asteroseismology
\cite[e.g.,][]{yang_2012_aa,burkart_2012_aa,moravveji_2012_aa},
rotationally induced mixing
\citep[e.g.,][]{denissenkov_2010_aa,chatzopoulos_2012_aa,chatzopoulos_2012_ab}
and enabled the discovery of new features in the evolution of low-mass
stars \citep{denissenkov_2012_aa}.  In addition, these enhanced
capabilities have allowed for applications of \MESAstar \ that were
not initially envisioned, such as explorations of stars under modified
gravity \citep{2011ApJ...732...25C,davis_2012_aa}, and stellar
oscillations induced by tidal disturbances in double white dwarf
binaries
\citep{2012arXiv1211.0624F,2012ApJ...756L..17F,2012arXiv1211.1393B}.

As an open source ``instrument'' for stellar astrophysics, it is
difficult to predict all the ways in which future development of
\mesa\ will occur.  We do know, however, that future versions of
\mesa\ will include advances in physics modules,
features driven by the \MESA\ user community, and architectural
refinements.  For example, the plethora of asteroseismological data is
driving new initiatives to incorporate non-adiabatic pulsation codes,
where possible, into \MESA. The prevalence of interacting binary star
systems, especially for massive stars, has increased the pressure for
\MESA \ development efforts that would yield the capability to
simultaneously evolve two interacting stellar models.  Physics module
developments will likely include general relativistic corrections to
the stellar structure equations (e.g., difference between
gravitational and baryonic mass), the mass diffusion coefficients in
electron degenerate environments, phase separation in cooling white
dwarfs, and nuclear statistical equilibrium solvers.  We also expect
the transition from multicore systems (with order 10 cores) to
many-core architectures (with order 100 cores) to drive new
directions in \MESA's algorithmic and architectural development.

\acknowledgements

It is a pleasure to thank Falk Herwig for significant contributions to
the \mesa\ project and J\o rgen Christensen-Dalsgaard for kindly
providing the ADIPLS code for inclusion in \mesa\ and assisting with
its integration. We likewise thank Jared Brooks for documenting
the \mesa\ test suite, Tristan Guillot for providing the CEPAM
evolutionary tracks, Alexander Potekhin for an update to his EOS code,
Didier Saumon and Jim MacDonald for providing EOS tables, Ren\'e
Rohrman for providing atmosphere tables, Richard Freedman for
providing opacity tables, Evert Glebbeek, Alex Heger, and Norbert
Langer for providing code for implementing rotation, and Haili Hu for
providing code for implementing diffusion.

We also thank
David Arnett, 
Tim Bedding,
Kent Budge,
Phil Chang,
Pieter DeGroote,
Pavel Denisenkov,
Jonathan Fortney,
Chris Fryer,
Gustavo Hime,
Raphael Hirschi,
Sam Jones,
Steve Kawaler,
Phillip Macias,
Pablo Marchant,
Travis Metcalfe,
Kevin Moore,
Ehsan Moravveji,
Jean-Claude Passy,
Hideyuki Saio,
Josiah Schwab,
Aldo Serenelli,
Josh Shiode,
Steinn Sigurdsson,
Anne Thoul,
Roni Waldman,
Achim Weiss,
Stan Woosley,
Sung-Chul Yoon,
and Patrick Young
for providing valuable discussions and correspondence.
Some of the simulations for this work were made possible by the Triton
Resource, a high performance research computing system operated by San
Diego Supercomputer Center at UC San Diego.

We thank the participants of the 2012 MESA Summer School for their willingness
to experiment with the new capabilities:
Jeff Andrews,
Umberto Battino,
Keaton Bell,
Harshal Bhadkamkar,
Kristen Boydstun,
Emmanouil Chatzopoulos,
Eugene Chen,
Jieun Choi,
Alex Deibel, 
Luc Dessart,
Ian Dobbs-Dixon,
Tassos Fragos,
Samuel Harrold, 
Daniel Huber,
Joe Hughto,
Max Katz,
Agnes Kim, 
Io Kleiser, 
Shri Kulkarni,
Gongjie Li,
Christopher Lindner,
Jing Luan,
Mia Lundkvist,
Morgan MacLeod, 
Jo\~ao Marques,
Grant Newsham, 
Rachel Olson,
Richard O'Shaughnessy,
Kuo-Chuan Pan,
Ilka Petermann, 
Theodore Sande,
Ken Shen,
Natalia Shabaltas,
Dave Spiegel, 
Jie Su,
Tuguldur Sukhbold,
David Tsang,
Bill Wolf,
Angie Wolfgang,
Tsing Wai Wong,
and Alexey Zinger.

This project was broadly supported by the NSF under grants PHY 11-25915 and AST 11-09174.
M.C. acknowledges partial support from the ``Alberto Barlettani'' Prize 2012.
P.A. acknowledges support by NSF AST-0908873 and NASA NNX09AF98G.
L.B. acknowledges support from the Wayne Rosing, Simon and Diana Raab Chair 
in Theoretical Astrophysics at KITP.
E.F.B acknowledges support by the Joint Institute for Nuclear Astrophysics under NSF PHY grant 08-22648.
A.L.D received support from the Australian Research Council under grant FL110100012.
M.H.M acknowledges support by the NSF grant AST-0909107, the NASA grant NNX12AC96G, the Norman 
Hackerman Advanced Research Program under grant 003658-0252-2009, and the Delaware
Asteroseismic Research Center.
D.S. acknowledges support by the Australian Research Council.
F.X.T acknowledges support from the NSF under grants AST 08-06720, AST
6736821, AST 09-07919, AST 10-07977, PHY 08-22648, and from NASA under
grants 08-NAI5-0018 and NNX11AD31G.
R.T. acknowledges support by NSF grants AST-0908688 and AST-0904607, and NASA grant NNX12AC72G.

\appendix
\section{Updates to Input Physics Modules}\label{s.input-physics}

There have been many updates and improvements to the physics modules since \mesaone.  In this appendix, 
we describe the changes that have been made to the microphysics modules \chem\ (\S~\ref{s.chem}), \eos\ (\S~\ref{s.EOS}), \kap\ (\S~\ref{s.opacities}), and \net\ (\S~\ref{s.reactions}).  We conclude by listing updates to the atmosphere boundary conditions (\S~\ref{s.atmospheres}).

\subsection{Atomic and Nuclear Data}\label{s.chem}

The \chem\ module now has the latest version (v2.0) of the JINA \reaclib\ nuclide data \citep{Cyburt2010The-JINA-REACLI}.  This contains updated mass evaluations, and now includes 7853 nuclides up to \copernicum[337].  For precision work, the \chem\ module now distinguishes between the 
atomic mass number $A_{i}$---the number of nucleons in a given isotope---and the atomic mass $W_{i}$.  The abundance of a species $i$ is defined as
\begin{equation}\label{e.abundance-def}
 Y_{i} \equiv \frac{n_{i}}{\nB},
\end{equation}
where \nB\ is the baryonic number density.  The \emph{baryon fraction} $X_{i}$ is then
\begin{equation}\label{e.baryon-fraction}
X_{i} = Y_{i}A_{i} = \frac{n_{i}A_{i}}{\nB},
\end{equation}
Note that $\sum_{i}X_{i} = \nB/\nB = 1$ and is invariant under nuclear reactions.  
We then define the baryon density (in mass units) as
\begin{equation}\label{e.baryon-density-def}
\rho = \nB\mb,
\end{equation}
where $\mb = 1.660538782\ee{-24}\nsp\gram$ is the atomic mass unit
\citep[CODATA 2006 value;][]{Mohr:08-CODATA}.  Note that the numerical
value \mb, along with other physics constants, are defined in the
\const\ module. 
The atomic mass of isotope $i$ is defined in \mesa\ as
\begin{equation}\label{e.atomic-weight-def}
W_{i} = A_{i} + \frac{\Delta_{i}}{\mb c^{2}} \ ,
\end{equation}
where $\Delta_{i}/c^{2}$ is the mass excess of isotope $i$.  This
treatment neglects the electronic binding energy, and $\Delta$ is
therefore independent of the ionization state of a given species.  The
electron rest masses are, however, included in this definition, since
the $W_{i}$ are atomic masses.

The \MESA\ microphysics modules---\kap, \eos, \neu, and \net---use
$\rho$, $T$, and $\left\{X_{i}\right\}$ as inputs.
\mesastar\ multiplies $\rho$ by a mass correction factor
$\bar{W}/\bar{A} = \sum_{i}W_{i}Y_{i}/\sum_{i}A_{i}Y_{i}$ to
distinguishes between $A_{i}$ and $W_{i}$ before starting the
calculation for a timestep.  A call to the routine
\code{composition\_info} in the \chem\ module returns the following
averaged quantities: the mean atomic mass number, $\bar{A} \equiv
\sum_{i} Y_{i}A_{i}/\sum_{i}Y_{i}$, mean atomic charge number,
$\bar{Z}\equiv \sum_{i}Z_{i} Y_{i}/\sum_{i}Y_{i}$, mean square atomic
charge number, $\sum_{i}Z_{i}^{2}Y_{i}/\sum_{i}Y_{i}$, the electron
abundance, $Y_{e} = \bar{Z}/\bar{A}$, and the mass correction term,
$\bar{W}/\bar{A}$. In addition, the routine returns the derivatives of
$\bar{A}$, $\bar{Z}$, and $\bar{W}/\bar{A}$ with respect to the baryon
fractions $X_{i}$:

\begin{eqnarray}
\left.\frac{\partial \bar{A}}{\partial X_{i}} \right|_{\rho, X_{j\ne i}} &=& \frac{\bar{A}}{A_{i}}\left(A_{i}-\bar{A}\right)\frac{1}{\sum_{i}X_{i}}; \label{e.dAbardx}\\
\left.\frac{\partial \bar{Z}}{\partial X_{i}} \right|_{\rho, X_{j\ne i}} &=& \frac{\bar{A}}{A_{i}}\left(Z_{i} - \bar{Z}\right) \frac{1}{\sum_{i}X_{i}}; \label{e.dZbardx}\\
\left.\frac{\partial (\bar{W}/\bar{A})}{\partial X_{i}} \right|_{\rho, X_{j\ne i}} &=& \left(\frac{W_{i}}{A_{i}} - \frac{\bar{W}}{\bar{A}}\right) \frac{1}{\sum_{i}X_{i}}. \label{e.dWoAdx}
\end{eqnarray}
Note that the routine does not make any assumption in these derivatives that $\sum_{i}X_{i} \equiv \sum_{i}A_{i}Y_{i} = 1$; in this formulation, $\sum_{i}X_{i}$ is not explicitly set to unity. 

At the beginning of each Newton iteration, the abundances are checked.  A mass fraction is considered good if its value exceeds \texttt{min\_xa\_hard\_limit}.  If all mass fractions meet this standard, then the mass fractions are clipped to range from 0 to 1, and the mass fractions are summed. If the sum differs from unity by less than a value \texttt{sum\_xa\_tolerance}, then the mass fractions are renormalized to sum to unity; otherwise, the 
code reports an error. 
Currently composition derivatives are ignored in the \eos\ and \kap\ routines. Equations~(\ref{e.dAbardx})--(\ref{e.dWoAdx}) allow, however, future additions to these routines to compute these derivatives analytically.

\subsection{Equation of State}\label{s.EOS}

The only significant change to the \eos\ module since \mesaone\ is the addition of tables 
for $Z > 0.04$, where $Z$ is the mass fraction of all elements heavier than He. The \eos\ module as described in \mesaone\ supplied equation of state (EOS) tables 
for $Z=0.0$, 0.02, and 0.04 at temperatures and densities for which neutral and partially-ionized species
are present (see \mesaone, Figure 1). For $Z > 0.04$ \mesa\ switched to the \code{HELM EOS} \citep{HELM}, which 
assumes full ionization. In order to rectify the inconsistent treatment of the partially-ionized
region at high $Z$, new EOS tables have been computed (J. MacDonald, priv.\ comm.) using the 
MacDonald EOS code \citep{MacDonald:2012} for $Z=0.2$ (scaled-solar), and two $Z=1.0$ compositions: one with 49.5\% C, 49.5\% O, and 1\% scaled-solar by mass; and  one with 50\% C and 50\% O by mass. Here 
``scaled-solar'' refers to the \citet{GN93} solar heavy element distribution adopted in the OPAL EOS 
tables \citep{OPAL2002}.

\subsection{Opacities}\label{s.opacities}

The \kap\ module now divides the opacity tables into a high-temperature domain, $\log(T/\K) \gtrsim 4$, and a low-temperature domain, $\log(T/\K)\lesssim 4$; the exact 
range of $\log T$ over which the tables are blended can be adjusted at runtime.  This treatment differs from the opacity tables described in \mesaone, which combined high- and low-temperature 
opacities into a single set of tables.  The motivation for separating the tables is to facilitate using different sources of low-$T$ opacity data. The \kap\ module now supports low-$T$ opacities from 
either \citet{Ferguson:2005} or \citet{Freedman:2008} with updates to the molecular hydrogen pressure-induced opacity \citep{frommhold2010} and the ammonia opacity \citep{yurchenko2011}. Either set may be selected at run time. 
The electron conduction opacity tables, based on \citet{Cassisi:2007}, have been expanded (Potekhin 2011, priv.\ comm.) to cover 
higher temperatures (up to $10^{10}\nsp\K$, originally $10^9\nsp\K$) and densities (up to 
$10^{11.5}\nsp\grampercc$, originally $10^{9.75}\nsp\grampercc$).

\subsection{Nuclear Reactions}\label{s.reactions}

Substantial improvements to the \code{net} module have been made since
\mesaone\ to increase the flexibility of the nuclear reaction networks
(see \S~\ref{s.nuts-and-bolts} for working details).  One such
improvement is the standalone one-zone burn routines.  These now
operate on a user-defined initial composition, nuclear network, and a
thermodynamic trajectory. Choices for the thermodynamic trajectory
include a burn with density and temperature held fixed,
a burn with pressure held fixed, and a burn with the density and temperature following an arbitrary, user-specified
profile. This last option is activated by setting
\code{read\_T\_Rho\_history=.true.} and specifying the file name
containing the profile through the variable
\code{T\_Rho\_history\_filename}.  The \mesa\ one-zone burn routines now
include user-specified options for the family of stiff ordinary
differential equation integrators from \citet{HairerWanner}.
In addition, three user-defined
switches are provided to switch between using dense matrix linear
algebra solvers, for smaller networks, and sparse matrix linear
algebra solvers for larger ones.  The option \code{decsol\_switch} sets the number of isotopes at which the switch occurs; options \code{small\_mtx\_decsol} and \code{large\_mtx\_decsol} specify the dense and sparse solvers, respectively.

\begin{figure}[htbp]
\centering
 \includegraphics[width=\figwidth]{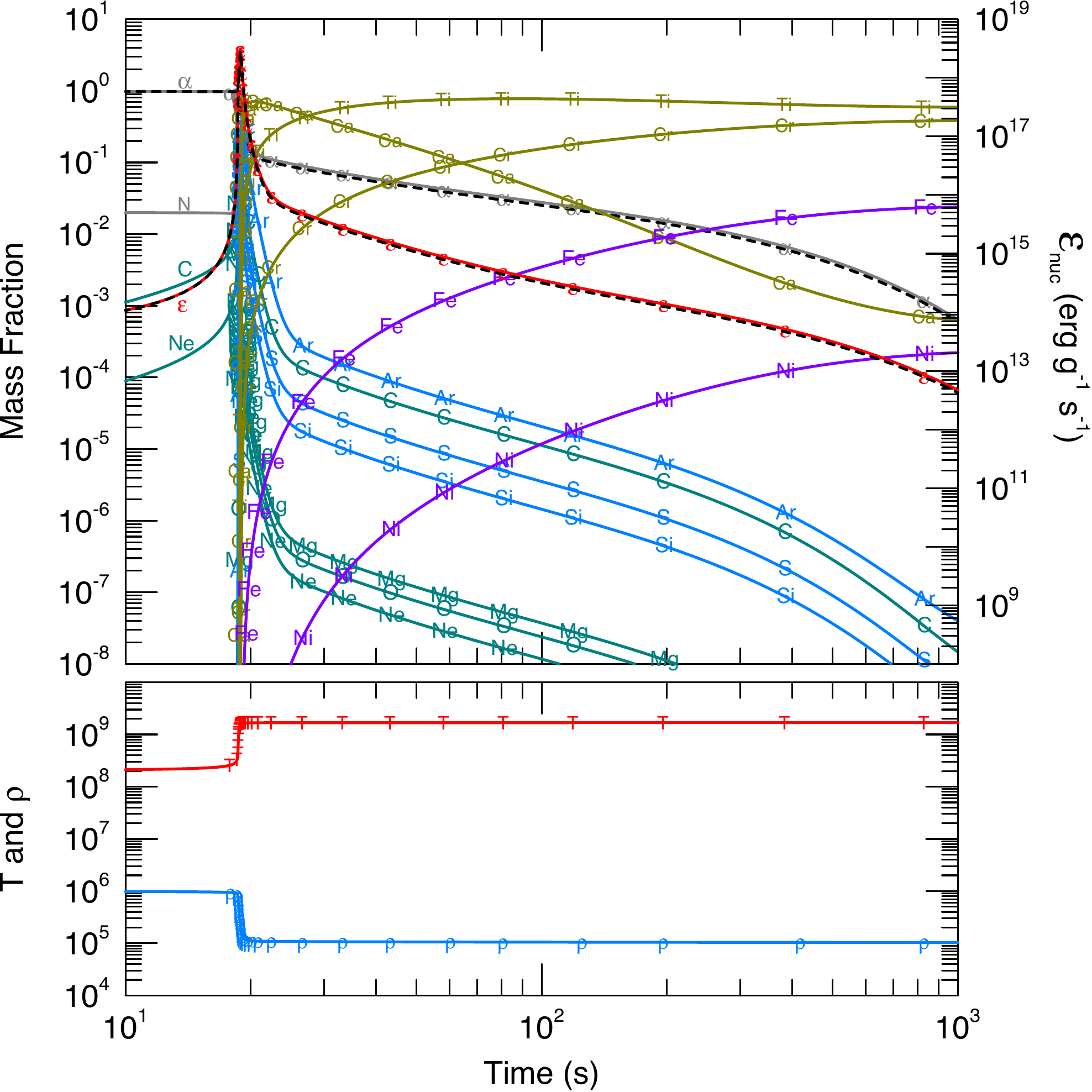}
\caption{
A one-zone helium and nitrogen burn at constant pressure, $P = 
3.1\times 10^{22}\nsp\ergs\usp\cm^{-3}$, starting from an initial 
temperature of $T= 2\times 10^8\nsp\K$. Evolution of the 
temperature and density are shown in the lower panel, while
the upper panel shows the mass fraction of key isotopes (right axis)
and the energy generation rate per unit mass (left axis; red curve).
\mesa\ results are shown by the colored and labelled curves, 
and the results from an independent one-zone burner 
\citep{timmes_1999_ab} are shown by the dashed black curves.
        }
\label{one_zone_burn}
\end{figure}

Figure \ref{one_zone_burn} shows the constant pressure option of these
routines operating on conditions that might be encountered for helium
burning on the surface of a white dwarf. The initial pressure is
$3.1\ee{22}\nsp\erg\usp\cm^{-1}$, the initial temperature is
$2\ee{8}\nsp\K$, the initial composition is $X(\helium)=0.98$ and
$X(\nitrogen)=0.02$, and the system was evolved for
$10^{4}\nsp\second$ with a 19-isotope network. Evolution of the
density and temperature under the constant burn conditions are shown
in the lower panel of Figure \ref{one_zone_burn}.  The temperature
slowly increases and the density slowly decreases as the material
begins to burn and release energy at a rate of $\epsnuc = c_P
\ dT/dt$. When the temperature crosses a critical threshold at
$\approx 20\nsp\second$, a runaway occurs as the temperature rapidly
rises and the composition burns to heavier elements. The material then
establishes a final equilibrium state, no energy from nuclear burning
is injected into the system, and the temperature reaches a plateau.

The upper panel of Figure \ref{one_zone_burn} compares the evolution
of key isotopes and the energy generation rate per unit mass of the
\mesa\ one-zone burner (colored and labeled curves) with an
independent one-zone burner (dashed black curves) based on
\citet{timmes_1999_ab}.  These comparisons indicate that both one-zone
burns produce a final composition that is mostly \titanium[44] and 
\chromium[48]. Over most of the evolution, the two
calculations give mass fractions of various isotopes that agree to
within 2--3 significant digits. Larger differences in some of the
heavier isotopes at the end of the calculation are due to differences
in the adopted nuclear reaction rates.

Another improvement is the \code{net} module now accesses reactions
from both \code{weaklib} and \code{reaclib}. Rather than evaluating
the standard seven-parameter fit for $\NA\langle\sigma v\rangle$ for
the \code{reaclib} rates \citep{Cyburt2010The-JINA-REACLI} every time
a reaction rate is needed, the \code{net} module caches separate rate
tables for each reaction.  Inverse rates are calculated directly from
the forward rates (those with positive $Q$-value) using detailed
balance, rather than using fitted rates. This is important for
explosive nucleosynthesis approaching nuclear statistical equilibrium
\citep[see][]{Calder2007Capturing-the-F}. The nuclear partition
functions used to calculate the inverse rates are taken from
\citet{rauscher00}.

\subsection{Atmosphere Boundary Conditions}\label{s.atmospheres}
The \atm\ module provides the surface boundary condition for the interior model. A collection of four new options that extend the set described in \mesaone\ are now available.
\begin{enumerate}

\item \code{solar\_Hopf\_grey}: Implements the solar-calibrated Hopf $T(\tau)$ relation, where
\begin{equation}
T^4(\tau) = \frac{3}{4} \Teff^4 \left[\tau + q(\tau)\right],
\end{equation}
and
\begin{equation}
q(\tau) = q_1 + q_2 \exp(-q_3 \tau) + q_4 \exp( -q_5 \tau).
\end{equation}
The $q_i$ are fit to the solar atmosphere with resulting values $q_1=1.0361$, $q_2=-0.3134$, 
$q_3 = 2.448$, $q_4 = -0.2959$, and $q_5 = 30.0$ (J.\ Christensen-Dalsgaard, 2011, priv.\ comm.).

\item \code{grey\_and\_kap}: Expands on the simple assumption that $P\simeq\tau g/\kappa$ 
by iterating to find a consistent solution among $P$, $T$, and $\kappa(\rho,T)$.

\item \code{grey\_irradiated}: 
Implements the \citet{2010A&A...520A..27G} $T(\tau)$ relation that includes both
external irradiation by the star and cooling flux from the interior; see
\citet[][eq.\ 49]{2010A&A...520A..27G} along with the discussion and results
in \S \ref{sec:irradiation}.
In addition to the external and internal fluxes, this boundary condition
requires two constant opacity values: $\kappav$ for the external radiation, and $\kappath$ for the thermal radiation
generated within the atmosphere.
This boundary condition is unique in that it is applied at a specified pressure level,
as opposed to optical depth. This pressure must be chosen sufficiently high to capture
any heating of the atmosphere by the irradiation.

\item \code{WD\_tau\_25\_tables}: Provides as outer boundary
  conditions the values of $\Pgas$ and $T$ at $\log(\tau)=1.4$ as
  extracted from pure hydrogen model atmospheres of WDs
  \citep{Rohrmann12,Rohrmann01}. The tables span a range of effective
  temperatures and surface gravities: $2,000\nsp\K \leq \Teff \leq
  40,000\nsp\K$ and $5.5 \leq \logg \leq 9.5$. See \S~\ref{s.completing-evolution} for
  an example of the use of these tables.

\end{enumerate}

\section{Nuts and Bolts}\label{s.nuts-and-bolts}

We now briefly describe the primary components of evolution calculations.
\MESAstar\ first reads the input files and initializes the physics
modules to create a nuclear reaction network and access the EOS and
opacity data.  The specified starting model is then loaded into memory
and the evolution loop is entered.

\subsection{Evolve a Step}\label{s.evolve_a_step}

The top level routine for evolving a star for a single timestep is
\code{do\_evolve\_step}.  If this is the first attempt to do a step
starting from the current state, the model is remeshed (see \S
\ref{s.mesh-controls}), and information for MLT++ is prepared by the
routine \code{set\_gradT\_excess\_alpha} (see \S
\ref{s.superadiabatic}).  Sufficient information is saved so that if
necessary it will be possible to make other attempts (i.e., after a
redo, a retry, or a backup).  In addition to the current state, we
keep the previous state (called ``old''), and the one that came before
``old'' (called ``older'').  During the step, the current state is
modified, and the old one holds the state at the start of the step.
If we do a redo or a retry, we copy old to current to restore the
starting state.  If we do a backup, we copy older to old before
copying old to current, making us start at the state prior to the
current one.  Note that the duration of the timestep is determined
before the call on \code{do\_evolve\_step} by the process described in
\S \ref{s.timestep-controls}.

After remeshing and the other initial preparations,
\code{do\_evolve\_step} begins the operations that are done on every
attempt.  It first calls the routine \code{do\_winds} which sets
\Mdot\ based on the current radius, luminosity, mass, metallicity, and
other properties as needed.  During the evalution of \code{do\_winds}
there is a call on the user-defined \code{other\_wind} routine giving
users an easy way to define different schemes for setting \Mdot.

Information for evaluating the Lagrangian time derivatives is stored
by a call to \code{save\_for\_d\_dt}. The ensuing call to
\code{do\_adjust\_mass} adds or removes mass without changing the
number of grid points (see \S \ref{s.mass-adjustment}).  Information
for evaluating the Lagrangian time derivatives is updated at this
point.  Variables for the model are evaluated to reflect the changes
made by remeshing and changing mass.  This includes evalution of the
Brunt-V\"ais\"al\"a frequency (see \S \ref{s.ledoux}), and the
diffusion coefficients for the mixing of composition (see \S
\ref{s.semiconvection} and \ref{s.thermohaline}).  The user-defined
routines \code{other\_brunt}, \code{other\_mlt}, and
\code{other\_mixing} are called as part of this.  If rotation is
enabled, there is a call to \code{set\_rotation\_mixing\_info} (see \S
\ref{s.rotation}) which in turn calls \code{other\_am\_mixing}.  If
element diffusion from gravitational settling and chemical diffusion
is active, the routine \code{do\_element\_diffusion} adjusts the
composition and includes a call on \code{other\_diffusion} The ensuing
call to \code{do\_struct\_burn\_mix} solves for the new structure and
composition of the star through repeated Newton iterations (see \S
\ref{s.solve-burn-and-mix}).  Non-convergence causes
\code{do\_evolve\_step} to return with a result indicating a failure.
Convergence is followed by a call to the routine
\code{do\_solve\_omega\_mix} which adjusts the total angular momentum
by solving a diffusion equation (see \S \ref{s.nuts-rotation}); it
calls \code{other\_torque}.  There is an option to repeat the
operations described in this paragraph in case rotationally enhanced
mass loss (see \S \ref{s.rotation-mass-loss}) has not been sufficient
to eliminate super-critical surface velocities.  In such a situation,
the mass loss is adjusted iteratively until slightly sub-critical
velocities result.  In effect, this is an implicit solution for the
appropriate \Mdot\ when super-critical rotation occurs.

Next, if specified by the user,
\code{smooth\_convective\_bdy} is called to smooth abundances behind retreating
convection boundaries.   Finally, a call to
\code{do\_report} gathers information and metrics about the
timestep for the user.  This information will then be available to the user's
\code{extras\_check\_model} routine.

\subsection{Solving the Coupled Structure, Burn, and Mix Equations}\label{s.solve-burn-and-mix} 

A call to \code{do\_struct\_burn\_mix} invokes a Newton method---an
N-dimensional root find---to solve a system of $N$ nonlinear 
differential-algebraic equations for the new structure and composition
of the stellar model.  Here $N$ is the number of zones in the current
model times the number of basic variables per zone and can exceed
100,000.

The equations to be solved are written as the relation
$F(\code{basic\_vars})=0$, where $F$ is the vector-valued function of
the residuals.  If the \code{basic\_vars} were a perfect solution to
the equations, we would have $F=0$; in practice, the solution is never
perfect.  The solution strategy is to iteratively adjust the values of
the \code{basic\_vars} to reduce $F$ towards zero.  An approximate
solution is accepted depending on both the magnitude of F and the
relative size of the adjustments to \code{basic\_vars}.  Adjustments
are chosen using the Jacobian matrix of partial derivatives of all the
$F$ equations with respect to all the \code{basic\_vars}.

\begin{figure*}[htbp]
\centering \includegraphics[width=\doublewide]{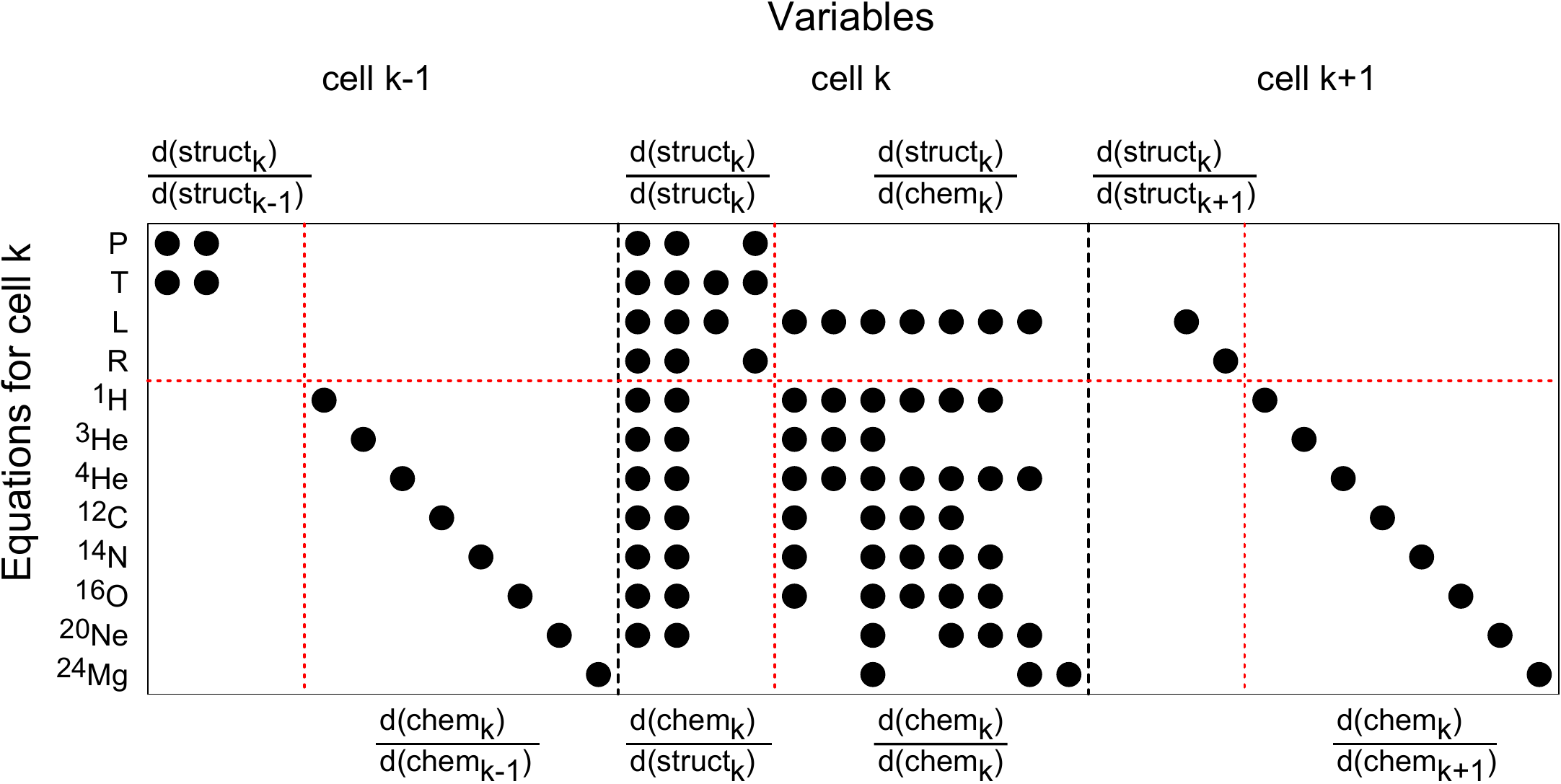}
\caption{
One row of the block tridiagonal Jacobian matrix for a $2.5\,\Msun$ ZAMS model, with black dots
showing the locations of non-zero entries.}
\label{fig.jacobian}
\end{figure*}

Figure \ref{fig.jacobian} shows the three blocks making up the
row of the block tridiagonal Jacobian matrix for the tenth from the
center cell of a non-rotating $2.5\,\Msun$ ZAMS model, with black dots showing
non-zero entries.  The partial derivatives of the equations for cell
$k$ form the rows of the blocks.  In this case, we have 4 equations
for the structure of the model ($P$, $T$, $L$, and $r$) and 8
equations for the chemical abundances ($^1$H through $^{24}$Mg).  Each
block of the tridiagonal matrix is demarcated by dashed black vertical
lines. The block matrix on the left shows the dependencies of the
equations for cell $k$ on the variables of cell $k-1$, the one in the
middle shows the dependencies of the equations for cell $k$ on the
variables of cell $k$, and the one on the right shows the dependencies
of the equations for cell $k$ on the variables of cell $k+1$. The
dashed lines partition each each block into four sub-blocks to
highlight the structure and abundance portions of each block.

The structure of the lower-right subblocks in the left and right
blocks shows that the chemical abundance of a particular species in
cell $k$ depends on the chemical abundances of that species in cells
$k-1$ and $k+1$; this is because of mixing between neighboring
cells. In this specific case of a non-rotating $2.5\,\Msun$
ZAMS model the mixing of chemical elements between cells is only due
to the treatment of convection.  The lower-right subblock of the
center block also shows the interdependencies of abundances due to
nuclear reactions in the cell.  The bottom-left subblocks are zero in
the left and right blocks but show dependencies on the $P$ and $T$
variables of the center block.  This is because the nuclear reactions
that change the abundances depend of $P$ and $T$ of that cell but do
not depend on $P$ and $T$ in the neighboring cells.  The columns for
$L$ and $r$ are zero in the center lower-left subblock because the
equations for the abundances do not directly depend on those
variables.  The upper-right subblocks are zero in the left and right
blocks but show that the equation for $L$ depends on the abundance
variables in the center block.  This is because the $L$ equation
includes results from nuclear burning, and that depends directly on
the composition of cell $k$ but not on the composition in neighboring
cells. The other rows in the center upper-right subblock are zero
because the equations for $P$, $T$, and $r$ do not directly depend on
composition.

Finally, consider the upper-left subblocks that show the dependencies
of structure equations on structure variables.  The upper-left
subblock in the center shows that each structure equation in $k$
depends on 3 or 4 of the structure variables in $k$, The $P$ and $T$
equations for cell $k$ also depend on both of the variables $P$ and
$T$ in $k-1$.  while the $L$ and $r$ equations for $k$ depend on the
corresponding variables in $k-1$.  This pattern reflects the form of
the finite differences in the implementation of the structure
equations: $P$ and $T$ differences use the outer neighbor ($k-1$)
while $L$ and $r$ differences use the inner neighbor ($k+1$).  The $L$
and $r$ equations for innermost cell $k=n$ use $L_{\rm center}$ and
$R_{\rm center}$; the $P$ and $T$ equations for the outermost cell
$k=1$ use the surface boundary conditions.

The structure variables for each zone always include the zone average
of the natural logarithm of the temperature, $\ln T$, the luminosity
at the outer edge of the zone, $L$, the natural logarithm of the
radius at the outer edge of the zone, $\ln r$, and a second
thermodynamic variable---either the zone average of the natural
logarithm of the mass density, $\ln \rho$, or the the zone average of
the natural logarithm of the gas pressure, $\ln\Pgas$.  Ideally it
would not matter whether $\ln \rho$ or $\ln \Pgas$ was used as the
second thermodynamic variable---for a given temperature and
composition the equation of state permits going back and forth between
the two.  Microphysics packages tend to use mass density as a primary
input (i.e., they use a Helmholtz free energy basis) leading to the
common choice of $\ln \rho$.  However, the structure equations are
solved only to within a finite but non-zero residual (see above).
Approximately correct values for the density and temperature can then
lead to anomalous pressure profiles, with tiny violations of
hydrostatic balance.  These local violations tend to appear near large
jumps in density, such as at a sharp H/He boundary.  Using $\Pgas$ as
the second thermodynamic variable (effectively using a Gibbs free
energy basis) removes these anomalous pressure profiles.  For example,
in stellar models without overshooting or semiconvection, the H/He
boundary is extremely sharp. Using the gas pressure as the second
thermodynamic variable results in single zone step function
transitions in the abundances and in the density, while the
temperature and pressure are smooth across the transition.
Applications that demand smooth pressure profiles, such as pulsation
analysis (see \S \ref{s.astroseismology}), should generally specify
the gas pressure as the second thermodynamic variable.

$\MESAstar$ treats convective mixing as a time-dependent, diffusive
process with a diffusion coefficient, $D$, determined by the
\code{MLT} module.  See \mesaone\ for the implementaton details of
standard mixing length treatment.  In addition to this standard MLT
treatment, the \code{MLT} module includes the option to use the
modified MLT++ prescription described in \S\ref{s.massive-evol} and
\S\ref{s.superadiabatic}.  After the convective mixing calculations
calculations have been performed, $\MESAstar$ calculates the
overshoot mixing diffusion coefficient as described in \mesaone.
During the solution of the coupled structure, burning, and mixing
equations the equation for mass fraction $X_{i,k}$ of species $i$ in
cell $k$ is determined by
\begin{eqnarray}
X_{i,k}(t + \delta t) - X_{i,k}(t) 
&=& dX_{\rm burn} + dX_{\rm mix} \nonumber \\
&=&  \frac{dX_{i,k}}{dt} \delta t + (F_{i,k+1} - F_{i,k}) \frac{\delta t}{dm_k},
\end{eqnarray}
where $dX_{i,k}/dt$ is the rate of change
from nuclear reactions, $F_{i,k}$ is the mass of species $i$ flowing across face $k$
\begin{equation}
F_{i,k} = \left ( X_{i,k} - X_{i,k-1} \right ) \frac{\sigma_k}{\overline{dm}_k}
\enskip,
\label{e.massfractions}
\end{equation}
where $\sigma_k$ is the Lagrangian diffusion coefficient from the
combined effects of convection and overshoot mixing
and $\overline{dm}_k = 0.5 (dm_{k-1} + dm_k)$.
For numerical
stability, $\sigma_k$ is calculated at the beginning of the timestep
and held constant during the implicit solver iterations.  This
assumption accommodates the non-local overshooting algorithm and
significantly improves the numerical convergence.  The structure of
the lower-right subblocks in the left and right blocks in Figure
\ref{fig.jacobian} shows that the dependency of the chemcial
abundances in cell $k$ depends on the chemical abundances of that
species in cells $k-1$ and $k+1$ as a result of convective mixing.

If the optional hydrodynamic mode is activated, then the radial
velocity at the outer edge of the cell, $v$, is added to the structure
variables.  Figure \ref{fig.jacobian} shows the order of the model
variables in the Jacobian: each cell includes the structure variables
followed by the mass fraction $X_i$ of each isotope.  Mass and the
local angular velocity $\omega$ are not treated as structure
variables because they are held constant during the Newton iterations.
The mass is set before the iterations, while $\omega$ is determined
after convergence. This is computed taking into account loss/gain 
of angular momentum during the time step, the new stellar structure and
internal transport of angular momentum calculated by a diffusion equation
(see \S~\ref{s.nuts-rotation}). 

The program flow to solve the coupled structure, burning, and mixing
equations is to first create the matrix of partial derivatives using
the current candidate solution, solve the block tridiagonal system of
linear equations for the corrections to the basic variables, apply the
possibly damped corrections (see next paragraph) to update the
candidate solution, and calculate the residual $F$. Then, if the
residual is small enough, we declare victory, otherwise we repeat the
general flow with the updated candidate solution. 

Each iteration of the Newton solver uses a linear approximation to
create a vector of corrections to the model.  These corrections do not
include the physical requirement that the abundance mass fractions
need to remain positive. To reduce the possible occurrence of negative
abundances \MESAstar\ now uses a damped Newton scheme.  This 
checks for proposed corrections that would produce negative abundances
and multiplies the entire correction vector by a factor less than one,
so that only part of the the full correction is applied.  In many
cases, this is sufficient to significantly improve the convergence
properties of a model.  In other cases, the damped correction scheme
may force so many small corrections that the Newton solver cannot
converge within the user-specified maximum number of iterations,
forcing the previous model to be attempted again with a smaller
timestep (termed ``a backup'').  On balance, this is usually a
small price to pay for an improved conservation of species and
more accurate solutions.

The modules in \code{star} provide routines to evaluate the residual
equations and create the Jacobian matrix.  Given a candidate solution
(i.e., the set of basic variables for each cell), the microphysics for
each cell (EOS, thermal neutrino loss, opacity, nuclear reaction
rates) are evaluated in parallel (see \S \ref{s.numerics}).  The
Jacobian matrix is then further populated with elements from rotation,
artificial viscosity, and mixing length theory for the temperature
gradient, and these are also evaluated in parallel.  Each of the
routines that evaluate these components returns output values and
partial derivatives of the output values with respect to the input
values. Analytic partial derivaties are used whenever feasible,
otherwise numerical partial derivatives are supplied.

\subsection{Timestep Controls}\label{s.timestep-controls}

Control of the timestep is a critical part of stellar evolution and
requires careful trade-offs. The timesteps must be small enough to
allow convergence in comparatively few iterations but large enough to
allow sufficiently efficient evolutions.  Changes to the timestep must respond
rapidly to varying structure or composition conditions, but they 
need to be  controlled to avoid large jumps
that can reduce the convergence rate or the accuracy of the results.  The routine
\code{pick\_next\_timestep} performs timestep control as a
two-stage process.  The first stage proposes a new timestep using the
H211B low-pass filter \citep{soderlind_2006_aa}, a scheme based on
digital control theory.  The second stage implements a wide range of
tests that can reduce the proposed timestep if certain selected
properties of the model are changing too much in a single timestep.

For the first stage, routine
\code{hydro\_timestep} sets the variable for the next timestep, \code{dt\_next},  according to the relative magnitude of
changes to the \code{basic\_vars}.  The variable reflecting the size of these changes is called \code{varcontrol} and is
calculated by the routine \code{eval\_varcontrol}.    For improved stability and response, the low-pass
controller uses previous and current values of \code{varcontrol} to
make the next timestep match the \code{varcontrol\_target}, $w_t$,
which is $10^{-4}$ by default.  To make this explicit, let
$\timestep_{i-1}$, $\timestep_{i}$, and $\timestep_{i+1}$ be the
previous, current, and next timestep, respectively, while $w_{c,i-1}$
and $w_{c,i}$ are the previous and current values of \code{varcontrol}.  The
maximum timestep for model $i+1$ is then determined by
\begin{equation}
\timestep_{i+1} = 
\timestep_{i} f \left[ 
\frac{f(w_{t}/w_{c,i}) f(w_{t}/w_{c,i-1})}{f(\timestep_{i}/\timestep_{i-1})} \right]^{1/4},
\label{filterA}
\end{equation}
where $f(x) = 1 + 2\tan^{-1}[0.5(x-1)]$.
This control scheme allows rapid changes in the timestep 
without undesirable fluctuations.

The timestep proposed by this low-pass filtering scheme
can be reduced in the second stage according to a
variety of special tests that have hard and soft limits.  
If a change exceeds its specified hard limit, the
current trial solution for the new step is rejected, and the code is forced to do a retry or
a backup.  If a change exceeds its specified soft limit, the next
timestep is reduced proportionally.  The current classes of special
cases that can reduce the next timestep are limits based on:
\begin{enumerate}
\item Number of Newton iterations required to converge.
\item Maximum absolute change in the  mass fraction of hydrogen or helium in any cell.
\item Maximum relative change in any mass fraction at any cell.
\item Magnitude in the relative change in the structure variables in each cell.
\item Nuclear energy generated in each cell for several categories of nuclear reactions.
\item Changes in the luminosity resulting from nuclear burning.
\item Changes at the photosphere in $\ln L$ and $\Teff$.
\item Changes in $\ln \rho_{\mathrm{center}}$, $\ln T_{\mathrm{center}}$, $X(\mathrm{H})_{\mathrm{center}}$, $X(\mathrm{He})_{\mathrm{center}}$.
\item Magnitude of the change in $\log (M/\Msun)$ due to winds or accretion.
\item Mass accreted so that compressional heating is correct (see \S\ref{s.compressional}).
\item Changes in the logarithm of the total angular momentum.
\item Distance moved in the HR diagram.
\item Maximum allowed timestep under any circumstance.
\item Any user specifed timestep limit, accomplished by setting \code{max\_years\_for\_timestep}, 
in the optional routine \code{extras\_check\_model}.
\end{enumerate}

For convergence studies with respect to the timestep it is vital to
change the control parameters that are actually setting the
timestep. Often, this is just \code{varcontrol\_target}, but in
many situations the timestep will be set by one of the 
special timestep control parameters.

\subsection{Mesh Controls}\label{s.mesh-controls} 

Control of the spatial mesh is a key ingredient of a stellar evolution
instrument, and requires careful trade-offs. The mesh must respond to
gradients in the structure, chemical composition, and energy generation,
in order to give an accurate result,
but it should not be overly dense since that will unnecessarily increase 
the cost of the calculation.

Since $\MESAstar$ allows for simulations with a fixed inner core mass,
\Mc, the total mass $M$ is $\Mc+\Mm$ where $\Mm$ is the modeled mass.
For cell $k$, $\MESAstar$ stores the relative cell mass $\dif q_k =
\dif m_k/\Mm$ where $\dif m_k$ is the mass contained in cell $k$.
The relative mass interior to the outer cell face is $q_k = 1 - \sum_{i=1}^{i=k-1} \dif q_i$,
and the total mass interior to the outer cell face is $m_k = q_k*\Mm + \Mc$.
In all cases, $m_1$ = $M$ and $q_1$ = 1.
We explicitly keep $\dif q_k$ in addition to $q$ and define $q$ in terms of $\dif q_k$
to avoid the need for evaluating $q_k-q_{k+1}$
since that can involve the subtraction of almost equal numbers
leading to an undesirable loss of precision \citep{lesaffre_2006_aa}.
For example, in the outer envelope of a star where the $q_k$ approach 1,
the $\dif q_k$ can be $10^{-12}$ or smaller.  By storing $\dif q_k$
we have 16 digit precision, whereas in this case,  $q_k-q_{k+1}$
would only give us 4 digits at best for the relative cell mass.

$\MESAstar$ checks the structure and composition profiles of the model
at the beginning of each timestep and, if necessary, adjusts the
mesh. A single cell can be split into two or more cells and two or more adjacent cells
can be merged.  In practice, only a small fraction of the cells are changed
during a remesh. This minimizes numerical diffusion, 
aids convergence, and keeps the cost of remeshing relatively small.
Remeshing is divided into a planning stage and an adjustment stage.

The planning stage determines which cells to split or merge based on
the magnitude of allowed cell-to-cell changes in a variety of mesh functions.  
Built-in mesh functions include gradients of the mass, radius, pressure,
temperature, adiabatic gradient, angular velocity and mass fractions
above some threshold.   Users can add others by defining their own \code{other\_mesh\_functions} routine.

Other controls are provided to increase the
sensitivity in regions selected by the user. Examples include
increasing the spatial resolution in regions with changes in 
user-specified abundances with respect to pressure, changes in the
energy generation rate with respect to pressure for different types of
burning (e.g., the pp chains, CNO cycles, triple-$\alpha$, and
others), for regions near burning or non-burning convective
boundaries, and others.

After the  mesh functions are evaluated,  the relative
magnitude of the changes between adjacent cells
are determined.  The magnitude of change is multiplied by
\code{mesh\_delta\_coeff} to obtain a weighted mesh function.  Cells
where the weighted changes are ``too large'' are marked for splitting, and
cells where the changes are ``too small'' are marked for merging.  For
example, if the weighted changes in all mesh functions from cells $k$ to
$k+n$ are  less than 1, the series of cells from $k$ to $k+n$ are
marked for merging.  If any weighted mesh function changes from cell $k$ to
$k+1$ by an amount greater than 1, the larger of cell $k$  and cell $k+1$ is
marked for splitting.  Finally, if adjacent cells have too large of a
relative size difference (as defined by
\code{mesh\_max\_allowed\_ratio} which defaults to 2.5), the larger cell is marked for
splitting and the check for excessive ratios is repeated.  This can lead to a cascade
of splitting in order to ensure that cells sizes do not have excessive jumps.

The adjustment stage executes the remesh plan by performing the merge
and split operations to calculate new values for basic variables.
Special care is taken to use physical knowledge whenever possible when
setting new values. For example, conservation of mass is accounted for
when determining new densities, and species conservation is used when
setting new mass fractions.  Energy conservation is used when setting the 
temperature (see \mesaone), and conservation of angular
momentum plays a role in determining the angular velocity.  Cells to be split are
constructed by first performing a monotonicity-preserving cubic
interpolation \citep{steffen_1990_aa} in mass to obtain the
luminosities and $\ln r$ values at the new cell boundaries.  The new
densities are then calculated from the new cell masses and volumes.
Next, new composition mass fraction vectors are calculated.  For cells
being merged, the mass averaged abundances are used. For cells being
split, neighboring cells are used to form a linear approximation of
mass fraction for each species as a function of mass coordinate within
the cell.  The slopes are adjusted so that the mass fractions sum to
one everywhere, and the functions are integrated over the new cell
mass to determine the abundances.

\subsection{Mass Adjustment}\label{s.mass-adjustment}

Mass adjustment for mass loss or accretion is performed at each timestep
when  \code{do\_evolve\_step} calls the routine \code{do\_adjust\_mass} (see \S \ref{s.evolve_a_step}). 
$\MESAstar$ offers several ways to set the rate of mass change \Mdot.  A
constant mass accretion rate (positive \Mdot) or mass loss rate
(negative \Mdot) can be specified in the input files, a wind can produce a mass loss,
the user can set \Mdot\ in an \code{other\_wind} routine or in an \code{other\_check\_model} routine.  
When \code{do\_adjust\_mass} is called, the
timestep $dt$ and the rate of mass change $\dot{M}$ are known, and thus the change
in mass, $\delta M = \Mdot\ \ \timestep$. 

When there is a change in mass, instead of adding or removing cells,
the total mass is changed by modifying the modeled mass $\Mm$, and
cell mass sizes are changed by revisions to $\dif q_k$ which in turn
changes cell mass locations $q_k$ (see \S \ref{s.mesh-controls}).
The mass structure is divided into an inner region where the
$m_k$ and $\dif m_k$ are unchanged but the $q_k$ and $\dif q_k$ change,  an 
outer region  where the $q_k$ and
$\dif q_k$ are unchanged but the $m_k$ and $\dif m_k$ change, and an
intermediate blending region where all of these change.
The selection of the region boundaries is discussed
in detail in \mesaone.  The implementation of $\epsgrav$ in 
the newly accreted matter is described in \S\ref{s.compressional}.

Once the three regions have been defined, the $\dif q_k$ are updated.  In the
inner region they are rescaled by $M/(M + \delta M$).  Thus,
$\dif m_k$, $m_k$, and $X_k$ have the same value before and after a
change in mass.  This eliminates the possibility of unwanted numerical
diffusion causing unphysical mixing in the center region.  In the outer region, cells retain the same
value of $\dif q_k$ to improve convergence  in the
high entropy parts of the star \citep{sugimoto_1981_aa}.  
In the intermediate region, the $\dif q_k$  are uniformly scaled to make $\sum \dif q_k = 1$.

The chemical mass fractions of cells in the intermediate and outer regions are
then updated by summing the abundances between the new cell mass
boundaries.  
This step is not necessary for the inner region since those cells have not changed mass location.
In the case of mass accretion, the composition of the
outermost cells whose enclosed mass totals $\delta M$ is set to match
the specified accretion abundances.  The single cell that is part old material
and part newly accreted material is given an appropriately mixed composition.

Finally, to create a somewhat better
starting model for the Newton iterations (see \S \ref{s.solve-burn-and-mix}),	
the $\ln T$ and $\ln \rho$ and $\ln \Pgas$ 
values are revised by monotonic cubic interpolating to the
cell center by mass from the values prior to mass adjustment.
The $\ln r$ and material speed $v$ are also set by monotonic cubic
interpolation to the value at the new outer mass boundary.  The angular
velocity is set by integrating the angular momentum between the new
cell mass boundaries and using the new $\ln r$ values,
conserving the total angular momentum to the floating point
limit of the arithmetic.

\subsection{Evolving the Angular Velocity}\label{s.nuts-rotation}

Initialization of rotation in \MESAstar\ begins from a non-rotating model. The angular velocity
$\omega$ is added to the set of model variables
and initialized to a constant value throughout the model (i.e.,
solid body rotation). The initial value of $\omega$ can be specified
as a surface rotational velocity (in km/s) or as a fraction of the surface critical
rotation rate (see \S\ref{s.rotation}).  
During the subsequent evolution, $\omega$ is changed at each timestep by
remeshing, mass adjustment, radius adjustment (as part of the
structure evolution), optional extra angular momentum removal in the outer
layers, and the transport of angular momentum optionally with user-defined
source terms for external torques.

The angular velocity $\omega$ is defined at cell boundaries.  Thus
\code{omega(k)} is at the outer boundary of cell \code{k}, which is
the same location as the radius, \code{r(k)},
the specific moment of inertia, \code{i\_rot(k)}, and 
the specific angular momentum, \code{j\_rot(k)}.  The mass associated
with \code{omega(k)}  spans the range from the center  by mass of cell \code{k} outward to the center by mass of cell \code{k-1}
and is referred to as \code{dm\_bar(k)} to distinguish it from the cell mass  \code{dm(k)}.

The remeshing operation splits and merges cells but does not change the physical stellar structure  (see \S\ref{s.mesh-controls}).  For regions 
where there has been a change in the mesh, the values of $\omega$ are
adjusted to give the same angular momentum  as
before. More specifically,  the angular momentum from the original model is summed 
over the mass range encompassed by the new \code{dm\_bar(k)}, and
\code{omega(k)} is adjusted to give the same total for the new model.
      
During the mass adjustment operation, when mass is added or removed
from the model, cells in the outer layers are moved to new mass
locations (see \S\ref{s.mass-adjustment}).  As part of this process, the angular
velocity values are updated to conserve angular momentum using the
same scheme as for remeshing: sum the angular momentum in the
original model and set \code{omega(k)} in the new model to conserve it. 
Newly added material from accretion is given the current
surface angular velocity.  In the case of
mass loss, this operation removes the amount of angular momentum 
contained in the lost mass at the start of the timestep; it does not
deal with possible transport of angular momentum into the lost mass
during the timestep.   That  is dealt with by an optional, 
user-specified removal prior to the angular momentum transport.
   
\MESAstar\ performs the transport of
angular momentum as a separate operation from the evolution of structure and composition.
This is done in order to obtain high accuracy in the angular momentum transport by
using substeps and quad-precision linear algebra.  It does not
introduce additional operator splitting errors since $\omega$ is not used in
the structure and abundance equations.  So we solve for the new
structure and composition after any mass change and before
the transport of angular momentum.   Calculation of the new stellar structure changes the radii but does
not change the mass partitioning of the model (see
\S\ref{s.solve-burn-and-mix}).  Given the new radius \code{r(k)}, we calculate 
the new \code{i\_rot(k)}.  Then using the unchanged \code{j\_rot(k)},
\code{omega(k)} is set to \code{j\_rot(k)/i\_rot(k)} to conserve specific angular momentum.  
Since \code{dm\_bar(k)} has not changed, this also conserves total angular momentum.

Next, \MESAstar\ applies an optional, user-specified amount of angular
momentum loss in the outer surface layers. This is to account for possible
transport of angular momentum during the timestep from these outer layers into the mass 
removed by the mass adjustment operation.  

The final operation is the transport of angular momentum within the
star, which is treated with a diffusion approximation 
\citep{Endal:1978,Pinsonneault:1989,Heger:2000}
\begin{equation}\label{e.jdiff}
  \dxdycz{\omega}{t}{m}=\frac{1}{i}\,\dxdycz{}{m}{t}\, 
  \SBrak{(4\pi r^2 \rho)^2 \, i\nu \,\dxdycz{\omega}{m}{t}}-\frac{2\omega}{r}\,\dxdycz{r}{t}{m} 
  \, \Brak{\frac{1}{2}\frac{\dif\ln i}{\dif\ln r}} \ ,
\end{equation}
where $i$ is the specific moment of inertia of a shell at mass
coordinate $m$, and $\nu$ is the turbulent viscosity
determined as the sum of the diffusion coefficients for convection, double diffusion, 
overshooting and rotationally-induced instabilities (see \S\ref{s.rotation}).   The diffusive transport is carefully implemented to
accurately conserve angular momentum.  The angular momentum associated with 
location $k$ is \code{dm\_bar(k)*i\_rot(k)*omega(k)}.  The change
in angular momentum for $k$ is  determined by the flux
in angular momentum from $k-1$ to $k$ and from $k+1$ to $k$.  The flux
from $k-1$ to $k$ is set by $\nu(k-1)$
and the difference between \code{omega(k)} and \code{omega(k-1)}.  The flux
from $k+1$ to $k$ is found similarly using $\nu(k)$ and the difference
between \code{omega(k)} and \code{omega(k+1)}. Source terms for location
$k$ are applied by user-supplied values for \code{extra\_jdot(k)} or
\code{extra\_omegadot(k)}.  The finite difference equation for the effects of the transport and source terms is solved over the
stellar timestep with an implicit time integration that uses multiple
smaller timesteps. The sizes of these substeps are determined by the
timescale set by the diffusion coefficients and the differences in
$\omega$.  It is not unusual to use 10 or more
substeps to evolve \code{omega(k)} over the stellar timestep. Each
implicit substep is solved using a quad-precision tridiagonal
matrix routine.  The conservation of total angular momentum is
monitored and the stellar timestep is rejected if there is any
deviation from conservation by more than a user-specified factor.  In
practice, we find the total angular momentum is conserved over the
stellar timestep to within a few digits of the floating point limit of
the arithmetic.


\subsection{Free Parameters}\label{s.nuts-free-parameters}
Stellar evolution calculations involve the choice of a number of free parameters.
The values of these parameters are not determined by first principles, 
and in the literature one can find a range of possibilities. 
In some cases the parameters can be constrained by matching a restricted set of observations; 
in other cases they represent common choices. Users need to be aware that their results will 
depend on these values, and that in some cases the sensitivity can be large. 
Below we illustrate this by discussing some of the main parameters involved in
the mixing of stellar interiors.

\subsubsection{Convection}
In the literature the value of the mixing length parameter $\alphaMLT$ (see e.g. Paper I for a definition) 
is usually found to vary within the range $1.0 \lesssim \alphaMLT \lesssim 2.0$.
Efforts are ongoing  to eliminate this free parameter  
\citep[e.g.,][]{Arnett:2010}.

\subsubsection{Overshooting}
In the literature the adopted value for the convective core overshooting parameter 
is in the range $0.1 \lesssim \fov \lesssim 0.6$, in units of the pressure scale height 
$\lambda_P$, when the overshoot zone is considered to be fully mixed 
\citep{MaederMeynet:1987,Dupret:2004,Straka:2005,Claret:2007,Briquet:2007}. 
 When overshoot mixing is treated as an exponential decay process the free parameter should be smaller, 
 $\fov\sim0.016$, \citep[see the discussion by][]{Herwig:2000}.  \MESA\ has the ability to treat overshoot mixing zones as either fully mixed or in the exponential decay formalism.

It has been suggested that the overshooting parameter is a function of both mass and metal abundance, in that it 
should transition smoothly from zero to a maximum value over a small range of stellar mass
 where a convective core is present on the main sequence \citep{Woo:2001,VandenBerg:2006}, but see also \citet{Claret:2007}. 
 A dependency on the evolutionary stage seems also likely \citep{Herwig:1997,Meakin:2007,Tian:2009}.
See Fig.~13, 14 and 15 in \S\ref{s.mixing} for an example of the sensitivity of the calculations to 
changes in the $\fov$ parameter.

\subsubsection{Semiconvection}
Semiconvection, as implemented in \MESA, requires a choice of the free parameter  $\alphasc$ 
(see \S\ref{s.semiconvection}). 
In the literature this spans the range  $0.001 \lesssim \alphasc \lesssim 1.0$  
\citep{Langer:1991,Yoon:2006}.  
See Fig.~13, 14 and 15 in \S\ref{s.mixing} for an example of the sensitivity of the calculations to changes 
in the $\alphasc$ parameter.
Research is ongoing to eliminate this free parameter \citep{Woods:2013,Spruit:2013}.

\subsubsection{Thermohaline Mixing}
The implemented formulation for thermohaline mixing requires the adoption of  the free parameter 
$\alphath$  (see \S\ref{s.thermohaline}). 
In the literature this parameter can be usually found within the range $1 \lesssim \alphath \lesssim 667$ 
\citep{Kippenhahn:1980,Charbonnel:2007,cantiello_2010_aa,Stancliffe:2010,Wachlin:2011}.  
Research is ongoing to eliminate this free parameter \citep{Traxler:2011,Brown:2013}.

\subsection{Nuclear Reactions}\label{s.nuts-nuclear-reactions}

A reaction network is defined by a set of isotopes and a set of
reactions; these sets are specified in a reaction network definition
file. \mesa\ comes with many predefined reaction networks in
\code{data/net\_data/nets} and can also incorporate user-defined
networks. To use a custom network, a user creates a reaction network
definition file containing the command
\code{add\_isos\_and\_reactions(isos\_list)}, which will automatically
add all reactions linking the isotopes in \code{isos\_list}.  The
sequence of isotopes in \code{isos\_list} may be specified by the name
of the isotope: for example, \code{add\_isos\_and\_reactions(he4)}
adds \helium[4].  Alternatively, one can specify the name of element
followed by the desired minimum and maximum nucleon number. For
example, the command \code{add\_isos\_and\_reactions(o 16 18)} adds
\oxygen[16], \oxygen[17], and \oxygen[18].  Note that because many of
the predefined networks may use effective rates---that is, using one
reaction rate to represent a reaction sequence or group of reaction
sequences---it is not recommended that the user extend one of the
pre-existing networks with this command.

\mesa\ creates and stores reaction rate tables for each reaction whose
entries are derived from evaluating standard analytic fitting
formulas (see \S\ref{s.reactions}), but these reaction rates may be
replaced with user-specified values. To change a rate for a given reaction,
\begin{enumerate}
\item create a file with two columns: the temperature in units of $10^{8}\nsp\K$ and the
rate $\NA\langle\sigma v\rangle$ in units of
$\cm^{3}\usp\gram^{-1}\usp\second^{-1}$;

\item list the file name in a local file \code{rate\_list.txt} along with 
its ``handle'' for the reaction rate in question (see discussion below); and

\item set the parameter \code{rate\_tables\_dir} in the namelist \code{star\_job} to the 
name of the directory in which \code{rate\_list.txt} is located; by
default this is \code{data/rates/rate\_tables}.
\end{enumerate}
The handle for a reaction is derived from the input and output channel isotopes according to a few rules.  
Capture reactions, such as $x(\pt,\gamma)y$, have handles of the form \code{r\_x\_pg\_y} and 
exchange reactions, such as $x(\alpha,\pt)y$, have handles of the form \code{r\_x\_ap\_y}.
Other arbitrary reactions may be added by 
listing them in a form \code{r\_inputs\_to\_outputs}
where \code{inputs} and \code{outputs} are isotopes separated by ``\_''.
If the same isotope appears two or more times, the isotope name may be repeated.
For example, the triple-$\alpha$ reaction is specified as 
\code{r\_he4\_he4\_he4\_to\_c12}.
Isotopes are ordered by increasing $Z$ and $N$, e.g., 
\code{r\_h3\_be7\_to\_neut\_h1\_he4\_he4}.  To see a list of reactions used, the parameter
\code{show\_net\_reactions\_info} in namelist \code{star\_job} should be set to  \code{.true.}.

\subsection{Multicore Performance}\label{s.numerics}

\mesa\ implements shared memory multiprocessing via
OpenMP\footnote{\url{http://www.openmp.org}}.  \mesaone\ explored the
runtime scaling of \mesastar\ which at that time used a banded matrix linear algebra
solver that did not benefit from multiple cores.  A large part of the
performance improvement in \mesastar\  since \mesaone\ comes from converting to
a parallel block tridiagonal linear algebra  solver derived from
BCYCLIC \citep{Hirshman10}.  This improved  solver is
particularly important since linear algebra is typically the
largest part of the runtime in \mesastar.  In addition, the new algorithm 
has the desirable property of producing numerically
identical results independent of the number of cores, an attribute
that is not generally true of parallel matrix solvers.

Our test case is a $1.5\,\Msun$ model with $Z=0.02$ that is evolved from
the ZAMS until the central H mass fraction falls to 0.35. This model
includes 25 isotopes and 4 structure variables per cell with a
variable number of zones typically exceeding 1700.  
The test takes  $\sim$55 time steps to cover $\sim1.4\,\Gyr$ and 
uses the default amount of I/O.

\begin{figure}[htbp]
\centering{
\includegraphics[width=\figwidth]{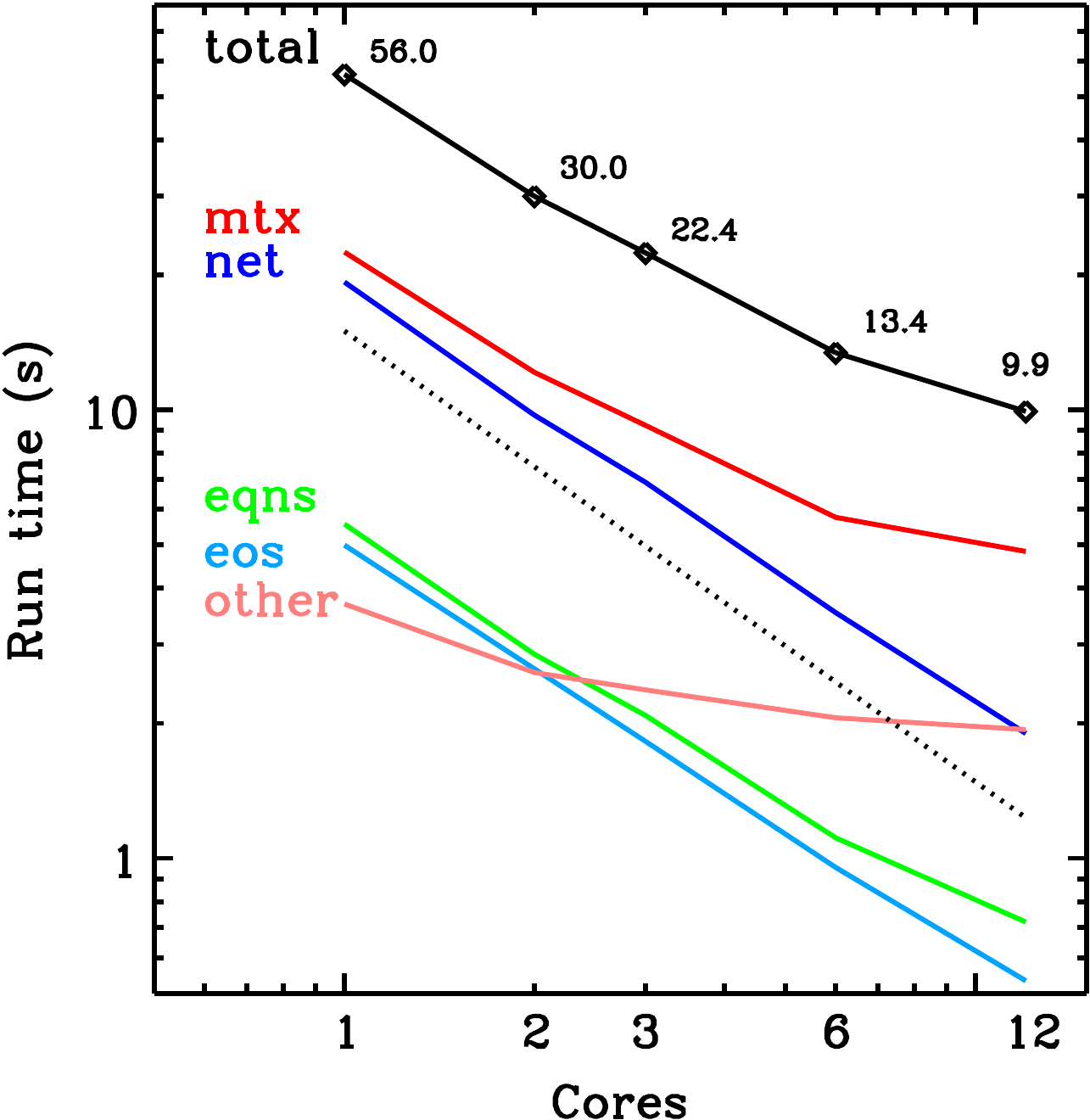}
\caption{\label{timing}Scaling behavior of various components of \mesastar\ 
using 1, 2, 3, 6, and 12 cores. 
The top curve shows the total run time, and the lower solid curves
show the run times for the components of the total.
The dotted line shows the ideal scaling relation.}
}
\end{figure}

Figure~\ref{timing} shows the scaling behavior of some key components
of \mesastar\ under GFORTRAN 4.7.2. on a 12 core 2010 Apple MacPro.  The
dotted line shows the ideal scaling relation where 
doubling the number of cores cuts the run time in half.  The linear algebra,
labeled ``mtx,'' dominates the total run time as the number of cores increases.  For example, in the case of 12 cores
it accounts for about half the total and is 2.5 times larger than ``net'',
the evaluation of the  nuclear reaction network.  The net evaluations
closely approach the ideal scaling behavior because they
 can be done in parallel, each cell independent of the others, with one core working on one cell at a time.  The equation
of state component, labeled ``eos,'' also closely approaches the ideal scaling
law while consuming less than a third of the run time for the net.  The
component labeled ``eqns,'' which includes the evaluation of the structure
equations and the creation of the block tridiagonal matrix, also 
is close to  the ideal scaling law and costs about the same as the eos.  
The ``other'' component is everything else.  It is dominated
by processes that currently are not efficient to parallelize because 
of the relatively large overhead for OpenMP operations.
Consequently it remains at roughly a constant run time independent of the
number of cores.  When a significantly larger number of cores per processor 
becomes available, the larger operations in this category will have to be
reworked or they will dominate the total run time.

The run time also depends on the hardware, the quality of the
compiled code, and the efficiency of the OpenMP implementation.  For
example,
we ran the test case under GFORTRAN 4.7.2 on a
40-core server.  While we obtained a speedup of 5.6 in going from 1 to
12 cores for the 12 core machine (see Figure~\ref{timing}), we find a
speedup of only 4.8 on the 40-core server in going from 1 to 12 cores.
Moreover, the speedup per core dropped steeply beyond 8--12 cores on the 40-core server,
confirming the expectation that much work will be required to make  full use of
machines with many cores.

\subsection{Visualization}\label{s.visualization}

\MESAstar\ provides alphanumeric output at user-specified regular intervals.  
In addition, the routines in module \code{star/public/pgstar.f} provide
 an option for
concurrent graphical output with the
PGPLOT\footnote{\url{http://www.astro.caltech.edu/~tjp/pgplot/}.}
library to create on-screen plots that can be saved for
post-processing into animations of an evolutionary sequence.  A
variety of options are provided and are all configurable
through the \code{PGstar} inlist. For example, a \code{PGstar} X11
window can simultaneously hold an H-R diagram, a $\Tc\textrm{--}\rhoc$ diagram,
and interior profiles of physical variables, such as nuclear energy
generation and composition.   The \code{PGstar} inlist is read at each timestep,
so the display options can be changed without have to stop  \MESAstar.

Since \mesaone, a number of \MESAstar\ users have developed and released 
toolkits\footnote{See \url{http://mesastar.org/tools-utilities}.}
to visualize the alphanumeric output with common graphical packages
including: Mathematica scripts (contributed by Richard O'Shaughnessy)
and the intuitive and efficient graphical user interface MESAFace
\citep{giannotti_2012_aa}; MatLab utilities (contributed by Dave
Spiegel and Gongjie Li); IDL functions (contributed by Rich Townsend);
Python scripts (contributions from Falk Herwig and the NuGrid
collaboration, David Kaplan, Alfred Gautschy, William Wolf); and Tioga
scripts (contributed by Christopher Mankovich and Bill Paxton).

\subsection{Operating System and Compiler Considerations}\label{s.compiler_OS}

We next consider the implications of running \mesa\ compiled 
with different compilers and on different operating systems. The 
operating systems examined are Linux (Gentoo 2.1; kernel 3.6.11) and Apple 
OS X (10.7.5), both 64-bit; the compilers are GNU gfortran 4.9.0 on Linux 
and OS X and Intel ifort 13.1.0 on Linux.
We used OpenMP in all cases. For optimization we used \code{-O2} with 
gfortran and \code{-O1} with ifort. The models described here were 
computed with \mesa\ revision 4942. The comparison case is the 
\code{example\_solar\_model} from the \code{test\_suite}. It evolves a 
calibrated solar model from the pre-main sequence to the solar age, 4.57 Gyr.

\begin{figure}[htbp]
\centering{
\includegraphics[width=\figwidth]{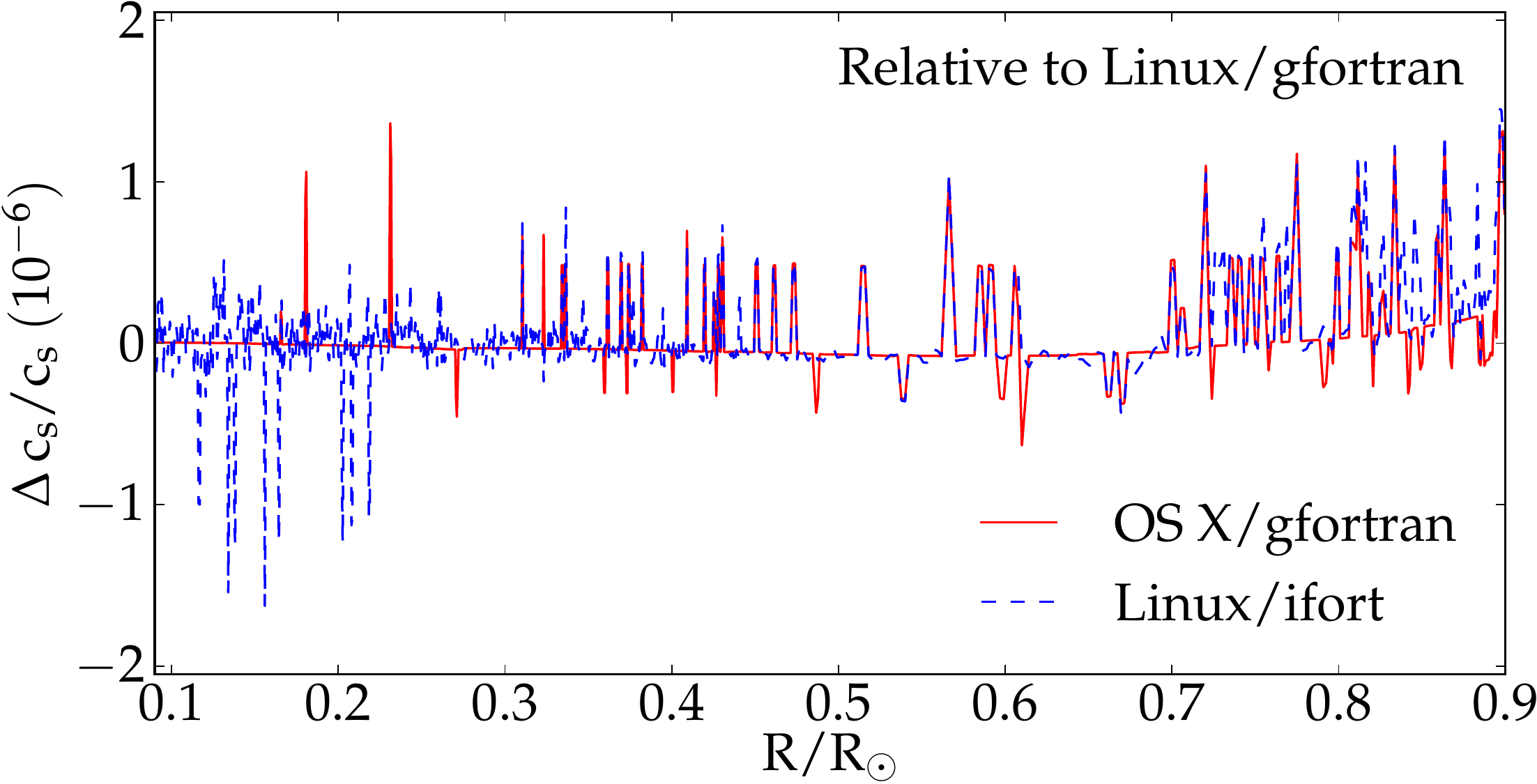}
\caption{\label{compOSsolar} Relative difference in the solar sound speed
profile from three models run using the same inlist but on different 
operating systems and/or compilers.}
}
\end{figure}

In Figure \ref{compOSsolar} we compare the sound speed profiles of these 
solar models in the same way that standard solar models are compared with
the solar sound speed profile in Section \ref{s.solar}, Figure 
\ref{astero:solar}. The reference was chosen to be the Linux/gfortran model;
the other two models were compared with it by interpolating their sound
speed profiles onto the radial grid of the reference model and then taking the 
relative sound speed difference with respect to that reference model. Figure
\ref{compOSsolar} indicates that models computed on different platforms are
consistent with one another at the level of the numerical tolerances with
which the equations are solved. The differences shown here are about one 
part per million or less, whereas the differences between standard solar 
models and the helioseismic data are in the parts per thousand (Figure
\ref{astero:solar})---a factor of 1000 difference. While this result is
reassuring, it is worthwhile to close this discussion with the comment that 
the consistency found in a low-mass, non-rotating model evolved about half 
way through the main sequence phase will not be representative of other cases 
dealing with different physics, stellar masses, and evolutionary phases.  
Projects using \mesa\ with heterogeneous architectures should perform their 
own consistency checks.

\section{The \mesa\ Software Development Kit}\label{s.mesasdk}

\mesa\ is provided as source code, allowing users access to all of the implementation details. Installation necessarily
involves building the code from source, which is a non-trivial task.
A successful build requires cooperation between the operating system, compiler,
libraries, and utilities.

\begin{table}
\caption{\label{tab:sdk} Principal components of the \MESA\ Software Development Kit}
\centering
\renewcommand{\arraystretch}{1.0}
\begin{tabular}{llll}
\hline\hline
	Name & Purpose & Version & License\tablenotemark{a}\\
\hline
GFORTRAN & Compiler             & 4.7.2       & Open source (GPL ver.\ 2)  \\
BLAS         & Matrix algebra       & 2011-04-19  & Open source (other) \\
LAPACK       & Matrix algebra       & 3.4.2       & Open source (other) \\
HDF5         & File storage         & 1.8.9       & Open source (other) \\
NDIFF        & Numerical comparison & 2.00        & Open source (GPL ver.\ 2) \\
PGPLOT       & Plotting             & 5.2.2       & Open source (non-commercial) \\
SE           & File storage    & 1.2.1       & Open source (other)\\
\hline\\
\end{tabular}
\tablenotetext{1}{``GPL'' denotes the GNU General Public License (with the version in
  parentheses); ``non-commercial'' denotes an open-source license with
  restrictions on commercial distribution; and ``other'' denotes to a
  variety of open-source licenses which permit largely unrestricted
  distribution.}
\end{table}

To address this issue we have created the \mesa\ Software Development
Kit (\SDK), which packages everything necessary to establish
a unified and maintained build environment.\footnote{Avaliable from \mbox{\url{http://www.astro.wisc.edu/~townsend/static.php?ref=mesasdk}}.}
The principal
components of the \SDK\ are summarized in Table~\ref{tab:sdk}; all of
these are distributed under an open-source license (detailed in the table), permitting their redistribution
without financial or copyright encumbrances. Perhaps the most
important component is the GFORTRAN compiler, part of the GNU Compiler
Collection. GFORTRAN implements almost all of the Fortran 2003 (F2003)
standard, and benefits from a high level of community support.

The \SDK\ is available for Intel x86 and x86-64 CPU
architectures running the Linux and Mac OS X operating systems (these
platforms comprise most of the \mesa\ user base). Installation
of the kit is straightforward, requiring a tar archive to be
unpacked (Linux) or an application folder to be copied (OS X),
followed by the initialization of a few environment
variables. By default, \mesa\ is configured to compile
``out-of-the-box'' with the \SDK.  \mesa\ can also be compiled without the \SDK, using any alternate compiler
which supports the F2003 standard. In this respect, GFORTRAN should
not be viewed as \emph{the} \mesa\ compiler (nor the full \SDK\ as
\emph{the} \mesa\ build environment). \mesa\ will adhere to Fortran standards
rather than rely on vendor-specific extensions.

Uptake of the \SDK\ has been very rapid: at the time of writing, we
estimate over 90\% of the \mesa\ community (over 500 users) are using
the \SDK.  This
growth has been matched by a significant decline in the number of
installation support requests, and a corresponding reduction in the
time taken to resolve these requests. With these maintenance overheads curbed, the
\mesa\ developers are able to devote more of their time to
refining and extending the code.

\bibliographystyle{apj}

\end{document}